\documentclass[structabstract]{aa}
\usepackage{graphicx}
\usepackage{txfonts}
\usepackage{natbib}
\bibpunct{(}{)}{;}{a}{}{,} 
\usepackage{aalongtable}
\begin{document}
\title{Optical spectroscopic variability of Herbig Ae/Be stars\thanks{Based
       on observations carried out by the EXPORT consortium}}

\author{I. Mendigut\'{\i}a\inst{1}
          \and
         C. Eiroa\inst{2}
         \and
         B. Montesinos\inst{1}
         \and
         A. Mora\inst{3}
         \and
         R.D. Oudmaijer\inst{4}
         \and
         B. Mer\'{\i}n\inst{5}
         \and
         G. Meeus\inst{2}}

\offprints{Ignacio Mendigut\'{\i}a\\
              \email{Ignacio.Mendigutia@cab.inta-csic.es}}
\institute{$^{1}$Centro de Astrobiolog\'{\i}a, Departamento de
     Astrof\'{\i}sica (CSIC-INTA), ESAC Campus, P.O. Box 78, 
     28691 Villanueva de la Ca\~nada, Madrid, Spain.\\
     $^{2}$Departamento de F\'{\i}sica Te\'{o}rica, M\'{o}dulo 15,
     Facultad de Ciencias, Universidad Aut\'{o}noma de Madrid, PO Box
     28049, Cantoblanco, Madrid, Spain.\\
     $^{3}$GAIA Science Operations Centre, ESA, European Space Astronomy Centre, PO Box 78, 28691, Villanueva de la Ca\~nada, Madrid,
     Spain.\\
     $^{4}$School of Physics \& Astronomy, University of Leeds, Woodhouse
     Lane, Leeds LS2 9JT, UK.\\
     $^{5}$Herschel Science Centre, ESA, European Space Astronomy Centre, P.O. Box 78, 28691, Villanueva de la Ca\~nada, Madrid,
     Spain.\\}

     \date{Received 2010, September 27; accepted 2011, February 15}
 
  \abstract{}
{In order to gain insights into the variability behaviour of the circumstellar (CS) atomic gas, we have analysed 337 multi-epoch optical spectra of 38 Herbig Ae/Be (HAeBe) stars.}
{Equivalent widths (EWs) and line fluxes of the H$\alpha$, [\ion{O}{i}]6300, \ion{He}{i}5876 and \ion{Na}{i}D lines were obtained for each spectrum; the H$\alpha$ line width at 10$\%$ of peak intensity (W$_{10}$) and profile shapes were also measured and classified. The mean line strengths and relative variabilities were quantified for each star. Simultaneous optical photometry was used to estimate the line fluxes.}
{We present a homogeneous spectroscopic database of HAeBe stars. The lines are variable in practically all stars and timescales, although 30 \% of the objects show a constant EW in [\ion{O}{i}]6300, which is also the only line that shows no variability on timescales of hours. The \ion{He}{i}5876 and \ion{Na}{i}D EW relative variabilities are typically the largest, followed by those in [\ion{O}{i}]6300 and H$\alpha$. The EW changes can be larger than one order of magnitude for the \ion{He}{i}5876 line, and up to a factor 4 for H$\alpha$. The [\ion{O}{i}]6300 and H$\alpha$ EW relative variabilities are correlated for most stars in the sample. The H$\alpha$ mean EW and W$_{10}$ are uncorrelated, as are their relative variabilities. The H$\alpha$ profile changes in $\sim$ 70 $\%$ of the objects. The massive stars in the sample (M$_{*}$ $>$ 3 M$_{\sun}$) usually show more stable H$\alpha$ profiles with blueshifted self-absorptions and less variable 10$\%$ widths.}
{Our data suggest multiple causes for the different line variations, but the [\ion{O}{i}]6300 and H$\alpha$ variability must share a similar origin in many objects. The physical mechanism responsible for the H$\alpha$ line broadening does not depend on the amount of emission; unlike in lower-mass stars, physical properties based on the H$\alpha$ luminosity and W$_{10}$ would significantly differ. Our results provide additional support to previous works that reported different physical mechanisms in Herbig Ae and Herbig Be stars. The multi-epoch observations we present are a useful tool for understanding the origin of the CS lines and their variability, and to establish distinctions in the physical processes operating in pre-main sequence stars.}

\maketitle
%
\section{Introduction}
\label{Section:Introduction}

Herbig Ae and Be (HAeBe) stars are intermediate mass, pre-main
sequence (PMS) objects, which are considered as the possible progenitors of
Vega-like stars surrounded by circumstellar (CS) debris disks and,
eventually, planets. The spectroscopic
monitoring of some sources \citep[see e.g.][]{Praderie86,Pogodin94,Rodgers02,Mora02,Mora04} revealed that the spectra of HAeBe objects are not only characterized by the presence of emission lines, but also
by the complex variations observed in both the emission and absorption
features. This variability is also characteristic of T-Tauri stars \citep[see e.g.][and references therein]{JohnsBasri95,Schisano09}. 

The optical spectra of PMS stars show several important features that have been related to different physical processes. Magnetospheric accretion models have succeeded in reproducing the profiles and
strengths of the H$\alpha$ and \ion{Na}{i}D lines \citep{Hartmann94,Muzerolle98a,Muzerolle98b,Muzerolle01,Muzerolle04}. Despite the unknown nature of the magnetic fields in HAeBe stars \citep[see e.g.][]{Wade05,Alecian07,Wade07,Hubrig09}, magnetospheric accretion has been shown to act in several HAe objects \citep{Muzerolle04,Mottram07}. The H$\alpha$ line has also been associated with winds \citep{Cabrit90} or with combined
magnetopheric accretion-wind models \citep{Kurosawa06}. The H$\alpha$ line width at 10$\%$ of peak intensity is used to estimate accretion rates in lower-mass PMS stars \citep[see e.g.][]{White03,Natta04,Jayawardhana06}. The [\ion{O}{i}]6300
line has been associated with accretion-powered outflows and winds
\citep{Finkenzeller85,Bohm94,Bohm97,Corcoran97,Corcoran98}, and
with the stellar UV-luminosity and disk-shape
\citep{Acke05,vanderplas08}. The high temperature close to the
stellar surface, which is generated in the accretion shock, has been suggested to be responsible for the \ion{He}{i}5876 line \citep{Tambovtseva99, Grinin01}. \ion{Na}{i}D lines seen in absorption describe the complex gas motions characterizing the
CS medium around PMS stars \citep{Mora02,Mora04}. 

Most spectroscopic studies are based on
isolated spectra of different sources or have been focused on particular
objects. The different approaches explaining the physical origin of the lines can profit from multi-epoch spectroscopic data. The EXPORT consortium \citep{Eiroa00} monitored a large
number of intermediate-mass PMS stars, allowing for an
extensive and homogeneous study of their spectroscopic variability. In this paper we analyse EXPORT multi-epoch optical spectra of HAeBe stars, specifically the lines H$\alpha$, [\ion{O}{i}]6300,
\ion{He}{i}5876, \ion{Na}{i} D$_{2}$5890 and D$_{1}$5896. The photometric and spectroscopic variability of those objects requires simultaneous measurements to obtain
accurate values for the line fluxes. These are derived using the already published simultaneous optical photometry of the stars in our sample \citep{Oudmaijer01}. We will present a dataset comprised of multi-epoch line equivalent widths, fluxes, H$\alpha$ profiles and widths, which is provided as a valuable observational legacy of the spectroscopic behaviour of HAeBe stars. We analyse and quantify the observed changes in the different lines and look for general trends, mainly focused on the equivalent width variability. A more specific analysis will be made in subsequent papers.

Sect. \ref{Section:Sample} describes the sample and spectra. Sect. \ref{Section:Results} presents our results. They are analysed in Sect. \ref{Sect:analisis}, which describes the spectroscopic variability shown by the different lines (Sect. \ref{Subsection:halfa} for H$\alpha$, Sect. \ref{Subsection:OI} for [\ion{O}{i}]6300, Sect. \ref{Subsection:HeI} for \ion{He}{i}5876, Sect. \ref{Subsection:NaD} for \ion{Na}{i}D and Sect. \ref{Subsubsection:EW summary} for a compendium of the spectroscopic characterization). Sect. \ref{Section:Discussion} discusses the results and Sect. \ref{Section:Conclusions} includes a brief summary and conclusions.

\section{Sample properties and observations}
\label{Section:Sample}

Table \ref{Table:sample} (left side) shows the 38 stars in the sample. Cols. 1 to 3 indicate the name of the star, the spectral type, and the stellar mass. While most of the objects are HAeBe stars \citep[covering almost all these objects in the Northern hemisphere from the catalogue of][]{The94}, 10 of them have spectral types ranging from F3 to G1. The stellar masses range between $\sim$ 1-- 6 M$_{\sun}$. As a selection criterion, all objects show the H$\alpha$ line in emission. The sample span the 1 -- 15 Myr age-range \citep{Manoj06,Montesinos08}, which is the period of the evolution from protoplanetary to young debris disks and the epoch of planet formation. There is a good balance between variable and non-variable objects, according to their simultaneous photopolarimetric behaviour \citep{Oudmaijer01,Eiroa01,Eiroa02}. 

The spectra were obtained by the EXPORT consortium \citep{Eiroa00} with the long
slit Intermediate Dispersion Spectrograph (IDS) on the 2.5 m Isaac
Newton Telescope (INT). The typical spectral resolution is R $\sim$
5500, covering the wavelength range $\lambda \lambda$ 5800--6700
\AA. The slit width was 1'' projected on the sky \citep[i.e. narrow enough to avoid confusion in most binaries of the sample; see e.g.][and references therein]{Wheelwright10}. Details of the observations and data reduction are given by
\citet{Mora01}. The stars were observed over one or more of the
four EXPORT runs. The right side of Table \ref{Table:sample} shows the log of the
observations, illustrating the monitoring timescales: from days to
months in general, and for a few stars even hours. A total of 337 spectra
were obtained, ranging from 3 to 18 per star, typically 6--10 spectra
per object. In order to study the non-photospheric contribution (see
Sect. \ref{Section:Results}), additional spectra of 28 spectroscopic
standard MS stars were also taken in the same campaigns and with the
same configuration. 
\begin{table*}
\renewcommand\arraystretch{1.013}
\renewcommand\tabcolsep{3.6pt}
\centering
\caption{Sample of stars and log of the INT/IDS spectra analysed.}
\label{Table:sample}
\begin{tabular}{lll||cccc|cccc|ccccc|ccc|c}
\hline\hline
Star & Spectral type & M$_{*}$ &\multicolumn{4}{c}{1998 May} & \multicolumn{4}{c}{1998 Jul} & \multicolumn{5}{c}{1998 Oct} & \multicolumn{3}{c}{1999 Jan} & Ntot \\
 &  & (M$_{\sun}$) &14 & 15 & 16 & 17 & 28 & 29 & 30 & 31 & 24 & 25 & 26 & 27 & 28 & 29 & 30 & 31 & \\
\hline
HD 31648 & A5 Vep & 2.0               & ... & ... & ... & ... & ... & ... & ... & ... & 1 & ... & 1 & 1 & 1 & 1 & 1 & 1 & 7 \\
HD 34282 & A3 Vne & $<$2.1$^{A}$  & ... & ... & ... & ... & ... & ... & ... & ... & 1 & 1 & 1 & 1 & 1 & 1 & 1 & 1 & 8 \\
HD 34700 & G0 IVe & 2.4$^{B}$         & ... & ... & ... & ... & ... & ... & ... & ... & 1 & 1 & 1 & 1 & ... & 1 & 1 & 1 & 7 \\
HD 58647 & B9 IVep & 6.0              & ... & ... & ... & ... & ... & ... & ... & ... & ... & ... & ... & ... & ... & 1 & 1 & 1 & 3 \\
HD 141569& B9.5 Vev & 2.2$^{A}$       & 1 & 1 & 1 & 1 & ... & 1 & 1 & 1 & ... & ... & ... & ... & ... & 1 & 1 & 1 & 10 \\
HD 142666& A8 Ve & 2.0$^{A}$          & 1 & 1 & 1 & 1 & 1 & 1 & 1 & 1 & ... & ... & ... & ... & ... & 1 & 1 & 1 & 11 \\
HD 144432& A9 IVev & 2.0$^{A}$        & 1 & 1 & 1 & 1 & 1 & 1 & 1 & 1 & ... & ... & ... & ... & ... & 1 & 1 & 1 & 11 \\
HD 150193& A2 IVe & 2.2                       & ... & 1 & 1 & 1 & ... & ... & ... & ... & ... & ... & ... & ... & ... & ... & ... & ... & 3 \\
HD 163296& A1 Vepv & 2.2                      & 1 & 1 & 1 & 1 & 1 & 2 & 2 & 2 & ... & ... & ... & ... & ... & ... & ... & ... & 11 \\
HD 179218& A0 IVe & 2.6                       & ... & 1 & 1 & 1 & ... & ... & ... & ... & ... & ... & ... & ... & ... & ... & ... & ... & 3 \\
HD 190073& A2 IVev & 5.1              &1 & 1 & 1 & 1 & 1 & 1 & 1 & 1 & ... & ... & ... & ... & ... & ... & ... & ... & 8 \\
AS 442   & B8 Ve & 3.5                & ... & 1 & 1 & 1 & 1 & 1 & ... & ... & ... & ... & ... & ... & ... & ... & ... & ... & 5 \\
VX Cas   & A0 Vep & 2.3                       & ... & ... & ... & ... & 1 & 1 & 1 & 1 & 1 & 1 & 1 & 1 & 1 & ... &     1 & 1 & 11 \\
BH Cep   & F5 IIIev & 1.7$^{A}$       & ... & 1 & 1 & 1 & 1 & 1 & 1 & 1 & 2 & 2 & 2 & 2 & 1 & ... & 1 & ... & 17 \\
BO Cep   & F5 Ve & 1.5$^{A}$          & ... & 1 & 1 & 1 & 1 & 1 & 1 & 1 & 1 & 1 & 1 & 1 & 1 & ... & ... & 1 & 13 \\
SV Cep   & A2 IVe & 2.4               & 1 & 1 & 1 & 1 & 1 & 1 & 1 & 1 & 1 & 1 & 1 & 1 & 1 & 1 & ... & ... & 14 \\
V1686 Cyg& A4 Ve & $>$3.5$^{A}$  & 1 & 1 & 1 & ... & 1 & 1 & 1 & 1 & 1 & 1 & 1 & 1 & 1 & ... & ... & ... & 12 \\
R Mon    & B8 IIIev & $>$5.1$^{A}$ & ... & ... & ... & ... & ... & ... & ... & ... & ... & ... & ... & ... & ... & 1 & 1 & 1 & 3 \\
VY Mon   & A5 Vep & $>$5.1$^{A}$ & ... & ... & ... & ... & ... & ... & ... & ... & ... & ... & ... & ... & ... & 1 & 1 & 1 & 3 \\
51 Oph   & B9.5 IIIe & 4.2       	 & 1 & 1 & 1 & 1 & 1 & 1 & 1 & 1 & ... & ... & ... & ... & ... & ... & ... & ... & 8 \\
KK Oph   & A8 Vev & 2.2$^{A}$    	 & 1 & 1 & 1 & 3 & 1 & 1 & 1 & 1 & ... & ... & ... & ... & ... & ... & ... & ... & 10 \\
T Ori    & A3 IVev & 2.4         		 & ... & ... & ... & ... & ... & ... & ... & ... & 1 & 1 & ... & 1 & 1 & 1 &1 & 1 & 7 \\
BF Ori   & A2 IVev & 2.6         	 & ... & ... & ... & ... & ... & ... & ... & ... & 1 & 1 & 1 & 1 & 1 & 1 & 1 & 2 & 9 \\
CO Ori   & F7 Vev & $>$3.6$^{A}$ 	 & ... & ... & ... & ... & ... & ... & ... & ... & 1 & 1 & 1 & 1 & 1 & 1 & 1 & 1 & 8 \\
HK Ori   & G1 Ve & 3.0$^{A}$     		 & ... & ... & ... & ... & ... & ... & ... & ... & 1 & 1 & 1 & 1 & 1 & 1 & 1 & 1 & 8 \\
NV Ori   & F6 IIIev & 2.2        		 & ... & ... & ... & ... & ... & ... & ... & ... & 1 & 1 & 1 & 1 & 1 & 1 & 1 & 1 & 8 \\
RY Ori   & F6 Vev & 2.5$^{A}$    	 & ... & ... & ... & ... & ... & ... & ... & ... & ... & 1 & 1 & 1 & 1 & 1 & 1 & 1 & 7 \\
UX Ori   & A4 IVe & 2.3          		 & ... & ... & ... & ... & ... & ... & ... & ... & 1 & 2 & 2 & 2 & 1 & 1 & 2 & 4 & 15 \\
V346 Ori & A2 IV & 2.5           	 & ... & ... & ... & ... & ... & ... & ... & ... & 1 & 1 & 1 & 1 & 1 & 1 & 1 & 1 & 8 \\
V350 Ori & A2 IVe & 2.2          	 & ... & ... & ... & ... & ... & ... & ... & ... & 1 & ... & 1 & 1 & 1 & 1 &1 & 1 & 7 \\
XY Per   & A2 IV & 2.8           		 & ... & ... & ... & ... & 1 & 1 & 1 & 1 & ... & 1 & 1 & 1 & 1 & 1 & 1 & 1 & 11 \\
VV Ser   & A0 Vevp & 4.0         	 & ... & 1 & 1 & 1 & 1 & 3 & 3 & 3 & 1 & 1 & 1 & 1 & 1 & ... & ... & ... & 18 \\
CQ Tau   & F5 IVe & 1.5$^{B}$    	 & ... & ... & ... & ... & ... & ... & ... & ... & 1 & 1 & 1 & 1 & 1 & 1 & 1 & 1 & 8 \\
RR Tau   & A0 IVev & 5.8         	 & ... & ... & ... & ... & ... & ... & ... & ... & 1 & 1 & 1 & 1 & 1 & 1 & 1 & 1 & 8 \\
RY Tau   & F8 IIIev & 1.3$^{C}$  	 & ... & ... & ... & ... & ... & ... & ... & ... & 1 & 1 & 1 & 1 & ... & 1 & 1 & ... & 6 \\
PX Vul   & F3 Ve & 1.5$^{A}$     & ... & ... & ... & ... & 1 & 1 & 1 & 1 & 1 & 1 & 1 & 1 & 1 & ... & ... & ... & 9 \\
WW Vul   & A2 IVe & 2.5          	 & 1 & 1 & 1 & 1 & 1 & 1 & 1 & 1 & 1 & 1 & 1 & 1 & 1 & ... & ... & ... & 13 \\
LkHa 234 & B5 Vev & $>$5.3$^{A}$ & ... & ... & ... & ... & 1 & 1 & 1 & 1 & 1 & 1 & 1 & 1 & 1 & ... & ... & ... & 9\\
\hline
\end{tabular}

\begin{minipage}{18cm}

  \underline{References and notes to Table \ref{Table:sample}}: Left side: spectral types are from \citet{Mora01}. The non-flagged stellar masses are from \citet{Montesinos08} and references therein. Flagged numbers are from $^A$\citet{Manoj06}, $^B$ \citet{AlonsoAlbi09}, $^C$\citet{Merin04}. Uncertainties for the stellar mass can be found in some of the references and are around 12$\%$. Right side: The number of spectra per star is given for each observing day. The last column indicates the total number of spectra per object.
\end{minipage}
\end{table*}

\section{Results}
\label{Section:Results}

The contribution of the CS component to the spectra of the PMS sample
was estimated from the spectra of the standard MS stars with similar
spectral types. These were rotationally broadened using the projected rotational velocities
derived by \citet{Mora01} for the PMS objects. The non-photospheric contribution is given by the residuals obtained from the subtraction of the broadened standard spectra from the observed PMS
spectra. The results refer to the normalized non-photospheric spectra.

Table \ref{Table:extract} gives estimates of the equivalent widths (EWs) of the H$\alpha$, [\ion{O}{i}]6300, \ion{He}{i}5876,
\ion{Na}{i}D$_2$ and \ion{Na}{i}D$_1$ lines for all spectra of our
sample. The H$\alpha$ widths at 10$\%$ of peak intensity (W$_{10}$) and profile types (see below) are also listed. As usual, positive and negative EW values correspond to absorption and emission features, respectively. When a spectral line is
seen partly in absorption and partly in emission, only the strongest
component (i.e., that with the largest $|$EW$|$) is included. This
double contribution appears mainly in the \ion{He}{i}5876 line, which usually
shows both a redshifted absorption and a blueshifted
emission. Although a specific analysis of this behaviour is beyond the
scope of this work, it probably indicates the presence of hot infalling
gas close to the stellar surface. In addition, the \ion{He}{i}5876 and \ion{Na}{i}D
lines change from absorption to emission in several objects.

Conservative EW uncertainties were estimated by determining the maximum and minimum EW values from two
different continuum levels that are located below and above unity ($\sim$ 1 $\mp$ 1.5$\sigma$), respectively. Both continuum levels bracket the adjacent noise to each spectral line. Typical median
uncertainties are 3$\%$, 8$\%$, 10$\%$, 20$\%$ and 30$\%$ of the EW
values in the H$\alpha$, \ion{Na}{i}D$_{2}$, \ion{Na}{i}D$_{1}$, \ion{He}{i}5876 and [\ion{O}{i}]6300
lines, respectively. Typical EW-uncertainties are larger than, or similar to, the typical strength of telluric absorption lines \citep[see e.g.][]{Caccin85,Lundstrom91,Reipurth96}. Telluric variability is also in general negligible compared with that reported for the lines (see below). Uncertainties for W$_{10}$ are typically $\sim$ 1$\%$ of its value.

We follow the two-dimensional scheme by
\citet{Reipurth96} to classify the observed H$\alpha$ profiles: type I are symmetric with no or
very weak dips; type II are double-peaked profiles with the secondary
peak higher than half the strength of the primary; type III have a secondary peak less than half the strength of the primary and type IV are P Cygni-type profiles. The scheme incorporates
the designations ``R'' or ``B'' to indicate if the secondary peak is
redshifted or blueshifted with respect to the primary. If both peaks have equal strengths, only the
profile type is indicated. For type IV, the letters
indicate a P-Cygni profile (``B'') or inverse P Cygni (``R''). The letter
``m'' is added when more than one absorption appears,
and also when small dips are apparent in the type I profiles.

An object is considered as `variable' in the EW of a line when its value differs in two or more spectra, taking into account the individual uncertainties. We use the relative
variability $|$$\sigma$(EW)/$<$EW$>$$|$ as a reliable measurement of
the strength of the EW-changes, where $\sigma$ is the standard deviation from the individual EW
measurements of a given star\footnote{i.e. for N spectra, $\sigma$ = [$\frac{1}{N-1}$$\cdot$$\sum$$_{i=1}^N${(EW$_i$ -
    $<$EW$>$)$^2$}]$^{1/2}$} and $<$EW$>$ is the mean line equivalent width. \citet{JohnsBasri95} used an equivalent parameter to characterize the variability of the H$\alpha$ intensity in several T-Tauri stars. The non-variable objects tend to show the lowest values of $|$$\sigma$(EW)/$<$EW$>$$|$; in these stars this ratio only contains information
about the precision of the measurements. $\sigma$(W$_{10}$)/$<$W$_{10}$$>$ and $\sigma$(L)/$<$L$>$ are
used to assess the relative variability of the H$\alpha$ line width and that of the line luminosities.

Table \ref{Table:meanresults} gives the mean line equivalent widths and their relative variabilities. Mean values and relative variabilities
of W$_{10}$(H$\alpha$) are also given, as well as the number of
spectra displaying the corresponding H$\alpha$ profiles. Uncertainties are listed for the non-variable stars in a given spectral line.

\begin{table*}
\caption{Mean equivalent and line widths, relative variabilities, and H$\alpha$ profiles.}
\label{Table:meanresults}
\centering
\renewcommand\arraystretch{1}
\renewcommand\tabcolsep{2.7pt}
\begin{tabular}{lrrlrrrr}
\hline\hline
Star & $<$EW$>$  & $<$W$_{10}$$>$  & Profile type (N$_{\rm spectra}$) & $<$EW$>$
 & $<$EW$>$  & $<$EW$>$  & $<$EW$>$  \\ 
 & H$\alpha$ & H$\alpha$ &  H$\alpha$ & [\ion{O}{i}]6300 & \ion{He}{i}5876 & \ion{Na}{i}D$_{2}$ & \ion{Na}{i}D$_{1}$ \\
 & ({\AA}) & (km/s) & & ({\AA}) & ({\AA}) & ({\AA}) & ({\AA})  \\
\hline
HD 31648  & -19.4 [0.10]    & $>$557 [0.19]& IIIB(7) & ... & -0.41 [0.34] & -0.64 [0.16] & -0.61  [0.15]\\
HD 34282  & -5.0 [0.22]     & $>$628 [0.19]& IIIR(3) IIIB(3) IIIRm(2) & ... & 0.18 [0.28] & 0.10 [0.20] & 0.06
[0.33]\\ 
HD 34700  & -2.4 [0.12]     & 355 [0.06]    & I(6) IIIR(1) & ... & ... & 0.09 [0.22] & 0.08 $\pm$ 0.01\\
HD 58647  & -11.4 $\pm$ 0.3 & 616 $\pm$ 6 & IIR(3) & -0.06 [0.50] & -0.13$^{f}$ $\pm$ 0.04 & 0.15 [0.20] & 0.13 $\pm$ 0.02\\
HD 141569 & -6.0 $\pm$ 0.1  & $>$641 [0.02] & IIR(10) & -0.14 [0.21] & ... & 0.1  [0.30] & 0.1 [0.50]\\
HD 142666 & -4.0 [0.25]     & 540 [0.24]   & IIIR(4) IIIRm(4) IIB(1) IIR(1) & -0.010$^{*}$ $\pm$ 0.009 & 0.19 [0.47]
&0.21 [0.19] & 0.18 [0.22]\\
HD 144432 & -11.8 [0.10]    & $>$441 [0.23]& IIIB(11) & -0.013$^{*}$ $\pm$ 0.006 & -0.29$^{f}$ [0.55] & (-0.27 [0.63])
& -0.26$^{*}$ [0.46]\\ 
HD 150193 & -13.8 [0.06]    & $>$481 [0.11] & IIIB(3) & ... & -0.21$^{f}$ $\pm$ 0.07 & -0.36 $\pm$ 0.06 & -0.28 $\pm$ 0.06\\ 
HD 163296 & -22.8 [0.08]    & 690 [0.09]    & IIIB(8) I(3) & -0.03$^{*}$ [0.67] & (0.54 [0.93]) & -0.47 [0.36] &
-0.46 [0.33]\\ 
HD 179218 & -13.6 [0.05]    & 475 [0.02]    & I(3) & -0.03 [0.33] & (0.04 [1.75]) & -0.05 $\pm$ 0.01 & -0.03
[0.67]\\
HD 190073 & -25.6 [0.05]    & 391 [0.03]    & IVB(5) I(3) & -0.03$^{*}$ [1.00] & -0.47$^{f}$ [0.13] & -1.02  [0.04] & -0.85 $\pm$ 0.02\\
AS 442    & -32.7 [0.12]    & 646 [0.05]    & IIIB(5) & -0.09 $\pm$ 0.02 & 0.19 [0.37] & 0.76 [0.22] & 0.63
[0.19]\\
VX Cas    & -22.1 [0.51]    & 650 [0.15]   & IIR(7) IIB(3) IIIR(1) & -0.21 [0.76] & 0.61 [0.41] & 0.53 [0.53] &
0.47 [0.45]\\
BH Cep    & -6.2 [0.37]     & 700 [0.09]    &IIR(5) IIIR(5) IIIBm(4) IIRm(2) IIBm(1) & -0.03$^{*}$ [0.67] & 0.36
[0.53] & 0.44 [0.36] & 0.37 [0.38]\\
BO Cep    & -7.5 [0.21]     & 645 [0.12]    & IIR(7) IIIRm(2) IIIR(2) IIB(1) IIRm(1) & -0.17 $\pm$ 0.02 & 0.31
[0.64] & 0.26 [0.35] & 0.21 [0.33]\\
SV Cep    & -11.7 [0.08]    & 705 [0.07]    & IIR(7) IIB(5) IIIR(2) & -0.13 $\pm$ 0.01 & 0.47 [0.45] & 0.31
[0.39] & 0.26 [0.38]\\
V1686 Cyg & -22.7 [0.42]    & $>$458 [0.10] & IIIB(7) IIIBm(5) & -0.30 [0.77] & (0.25$^{*}$ [0.88]) & 1.57
[0.47] & 0.86 [0.69]\\
R Mon     & -114 $\pm$ 2    & 832 [0.01]     & Im(3) & -5.0 $\pm$ 0.1 & 1.2$^{f}$ $\pm$ 0.2 & -0.15 $\pm$ 0.02 & -0.47 $\pm$ 0.06\\
VY Mon    & -41 $\pm$ 1     & 711 [0.03]    & IIIB(3) & -1.9 $\pm$ 0.1 & ... & 2.32 $\pm$ 0.07 & 1.53 [0.05]\\
51 Oph    & -3.3 $\pm$ 0.1  & 521 [0.03]    & IIB(8) & ... & -0.1 [0.30] & (-0.06$^{*}$ [0.83]) & (-0.14 [0.57])\\
KK Oph    & -74.1 [0.15]    & 607 [0.08]    & I(3) Im(3) IIR(3) IIB(1) & -2.41 [0.22] & (0.46 [0.56]) & (-0.4
[1.12]) & (-0.22 [1.68])\\
T Ori     & -21.5 [0.27]    & 675 [0.05]    & IIR(7) & -0.13 [0.31] & 0.58 [0.28] & 0.42 [0.40] & 0.30 [0.40]\\
BF Ori    & -9.9 [0.06]     & 752 [0.09]    & IIR(7) IIB(2) & -0.04 [0.25] & 0.67 [0.31] & 0.60 [0.33] & 0.50
[0.30]\\
CO Ori    & -21.1 [0.11]    & 549 [0.06]    & I(5) Im(3) & -0.29 [0.24] & 0.07$^{*}$ [1.14] & 0.31 [0.55] & 0.27
[0.41]\\
HK Ori    & -57.6 [0.27]    & 582 [0.06]    & IIB(4) IIR(3) I(1) & -1.29 [0.26] & 0.40$^{*}$ [0.48] & (0.35
[0.80]) & (0.34 [0.65])\\
NV Ori    & -4.0 [0.05]     & 608 [0.11]    & IIIR(3) IIR(3) Im(1) IIB(1) & -0.012$^{*}$ $\pm$ 0.006  & 0.20$^{*}$
[0.50] & (-0.21 [0.86]) & (-0.22 [0.73])\\
RY Ori    & -15.8 [0.23]    & 611 [0.06]    & IIR(6) IIB(1) & -0.12 [0.50] & 0.59 [0.44] & 0.98 [0.30] & 1.01
[0.30]\\
UX Ori    & -9.3 [0.33]     & $>$659 [0.14] & Im(8) IIIRm(3) IIRm(2) IIBm(1) IVR(1) & -0.05 $\pm$ 0.01 & (0.27
[0.78]) & (-0.25 [0.92]) & (-0.13 [1.23])\\
V346 Ori  & -3.8 [0.24]     & 908 $\pm$ 55& IIIRm(5) IIIBm(3) & ... & ... & (0.15 [0.80]) & (0.11 [0.73])\\
V350 Ori  & -12.3 [0.17]    & 722 [0.06]    & IIR(4) IIIR(2) IIIRm(1) & -0.13 [0.31] & 0.43 [0.37] & 0.71 [0.18]
& 0.60 [0.27]\\
XY Per    & -9.8 [0.11]     & 726 [0.02]    & IIB(6) IIR(3) IIRm(1) IIBm(1) & -0.04$^{*}$ [0.75] & 0.23 [0.43] &
0.55 [0.16] & 0.44 [0.11]\\
VV Ser    & -49.7 [0.05]    & 695 [0.04]    & IIR(8) IIB(8) II(2) & -0.54 [0.11] & (0.57 [0.47]) & 0.83 [0.19] &
0.71 [0.24]\\
CQ Tau    & -4.8 [0.31]     & $>$531 [0.24]& IIIR(5) IIR(2) Im(1) & -0.06 [0.50] & 0.10$^{*}$ [1.00] & 0.39
[0.26] & 0.33 [0.24]\\
RR Tau    & -25.6 [0.34]    & 683 [0.02]    & IIR(5) IIB(3) & -0.39 [0.23] & 0.43 [0.18] & 0.76 [0.21] & 0.66
[0.20]\\
RY Tau    & -15.3 [0.19]    & 680 [0.12]    & IIB(3) IIIB(1) IIBm(1) IIR(1) & -0.75 $\pm$ 0.05 & (-0.24 [1.12])
& (0.48 [0.90]) & (0.41 [0.98])\\
PX Vul    & -14.4 [0.08]    & 626 [0.05]    & IIIB(5) Im(2) IIB(2) & -0.08 [0.25] & 0.19$^{*}$ [0.74] & 0.17
[0.23] & 0.16 [0.19]\\
WW Vul    & -19.1 [0.10]    & 744 [0.06]    & IIB(9) IIR(4) & -0.09 [0.22] & 0.69 [0.35] & 0.66 [0.35] & 0.50
[0.44]\\
LkHa 234  & -69.9 [0.07]    & 762 [0.04]    & IIB(6) IIIB(3) & -0.57 [0.12] & 0.81 [0.23] & (-0.39 [2.51]) &
(-0.22 [3.30])\\
\hline
\end{tabular}
\begin{minipage}{18cm}
  \underline{Notes to Table \ref{Table:meanresults}}: Values from the data in Table \ref{Table:extract}. Numbers in brackets are the EW relative variabilities. Typical uncertainties are $\delta$($|$$\sigma$/$<$EW$>$$|$) $\sim$
  0.03, 0.15, 0.16, 0.14 and 0.12 for the H$\alpha$, [\ion{O}{i}]6300, \ion{He}{i}5876
  and \ion{Na}{i}D$_{2}$ and D$_{1}$ lines, respectively. $\delta$($\sigma$/$<$W$_{10}$$>$) $\sim$ 0.01. Propagated uncertainties ($\pm$) are indicated for the
  non-variable objects. Values in parentheses indicate those
  lines shown both in absorption and emission. Only the values of the
  most frequent absorption/emission behaviour are considered in these
  cases to derive the mean values. $^{*}$ indicates that there is at least one spectrum not showing the line. ``...'' indicates that the line is
  not detected. $^{f}$ indicates that the \ion{He}{i}5876 line has only one component (absorption or emission) in all spectra.
\end{minipage}
\end{table*}

Table \ref{Table:fluxes} shows the line fluxes for the spectra with
simultaneous optical photometry \citep{Oudmaijer01}. They were
estimated using the EWs in Table \ref{Table:extract}, and the $V$ and
$R$ magnitudes obtained during the same night, taken within a time span of less than 2-3 hours. Fluxes are derived for 137 spectra (i.e. 41\% of
the initial 337 spectra) of 36 stars (the initial sample excluding
\object{AS 442} and \object{R Mon}). We used the expression F =
F$_{0}$ $\times$ $|$EW$|$/10$^{0.4m}$, with F$_{0}$ the
zero-magnitude fluxes \citep{Bessell79} and $m$ the $V$ (for the
\ion{Na}{i}D and \ion{He}{i}5876 lines) or $R$ (for the H$\alpha$ and [\ion{O}{i}]6300
lines) magnitudes. We also computed fluxes for the \ion{Na}{i}D and \ion{He}{i}5876 lines that showed positive EWs, meaning the stellar flux absorbed by the lines. These fluxes are related to the amount of gas in the line of sight, traced by these species. 

Table \ref{Table:luminosities} shows the typical (mean) line
luminosities of each star, obtained by averaging the fluxes in Table
\ref{Table:fluxes} and assuming spherical symmetry. The distances considered are indicated in Col. 3. The line luminosity relative
variabilities are also given, although the statistics is poorer than for the EWs,
because of the lower number of spectra with simultaneous photometry (given in Col. 2). 

\begin{table*}
\caption{Mean line luminosities.}
\label{Table:luminosities}
\centering
\begin{tabular}{lrlrrrrr}
\hline\hline
Star & N$_{\rm spectra}$ & d  & $<$L$>$  &
$<$L$>$  & $<$L$>$  &
$<$L$>$  & $<$L$>$ \\
 & &      & H$\alpha$ & [\ion{O}{i}]6300 & \ion{He}{i}5876 & \ion{Na}{i}D$_{2}$ & \ion{Na}{i}D$_{1}$ \\  
 & & (pc) & ($\times$10$^{31}$ erg s$^{-1}$)  & ($\times$10$^{29}$ erg s$^{-1}$) & ($\times$10$^{30}$ erg s$^{-1}$) &
 ($\times$10$^{30}$ erg s$^{-1}$) & ($\times$10$^{30}$ erg s$^{-1}$) \\
\hline		   
HD 31648  &3&146        &  10.4 [0.15] & ... 	        & 2.35 [0.45]  &  3.91 [0.21]&  3.86 [0.15] \\
HD 34282  &2&164$^{1}$  & 0.50$\pm$0.08        & ... 	        & \emph{0.195} [0.27]  & \emph{0.100} [0.36]&\emph{0.0644} [0.50] \\
HD 34700  &1&336$^{4}$  &  2.0$\pm$0.3        & ... 	        & ...		& \emph{0.7}$\pm$0.2  & \emph{0.6}$\pm$0.2   \\
HD 58647  &1&543        &   179$\pm$10        & 81$\pm$30        &  22$\pm$9 	&  \emph{31}$\pm$4  &  \emph{26}$\pm$4   \\
HD 141569 &4&99$^{1}$   &  2.51$\pm$0.09        & 6$\pm$1        & ...		& \emph{0.47}$\pm$0.05  & \emph{0.481} [0.47] \\
HD 142666 &3&145$^{1}$  & 0.81$\pm$0.08        &    ...	&  \emph{0.37}$\pm$0.07 	& \emph{0.470} [0.15]& \emph{0.40}$\pm$0.04   \\
HD 144432 &3&145$^{1}$  &  4.79 [0.11] &0.4$^{*}$$\pm$0.4  &  0.860 [0.53]&  0.928 [0.68]&   1.0$\pm$0.1   \\
HD 150193 &2&203        &  6.2$\pm$0.3        &    ...	&  0.9$\pm$0.2 	&  1.4$\pm$0.2  &  1.1$\pm$0.2   \\
HD 163296 &3&130        &  21.3 [0.05] & 1.50$^{*}$ [1.73]&\emph{5.68} [0.41]& 2.96 [0.39]&   3.13 [0.48] \\
HD 179218 &1&201        &  18$\pm$1        & 4$\pm$1        &  \emph{0.7}$\pm$0.2 	& 0.6$\pm$0.1  & 0.3$\pm$0.1   \\
HD 190073 &1&767        &   360$\pm$12        & 82$\pm$10        &   68$\pm$10 	&   \emph{146}$\pm$10  &   \emph{128}$\pm$10   \\
VX Cas    &6&619        &  4.88 [0.18] & 3.83 [0.16] &  \emph{1.69} [0.62]	&  \emph{1.11} [0.41]&  \emph{1.02} [0.40] \\
BH Cep    &7&450$^{1}$  &  1.26 [0.16] &0.784 [0.60] &  \emph{0.537} [0.61] & \emph{0.669} [0.38]&  \emph{0.578} [0.53] \\
BO Cep    &5&400$^{1}$  &  1.11 [0.24] & 2.6$\pm$0.5        & \emph{0.464} [0.55]  & \emph{0.277} [0.51]& \emph{0.232} [0.46] \\
SV Cep    &5&596        &  6.52 [0.08] & 7$\pm$1        &  \emph{2.46} [0.41]  &  \emph{1.48} [0.49]&  \emph{1.21} [0.42] \\
V1686 Cyg &6&980$^{1}$  &  6.19 [0.44] & 7.85 [0.35] & (\emph{0.471} [0.90])&  \emph{2.64} [0.81]&  \emph{1.47} [0.97] \\
VY Mon    &1&800$^{1}$  &  8.2$\pm$0.2        & 39$\pm$4        & ...		&  \emph{1.5}$\pm$0.1  &  \emph{1.02}$\pm$0.08   \\
51 Oph    &3&142        &  23$\pm$2        & ... 	        &  8$\pm$2 	&  2.49$^{*}$ [1.06]&  8$\pm$2   \\
KK Oph    &3&160$^{1}$  &  1.03$\pm$0.03        & 3.54 [0.09] &(\emph{0.0630} [0.42])&(\emph{0.0227} [0.31])&(\emph{0.0282} [0.28]) \\
T Ori     &3&472        &  12.6 [0.13] & 7.49 [0.27] &   \emph{1.972} [0.22]&  \emph{2.18} [0.30]&  \emph{1.52} [0.34] \\
BF Ori    &4&603        &  14.8$\pm$0.6        & 8$\pm$1        &  \emph{9.86} [0.57]  &  \emph{9.66} [0.49]&  \emph{9.38} [0.23] \\
CO Ori    &5&450$^{1}$  &  5.27 [0.07] & 6.58 [0.22] &\emph{0.0573}$^{*}$ [2.24]&  \emph{0.469} [0.94]& \emph{0.397} [0.72] \\
HK Ori    &5&460$^{1}$  &  9.98 [0.41] & 23.9 [0.4]  &\emph{0.590}$^{*}$ [0.68]& (\emph{0.565} [0.20])& (\emph{0.566} [0.27]) \\
NV Ori    &5&450$^{2}$  &  3.5$\pm$0.2        &0.457$^{*}$ [2.24]&\emph{1.22}$^{*}$ [0.68]&  1.75 [0.16]&  1.71 [0.21] \\
RY Ori    &4&460        &  1.99 [0.12] & 1.43 [0.44] & \emph{0.460} [0.22]  &  \emph{0.993} [0.33]&  \emph{1.01} [0.33] \\
UX Ori    &5&517        &  10.8 [0.21] & 6$\pm$1        &  (\emph{2.73} [0.65])  &  (2.36 [0.30])&  (1.08 [0.69]) \\
V346 Ori  &3&586        &  3.2$\pm$0.4        & ... 	        &  ...  	&  (\emph{1.55} [0.42])&  (\emph{1.18} [0.43]) \\
V350 Ori  &2&735        &  4.8$\pm$0.2        & 5$\pm$1        &  \emph{1.3}$\pm$0.2 	&  \emph{2}$\pm$1  &  \emph{2.6}$\pm$0.2   \\
XY Per    &4&347        &  9.00 [0.28] & 4.37$^{*}$ [0.87] &  \emph{2.35} [0.41]  &  \emph{4.51} [0.19]&  \emph{3.58} [0.14] \\
VV Ser    &7&614        &  17.3 [0.06] & 19.8 [0.18] &  \emph{1.28} [0.25]  &  \emph{1.67} [0.20]&  \emph{1.41} [0.26] \\
CQ Tau    &5&130$^{3}$  & 0.501 [0.19] & 0.49$\pm$0.08       & \emph{0.10}$^{*}$$\pm$0.02 	& \emph{0.409} [0.18]& \emph{0.316} [0.11] \\
RR Tau    &5&2103       &   169 [0.18] &  253 [0.14] &  \emph{24.4} [0.41]  &  \emph{42.4} [0.33]&  \emph{37.7} [0.33] \\
RY Tau    &5&134$^{2}$  & 0.850 [0.09] & 4.4$\pm$0.3        &(0.0694 [0.26])& (0.0993 [0.47])& (0.103 [0.50]) \\
PX Vul    &5&420$^{1}$  &  2.76$\pm$0.09        & 1.71 [0.28] & \emph{0.210} [0.62]  & \emph{0.227} [0.27]& \emph{0.218} [0.22] \\
WW Vul    &6&696        &  15.3 [0.07] & 8.13 [0.22] &  \emph{5.83} [0.29]  &  \emph{6.00} [0.42]&  \emph{4.79} [0.50] \\
LkHa 234  &4&1250$^{1}$ &  43.9$\pm$0.6        & 36$\pm$3        &  \emph{2.71} [0.19]  &  1.18 [0.62]&  0.8$\pm$0.2   \\
\hline
\end{tabular}
\begin{minipage}{18cm}

  \underline{Notes to Table \ref{Table:luminosities}}: Values from the data in Table \ref{Table:fluxes}. Numbers in brackets are the line luminosity relative variabilities. Typical uncertainties are $\delta$($\sigma$/$<$L$>$) $\sim$ 0.06, 0.1, 0.2, and 0.1 for the H$\alpha$, [\ion{O}{i}]6300, \ion{He}{i}5876 and \ion{Na}{i}D lines,
  respectively. Propagated uncertainties ($\pm$; distance errors not considered) are indicated for the
  non-variable objects. Italic numbers refer to lines seen in absorption and values in parentheses are those lines shown both in absorption and emission. Only
  the values of the most frequent absorption/emission behaviour are
  considered in these cases to derive the mean values. $^{*}$ indicates that
  there is at least one spectrum not
  showing the line. ``...''  indicates that the line is not
  detected. The distances are from \citet{Montesinos08} by default, the remaining ones are taken from $^{1}$\citet{Manoj06}, $^{2}$\citet{Blondel06}, $^{3}$\citet{GarciaLopez06}, $^{4}$\citet{Acke05}.
\end{minipage}
\end{table*}

The reddening towards the objects is low for most stars, with very few exceptions \citep[E($B$-$V$) $\leq$ 0.1 for almost half of the objects, and E($B$-$V$) $\sim$ 1 for the most reddened sources -\object{V1686 Cyg} and \object{LkHa 234}-,][]{Merin04}. The fact that our multi-epoch line fluxes and luminosities are not de-reddened avoids introducing additional uncertainties and does not affect the analysis and discussion in the following sections. 

\section{Analysis}
\label{Sect:analisis}

\subsection{The H$\alpha$ line}
\label{Subsection:halfa}

Our data show that the H$\alpha$ emission line remains constant, at
least on timescales of days-months, for \object{HD 141569} and
\object{51 Oph}, based on 8 and 10 spectra, respectively. In addition,
H$\alpha$ is non-variable in \object{HD 58647}, \object{R Mon}
and \object{VY Mon}; however, there are only three spectra per object, so that variability at timescales longer than
days cannot be excluded. The remaining stars show EW(H$\alpha$)
variability.

The typical H$\alpha$ relative variability, $|$$\sigma$/$<$EW$>$$|$, is 0.19 (mean) and 0.15 (median)\footnote{In order to improve the reliability of the typical values and ranges derived, the 33 objects with 5 or more spectra are usually taken as a reference.}. H$\alpha$ EW variations up to a factor EW$_{max}$/EW$_{min}$ $\sim$ 4 can be observed for individual stars (e.g. \object{V1686 Cyg}, \object{VX Cas}). The $<$EW$>$(H$\alpha$) ranges between $-2$ and $-74$
\AA{}, with typical values of $-19.4$ \AA{} (mean) and $-14.4$ \AA{}
(median)

All objects change their W$_{10}$(H$\alpha$) but \object{HD 58647} with only three
measurements and \object{V346 Ori} with very large uncertainties. The typical relative
variability is only $\sigma$(W$_{10}$)/$<$W$_{10}$$>$ = 0.09 (mean), 0.07 (median). Some stars show W$_{10}$(H$\alpha$) changes up to a factor $\sim$ 2.5 (e.g. \object{CQ Tau}). The $<$W$_{10}$$>$(H$\alpha$) ranges from 355 km s$^{-1}$ to 908
km s$^{-1}$ and is typically around 640 km s$^{-1}$

Changes of both EW and W$_{10}$ are seen on timescales as short as
hours from the five objects with more than one spectrum
per night. Our data show typical changes of a
factor $\sim$ 1.1 in both parameters on this timescale. The strongest
variations occur over longer timescales however, i.e. weeks-months,
at least in our database.

Variations in the line profile are very frequent --only
18$\%$ of the stars display the same type in all spectra-- and are common on timescales of days; $\sim\!73$\% show variations from
one day to the next one. Some extreme cases are \object{VV Ser},
changing from type IIR to type IIB profiles practically from spectrum
to spectrum, and \object{UX Ori}, \object{BO Cep} and \object{BH Cep}, which show five different
profile types. \object{BH Cep} also shows the fastest variation, changing from IIB to IIIB in less than one hour. The high-mass objects in our sample tend to show the same H$\alpha$ profile type during a given observing run, i.e. on timescales of a week, and all objects showing stable
H$\alpha$ profiles only on shorter timescales have M$_{*}$ $<$ 3 M$_\odot$. 

All H$\alpha$ profiles classified by \citet{Reipurth96} are
observed in our spectra with a similar distribution \citep[see also][]{Vieira03}: type II
profiles are the most frequent (52$\%$ of the observations), followed
by type III (32$\%$) and I (13$\%$). P-Cygni type profiles are the
least common (less than 3$\%$). ``R'' and ``B''
profiles appear almost in the same proportion (44$\%$ versus
43$\%$). However, B profiles are mainly shown by the most
massive objects in our sample (the stars with M$_{*}$ $\geq$ 3 M$_{\sun}$ have ``B''
profiles in 60$\%$ of the spectra; in fact only they show
P-Cygni signatures). ``R'' profiles dominate in the HAe and lower-mass
range of our sample (47$\%$ of the spectra, against 37$\%$ in
``B'').

\subsection{The [\ion{O}{i}]6300 line}
\label{Subsection:OI}
Thirty-two out of the 38 stars show the [\ion{O}{i}]6300 emission line. The line is variable in $\sim\!70$\% of the stars;
typical values of $|$$\sigma$/$<$EW$>$$|$ are 0.51 (mean) and 0.26
(median). The largest EW variations -a factor $\sim$ 7--9- are seen, as for the H$\alpha$ line, in \object{V1686 Cyg} and \object{VX Cas}. [\ion{O}{i}]6300 variability is not detected on timescales of hours. For seven objects, the line is seen only in several spectra. The emission is very faint in these
stars and their $|$$\sigma$/$<$EW$>$$|$ values
should be taken with caution. They could be affected by telluric
variability or by artefacts from
the telluric emission subtraction. Higher SNR spectra of the objects with
the weakest [\ion{O}{i}]6300 line would be necessary to better estimate their line relative variability. Typical
$<$EW([\ion{O}{i}]6300)$>$ values are $-0.29$ \AA{} (mean) and $-0.12$ \AA{}
(median). The difference between the mean and median values comes from
the few objects showing $|$$<$EW([\ion{O}{i}]6300)$>$$|$ $>>$ 1 \AA{}. 

Most of the stars display a single-peaked almost symmetric low-velocity component. For the few objects with profiles different from single-peaked, the emission tends to be faint and the SNR low, probably distorting their profiles.

\subsection{The \ion{He}{i}5876 line}
\label{Subsection:HeI}

Only 4 out of the 38 stars do not show the \ion{He}{i}5876 line in any of
their spectra. The line is present in the rest of the stars, although
five objects do not show it in all their spectra. In seven stars the
line appears either in absorption or in emission, depending on the
observing date. In most cases, 84$\%$ of the objects, the line is
present either fully in absorption, or the absorption is
dominant. All objects show variations in the EW of the
line, with the exception of \object{HD 150193}, \object{HD 58647}, and \object{R
  Mon}. There are only three spectra available for each object, thus we cannot exclude that this result
is due to the comparatively poor spectroscopic coverage.

Changes in EW(\ion{He}{i}5876) take place in all timescales. The relative variability range is 0.12 $<$
$|$$\sigma$/$<$EW$>$$|$ $<$ 1.22, with a typical value of 0.53 (mean)
and 0.46 (median). The variability is larger in those objects where the line is dominated by absorption. Line EW variations up to a factor EW$_{max}$/EW$_{min}$ $\sim$ 13 can be observed in these cases (e.g. \object{BO Cep}), compared to a factor $\sim$ 6 for the objects showing the line emission (e.g. \object{HD 144432}). The typical mean and median value of $|$$<$EW(\ion{He}{i}5876)$>$$|$ is 0.38
$\AA$, ranging between $-$0.47 $\leq$ $<$EW(\ion{He}{i}5876)$>$ $\leq$ 0.81
$\AA$. 

\subsection{The \ion{Na}{i}D lines}
\label{Subsection:NaD}

\ion{Na}{i}D lines are seen in absorption for most of the cases ($\sim$ 70 $\%$ of the stars). The \ion{Na}{i}D absorption lines can have a non-negligible
interstellar contribution owing to clouds in the line of sight of our
objects \citep[see e.g.][]{Redfield07}. Therefore, several EWs given
in Tables \ref{Table:extract} and \ref{Table:meanresults} should be
considered as upper limits to the CS absorption. The timescale variability of the
interstellar absorption is, however, much longer than that covered by
our spectra \citep[see e.g.][]{Lauroesch03}. Thus, the observed
variability of the \ion{Na}{i}D lines is caused by the CS gas component.

There are seven objects with a constant EW in the \ion{Na}{i}D lines. Again, apart from \object{HD 34700} and \object{HD 190073}, only three spectra are available for each one. The remaining objects show EW variability. As expected, the relative variability is equal in both \ion{Na}{i}D lines ($\sim$ 0.50), within the uncertainties. As for the \ion{He}{i}5876 line, the smallest  EW$_{max}$/EW$_{min}$ factors are shown by the objects with \ion{Na}{i}D in emission (up to a a factor $\sim$ 4 in e.g. \object{HD 163296}, against a factor $\sim$ 6 for objects with the lines in absorption such as \object{CO Ori}). \ion{Na}{i}D EWs do not usually change in hours. The only exception is \object{HD 163296}; its \ion{Na}{i}D emission changed a factor $\sim\!2$ in one night (29-Jul-1998), but no variations were detected the two following
nights. We note that variations on timescales of hours have been reported for \object{UX Ori} using higher resolution spectra \citep{Mora02}. The typical $<$EW(\ion{Na}{i}D)$>$ in our sample is $\sim$ 0.40 \AA.

The \ion{Na}{i}D ratio is a good indicator of the optical thickness at these wavelengths \citep[e.g.][]{Mora02,Mora04}. Changes in the optical depth of the CS medium in the line of sight are observed in almost all objects, however, averaged values indicate optically thick media for most of the stars ($<$EW(\ion{Na}{i}D$_{2}$)/EW(\ion{Na}{i}D$_{1}$)$>$ $\sim$ 1).

\subsection{Compendium of the EW variability}
\label{Subsubsection:EW summary}

Table \ref{Table:summary_halfa} summarizes typical values for the
equivalent widths and their relative variabilities, the minimum and maximum values, the percentage of objects showing
line variations, and the number of objects with variability on
a timescale of hours. We remark that this one refers only to a sample of
five stars with that spectroscopic timescale coverage. The percentage
of objects where the corresponding line is undetected is also given. The values in the last three rows of the bottom
panel are derived considering both \ion{Na}{i}D lines.
\begin{table*}
\centering
\caption{Summary of the typical equivalent and line widths and their relative variability.}
\label{Table:summary_halfa}
\begin{tabular}{l|cc|cc}
\hline\hline
 & EW(H$\alpha$) & $|$$\sigma$/$<$EW(H$\alpha$)$>$$|$ & W$_{10}$(H$\alpha$) & $\sigma$/$<$W$_{10}$(H$\alpha$)$>$ \\
 & ($\AA$) & & (km s$^{-1}$) & \\
\hline
mean   & 19.4    & 0.19         & 627        & 0.09         \\
median & 14.4    & 0.15         & 646        & 0.07         \\
range  & 2 -- 74 & 0.05 -- 0.51 & 355 -- 908 & 0.02 -- 0.24 \\
$\%$ var & \multicolumn{2}{c|}{94} & \multicolumn{2}{c}{97} \\
N$_{\rm stars}$ var (hours) & \multicolumn{2}{c|}{3} & \multicolumn{2}{c}{4} \\
\hline
\end{tabular}
\end{table*}
\begin{table*}
[!hbtp]
\centering
\begin{tabular}{l|cc|cc}
\hline\hline
 & EW([\ion{O}{i}]6300) & $|$$\sigma$/$<$EW([\ion{O}{i}]6300)$>$$|$ & EW(\ion{He}{i}5876) & $|$$\sigma$/$<$EW(\ion{He}{i}5876)$>$$|$ \\ 
 & ($\AA$) & & ($\AA$) \\
\hline
mean   & 0.29         & 0.51         & 0.38            & 0.53 \\
median & 0.12         & 0.26         & 0.38            & 0.46 \\
range  & 0.01 -- 2.41 & 0.11 -- 1.41 & $-0.47$ -- 0.81 & 0.12 -- 1.22 \\
$\%$ var & \multicolumn{2}{c|}{69} & \multicolumn{2}{c}{100} \\
N$_{\rm stars}$ var (hours) & \multicolumn{2}{c|}{0} & \multicolumn{2}{c}{3} \\
$\%$ no detection & \multicolumn{2}{c|}{15} & \multicolumn{2}{c}{9} \\
\hline
\end{tabular}
\end{table*}
\begin{table*}
[!hbtp]
\centering
\begin{tabular}{l|cccc}
\hline\hline
& EW(\ion{Na}{i}D${2}$) & $|$$\sigma$/$<$EW(\ion{Na}{i}D$_{2}$)$>$$|$ & EW(\ion{Na}{i}D$_{1}$) & $|$$\sigma$/$<$EW(\ion{Na}{i}D$_{1}$)$>$$|$ \\
 & ($\AA$) & & ($\AA$) \\ 
\hline
mean   & 0.40            & 0.53         & 0.38            & 0.57  \\
median & 0.47            & 0.37         & 0.33            & 0.41   \\
range  & $-1.02$ -- 1.57 & 0.15 -- 2.51 & $-0.85$ -- 1.01 & 0.13 -- 3.27  \\
$\%$ var & \multicolumn{4}{c}{100} \\
N$_{\rm stars}$ var (hours) & \multicolumn{4}{c}{1} \\
$\%$ no detection & \multicolumn{4}{c}{0} \\
\hline
\end{tabular}
\end{table*}

In general, CS absorption features show a larger EW variability than the emission lines. The EW relative variability is
significantly higher for the \ion{He}{i}5876, \ion{Na}{i}D and [\ion{O}{i}]6300 lines than for
H$\alpha$. In addition, approximately 30 \% of the
objects show a constant [\ion{O}{i}]6300 equivalent width, but the remaining lines are variable in practically all
stars in which these are detected. Considering the short
timescale variations, the number of variable stars is similar
for the \ion{He}{i}5876 and H$\alpha$ lines, a small percentage seems to present
\ion{Na}{i}D variability, while no star shows changes in
[\ion{O}{i}]6300. 

Finally, when simultaneous EWs and line fluxes are compared, the relative variability of the EWs tends to be equal to or an upper limit of that of the line fluxes for most of the stars. 

\section{Discussion}
\label{Section:Discussion}

The described results indicate that the physical conditions in the line-forming regions are highly complex and variable on practically any timescale, and that the use of individual EW or line flux measurements could lead to biased conclusions. Averaged EW values are likely more representative, which might be specially true for the absorption component of the \ion{He}{i}5876 line, where the variations can be larger than one order of magnitude. Our sample shows that there is no significant correlation between the mean strengths and their relative variabilities, therefore, both are necessary to completely characterize the line behaviour of the stars.

\begin{figure}
\centering
 \includegraphics[height=85mm, clip=true]{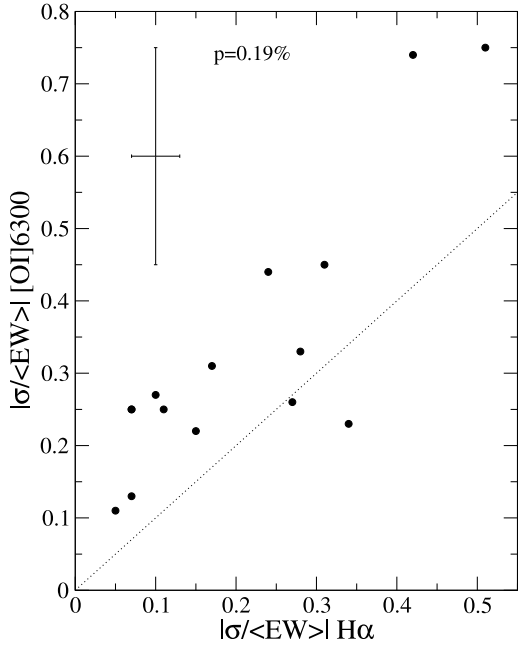}
\caption{Equivalent width relative variability of the [\ion{O}{i}]6300 line against that in
  H$\alpha$ for the objects showing clear [\ion{O}{i}] detections in all their
  spectra. The dotted line indicates equal values. The typical error bars and the Spearman's probability of false correlation are indicated.}
\label{Figure:varEW_alfa_OI}
\end{figure}

Although a detailed study of the physical origin of the lines and their variations is beyond the scope of this work, the observed differences between the typical variabilities of the features, both in strength and in timescale, suggest multiple causes for the different line variations. In addition, our results show that there are no clear correlations among the relative variabilities of the different lines, excluding the obvious relation between the \ion{Na}{i}D$_2$ and D$_1$ changes, and the one between H$\alpha$ and [\ion{O}{i}]6300. Fig. \ref{Figure:varEW_alfa_OI} shows the H$\alpha$ and [\ion{O}{i}]6300 EW relative variabilities of the sources with a clear detection of the [\ion{O}{i}] line in all their spectra (this avoids possible telluric/instrumental effects; see Sect. \ref{Subsection:OI}). The Spearman's probability of false correlation \citep[see e.g.][]{Conover80} is only 0.19$\%$. Although $|$$\sigma$/$<$EW$>$$|$([\ion{O}{i}]6300) tends to be larger than $|$$\sigma$/$<$EW$>$$|$(H$\alpha$), the strength of the corresponding EW variations are coupled in many stars. This suggests that the EW variations share a common origin. One possibility is that H$\alpha$ and [\ion{O}{i}]6300 are affected by accretion-wind variability \citep[see below and e.g.][]{Corcoran97,Corcoran98}. Variability in the UV radiation would have influence on the strength of the [\ion{O}{i}]6300 emission \citep{Acke05}. Changes in the stellar continuum level could also affect the EW variations of both lines simultaneously, as we outline below.

\begin{figure}
\centering
 \includegraphics[width=88mm,clip=true]{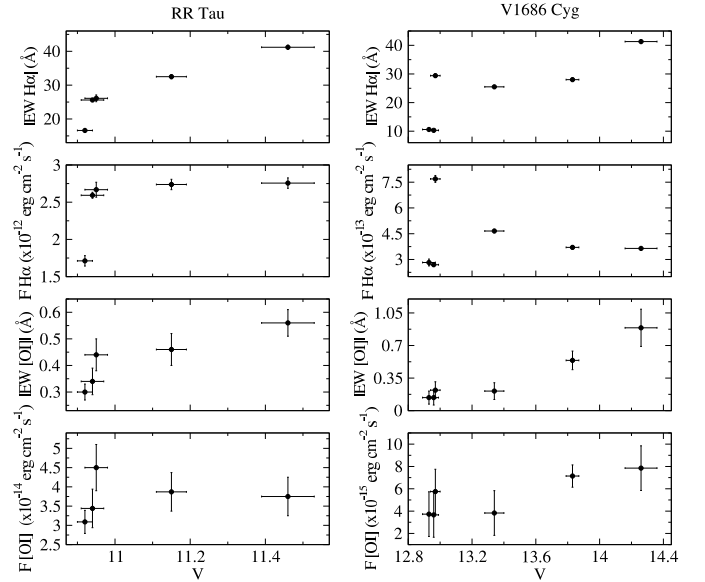}
\caption{Equivalent widths and line fluxes of the H$\alpha$ and
  [\ion{O}{i}]6300 lines against the simultaneous $V$ magnitude \citep{Oudmaijer01} for \object{RR
    Tau} and \object{V1686 Cyg}.}
\label{Figure:uxornouxor}
\end{figure}

Line EWs change because the conditions in the CS gas vary (producing variations in the line luminosity) and/or because there are
changes in the continuum level. Most objects with significant $V$-band variability ($\Delta$$V$ $\geq$ 0.4) increase their H$\alpha$ and [\ion{O}{i}]6300 EWs as the stellar brightness decreases, leaving the corresponding line luminosities almost constant or even decreasing \citep[see e.g. \object{RR Tau} on the left panel of Fig. \ref{Figure:uxornouxor}, and also][]{Rodgers02}. This behaviour has been explained as due to the coronographic effect caused by dusty clouds that occult the stellar surface \citep{Grinin94,Rodgers02}. The EW enhancement would result from the contrast between the continuum dimming and the almost constant line luminosity. This explanation has been suggested for stars showing the UXOr behaviour, but we note that objects such as \object{V1686 Cyg}, which is not classified as an UXOr from its simultaneous optical photopolarimetry \citep{Oudmaijer01}, show a similar pattern in the H$\alpha$ and [\ion{O}{i}]6300 lines (see right panel of Fig \ref{Figure:uxornouxor}).

\begin{figure}
\centering
 \includegraphics[width=88mm,clip=true]{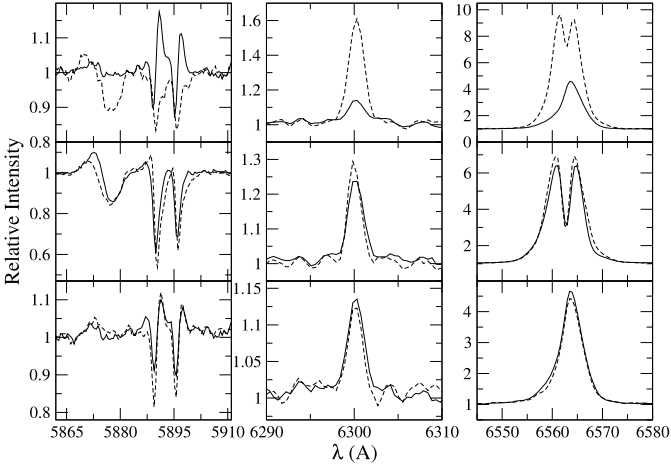}
\caption{\ion{Na}{i}D and \ion{He}{i}5876 (left panels), [\ion{O}{i}]6300 (middle panels) and
  H$\alpha$ (right panels) lines shown by \object{HK Ori} (top panels, 1998, October 25 and 26), \object{VV Ser} (middle panels, 1998, October 24 and 27) and \object{CO Ori} (bottom panels, 1998, October 25 and 26). The dashed and solid lines correspond to the first and second night.}
\label{Figure:hkori_lines}
\end{figure}

Continuum changes can be rejected as the origin of the EW variability in other cases. Several examples are given in Fig. \ref{Figure:hkori_lines}, where the remarkable line variations are not accompanied by significant changes in the simultaneous optical brightness \citep{Oudmaijer01}. The main spectroscopic features shown by \object{HK Ori} during two consecutive nights are
plotted in the top panels. The appearance of redshifted \ion{Na}{i}D emission lines is
accompanied by a decrease of the absorption component of the \ion{He}{i}5876
line. Simultaneously, the [\ion{O}{i}]6300 and H$\alpha$ lines reduce
their luminosities by a factor $\sim$ 3. H$\alpha$ changes from
double-peaked to a redshifted single-peaked profile, with W$_{10}$ remaining constant. The corresponding \ion{Na}{i}D ratios indicate that the CS gas changed from optically thick to
optically thin at these wavelengths. These findings are difficult to
interpret, but obscuring dusty screens in the line of
sight are clearly excluded. An alternative could be that the accretion and/or wind rate diminished from one night
to the other and produced the decrease of the H$\alpha$ and [\ion{O}{i}]
strengths and reduced the gas density, which explains the change to
optically thin. This would allow the detection of hot infalling gas
very close to the stellar surface, seen in emission. The examples in the middle and bottom panels of Fig. \ref{Figure:hkori_lines} show that small variations in the H$\alpha$ and [\ion{O}{i}]6300 lines are again related to each other.

The data show that the main origin for the EW-variability (gas or continuum changes) strictly depends on each star, each line considered and the epoch of observation. The complex behaviour of the lines and the
continuum requires their simultaneous characterization to
distinguish the origin of the spectroscopic variability.

\begin{figure}
\centering
 \includegraphics[width=88mm,clip=true]{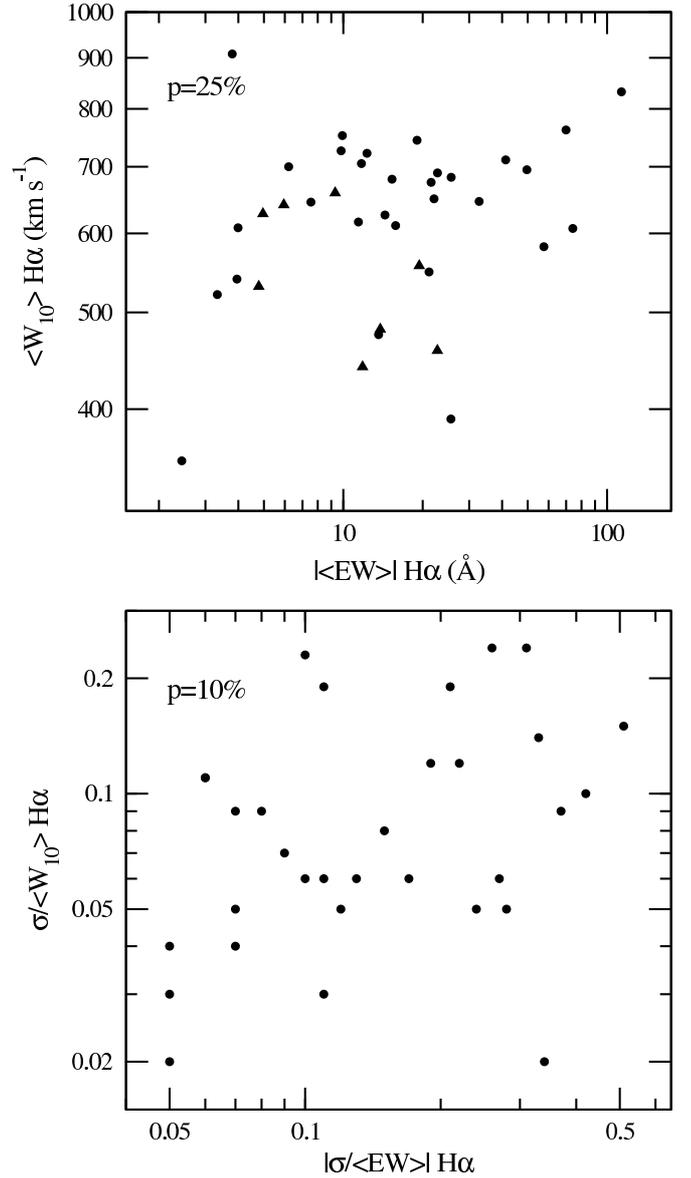}
\caption{Mean value of the width of the H$\alpha$ line against that of the equivalent width (top) and relative variabilities (bottom). Triangles are lower limits for $<$W$_{10}$(H$\alpha$)$>$. The Spearman's probabilities of false correlation are indicated.} 
 \label{Figure:eww10alpha}
\end{figure}

Regarding the H$\alpha$ line, its relative variability in W$_{10}$ is twice as low as that in the EW, and we find no significant correlation either between the mean widths and line strengths or between their relative variabilities. Fig. \ref{Figure:eww10alpha} shows that these parameters have a very high Spearman's probability of false correlation. This suggests that the physical mechanism responsible for the H$\alpha$ broadening does not depend on the column density of hydrogen atoms. A similar result was found for the [OI]6300 line by \citet{Acke05}. Both the H$\alpha$ luminosity and W$_{10}$ are used as empirical accretion tracers in lower-mass stars \citep[see e.g.][and references therein]{Fang09,Jayawardhana06}. Our result indicates that both H$\alpha$ measurements would typically produce different estimates in HAeBe stars.

\begin{figure}
\centering
 \includegraphics[width=88mm,clip=true]{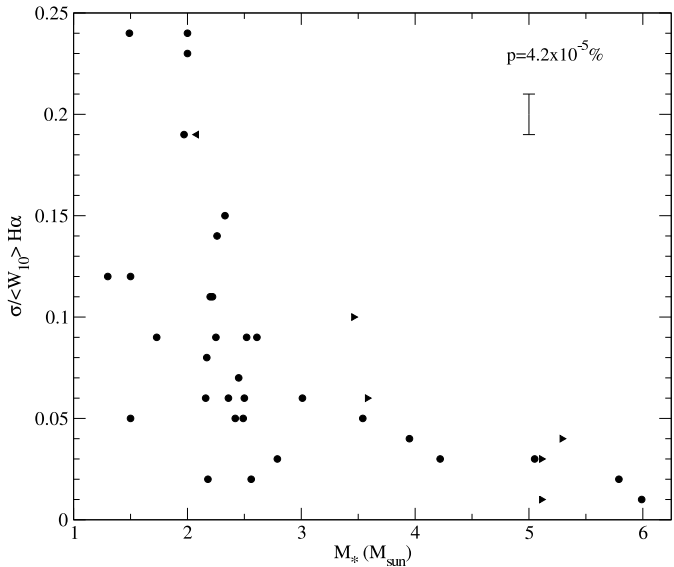}
\caption{Relative variability of the H$\alpha$ width at 10$\%$ of peak
  intensity against the stellar mass. Triangles are upper and lower limits for M$_{*}$. The typical
  $\sigma$/$<$W$_{10}$(H$\alpha$)$>$ uncertainty and the Spearman's probability of false correlation are indicated.}
\label{Figure:M_varW}
\end{figure}

Finally, we stress that the H$\alpha$ behaviour depends on the stellar mass. The more massive objects in our sample tend to have more stable H$\alpha$ line profiles with blueshifted self-absorptions, which could be indicative of a strong wind contribution \citep[see also e.g.][]{Finkenzeller84}. The less massive stars in our sample are slightly dominated by redshifted self-absorptions in their H$\alpha$ profiles, which suggests that they are influenced by accretion \citep{Muzerolle04}. In addition, our multi-epoch data suggest that the strength of the W$_{10}$(H$\alpha$) variability is significantly anti-correlated with the mass of the central object. Fig. \ref{Figure:M_varW} shows our $\sigma$/$<$W$_{10}$(H$\alpha$)$>$ estimates against the stellar mass. Despite the scatter at low masses, almost all stars with M$_{*}$ $\geq$ 3.0 M$_\odot$ show a low W$_{10}$(H$\alpha$) relative variability ($\leq$ 0.06). A more extended emitting region could produce more stability in the line width. Preliminary magnetospheric accretion modelling that we are currently applying on HAe stars suggests that changes in the size of the magnetosphere and/or in the gas temperature of this region could induce significant changes in W$_{10}$(H$\alpha$) (hundreds of km s$^{-1}$). The results found point to different physical processes operating in Herbig Ae and Herbig Be stars, which agrees with previous spectropolarimetric studies \citep[see e.g.][]{Vink02,Mottram07}.

\section{Summary and conclusions}
\label{Section:Conclusions}

The work presented here shows that multi-epoch spectroscopic observations together with simultaneous photometry are an extremely useful tool to better understand the variability of the circumstellar lines in PMS stars. By means of a large number of optical spectra the variability  of the H$\alpha$, [\ion{O}{i}]6300, \ion{He}{i}5876 and \ion{Na}{i}D lines has been analysed in a sample of 38 HAeBe stars. These spectra and the simultaneous photometry have allowed us to estimate line fluxes and to soundly assess if the observed EW variations are caused by changes in the stellar continuum or by variations of the circumstellar gas itself. Indeed, the spectra and photometry -and their simultaneous character- on which the  analysis is based constitute one of the largest existing data bases to study some of the variability properties in intermediate-mass PMS stars. Several specific results we obtained are:
\begin{itemize}
\item The EW variability of the different lines depends on the analysed timescale and is independent of the mean line strength. The \ion{He}{i}587 and \ion{Na}{i}D lines show the largest EW variations and can be seen either in absorption or in emission. In contrast, [\ion{O}{i}]6300 is the only line without variations on timescales of hours and is also the line with variability in a smaller percentage of the stars in the sample.
\item There is a correlation between the H$\alpha$ and [\ion{O}{i}]6300 relative variabilities, which  suggest a common origin. In  some stars the EW  variability of both  lines are due to variations of the continuum, but in other objects the EW variability reflects variations of the line luminosities and, consequently, in the CS gas properties.
\item Mean values and relative variabilities of the H$\alpha$ line width W$_{10}$ and EW are uncorrelated in our sample. The lack of correlation between both parameters suggests that H$\alpha$ broadening does not depend on the column density of hydrogen atoms. Thus, estimates of gas properties, such as accretion rates, based on the H$\alpha$ W$_{10}$ or EW would differ significantly, unlike in lower mass stars.
\item The H$\alpha$ behaviour differs depending on the stellar mass, which suggests different physical processes for Herbig Ae and Herbig Be stars. The massive stars tend to show stable H$\alpha$ profiles, mainly dominated by blueshifted self-absorptions. In addition, stars with  M$_{*}$ $\geq$ 3.0 M$_\odot$ show very low W$_{10}$(H$\alpha$) relative
variabilities ($\leq$ 0.06).
\end{itemize}
Finally, we point out that in addition to the mean spectra and relative variability distributions available as online
material, any of the spectra can be requested from the authors and will also be available from Virtual Observatory tools soon.

\begin{acknowledgements}
C. Eiroa, G. Meeus, I. Mendigut\'{\i}a and B. Montesinos are partially supported 
by grant AYA-2008 01727. We thank Enrique Solano and Mauro L\'opez for making the spectra available from VO tools.
\end{acknowledgements}

\clearpage
\begin{appendix}
\section{Tables with multi-epoch line EWs and fluxes}
\label{appendixonline}
\scriptsize
\renewcommand\arraystretch{1.3}
\renewcommand\tabcolsep{4pt}
\begin{longtable}{lp{1cm}p{0.7cm}p{0.7cm}p{0.7cm}p{0.7cm}cp{0.7cm}p{0.7cm}p{0.7cm}p{0.7cm}p{0.7cm}p{0.7cm}p{0.7cm}p{0.7cm}p{0.7cm}}
\caption{\label{Table:extract}Equivalent widths,  H$\alpha$ widths at 10$\%$ of peak intensity and H$\alpha$ profile types on different observing Julian Dates.}\\
\hline\hline
Star & JD  & EW & $\delta$EW & W$_{10}$ & $\delta$W$_{10}$ & profile type & EW & $\delta$EW & EW & $\delta$EW & EW & $\delta$EW & EW & $\delta$EW \\ 
   &(+2450000)  & H$\alpha$ & H$\alpha$ & H$\alpha$ & H$\alpha$ & H$\alpha$ & [\ion{O}{i}]6300 & [\ion{O}{i}]6300 & \ion{He}{i}5876 & \ion{He}{i}5876 & \ion{Na}{i}D$_{2}$ & \ion{Na}{i}D$_{2}$ &  \ion{Na}{i}D$_{1}$ & \ion{Na}{i}D$_{1}$ \\
 & & ({\AA}) & ({\AA}) & (km/s) & (km/s) & ... & ({\AA}) &  ({\AA}) & ({\AA}) & ({\AA}) & ({\AA}) & ({\AA}) & ({\AA}) & ({\AA}) \\
\hline
\endfirsthead
\caption{continued.}\\
\hline\hline
Star & JD  & EW & $\delta$EW & W$_{10}$ & $\delta$W$_{10}$ & profile type & EW & $\delta$EW & EW & $\delta$EW & EW & $\delta$EW & EW & $\delta$EW \\ 
   &(+2450000)  & H$\alpha$ & H$\alpha$ & H$\alpha$ & H$\alpha$ & H$\alpha$ & [\ion{O}{i}]6300 & [\ion{O}{i}]6300 & \ion{He}{i}5876 & \ion{He}{i}5876 & \ion{Na}{i}D$_{2}$ & \ion{Na}{i}D$_{2}$ &  \ion{Na}{i}D$_{1}$ & \ion{Na}{i}D$_{1}$ \\
 & & ({\AA}) & ({\AA}) & (km/s) & (km/s) & ... & ({\AA}) &  ({\AA}) & ({\AA}) & ({\AA}) & ({\AA}) & ({\AA}) & ({\AA}) & ({\AA}) \\
\hline
\endhead
\hline
\endfoot
HD 31648     &  1111.53&-17 & 1   &	$>$485 &	2  &	IIIB &  ...    &  ...   &  -0.35  &0.05   &-0.61  &0.03   &-0.68 &
0.04\\
     &  1113.62&-20.6 & 0.5 &	635 &	7  &	IIIB &  ...    &  ...   &  -0.58  &0.07   &-0.80   &0.05   &-0.73 &
0.02\\
     &  1114.60&-17.2 & 0.5 &	664 &	3  &	IIIB &  ...    &  ...   &  -0.26  &0.09   &-0.58  &0.02   &-0.56 &
0.03\\
     &  1115.64&-19.7 & 0.1 &	632 &	4  &	IIIB &  ...    &  ...   &  -0.52  &0.05   &-0.68  &0.02   &-0.70  &
0.03\\
     &  1208.43&-18.1 & 0.1 &	$>$440 &	2  &	IIIB &  ...    &  ...   &  -0.30   &0.05   &-0.49  &0.03   &-0.58 &
0.03\\
     &  1209.44&-20.6 & 0.6 &	627 &	9  &	IIIB &  ...    &  ...   &  -0.31  &0.05   &-0.63  &0.02   &-0.47 &
0.02\\
     &  1210.45&-22.6 & 0.5 &	$>$415 &	2  &	IIIB &  ...    &  ...   &  -0.6  &0.1    &-0.68  &0.03   &-0.56 &
0.02\\
HD 34282&  1111.64&-3.9  & 0.3 &	608 &	18 &	IIIR &  ...    &  ...   &  0.15   &0.04   &0.12   &0.02   &0.08  &
0.02\\
     &  1112.68&-5.4  & 0.2 &	660 &	6  &	IIIR &  ...    &  ...   &  0.13   &0.03   &0.10    &0.02   &0.06  &
0.02\\
     &  1113.67&-5.1  & 0.2 &	665 &	9  &	IIIR &  ...    &  ...   &  0.25   &0.05   &0.09   &0.02   &0.04  &
0.02\\
     &  1114.65&-6  & 2   &	$>$352 &	15 &	IIIRm&  ...    &  ...   &  0.19   &0.06   &0.09   &0.02   &0.05  &
0.02\\
     &  1115.77&-2.9  & 0.3 &	716 &	18 &	IIIRm&  ...    &  ...   &  0.17   &0.06   &0.08   &0.02   &0.05  &
0.02\\
     &  1208.45&-5.3  & 0.5 &	678 &	37 &	IIB  &  ...    &  ...   &  0.12   &0.03   &0.09   &0.02   &0.06  &
0.02\\
     &  1209.46&-5.7  & 0.4 &	621 &	23 &	IIB  &  ...    &  ...   &  0.24   &0.06   &0.13   &0.02   &0.09  &
0.02\\
     &  1210.46&-5.2  & 0.2 &	720 &	12 &	IIB  &  ...    &  ...   &  0.18   &0.06   &0.10    &0.02   &0.06  &
0.02\\
HD 34700     &  1111.64&-2.5  & 0.1 &	351 &	11 &	I    &  ...    &  ...   &  ...      &...      &0.09   &0.02   &0.09  &
0.03\\
     &  1112.19&-2.8  & 0.1 &	380 &	11 &	I    &  ...    &  ...   &  ...      &...      &0.09   &0.02   &0.08  &
0.02\\
     &  1113.00&-2.0    & 0.1 &	353 &	10 &	I    &  ...    &  ...   &  ...      &...      &0.05   &0.02   &0.06  &
0.02\\
     &  1114.65&-2.1  & 0.2 &	334 &	13 &	I    &  ...    &  ...   &  ...      &...      &0.09   &0.02   &0.08  &
0.02\\
     &  1208.46&-2.6  & 0.1 &	373 &	9  &	I    &  ...    &  ...   &  ...      &...      &0.10    &0.02   &0.08  &
0.02\\
     &  1209.46&-2.8  & 0.3 &	367 &	22 &	I    &  ...    &  ...   &  ...      &...      &0.11   &0.02   &0.08  &
0.02\\
     &  1210.46&-2.3  & 0.1 &	327 &	7  &	I    &  ...    &  ...   &  ...      &...      &0.08   &0.02   &0.07  &
0.02\\
HD 58647     &  1208.62&-11.3 & 0.4 &	610 &	10 &	IIR  &  -0.10 &  0.02&  -0.12  &0.05   &0.14   &0.02   &0.12  &
0.02\\
     &  1209.54&-11.1 & 0.6 &	619 &	14 &	IIR  &  -0.05&  0.02&  -0.13  &0.05   &0.18   &0.02   &0.15  &
0.02\\
     &  1210.62&-11.8 & 0.4 &	619 &	6  &	IIR  &  -0.04&  0.02&  -0.13  &0.05   &0.12   &0.02   &0.11  &
0.02\\
HD 141569    &  0948.49&-6.2  & 0.3 &	658 &	11 &	IIR  &  -0.15&  0.06&  ...      &...      &0.06   &0.01   &0.07  &
0.02\\
    &  0949.49&-6.0    & 0.3 &	649 &	14 &	IIR  &  -0.13&  0.06&  ...      &...      &0.12   &0.03   &0.13  &
0.04\\
    &  0950.42&-6.0    & 0.3 &	649 &	14 &	IIR  &  -0.14&  0.06&  ...      &...      &0.12   &0.03   &0.16  &
0.04\\
    &  0951.51&-6.0    & 0.5 &	638 &	17 &	IIR  &  -0.08&  0.04&  ...      &...      &0.12   &0.02   &0.09  &
0.02\\
    &  1024.40&-6.2  & 0.4 &	636 &	14 &	IIR  &  -0.13&  0.07&  ...      &...      &0.10    &0.01   &0.10   &
0.02\\
    &  1025.37&-6.1  & 0.5 &	641 &	25 &	IIR  &  -0.17&  0.06&  ...      &...      &0.12   &0.02   &0.18  &
0.05\\
    &  1026.37&-6.1  & 0.2 &	632 &	10 &	IIR  &  -0.19&  0.06&  ...      &...      &0.11   &0.02   &0.10   &
0.02\\
    &  1208.70&-5.3  & 0.6 &	623 &	13 &	IIR  &  -0.18&  0.04&  ...      &...      &0.09   &0.02   &0.06  &
0.02\\
    &  1209.73&-5.9  & 0.3 &	649 &	14 &	IIR  &  -0.14&  0.06&  ...      &...      &0.08   &0.01   &0.04  &
0.01\\
    &  1210.72&-5.7  & 0.3 &	636 &	14 &	IIR  &  -0.11&  0.06&  ...      &...      &0.05   &0.01   &0.04  &
0.01\\
HD 142666    &  0948.50&-2.8  & 0.3 &	$>$399 &	11 &	IIIR &  0    &  0.03   &  0.07   &0.03   &0.18   &0.02   &0.16  &
0.02\\
    &  0949.49&-2.7  & 0.3 &	$>$411 &	8  &	IIIR &  0    &  0.03   &  0.14   &0.06   &0.18   &0.03   &0.16  &
0.03\\
    &  0950.58&-3.1  & 0.3 &	$>$364 &	10 &	IIIR &  0    &  0.03      &0.15  & 0.04  & 0.25  & 0.02  & 0.19
&0.04\\
    &  0951.54&-5.0    & 0.3 &	$>$406 &	7  &	IIIR &  0    &  0.03      &0.30   & 0.05  & 0.21  & 0.03  & 0.21
&0.05\\
    &  1023.39&-3.9  & 0.3    &720    &15     &IIIRm  &0      &0.03      &0.25  & 0.05  & 0.21  & 0.01  & 0.19
&0.01\\
    &  1024.49&-3.3  & 0.3    &675    &18     &IIIRm  &0      &0.03      &0.22  & 0.06  & 0.22  & 0.02  & 0.21
&0.02\\
    &  1025.38&-3.1  & 0.2    &618    &15     &IIIRm  &-0.04  &0.02   &0.25  & 0.07  & 0.26  & 0.02  & 0.24
&0.02\\
    &  1026.38&-4.0    & 0.2    &698    &10     &IIIRm  &-0.04  &0.02   &0.37  & 0.06  & 0.28  & 0.02  & 0.25 &0.03\\
    &  1208.74&-5.5  & 0.4    &584    &15     &IIR    &0      &0.03      &0.11  & 0.03  & 0.18  & 0.02  & 0.14 &0.02\\
    &  1209.74&-4.8  & 0.2    &544    &11     &IIB    &-0.04  &0.03   &0.08  & 0.03  & 0.17  & 0.02  & 0.13 &0.02\\
    &  1210.23&-5.2  & 0.4    &517    &16     &IIB    &0      &0.03      &0.18  & 0.05  & 0.14  & 0.02  & 0.13 &0.02\\
HD 144432    &  0948.58&-13.9 & 0.4    &$>$386    &2      &IIIB   &0      &0.02      &-0.6 & 0.2   & -0.53 & 0.08 
& -0.42&0.05\\
    &  0949.52&-11.4 & 0.3    &$>$360    &1      &IIIB   &0      &0.02      &-0.19 & 0.07  & -0.28 & 0.05  & -0.25&0.05\\
    &  0950.59&-12.7 & 0.4    &$>$369    &1      &IIIB   &0      &0.02      &-0.4 & 0.1   & -0.41 & 0.06  & -0.31&0.06\\
    &  0951.55&-12.9 & 0.9    &$>$343    &4      &IIIB   &0      &0.02      &-0.4 & 0.1   & -0.32 & 0.06  & -0.25&0.06\\
    &  1023.42&-11.2 & 0.6    &576    &6      &IIIB   &-0.04  &0.02   &-0.16 & 0.07  & -0.16 & 0.04  & -0.45&0.09\\
    &  1024.40&-12 & 1      &534    &9      &IIIB   &-0.03  &0.02   &-0.16 & 0.07  & -0.08 & 0.03  & -0.31&0.09\\
    &  1025.39&-10.7 & 0.3    &583    &6      &IIIB   &-0.04  &0.02   &-0.2 & 0.1   & -0.04 & 0.02  & -0.17&0.05\\
    &  1026.38&-9.6  & 0.3    &568    &6      &IIIB   &-0.03  &0.02   &-0.11 & 0.08  & 0.08  & 0.02  & 0    &0.09\\
    &  1208.76&-12.5 & 0.4    &$>$381    &3      &IIIB   &0      &0.02      &-0.2 & 0.1   & -0.31 & 0.05  & -0.25&0.05\\
    &  1209.75&-11.7 & 0.4    &$>$352    &3      &IIIB   &0      &0.02      &-0.4 & 0.1   & -0.24 & 0.04  & -0.23&0.05\\
    &  1210.74&-11.2 & 0.4    &$>$395    &3      &IIIB   &0      &0.02      &-0.21 & 0.07  & -0.31 & 0.05  & -0.22&0.05\\
HD 150193    &  0949.53&-13.1 & 0.1    &$>$424    &5      &IIIB   &...      &...      &-0.3  & 0.1   & -0.33 & 0.08 
& -0.26&0.08\\
    &  0950.59&-14 & 1      &491    &8      &IIIB   &...      &...      &-0.22 & 0.08  & -0.43 & 0.07  & -0.32&0.08\\
    &  0951.56&-14.8 & 0.4    &528    &3      &IIIB   &...      &...      &-0.12 & 0.09  & -0.32 & 0.06  & -0.25&0.07\\
HD 163296    &  0948.59&-23 & 1      &731    &8      &IIIB   &0      &0.02      &0.70   & 0.05  & -0.47 & 0.03 
& -0.54&0.05\\
    &  0949.59&-24.3 & 1      &744    &6      &I      &0      &0.02      &0.36  & 0.05  & -0.43 & 0.04  & -0.49&0.05\\
    &  0950.62&-22.0   & 0.8    &700    &4      &I      &0      &0.02      &0.84  & 0.07  & -0.29 & 0.04  & -0.29&0.05\\
    &  0951.60&-21.9 & 0.2    &695    &8      &I      &0      &0.02      &-0.30  & 0.06  & -0.49 & 0.06  & -0.49&0.06\\
    &  1023.46&-23.1 & 0.7    &733    &6      &IIIB   &-0.03  &0.01   &-0.4 & 0.1   & -0.78 & 0.07  & -0.54&0.06\\
    &  1024.46&-22.5 & 0.6    &712    &6      &IIIB   &-0.04  &0.02   &0.48  & 0.07  & -0.40  & 0.04  & -0.33&0.04\\
    &  1024.53&-23.2 & 0.8    &733    &6      &IIIB   &-0.05  &0.02   &0.57  & 0.05  & -0.20  & 0.03  & -0.19&0.03\\
    &  1025.43&-25   & 1      &706    &6      &IIIB   &-0.05  &0.02   &-0.6 & 0.1   & -0.66 & 0.05  & -0.73&0.04\\
    &  1025.49&-25.4 & 0.7    &683    &6      &IIIB   &-0.05  &0.02   &-0.32 & 0.06  & -0.66 & 0.06  & -0.58&0.06\\
    &  1026.42&-20.6 & 0.8    &592    &6      &IIIB   &-0.02  &0.01   &0.45  & 0.05  & -0.39 & 0.04  & -0.44&0.05\\
    &  1026.48&-19.7 & 0.5    &557    &2      &IIIB   &-0.04  &0.01   &0.35  & 0.04  & -0.45 & 0.04  & -0.47&0.04\\
HD 179218    &  0949.64&-12.8 & 0.8    &464    &4      &I      &-0.03  &0.01   &0.05  & 0.01  & -0.04 & 0.01 
& -0.02&0.01\\
    &  0950.64&-14.0   & 0.2    &484    &3      &I      &-0.04  &0.01   &0.03  & 0.01  & -0.06 & 0.01  & -0.01&0.01\\
    &  0951.65&-14.0   & 0.2    &478    &3      &I      &-0.01  &0.01   &-0.08 & 0.03  & -0.04 & 0.01  & -0.06&0.01\\
HD 190073    &  0948.73&-25.3 & 0.9    &411    &2      &I      &0      &0.02      &-0.50  & 0.06  & -1.09 & 0.05 
& -0.85&0.04\\
    &  0949.64&-25.1 & 0.9    &407    &1      &I      &0      &0.02      &-0.51 & 0.07  & -1.02 & 0.05  & -0.85&0.04\\
    &  0950.65&-24.4 & 0.8    &397    &2      &I      &0      &0.02      &-0.53 & 0.08  & -1.00    & 0.04  & -0.82&0.04\\
    &  0951.65&-23.6 & 0.6    &379    &2      &IVB    &0      &0.02      &-0.52 & 0.06  & -1.02 & 0.05  & -0.85&0.04\\
    &  1023.53&-26 & 1    &381    &2      &IVB    &-0.07  &0.02   &-0.37 & 0.04  & -1.04 & 0.06  & -0.92&0.07\\
    &  1024.55&-26.0   & 0.8    &378    &2      &IVB    &-0.06  &0.01   &-0.45 & 0.06  & -0.96 & 0.04  & -0.84&0.05\\
    &  1025.51&-27.0   & 0.8    &389    &2      &IVB    &-0.06  &0.02   &-0.44 & 0.06  & -1.00    & 0.05  & -0.83&0.04\\
    &  1026.51&-27.6 & 0.7    &387    &1      &IVB    &-0.06  &0.02   &-0.42 & 0.07  & -1.01 & 0.04  & -0.85&0.04\\
AS 442 &0949.66&-36.1  &0.7   & 669    &5      &IIIB   &-0.08  &0.04   &0.11   &0.03  & 0.88  & 0.02  & 0.70   & 0.02\\
 &0950.66&-34.3  &0.6	& 649	 &5	 &IIIB   &-0.12  &0.05   &0.22   &0.09  & 0.86  & 0.02  & 0.72  & 0.02\\
 &0951.66&-36.2  &0.7	& 680	 &5	 &IIIB   &-0.08  &0.04   &0.29   &0.06  & 0.89  & 0.02  & 0.74  & 0.02\\
 &1023.56&-27.2  &0.5	& 614	 &6	 &IIIB   &-0.10   &0.04   &0.14   &0.05  & 0.53  & 0.03  & 0.49  & 0.01\\
 &1024.59&-30  &1	& 616	 &5	 &IIIB   &-0.06  &0.04   &0.20	 &0.06  & 0.62  & 0.03  & 0.52  & 0.02\\
VX Cas &1023.69&-50    &4     & 559    &8      &IIR    &-0.54  &0.05   &0.20    &0.05  & 0.49  & 0.02  & 0.41  & 0.02\\
 &1024.72&-37  &3	& 554	 &6	 &IIR	 &-0.44  &0.05   &0.4   &0.1	& 0.70	& 0.05  & 0.7  & 0.1 \\
 &1025.70&-24  &2	& 638	 &11	 &IIR	 &-0.33  &0.06   &0.84   &0.09  & 1.26  & 0.07  & 0.95  & 0.05\\
 &1026.69&-20  &1	& 635	 &5	 &IIR	 &-0.20   &0.06   &0.8   &0.1	& 0.60	& 0.04  & 0.5  & 0.1 \\
 &1111.52&-16.7  &0.6	& 738	 &4	 &IIB	 &-0.11  &0.04   &0.31   &0.07  & 0.30	& 0.02  & 0.26  & 0.02\\
 &1112.54&-18.0	 &0.7	& 671	 &7	 &IIB	 &-0.13  &0.02   &0.36   &0.06  & 0.31  & 0.02  & 0.28  & 0.04\\
 &1113.52&-16.0	 &0.8	& 707	 &9	 &IIR	 &-0.08  &0.03   &0.93   &0.07  & 0.34  & 0.02  & 0.33  & 0.03\\
 &1114.51&-15.9  &0.4	& 828	 &17	 &IIB	 &-0.13  &0.03   &0.67   &0.06  & 0.37  & 0.02  & 0.32  & 0.02\\
 &1115.50&-15.4  &0.6	& 757	 &9	 &IIR	 &-0.14  &0.02   &0.70	 &0.08  & 0.36  & 0.02  & 0.32  & 0.02\\
 &1209.36&-14.7  &0.7	& 535	 &7	 &IIR	 &-0.15  &0.02   &0.9	 &0.1	& 0.67  & 0.03  & 0.61  & 0.05\\
 &1210.36&-15.6  &0.8	& 529	 &7	 &IIIR   &-0.06  &0.02   &0.60	 &0.08  & 0.46  & 0.04  & 0.42  & 0.07\\
BH Cep &0949.70&-5.5   &0.2   & 779    &10     &IIR    &-0.04  &0.02   &0.20    &0.09  & 0.42  & 0.03  & 0.41  & 0.09\\
 &0950.70&-6.2   &0.1	& 619	 &2	 &IIIR   &0	 &0.03	 &0.4	 &0.1	& 0.52  & 0.04  & 0.41  & 0.04\\
 &0951.73&-5   &1	& 788	 &33	 &IIIR   &0	 &0.03	 &0.3   &0.2	& 0.60	& 0.05  & 0.5  & 0.2 \\
 &1023.60&-4.6   &0.1	& 645	 &13	 &IIR	 &0	 &0.03	 &0.4   &0.1	& 0.73  & 0.03  & 0.54  & 0.03\\
 &1024.65&-5.6   &0.5	& 599	 &23	 &IIIR   &-0.02  &0.01   &0.6   &0.1	& 0.64  & 0.04  & 0.7  & 0.3 \\
 &1025.59&-4.2   &0.5	& 603	 &33	 &IIIR   &-0.02  &0.01   &0.71   &0.08  & 0.54  & 0.03  & 0.42  & 0.02\\
 &1026.56&-4.0	 &0.3	& 678	 &9	 &IIIR   &0	 &0.03	 &0.53   &0.07  & 0.53  & 0.01  & 0.42  & 0.01\\
 &1111.41&-11.1  &0.6	& 650	 &6	 &IIR	 &-0.06  &0.03   &0.24   &0.08  & 0.30	& 0.03  & 0.25  & 0.02\\
 &1111.49&-12.6  &0.4	& 665	 &7	 &IIR	 &-0.03  &0.01   &0.3   &0.1	& 0.34  & 0.02  & 0.27  & 0.03\\
 &1112.36&-6.6   &0.2	& 766	 &9	 &IIRm   &-0.04  &0.02   &0.60	 &0.09  & 0.57  & 0.03  & 0.44  & 0.07\\
 &1112.46&-6.2   &0.2	& 801	 &8	 &IIRm   &-0.03  &0.01   &0.61   &0.09  & 0.57  & 0.02  & 0.43  & 0.08\\
 &1113.36&-4.7   &0.3	& 720	 &6	 &IIIBm  &-0.03  &0.01   &0.17   &0.04  & 0.32  & 0.02  & 0.25  & 0.02\\
 &1113.49&-4.5   &0.2	& 692	 &7	 &IIIBm  &-0.02  &0.01   &0.18   &0.07  & 0.28  & 0.02  & 0.23  & 0.02\\
 &1114.37&-6.5   &0.2	& 709	 &9	 &IIBm   &-0.07  &0.02   &0.10	 &0.03  & 0.26  & 0.02  & 0.21  & 0.02\\
 &1114.49&-6.8   &0.3	& 728	 &13	 &IIIBm  &-0.05  &0.02   &0.07   &0.03  & 0.24  & 0.01  & 0.20	& 0.01\\
 &1115.37&-6.6  &0.2	& 709	 &10	 &IIIBm  &-0.05  &0.02   &0.31   &0.05  & 0.22  & 0.01  & 0.20	& 0.01\\
 &1209.33&-5.3   &0.2	& 742	 &6	 &IIR	 &0	 &0.03	 &0.45   &0.06  & 0.34  & 0.02  & 0.37  & 0.05\\
BO Cep &0949.71&-9.6   &0.3   & 647    &4      &IIR    &-0.21  &0.09   &0.16   &0.04  & 0.30   & 0.05  & 0.22  & 0.02\\
 &0950.71&-7.6   &0.2	& 623	 &5	 &IIB	 &-0.19  &0.08   &0.26   &0.09  & 0.29  & 0.05  & 0.22  & 0.03\\
 &0951.72&-9.2   &0.4	& 614	 &15	 &IIR	 &-0.15  &0.07   &0.05   &0.03  & 0.23  & 0.03  & 0.18  & 0.03\\
 &1023.61&-3.9   &0.2	& 663	 &12	 &IIR	 &-0.15  &0.08   &0.22   &0.07  & 0.28  & 0.04  & 0.23  & 0.03\\
 &1024.67&-5.1   &0.2	& 568	 &12	 &IIIRm  &-0.11  &0.07   &0.22   &0.09  & 0.41  & 0.05  & 0.33  & 0.09\\
 &1025.60&-7.4   &0.3	& 476	 &4	 &IIIR   &-0.17  &0.08   &0.32   &0.08  & 0.34  & 0.04  & 0.23  & 0.04\\
 &1026.68&-7.4   &0.1	& 562	 &3	 &IIR	 &-0.10   &0.09   &0.67   &0.09  & 0.42  & 0.04  & 0.34  & 0.02\\
 &1111.42&-8.5   &0.2	& 658	 &9	 &IIR	 &-0.19  &0.08   &0.12   &0.06  & 0.20	& 0.02  & 0.18  & 0.02\\
 &1112.46&-9.2   &0.3	& 775	 &7	 &IIR	 &-0.19  &0.09   &0.57   &0.08  & 0.20	& 0.02  & 0.15  & 0.02\\
 &1113.45&-8.9   &0.3	& 729	 &7	 &IIR	 &-0.21  &0.08   &0.40	 &0.07  & 0.11  & 0.02  & 0.11  & 0.02\\
 &1114.43&-6.8   &0.3	& 696	 &11	 &IIIRm  &-0.22  &0.08   &0.62   &0.08  & 0.21  & 0.02  & 0.18  & 0.02\\
 &1115.44&-6.9   &0.3	& 660	 &7	 &IIIR   &-0.16  &0.07   &0.24   &0.06  & 0.19  & 0.02  & 0.16  & 0.02\\
 &1210.32&-7.5   &0.3	& 720	 &9	 &IIRm   &-0.13  &0.04   &0.14   &0.04  & 0.23  & 0.02  & 0.20	& 0.03\\
SV Cep &0948.71&-11.4  &0.3   & 715    &10     &IIR    &-0.16  &0.04   &0.16   &0.05  & 0.22  & 0.02  & 0.19  & 0.02\\
 &0949.73&-11.5  &0.5	& 694	 &11	 &IIR	 &-0.14  &0.04   &0.31   &0.07  & 0.23  & 0.02  & 0.2	& 0.02\\
 &0950.72&-10.8  &0.5	& 678	 &6	 &IIB	 &-0.16  &0.05   &0.25   &0.07  & 0.23  & 0.02  & 0.16  & 0.03\\
 &0951.71&-11.6  &0.4	& 719	 &6	 &IIR	 &-0.11  &0.05   &0.34   &0.04  & 0.23  & 0.02  & 0.16  & 0.02\\
 &1023.63&-11.5  &0.5	& 619	 &7	 &IIIR   &-0.11  &0.06   &0.72   &0.09  & 0.33  & 0.02  & 0.28  & 0.02\\
 &1024.68&-11.4  &0.6	& 708	 &14	 &IIIR   &-0.14  &0.04   &0.20	 &0.05  & 0.53  & 0.03  & 0.41  & 0.02\\
 &1025.62&-10.2  &0.4	& 645	 &8	 &IIR	 &-0.12  &0.04   &0.67   &0.09  & 0.55  & 0.03  & 0.50	& 0.06\\
 &1026.60&-9.9   &0.4	& 638	 &13	 &IIR	 &-0.16  &0.05   &0.45   &0.07  & 0.45  & 0.03  & 0.37  & 0.03\\
 &1111.44&-13.0	 &0.2	& 700	 &6	 &IIR	 &-0.14  &0.05   &0.52   &0.07  & 0.22  & 0.02  & 0.19  & 0.02\\
 &1112.49&-12.8  &0.5	& 756	 &10	 &IIB	 &-0.13  &0.04   &0.84   &0.09  & 0.22  & 0.02  & 0.21  & 0.02\\
 &1113.47&-13.4  &0.5	& 776	 &8	 &IIB	 &-0.12  &0.05   &0.55   &0.07  & 0.21  & 0.02  & 0.18  & 0.02\\
 &1114.45&-12.3  &0.3	& 714	 &5	 &IIB	 &-0.12  &0.04   &0.61   &0.07  & 0.42  & 0.02  & 0.32  & 0.02\\
 &1115.46&-12.1  &0.2	& 737	 &7	 &IIB	 &-0.11  &0.04   &0.28   &0.05  & 0.28  & 0.02  & 0.24  & 0.02\\
 &1208.32&-12.1  &0.6	& 774	 &16	 &IIR	 &-0.08  &0.04   &0.61   &0.07  & 0.28  & 0.02  & 0.22  & 0.02\\
V1686 Cyg      &0948.69&-25.7 & 0.4    &$>$473    &6      &IIIB   &-0.5  &0.2    &0.19  & 0.09  & 1.74  & 0.09  & 0.81&  0.06\\
      &0949.68&-28.0   & 0.5    &$>$474    &6      &IIIB   &-0.5  &0.1    &0.08  & 0.05  & 1.5  & 0.1   & 0.77&  0.05\\
      &0950.68&-41.3 & 0.7    &$>$528    &8      &IIIB      &-0.9  &0.2    &0     & 0.09     & 0.6   & 0.1   & 0.22&  0.03\\
      &1023.54&-25.4 & 0.4    &$>$466    &5      &IIIBm  &-0.20   &0.09   &-0.1  & 0.07  & 0.87  & 0.06  & 0.41&  0.04\\
      &1024.57&-25.5 & 0.4    &$>$482    &5      &IIIB   &-0.21  &0.09   &0.31  & 0.09  & 0.94  & 0.07  & 0.40 &  0.04\\
      &1025.53&-25.8 & 0.4    &$>$476    &5      &IIIB   &-0.20   &0.08   &-0.10  & 0.06  & 0.94  & 0.07  & 0.33&  0.05\\
      &1026.52&-26.5 & 0.4    &$>$502    &12     &IIIB   &-0.13  &0.06   &0     & 0.09     & 1.15  & 0.06  & 0.45&  0.05\\
      &1111.36&-29.4 & 0.6    &$>$476    &8      &IIIB   &-0.22  &0.09   &-0.10  & 0.06  & 1.12  & 0.07  & 0.44&  0.05\\
      &1112.41&-13.8 & 0.2    &$>$387    &6      &IIIBm  &-0.26  &0.08   &0.40   & 0.07  & 2.46  & 0.07  & 1.61&  0.08\\
      &1113.38&-10.6 & 0.6    &$>$374    &10     &IIIBm  &-0.14  &0.07   &0.38  & 0.08  & 2.53  & 0.06  & 1.55&  0.06\\
      &1114.39&-10.3 & 0.2    &$>$408    &5      &IIIBm  &-0.14  &0.08   &0.53  & 0.03  & 2.7  & 0.1   & 1.67&  0.07\\
      &1115.40&-10.4 & 0.7    &$>$444    &10     &IIIBm  &-0.24  &0.08   &0.33  & 0.09  & 2.28  & 0.07  & 1.7&  0.1\\
R Mon  &1208.63&-115 &3     & 824    &1      &Im     &-5.1  &0.2    &1.2   &0.2   & -0.15 & 0.03  & -0.46
& 0.07\\
  &1209.65&-111 &3	 & 841    &8	  &Im	  &-5.0  &0.2    &1.2   &0.3   & -0.16 & 0.02  & -0.44 & 0.08\\
  &1210.64&-115 &3	 & 832    &10	  &Im	  &-4.9  &0.2    &1.2   &0.2   & -0.13 & 0.02  & -0.51 & 0.07\\
VY Mon &1208.60&-42  &1     & 726    &3      &IIIB   &-2.0  &0.1    &...      &...     & 2.24  & 0.09  & 1.44  & 0.04\\
 &1209.63&-40  &1	& 719	 &5	 &IIIB   &-1.9   &0.2	 &...	 &...	& 2.34  & 0.08  & 1.56  & 0.05\\
 &1210.63&-41  &1	& 688	 &6	 &IIIB   &-1.9  &0.2	 &...	 &...	& 2.39  & 0.09  & 1.6	& 0.05\\
51 Oph &0948.62&-3.3   &0.1   & 537    &11     &IIB    &...      &...      &-0.10   &0.04  & -0.08 & 0.03  & -0.19 & 0.04\\
 &0949.59&-3.2   &0.2	& 529	 &14	 &IIB	 &...	 &...	 &-0.08  &0.04  & 0	& 0.03	& -0.08 & 0.04\\
 &0950.61&-3.3   &0.1	& 526	 &8	 &IIB	 &...	 &...	 &-0.11  &0.06  & -0.07 & 0.03  & -0.14 & 0.04\\
 &0951.59&-3.2   &0.2	& 542	 &11	 &IIB	 &...	 &...	 &-0.12  &0.04  & -0.04 & 0.03  & -0.12 & 0.04\\
 &1023.45&-3.5   &0.2	& 509	 &10	 &IIB	 &...	 &...	 &-0.14  &0.04  & -0.13 & 0.02  & -0.24 & 0.05\\
 &1024.45&-3.4   &0.2	& 511	 &8	 &IIB	 &...	 &...	 &-0.12  &0.05  & -0.03 & 0.01  & -0.10  & 0.02\\
 &1025.42&-3.3   &0.2	& 504	 &15	 &IIB	 &...	 &...	 &-0.06  &0.02  & 0.04  & 0.01  & 0.03  & 0.01\\
 &1026.41&-3.4   &0.1	& 509	 &6	 &IIB	 &...	 &...	 &-0.08  &0.04  & -0.04 & 0.01  & -0.11 & 0.02\\
KK Oph &0948.61&-69  &4     & 627    &6      &I      &-2.1  &0.2    &0.4   &0.1   & 0.14  & 0.04  & 0.09  & 0.03\\
 &0949.60&-67  &4	& 616	 &4	 &I	 &-2.0	 &0.1	 &0.6   &0.1	& 0.20	& 0.05  & 0.26  & 0.08\\
 &0950.61&-72  &3	& 589	 &2	 &Im	 &-2.8  &0.2	 &-0.22  &0.07  & -0.32 & 0.07  & -0.15 & 0.04\\
 &0951.58&-76  &5	& 594	 &3	 &Im	 &-2.7   &0.2	 &0.29   &0.09  & -0.4 & 0.1	& -0.21 & 0.06\\
 &0951.64&-67  &2	& 557	 &5	 &Im	 &-2.3  &0.2	 &0.4   &0.1	& -0.4 & 0.2	& -0.20  & 0.06\\
 &0951.68&-79  &3	& 611	 &2	 &I	 &-2.4  &0.1	 &0.4	 &0.1	& -0.3 & 0.1	& -0.14 & 0.04\\
 &1023.44&-96  &3	& 537	 &3	 &IIR	 &-3.3  &0.2	 &0.38   &0.09  & -0.5 & 0.1	& -0.40  & 0.09\\
 &1024.44&-89  &6	& 574	 &1	 &IIR	 &-3.0  &0.2	 &0.7	 &0.1	& 0.30	& 0.04  & 0.35  & 0.08\\
 &1025.41&-67  &5	& 702	 &5	 &IIR	 &-1.8  &0.1	 &0.36   &0.08  & 0.71  & 0.06  & 0.68  & 0.06\\
 &1026.41&-61  &5	& 664	 &11	 &IIB	 &-1.7  &0.1	 &0.6   &0.1	& 0.61  & 0.05  & 0.62  & 0.08\\
T Ori  &1111.72&-23  &1     & 712    &6      &IIR    &-0.14  &0.03   &0.56   &0.07  & 0.67  & 0.02  & 0.48  & 0.03\\
  &1112.73&-34  &2	 & 690    &6	  &IIR    &-0.22  &0.03   &0.50    &0.05  & 0.53  & 0.02  & 0.35  & 0.02\\
  &1114.72&-20.7  &0.8   & 639    &7	  &IIR    &-0.10   &0.02   &0.41   &0.05  & 0.42  & 0.02  & 0.29  & 0.02\\
  &1115.76&-23.0    &0.3   & 626    &9	  &IIR    &-0.16  &0.04   &0.61   &0.06  & 0.57  & 0.02  & 0.41  & 0.03\\
  &1208.53&-16.0    &0.2   & 711    &6	  &IIR    &-0.10   &0.03   &0.92   &0.06  & 0.28  & 0.02  & 0.19  & 0.02\\
  &1209.55&-17.1  &0.5   & 687    &5	  &IIR    &-0.11  &0.02   &0.56   &0.05  & 0.27  & 0.02  & 0.20   & 0.03\\
  &1210.55&-17.9  &0.4   & 659    &4	  &IIR    &-0.11  &0.02   &0.50    &0.05  & 0.22  & 0.02  & 0.16  & 0.02\\
BF Ori &1111.71&-8.9   &0.3   & 607    &4      &IIR    &-0.05  &0.02   &1.13   &0.06  & 0.24  & 0.02  & 0.56  & 0.04\\
 &1112.64&-9.6   &0.3	& 692	 &4	 &IIB	 &-0.05  &0.01   &0.53   &0.05  & 0.74  & 0.01  & 0.65  & 0.02\\
 &1113.69&-10.6  &0.4	& 776	 &9	 &IIR	 &-0.07  &0.02   &0.42   &0.07  & 0.58  & 0.02  & 0.46  & 0.01\\
 &1114.67&-11  &1	& 850	 &13	 &IIB	 &-0.04  &0.01   &0.47   &0.03  & 0.99  & 0.01  & 0.78  & 0.01\\
 &1115.66&-9   &1	& 733	 &11	 &IIR	 &-0.05  &0.02   &0.72   &0.08  & 0.66  & 0.02  & 0.48  & 0.03\\
 &1208.53&-9.7   &0.3	& 753	 &8	 &IIR	 &-0.04  &0.01   &0.76   &0.08  & 0.51  & 0.02  & 0.35  & 0.02\\
 &1209.54&-9.9   &0.2	& 815	 &8	 &IIR	 &-0.04  &0.01   &0.55   &0.07  & 0.65  & 0.03  & 0.49  & 0.05\\
 &1210.52&-10.6  &0.3	& 786	 &4	 &IIR	 &-0.04  &0.01   &0.67   &0.06  & 0.50	& 0.02  & 0.33  & 0.02\\
 &1210.58&-10.0	 &0.3	& 758	 &7	 &IIR	 &-0.02  &0.01   &0.75   &0.07  & 0.53  & 0.02  & 0.36  & 0.03\\
CO Ori &1111.57&-16.6  &0.5   & 566    &30     &Im     &-0.13  &0.03   &0      &0.07     & 0.64  & 0.04  & 0.46  & 0.04\\
 &1112.57&-21.0	 &0.6	& 548	 &6	 &Im	 &-0.26  &0.04   &0	 &0.07	& 0.22  & 0.02  & 0.18  & 0.02\\
 &1113.56&-23	 &1	& 562	 &8	 &I	 &-0.34  &0.06   &0	 &0.07	& 0.09  & 0.02  & 0.11  & 0.02\\
 &1114.54&-21  &1	& 502	 &6	 &I	 &-0.29  &0.04   &0.22   &0.07  & 0.17  & 0.01  & 0.21  & 0.02\\
 &1115.55&-21  &1	& 522	 &9	 &I	 &-0.35  &0.05   &0.11   &0.07  & 0.22  & 0.02  & 0.21  & 0.02\\
 &1208.48&-20.9  &0.6	& 518	 &5	 &I	 &-0.33  &0.05   &0.13   &0.06  & 0.37  & 0.04  & 0.32  & 0.06\\
 &1209.49&-25.1  &0.7	& 586	 &6	 &I	 &-0.34  &0.07   &0	 &0.07	& 0.37  & 0.04  & 0.34  & 0.05\\
 &1210.48&-21.4  &0.7	& 584	 &7	 &Im	 &-0.29  &0.07   &0.08   &0.04  & 0.39  & 0.03  & 0.36  & 0.03\\
HK Ori &1111.68&-63  &1     & 599    &4      &IIB    &-1.4  &0.1    &0.53   &0.05  & 0.53  & 0.04  & 0.52  & 0.05\\
 &1112.74&-61  &1	& 533	 &4	 &IIR	 &-1.4  &0.1	 &0.58   &0.05  & 0.47  & 0.04  & 0.44  & 0.06\\
 &1113.66&-21  &2	& 530	 &7	 &I	 &-0.5  &0.1	 &0	 &0.09	& -0.36 & 0.03  & -0.19 & 0.06\\
 &1114.71&-63  &1	& 578	 &1	 &IIR	 &-1.4  &0.1	 &0.53   &0.05  & 0.35  & 0.03  & 0.41  & 0.04\\
 &1115.68&-71  &3	& 567	 &4	 &IIR	 &-1.5  &0.1	 &0.30	 &0.09  & 0.30	& 0.05  & 0.27  & 0.06\\
 &1208.49&-58  &1	& 630	 &1	 &IIB	 &-1.33  &0.09   &0.4	 &0.1	& 0.14  & 0.02  & 0.16  & 0.03\\
 &1209.50&-59  &1	& 624	 &4	 &IIB	 &-1.4  &0.1	 &0.48   &0.06  & 0.22  & 0.03  & 0.21  & 0.04\\
 &1210.51&-65  &3	& 595	 &5	 &IIB	 &-1.29  &0.08   &0.36   &0.09  & 0.41  & 0.05  & 0.39  & 0.05\\
NV Ori &1111.77&-4.4   &0.2   & 473    &10     &Im     &0      &0.02      &0.19   &0.05  & -0.27 & 0.03  & -0.20 
& 0.02\\
 &1112.72&-4.0	 &0.3	& 546	 &10	 &IIIR   &0	 &0.02	 &0	 &0.07	& -0.26 & 0.04  & -0.23 & 0.03\\
 &1113.76&-3.9   &0.1	& 616	 &8	 &IIR	 &0	 &0.02	 &0.13   &0.05  & -0.19 & 0.03  & -0.28 & 0.06\\
 &1114.74&-3.9   &0.1	& 627	 &1	 &IIR	 &0	 &0.02	 &0.30	 &0.07  & -0.20  & 0.03  & -0.16 & 0.04\\
 &1115.74&-3.8   &0.1	& 633	 &7	 &IIIR   &0	 &0.02	 &0.26   &0.06  & -0.16 & 0.03  & -0.13 & 0.02\\
 &1208.52&-3.8   &0.2	& 697	 &16	 &IIB	 &-0.04  &0.02   &0.18   &0.05  & -0.18 & 0.02  & -0.25 & 0.03\\
 &1209.53&-4.3   &0.1	& 654	 &7	 &IIR	 &-0.03  &0.01   &0.20	 &0.04  & -0.24 & 0.02  & -0.27 & 0.03\\
 &1210.54&-3.8   &0.2	& 618	 &13	 &IIIR   &-0.03  &0.01   &0.30	 &0.06  & 0.28  & 0.02  & 0.22  & 0.04\\
RY Ori &1112.76&-19.4  &0.6   & 585    &4      &IIR    &-0.17  &0.06   &0.7    &0.1   & 1.00     & 0.04  & 1.10  
& 0.06\\
 &1113.74&-14.3  &0.4	& 588	 &8	 &IIB	 &-0.06  &0.03   &0.33   &0.08  & 0.76  & 0.03  & 0.73  & 0.03\\
 &1114.68&-11.0	 &0.3	& 571	 &6	 &IIR	 &-0.06  &0.04   &0.46   &0.09  & 0.95  & 0.04  & 0.96  & 0.06\\
 &1115.67&-21.4  &0.7	& 606	 &6	 &IIR	 &-0.11  &0.03   &0.32   &0.06  & 0.57  & 0.03  & 0.62  & 0.04\\
 &1208.51&-13.8  &0.5	& 616	 &7	 &IIR	 &-0.10   &0.04   &1.03   &0.08  & 1.11  & 0.03  & 1.08  & 0.04\\
 &1209.52&-13.2  &0.4	& 647	 &10	 &IIR	 &-0.13  &0.04   &0.5   &0.1	& 1.51  & 0.04  & 1.54  & 0.07\\
 &1210.53&-17.4  &0.5	& 662	 &6	 &IIR	 &-0.20   &0.07   &0.78   &0.08  & 0.98  & 0.03  & 1.03  & 0.04\\
UX Ori &1111.61&-13.2  &0.2   & 576    &6      &IIIB   &-0.06  &0.02   &-0.14  &0.05  & 0.15  & 0.03  & 0.27  & 0.05\\
 &1112.60&-13.0	 &0.4	& 772	 &6	 &Im	 &-0.05  &0.02   &0.15   &0.04  & -0.29 & 0.07  & -0.02 & 0.01\\
 &1112.71&-12.8  &0.4	& 740	 &6	 &Im	 &-0.07  &0.04   &0.24   &0.04  & -0.25 & 0.06  & -0.04 & 0.03\\
 &1113.61&-11.3  &0.3	& 687	 &17	 &Im	 &-0.07  &0.04   &0.21   &0.04  & -0.29 & 0.06  & -0.16 & 0.03\\
 &1113.71&-11.8  &0.2	& 744	 &9	 &Im	 &-0.05  &0.02   &-0.15  &0.04  & -0.36 & 0.07  & -0.17 & 0.03\\
 &1114.58&-11.8  &0.6	& 735	 &6	 &Im	 &-0.06  &0.02   &0.21   &0.04  & 0.22  & 0.01  & 0.15  & 0.02\\
 &1114.70&-11.5  &0.2	& 704	 &4	 &Im	 &-0.05  &0.02   &0.18   &0.03  & 0.28  & 0.02  & 0.18  & 0.01\\
 &1115.62&-10.7  &0.3	& 777	 &7	 &IIBm   &-0.04  &0.02   &0.48   &0.04  & 0.28  & 0.02  & 0.19  & 0.02\\
 &1208.44&-7.4   &0.4	& 730	 &15	 &IIRm   &-0.07  &0.02   &0.28   &0.05  & -0.24 & 0.05  & -0.16 & 0.03\\
 &1209.45&-6.6   &0.3	& 616	 &9	 &IVRm   &-0.06  &0.02   &0.54   &0.07  & -0.15 & 0.03  & -0.14 & 0.04\\
 &1209.57&-6.0	 &0.2	& 601	 &9	 &IVRm   &-0.03  &0.02   &0.49   &0.08  & -0.14 & 0.03  & -0.12 & 0.03\\
 &1210.31&-5.5   &0.5	& 532	 &21	 &IVR	 &-0.05  &0.02   &0.18   &0.05  & -0.23 & 0.04  & -0.16 & 0.03\\
 &1210.41&-5.8   &0.3	& 537	 &14	 &IVR	 &-0.04  &0.02   &-0.07  &0.04  & -0.25 & 0.05  & -0.16 & 0.03\\
 &1210.49&-6.3   &0.2	& 554	 &5	 &IIIRm  &-0.05  &0.02   &0.06   &0.03  & -0.30  & 0.08  & -0.15 & 0.03\\
 &1210.57&-5.9   &0.4	& $>$586	 &93	 &IIIRm  &-0.03  &0.02   &0.16   &0.05  & -0.22 & 0.06  & -0.10  & 0.03\\
V346 Ori       &1111.66&-3.8  & 0.7    &829    &91     &IIIRm  &...      &...      &...     & ...     & -0.13 & 0.04 
& -0.08&0.02\\
       &1112.70&-4.4  & 0.8    &1056   &171    &IIIRm  &...      &...      &...     & ...     & 0.09  & 0.02  & 0.06&	0.02\\
       &1113.72&-5  & 1    &972    &109    &IIIRm  &...      &...      &...     & ...     & 0.08  & 0.02  & 0.07&	0.02\\
       &1114.69&-4.0    & 0.8    &974    &141    &IIIBm  &...      &...      &...     & ...     & 0.10   & 0.02  & 0.08&  0.03\\
       &1115.73&-4.1  & 0.6    &880    &80     &IIIRm  &...      &...      &...     & ...     & 0.12  & 0.03  & 0.09&  0.03\\
       &1208.47&-2.9  & 0.5    &871    &101    &IIIRm  &...      &...      &...     & ...     & 0.21  & 0.02  & 0.15&  0.03\\
       &1209.48&-2.8  & 0.5    &864    &145    &IIIBm  &...      &...      &...     & ...     & 0.23  & 0.02  & 0.17&  0.03\\
       &1210.47&-2.9  & 0.8    &821    &198    &IIIBm  &...      &...      &...     & ...     & 0.21  & 0.02  & 0.16&  0.03\\
V350 Ori       &1111.76&-11.8 & 0.6    &729    &10     &IIR    &-0.12  &0.03   &0.32  & 0.09  & 0.67  & 0.03 
& 0.53&  0.05\\
       &1113.77&-11.4 & 0.5    &767    &8      &IIR    &-0.11  &0.03   &0.39  & 0.07  & 0.66  & 0.03  & 0.52&  0.05\\
       &1114.76&-12.8 & 0.7    &680    &10     &IIR    &-0.12  &0.04   &0.4  & 0.1   & 0.67  & 0.03  & 0.53&  0.04\\
       &1115.72&-15 & 1    &645    &7      &IIIR   &-0.14  &0.02   &0.40   & 0.08  & 0.63  & 0.04  & 0.54&  0.07\\
       &1208.54&-8.9  & 0.9    &770    &40     &IIR    &-0.20   &0.06   &0.23  & 0.08  & 0.56  & 0.04  & 0.45&  0.07\\
       &1209.58&-15   & 1    &731    &16     &IIIR   &-0.13  &0.04   &0.7  & 0.1   & 0.82  & 0.04  & 0.85&  0.07\\
       &1210.56&-11.3 & 0.8    &730    &14     &IIIRm  &-0.08  &0.04   &0.53  & 0.08  & 0.96  & 0.04  & 0.81&  0.06\\
XY Per &1023.72&-10.7  &0.7   & 733    &6      &IIRm   &-0.02  &0.01   &0.17   &0.02  & 0.46  & 0.01  & 0.38  & 0.01\\
 &1024.73&-8.5   &0.3	& 762	 &8	 &IIR	 &0	 &0.03	 &0.32   &0.05  & 0.54  & 0.02  & 0.43  & 0.02\\
 &1025.71&-9.0	 &0.3	& 744	 &6	 &IIB	 &-0.03  &0.01   &0.24   &0.04  & 0.69  & 0.02  & 0.52  & 0.02\\
 &1026.73&-10.7  &0.1	& 704	 &4	 &IIBm   &0	 &0.03	 &0.12   &0.02  & 0.65  & 0.02  & 0.48  & 0.01\\
 &1112.65&-10.1  &0.8	& 713	 &12	 &IIR	 &-0.06  &0.03   &0.32   &0.04  & 0.51  & 0.01  & 0.40	& 0.01\\
 &1113.57&-11  &1	& 729	 &12	 &IIR	 &-0.08  &0.03   &0.30	 &0.03  & 0.57  & 0.01  & 0.43  & 0.01\\
 &1114.56&-11  &1	& 729	 &15	 &IIB	 &-0.03  &0.01   &0.35   &0.03  & 0.62  & 0.01  & 0.50	& 0.01\\
 &1115.60&-11  &1	& 706	 &12	 &IIB	 &-0.08  &0.03   &0.11   &0.03  & 0.57  & 0.01  & 0.45  & 0.01\\
 &1208.37&-8.2   &0.7	& 743	 &14	 &IIB	 &-0.02  &0.01   &0.30	 &0.03  & 0.42  & 0.01  & 0.40	& 0.04\\
 &1209.38&-8.5   &0.7	& 708	 &11	 &IIB	 &-0.04  &0.01   &0.15   &0.03  & 0.56  & 0.02  & 0.48  & 0.02\\
 &1210.38&-9.4   &0.6	& 720	 &7	 &IIB	 &-0.10   &0.03   &0.10	 &0.02  & 0.43  & 0.01  & 0.34  & 0.01\\
VV Ser &0949.56&-50.0    &0.8   & 711    &6      &IIR    &-0.6  &0.1    &0.8   &0.1   & 0.65  & 0.04  & 0.47  & 0.04\\
 &0950.62&-50.2  &0.8	& 701	 &6	 &IIR	 &-0.46  &0.07   &0.5   &0.2	& 0.79  & 0.06  & 0.8  & 0.1 \\
 &0951.61&-47	 &2	& 726	 &6	 &IIR	 &-0.43  &0.07   &0.9	 &0.1	& 0.75  & 0.04  & 0.61  & 0.04\\
 &1023.47&-48.9  &0.9	& 704	 &6	 &IIR	 &-0.6  &0.1	 &0.47   &0.08  & 0.68  & 0.04  & 0.53  & 0.04\\
 &1024.48&-51.4  &0.8	& 728	 &6	 &IIR	 &-0.50   &0.08   &0.6   &0.1	& 0.98  & 0.05  & 0.93  & 0.09\\
 &1024.54&-49  &2	& 736	 &6	 &IIR	 &-0.58  &0.09   &0.8   &0.2	& 0.86  & 0.06  & 0.8  & 0.1 \\
 &1024.61&-48  &3	& 740	 &7	 &IIR	 &-0.59  &0.05   &0.81   &0.08  & 0.98  & 0.05  & 0.87  & 0.08\\
 &1025.44&-51.0	 &0.9	& 673	 &6	 &IIB	 &-0.54  &0.08   &0.44   &0.09  & 1.08  & 0.05  & 0.90	& 0.08\\
 &1025.50&-52.0	 &0.9	& 666	 &6	 &IIB	 &-0.52  &0.08   &0.3   &0.1	& 1.07  & 0.05  & 0.94  & 0.08\\
 &1025.57&-50.9  &0.9	& 667	 &6	 &IIB	 &-0.56  &0.08   &0.42   &0.06  & 1.07  & 0.05  & 0.95  & 0.09\\
 &1026.44&-52.0	 &0.8	& 685	 &6	 &II	 &-0.51  &0.07   &0.48   &0.05  & 0.81  & 0.05  & 0.69  & 0.08\\
 &1026.49&-53.9  &0.8	& 679	 &6	 &IIB	 &-0.56  &0.06   &0.45   &0.06  & 0.82  & 0.05  & 0.68  & 0.09\\
 &1026.56&-52.9  &0.8	& 676	 &6	 &IIR	 &-0.53  &0.06   &-0.3  &0.1	& 0.78  & 0.05  & 0.70	& 0.08\\
 &1111.30&-48  &3	& 671	 &6	 &IIB	 &-0.6  &0.1	 &0.79   &0.09  & 1.00	& 0.04  & 0.83  & 0.07\\
 &1112.31&-50  &3	& 706	 &6	 &IIB	 &-0.5   &0.1	 &0.4   &0.1	& 0.66  & 0.04  & 0.51  & 0.03\\
 &1113.30&-46	 &3	& 673	 &5	 &IIB	 &-0.53  &0.08   &0.4   &0.1	& 0.66  & 0.04  & 0.53  & 0.04\\
 &1114.31&-44  &3	& 646	 &5	 &II	 &-0.7  &0.1	 &0.65   &0.09  & 0.76  & 0.04  & 0.59  & 0.04\\
 &1115.30&-49	 &1	& 727	 &6	 &IIB	 &-0.5  &0.1	 &0.43   &0.09  & 0.61  & 0.04  & 0.50	& 0.05\\
CQ Tau &1111.74&-3.3   &0.1   & $>$272    &4      &IIIR   &-0.05  &0.02   &0.23   &0.06  & 0.53  & 0.03  & 0.39  & 0.03\\
 &1112.76&-5.1   &0.1	& 670	 &4	 &IIIR   &-0.04  &0.02   &0	 &0.08	& 0.52  & 0.03  & 0.39  & 0.03\\
 &1113.75&-3.6   &0.1	& 558	 &3	 &IIIR   &-0.03  &0.01   &0	 &0.08	& 0.30	& 0.02  & 0.22  & 0.02\\
 &1114.73&-3.7   &0.1	& 577	 &6	 &IIR	 &-0.03  &0.01   &0.11   &0.06  & 0.32  & 0.02  & 0.24  & 0.02\\
 &1115.69&-4.3   &0.1	& 636	 &1	 &IIR	 &-0.05  &0.02   &0	 &0.08	& 0.32  & 0.04  & 0.25  & 0.02\\
 &1208.56&-7.7   &0.2	& 519	 &7	 &IIIR   &-0.10   &0.02   &0.08   &0.04  & 0.35  & 0.02  & 0.33  & 0.04\\
 &1209.60&-6.1   &0.2	& 569	 &6	 &IIIR   &-0.07  &0.02   &0.26   &0.08  & 0.48  & 0.02  & 0.45  & 0.04\\
 &1210.59&-4.4   &0.2	& 445	 &6	 &Im	 &-0.08  &0.02   &0.15   &0.07  & 0.33  & 0.01  & 0.34  & 0.04\\
RR Tau &1111.73&-32.5  &0.6   & 667    &3      &IIR    &-0.46  &0.06   &0.20    &0.03  & 0.70   & 0.02  & 0.64  & 0.04\\
 &1112.77&-41.2  &0.6	& 685	 &4	 &IIR	 &-0.56  &0.05   &0.69   &0.03  & 0.64  & 0.02  & 0.60	& 0.04\\
 &1113.78&-26  &1	& 672	 &3	 &IIR	 &-0.44  &0.06   &0.40	 &0.04  & 1.00	& 0.03  & 0.92  & 0.05\\
 &1114.75&-25.6  &0.3	& 700	 &5	 &IIR	 &-0.34  &0.05   &0.64   &0.04  & 0.88  & 0.02  & 0.76  & 0.04\\
 &1115.71&-28.5  &0.3	& 660	 &1	 &IIR	 &-0.38  &0.05   &0.27   &0.04  & 0.66  & 0.03  & 0.55  & 0.04\\
 &1208.57&-18.0	 &0.2	& 703	 &1	 &IIB	 &-0.36  &0.07   &0.31   &0.05  & 0.71  & 0.05  & 0.53  & 0.04\\
 &1209.61&-16.6  &0.6	& 682	 &5	 &IIB	 &-0.30   &0.03   &0.42   &0.05  & 0.71  & 0.03  & 0.60	& 0.07\\
 &1210.60&-16.6  &0.4	& 694	 &5	 &IIB	 &-0.30   &0.03   &0.54   &0.06  & 0.77  & 0.05  & 0.68  & 0.07\\
RY Tau &1111.62&-14.1 & 0.4  &  634  &  2   &	IIIB &  -0.7 & 0.1  &  0.25  & 0.06 &  0.50   & 0.04 &  0.41  & 0.05 \\
 &1112.59&-12.0	& 0.1  &  641  &  2   &   IIBm &  -0.7 & 0.1  &  0.30	& 0.06 &  0.49  & 0.04 &  0.41  & 0.04 \\
 &1113.60&-13.0	& 0.3  &  624  &  4   &   IIB  &  -0.7 & 0.1  &  -0.19 & 0.06 &  -0.18 & 0.03 &  -0.19 & 0.03 \\
 &1114.59&-16.2 & 0.8  &  664  &  7   &   IIB  &  -0.7  & 0.1  &  -0.22 & 0.06 &  -0.32 & 0.06 &  -0.37 & 0.05 \\
 &1208.40&-19.9 & 0.4  &  837  &  18  &   IIR  &  -0.8 & 0.1  &  -0.33 & 0.06 &  0.44  & 0.02 &  0.42  & 0.02 \\
 &1209.41&-16.6 & 0.2  &  677  &  2   &   IIB  &  -0.9 & 0.1  &  -0.20  & 0.04 &  -0.39 & 0.05 &  -0.36 & 0.05 \\
PX Vul & 1023.51 & -15.8 & 0.2 & 598 & 4 & IIIB & -0.08 & 0.02	& 0 & 0.09 & 0.13 & 0.01 &	0.15 & 0.01 \\
 & 1024.51 & -15.1 & 0.3 & 588 & 4 & IIIB & -0.06 & 0.02 & 0.24 & 0.05 & 0.25 & 0.02 & 0.23 & 0.01 \\
 & 1025.47 & -15.1 & 0.2 & 613 & 2 & IIIB & -0.05 & 0.02 & 0.39 & 0.05 & 0.18 & 0.01 & 0.18 & 0.01 \\
 & 1026.47 & -14.0 & 0.9 & 671 & 7 & Im & -0.06 & 0.02 & 0.38 & 0.04 & 0.16 & 0.02 & 0.15 & 0.02 \\
 & 1111.34 & -13 & 1 & 651 & 9 & IIB & -0.11 & 0.03 & 0.09 & 0.02 & 0.13 & 0.01 & 0.14 & 0.01 \\
 & 1112.34 & -13 & 1 & 673 & 9 & IIB &-0.07 & 0.02 & 0.14 & 0.03 & 0.14 & 0.01 & 0.14 & 0.01 \\
 & 1113.34 & -14.0 & 0.4 & 630 & 6 & IIIB & -0.10 & 0.03 & 0.27 & 0.09 & 0.19 & 0.01 & 0.17 & 0.01 \\
 & 1114.35 & -15 & 1 & 598 & 10 & IIIB & -0.09 & 0.03 & 0.04 & 0.02 & 0.13 & 0.01 & 0.13 & 0.01 \\
 & 1115.35 & 15 & 1 & 616 & 7 & Im & -0.10 & 0.02 & 0.18 & 0.04 & 0.21 & 0.01 & 0.18 & 0.01 \\
WW Vul & 0948.54 & -21.2 & 0.2 & 749 & 6 & IIB & -0.05 & 0.03 & 0.85 & 0.06 & 0.50 & 0.02 & 0.32 & 0.02 \\
 & 0949.53 & -20.8 & 0.2 & 713 & 6 & IIB & -0.07 & 0.01 & 0.88 & 0.04 & 0.69 & 0.02 & 0.42 &0.02 \\
 & 0950.54 & -19.8 & 0.2 & 706 & 6 & IIB & -0.08 & 0.01 & 0.52 & 0.06 & 0.49 & 0.02 & 0.34 & 0.02 \\
 & 0951.62 & -20.0 & 0.2 & 769 & 8 & IIB & -0.07 & 0.02 & 0.76 & 0.05 & 0.75 & 0.02 & 0.48 & 0.02 \\
 & 1023.49 & -16.5 & 0.1 & 747 & 6 & IIB & -0.10 & 0.03 & 0.20 & 0.03 & 0.50 & 0.01 & 0.36 & 0.01 \\
 & 1024.50 & -17.9 & 0.4 & 720 & 6 & IIR & -0.06 & 0.03 & 0.66 & 0.05 & 0.51 & 0.01 & 0.39 & 0.01\\
 & 1025.46 & -17.6 & 0.8 & 707 & 9 & IIR & -0.12 & 0.03 & 0.64 & 0.05 & 0.51 & 0.02 & 0.38 & 0.01\\
 & 1026.45 & -19.2 & 0.4 & 742 & 9 & IIR & -0.09 & 0.03 & 0.71 & 0.05 & 0.39 & 0.01 & 0.29 & 0.01\\
 & 1111.32 & -17.8 & 0.7 & 843 & 9 & IIR & -0.10 & 0.03 & 0.97 & 0.04 & 1.23 & 0.01 & 1.05 & 0.02\\
 & 1112.33 & -17.4 & 0.3 & 721 & 5 & IIB & -0.11 & 0.02 & 0.58 & 0.04 & 0.91 & 0.01 & 0.76 & 0.03\\
 & 1113.32 & -22.8 & 0.3 & 728 & 6 & IIB & -0.13 & 0.02 & 0.96 & 0.03 & 0.79 & 0.01 & 0.73 & 0.02\\
 & 1114.33 & -17.0 & 0.1 & 812 & 6 & IIB & -0.12 & 0.02 & 0.90 & 0.02 & 0.75 & 0.01 & 0.57 & 0.01\\
 & 1115.33 & -19.5 & 0.2 & 711 & 8 & IIB & -0.09 & 0.03 & 0.32 & 0.04 & 0.50 & 0.01 & 0.40 & 0.02\\
LkHa 234 & 1023.50 & -70 & 1 & 749 & 7 & IIB & -0.60 & 0.08 & 0.58 & 0.06 & 1.15 & 0.05 & 1.00 & 0.04 \\
 & 1024.50 & -69 & 1 & 812 & 7 & IIB & -0.61 & 0.08 & 0.91 & 0.06 & 1.08 & 0.05 & 0.91 & 0.05 \\
 & 1025.50 & -61.4 & 0.9 & 762 & 5 & IIB & -0.39 & 0.05 & 0.82 & 0.07 &	2.14 & 0.07 & 1.60 & 0.06 \\ 
 & 1026.50 & -74 & 2 & 776 & 5 & IIB & -0.64 & 0.09 & 1.19 & 0.07 & 0.97 & 0.06 & 0.83 & 0.07 \\
 & 1111.39 & -64 & 1 & 804 & 5 & IIIB & -0.60 & 0.08 & 0.81 & 0.07 & -0.2 & 0.1 & -0.25 & 0.09 \\
 & 1112.44 & -71 & 2 & 745 & 4 & IIIB & -0.54 & 0.08 & 0.74 & 0.06 & -0.09 & 0.05 & -0.13 & 0.07 \\
 & 1113.44 & -75 & 2 & 722 & 8 & IIB & -0.58 & 0.08 & 0.83 & 0.08 & -0.5 & 0.1 & -0.23 & 0.09 \\
 & 1114.42 & -73 & 2 & 718 & 6 & IIB & -0.57 & 0.07 & 0.54 & 0.07 & -0.5 & 0.1 & -0.24 & 0.07 \\
 & 1115.43 & -72 & 2 & 773 & 5 & IIIB & -0.58 & 0.08 & 0.87 & 0.06 & -0.7 & 0.1 & -0.3 &   0.1 \\
\end{longtable}
\normalsize
\begin{minipage}{18cm}

\underline{Notes to Table \ref{Table:extract}}: Equivalent widths and their uncertainties (EW and $\delta$EW,
respectively) for the indicated spectroscopic lines. For the H$\alpha$
line, the profile shape according to the \citet{Reipurth96}
classification scheme, the width of the wings at 10$\%$ of peak
intensity and its uncertainty (W$_{10}$ and $\delta$W$_{10}$) are also
given. ``...'' indicates that the spectroscopic line is not
detected in any of the spectra of the star considered. When a line is measured only in several spectra of a given object, upper limits are shown for the non-detections.
\end{minipage}
\scriptsize
\renewcommand\arraystretch{1.3}
\renewcommand\tabcolsep{4pt}
\begin{longtable}{lp{1.5cm}p{1cm}p{1cm}p{1cm}p{1cm}cp{1cm}p{1cm}p{1cm}p{1cm}p{1cm}p{1cm}p{1cm}p{1cm}p{1cm}}
\caption{\label{Table:fluxes}Line fluxes on several observing Julian Dates for a subsample of the stars.}\\
\hline\hline
Star & JD  & F & $\delta$F & F & $\delta$F & F & $\delta$F & F & $\delta$F & F & $\delta$F \\ 
   &(+2450000)  & H$\alpha$ & H$\alpha$ & [\ion{O}{i}]6300 & [\ion{O}{i}]6300 & \ion{He}{i}5876 & \ion{He}{i}5876 & \ion{Na}{i}D$_{2}$ & \ion{Na}{i}D$_{2}$ & 
   \ion{Na}{i}D$_{1}$ & \ion{Na}{i}D$_{1}$ \\
 & & ($\times$10$^{-12}$) & ($\times$10$^{-12}$) & ($\times$10$^{-14}$) &  ($\times$10$^{-14}$) & ($\times$10$^{-14}$) & ($\times$10$^{-14}$) & ($\times$10$^{-14}$) & ($\times$10$^{-14}$) & ($\times$10$^{-14}$) & ($\times$10$^{-14}$) \\
\hline
\endfirsthead
\caption{continued.}\\
\hline\hline
Star & JD  & F & $\delta$F & F & $\delta$F & F & $\delta$F & F & $\delta$F & F & $\delta$F \\ 
   &(+2450000)  & H$\alpha$ & H$\alpha$ & [\ion{O}{i}]6300 & [\ion{O}{i}]6300 & \ion{He}{i}5876 & \ion{He}{i}5876 & \ion{Na}{i}D$_{2}$ & \ion{Na}{i}D$_{2}$ & 
   \ion{Na}{i}D$_{1}$ & \ion{Na}{i}D$_{1}$ \\
 & & ($\times$10$^{-12}$) & ($\times$10$^{-12}$) & ($\times$10$^{-14}$) &  ($\times$10$^{-14}$) & ($\times$10$^{-14}$) & ($\times$10$^{-14}$) & ($\times$10$^{-14}$) & ($\times$10$^{-14}$) & ($\times$10$^{-14}$) & ($\times$10$^{-14}$) \\
\hline
\endhead
\hline
\endfoot

HD 31648   &    1111.53&37    & 3   &   ...&  ...&    78 &   13&136  &14  &152  &17   \\
           &    1113.62&48    & 1   &   ...&  ...&    138&   17&191  &13  &174  &7    \\
           &	1114.60&38    & 1   &   ...&  ...&    59 &   21&132  &6   &128  &8    \\
HD 34282   &	1114.65&1.5   & 0.5 &   ...&  ...&\emph{5}&	2&2.3	&0.5 &\emph{1.3}   &0.5  \\
           &	1209.46&1.6   & 0.1 &   ...&  ...&\emph{7}&	2&3.9	&0.6 &\emph{2.7}   &0.6  \\
HD 34700   &	1114.65&1.4   & 0.2 &   ...&  ...&	...&  ...&\emph{5}	&1   &\emph{5}     &1    \\
HD 58647   &	1209.54&51    & 3   &   23&	 9&	63&   24&\emph{87}	&10  &\emph{73}    &10   \\
HD 141569  &	0949.49&21    & 2   &   46&	20&	...&  ...&\emph{45}	&11  &\emph{49}    &15   \\
  	   &	0950.42&21    & 2   &   50&	20&	...&  ...&\emph{45}	&12  &\emph{61}    &15   \\
  	   &	1024.40&22    & 2   &   46&	30&	...&  ...&\emph{39}	&5   &\emph{39}    &8    \\
  	   &	1209.73&21    & 1   &   49&	20&	...&  ...&\emph{31}	&4   &\emph{15}    &4    \\
HD 142666  &	0949.49&2.9   & 0.5 &   ...&  ...&	 \emph{13}&	5&\emph{16}	&3   &\emph{14}    &3    \\
           &    0950.58&3.3   & 0.5 &   ...&  ...&	 \emph{13}&	4&\emph{22}	&2   &\emph{17}    &4    \\
           &    1024.49&3.4   & 0.5 &   ...&  ...&	 \emph{18}&	5&\emph{18}	&2   &\emph{17}    &2    \\
HD 144432  &	0949.52&17.5  & 0.6 &     0&	 3&	27&   10&40   &7   &35   &7    \\
           &    0950.59&21.3  & 0.9 &     0&	 3&	55&   15&61   &9   &46   &9    \\
           &    1024.40&18    & 2   &    4&	 3&	21&	9&11   &4   &41   &12   \\
HD 150193  &	0949.53&12.2  & 0.4 &   ...&  ...&	22&	7&24   &6   &19   &6    \\
           &    0950.59&13    & 1   &   ...&  ...&	16&	6&32   &5   &24   &6    \\
HD 163296  &	0949.59&111   & 5   &     0&	 9&	\emph{173}&   24&206  &20  &235  &25   \\
           &    0950.62&102   & 4   &     0&	 9&	\emph{404}&   37&139  &20  &139  &25   \\
           &    1024.53&103   & 4   &   22&	 9&	\emph{266}&   29&93   &15  &89   &15   \\
HD 179218  &	0949.64&36    & 2   &    8&	 3&	 \emph{15}&	3&11   &3   &6    &3    \\
HD 190073  &	1024.55&51    & 2   &   12&	 2&	97&   14&207  &14  &181  &15   \\
VX Cas	   &	1024.72&0.76  & 0.06&  0.9&  0.1&	\emph{0.7}&  0.2&\emph{1.2}	&0.1 &\emph{1.2}   &0.2  \\
           &    1111.52&1.22  & 0.05&  0.8&  0.3&	\emph{2.2}&  0.5&\emph{2.1}	&0.2 &\emph{1.8}   &0.1  \\
     	   &	1112.54&1.21  & 0.08&  0.9&  0.2&	\emph{2.4}&  0.4&\emph{2.0}	&0.2 &\emph{1.8}   &0.3  \\
     	   &	1113.52&1.16  & 0.07&  0.6&  0.2&	\emph{6.6}&  0.6&\emph{2.4}	&0.2 &\emph{2.3}   &0.2  \\
     	   &	1114.51&1.12  & 0.04&  0.9&  0.2&	\emph{4.7}&  0.4&\emph{2.6}	&0.2 &\emph{2.2}   &0.2  \\
     	   &	1209.36&0.90  & 0.04& 0.92&  0.1&	\emph{5.6}&  0.6&\emph{4.2}	&0.2 &\emph{3.8}   &0.3  \\
BH Cep	   &	1024.65&0.48  & 0.07& 0.17& 0.09&     \emph{3.9}&  0.8&\emph{4.4}   &0.5 &\emph{5}	        &2    \\
     	   &	1111.41&0.57  & 0.04&  0.3&  0.2&     \emph{0.9}&  0.3&\emph{1.1}   &0.1 &\emph{0.9}	&0.1  \\
     	   &	1112.36&0.46  & 0.07&  0.3&  0.1&     \emph{3.2}&  0.5&\emph{3.0}   &0.3 &\emph{2.3}	&0.4  \\
     	   &	1112.46&0.48  & 0.07& 0.23& 0.08&     \emph{3.7}&  0.6&\emph{3.5}   &0.3 &\emph{2.6}	&0.5  \\
     	   &	1113.36&0.50  & 0.06& 0.3 &  0.1&     \emph{1.5}&  0.3&\emph{2.7}   &0.2 &\emph{2.1}	&0.2  \\
     	   &	1113.49&0.46  & 0.06& 0.2 &  0.1&     \emph{1.5}&  0.6&\emph{2.3}   &0.2 &\emph{1.9}	&0.2  \\
     	   &	1114.37&0.70  & 0.05&  0.8&  0.2&     \emph{0.9}&  0.3&\emph{2.2}   &0.2 &\emph{1.8}    &0.2  \\
BO Cep	   &	1024.67&0.38  & 0.04&  0.8&  0.5&   \emph{1.4}&  0.6&\emph{2.7	  }&0.4 &\emph{2.1  } &0.6  \\
     	   &	1111.42&0.68  & 0.04&  1.5&  0.6&   \emph{0.8}&  0.4&\emph{1.3	  }&0.1 &\emph{1.2  } &0.1  \\
     	   &	1112.46&0.69  & 0.07&  1.4&  0.7&   \emph{3.6}&  0.6&\emph{1.3	  }&0.2 &\emph{1.0  } &0.1  \\
     	   &	1113.45&0.65  & 0.06&  1.5&  0.6&   \emph{2.4}&  0.4&\emph{0.7	  }&0.1 &\emph{0.7  } &0.1  \\
     	   &	1114.43&0.50  & 0.03&  1.6&  0.6&   \emph{3.8}&  0.5&\emph{1.3	  }&0.1 &\emph{1.1  } &0.1  \\
SV Cep	   &	1024.68&1.4   & 0.1 &  1.7&  0.5&   \emph{2.3}&  0.6&\emph{6.1	  }&0.5 &\emph{4.7   }&0.4  \\
     	   &	1111.44&1.70  & 0.06&  1.8&  0.7&   \emph{5.7}&  0.8&\emph{2.4	  }&0.2 &\emph{2.1   }&0.2  \\
     	   &	1112.49&1.5   & 0.1 &  1.6&  0.5&   \emph{  9}&    1&\emph{2.3	  }&0.3 &\emph{2.2   }&0.3  \\
     	   &	1113.47&1.60  & 0.09&  1.4&  0.6&   \emph{5.8}&  0.8&\emph{2.2	  }&0.2 &\emph{1.9   }&0.2  \\
     	   &	1114.45&1.44  & 0.05&  1.4&  0.5&   \emph{6.3}&  0.7&\emph{4.3	  }&0.2 &\emph{3.3   }&0.2  \\
V1686 Cyg  &	0949.68&0.47  & 0.01&  0.9&  0.2&      \emph{0.06}& 0.04&\emph{1.2	}&0.1 &\emph{0.60  }&0.05 \\
  	   &	0950.68&0.47  & 0.01&  1.0&  0.2&	  0&  0.1&\emph{0.3	}&0.1 &\emph{0.12  }&0.02 \\
  	   &	1024.57&0.60  & 0.02&  0.5&  0.2&	\emph{0.4}&  0.1&\emph{1.2	}&0.1 &\emph{0.49  }&0.06 \\
  	   &	1111.36&0.98  & 0.02&  0.7&  0.3&       0.2&  0.1&\emph{1.9	}&0.1 &\emph{0.76  }&0.09 \\
  	   &	1113.38&0.36  & 0.03&  0.5&  0.2&	\emph{0.7}&  0.1&\emph{4.5	}&0.2 &\emph{2.8   }&0.1  \\
  	   &	1114.39&0.34  & 0.01&  0.5&  0.3&	\emph{0.9}&  0.1&\emph{4.7	}&0.2 &\emph{2.9   }&0.1  \\
VY Mon	   &	1209.63&1.07  & 0.03&  5.1&  0.5&	...&  ...&\emph{2.0}	&0.2 &\emph{1.3}   &0.1  \\
51 Oph	   &	0949.00&92    & 10  &   ...&  ...&    249&  127&0	&96  &249  &127  \\
           &    0950.00&96    & 10  &   ...&  ...&    343&  190&218  &96  &436  &131  \\
           &    1024.00&95    & 10  &   ...&  ...&    367&  157&92   &32  &306  &67   \\
KK Oph	   &	0949.60&3.6   & 0.2 & 10.7&  0.7&	\emph{2.5}&  0.4&\emph{0.8}	&0.2 &\emph{1.0}   &0.3  \\
           &    0950.61&3.2   & 0.1 & 13  &    1&    0.8&  0.2&1.1	&0.3 &0.5  &0.1  \\
           &    1024.44&3.3   & 0.2 & 11.2&  0.9&	\emph{1.6}&  0.3&\emph{0.7}	&0.1 &\emph{0.8}   &0.2  \\
T Ori	   &	1111.72&4.7   & 0.2 &  2.9&  0.6&   \emph{9.2}&    1&\emph{11.1    }&0.4 &\emph{7.9   }&0.5  \\
      	   &	1112.73&5.4   & 0.3 &  3.5&  0.6&   \emph{6.3}&  0.7&\emph{6.6	  }&0.5 &\emph{4.4   }&0.4  \\
      	   &	1114.72&4.1   & 0.2 &  2.0&  0.4&   \emph{6.7}&  0.8&\emph{6.9	  }&0.4 &\emph{4.7   }&0.4  \\
BF Ori	   &	1111.71&3.2   & 0.2 &  1.8&  0.7&   \emph{42}&	 3&\emph{8.8   }&0.8 &\emph{21    }&2    \\
     	   &	1112.64&3.4   & 0.3 &  1.8&  0.4&   \emph{19}&	 2&\emph{27    }&2   &\emph{24    }&2    \\
     	   &	1113.69&3.4   & 0.2 &  2.2&  0.6&   \emph{14}&	 2&\emph{18.7  }&0.9 &\emph{15.0  }&0.6  \\
     	   &	1114.67&3.6   & 0.3 &  1.3&  0.3&   \emph{16}&   1&\emph{34    }&1   &\emph{26.6  }&0.8  \\
CO Ori	   &	1111.57&2.21  & 0.08&  1.7&  0.4&	  0&  0.4&\emph{5.1	}&0.3 &\emph{3.6   }&0.3  \\
     	   &	1112.57&2.3   & 0.1 &  2.9&  0.5&	  0&  0.4&\emph{1.4	}&0.2 &\emph{1.1   }&0.1  \\
     	   &	1113.56&2.3   & 0.1 &  3.4&  0.6&	  0&  0.4&\emph{0.5	}&0.1 &\emph{0.6   }&0.1  \\
     	   &	1114.54&2.0   & 0.1 &  2.8&  0.4&      \emph{1.2} &  0.4&\emph{0.9	}&0.1 &\emph{1.1   }&0.1  \\
     	   &	1209.49&2.11  & 0.06&  2.9&  0.6&      0.0 &  0.4&\emph{1.8	}&0.2 &\emph{1.7   }&0.2  \\
HK Ori	   &	1111.68&4.61  & 0.08& 10.3&  0.8&    \emph{3.0} &  0.3&\emph{3.0}	&0.2 &\emph{2.9}   &0.3  \\
     	   &	1112.74&4.6   & 0.1 &   11&    1&    \emph{3.2} &  0.3&\emph{2.6}	&0.3 &\emph{2.4}   &0.4  \\
     	   &	1113.66&1.6   & 0.2 &  3.8&  0.8&    0.0 &  0.4&2.0	&0.2 &1.1  &0.3  \\
     	   &	1114.71&4.38  & 0.07& 10.1&  0.8&    \emph{2.7} &  0.3&\emph{1.8}	&0.2 &\emph{2.1}   &0.2  \\
     	   &	1209.50&5.19  & 0.09& 12.1&  0.9&    \emph{3.4} &  0.4&\emph{1.6}	&0.2 &\emph{1.5}   &0.3  \\
NV Ori	   &	1111.77&1.6   & 0.2 &     0&  0.3&    \emph{6} &    2&9     &1	&6.3  &0.7  \\
     	   &	1112.72&1.4   & 0.3 &     0&  0.3&    0 &    2&8     &1	&7    &1    \\
     	   &	1113.76&1.4   & 0.2 &     0&  0.3&    \emph{4} &    2&6     &1	&9    &2    \\
     	   &	1114.74&1.4   & 0.1 &     0&  0.3&    \emph{9} &    2&6     &1	&5    &1    \\
     	   &	1209.53&1.4   & 0.1 &  0.9&  0.3&     \emph{6} &    1&6.7   &0.6 &7.5  &0.9  \\
RY Ori	   &	1112.76&0.91  & 0.05&  0.8&  0.3&     \emph{2} &  0.3&\emph{3.1     }&0.2 &\emph{3.4   }&0.3  \\
     	   &	1113.74&0.79  & 0.04&  0.3&  0.2&     \emph{1} &  0.3&\emph{2.8     }&0.2 &\emph{2.7   }&0.2  \\
     	   &	1114.68&0.68  & 0.03&  0.4&  0.2&     \emph{2} &  0.4&\emph{4.0     }&0.2 &\emph{4.1   }&0.3  \\
     	   &	1209.52&0.76  & 0.03&  0.8&  0.2&     \emph{2} &  0.4&\emph{5.7     }&0.2 &\emph{5.8   }&0.3  \\
UX Ori	   &	1111.61&3.7   & 0.1 &  1.7&  0.6&	 4 &	1&\emph{3.9}	&0.8 &\emph{7}     &1    \\
     	   &	1112.60&3.8   & 0.3 &  1.5&  0.6&   \emph{4} &	 1&8	 &2   &0.6   &0.3  \\
     	   &	1113.61&3.6   & 0.2 &    2&    1&   \emph{7} &	 1&9	 &2   &5     &1    \\
     	   &	1114.58&3.5   & 0.2 &  1.8&  0.6&   \emph{6} &	 1&\emph{6.4}   &0.3 &\emph{4.3}   &0.6  \\
     	   &	1209.45&2.1   & 0.1 &  1.9&  0.6&   \emph{17}&	 2&5	 &1   &4     &1    \\
V346 Ori   &	1111.66&0.8   & 0.2 &   ...&  ...&    ... &  ...&2.9	&0.9 &1.8  &0.4  \\
   	   &	1114.69&0.9   & 0.2 &   ...&  ...&    ... &  ...&\emph{2.3}	&0.5 &\emph{1.8}   &0.7  \\
   	   &	1209.48&0.6   & 0.1 &   ...&  ...&    ... &  ...&\emph{5.3}	&0.5 &\emph{3.9}   &0.7  \\
V350 Ori   &	1113.77&0.72  & 0.05&  0.7&  0.2&  \emph{2.1} &  0.4&\emph{3.5}      &0.2 &\emph{2.7}   &0.3  \\
           &	1114.76&0.77  & 0.05&  0.7&  0.2&  \emph{2.1} &  0.5&\emph{3.4}      &0.2 &\emph{2.7}   &0.2  \\
XY Per	   &	1024.73&5.8   & 0.4 &     0&	 2&  \emph{21}&    4&\emph{35	  }&3	&\emph{28    }&2    \\
           &    1112.65&7.2   & 0.8 &    4&	 2&  \emph{19}&    3&\emph{31	  }&2	&\emph{24    }&2    \\
           &    1113.57&8.0   & 0.9 &    6&	 2&  \emph{19}&    2&\emph{36	  }&1	&\emph{28    }&1    \\
           &    1209.38&4.0   & 0.3 &  1.9&  0.5&    \emph{6} &    1&\emph{23.4    }&0.9 &\emph{20.1  }&0.9  \\
VV Ser	   &	0949.56&3.87  & 0.07&  4.6&  0.8&  \emph{3.5 }&  0.5&\emph{3.0	    }&0.2 &\emph{2.2   }&0.2  \\
     	   &	0950.62&3.89  & 0.07&  3.6&  0.6&  \emph{2.5 }&  0.9&\emph{3.7	    }&0.3 &\emph{3.6   }&0.5  \\
     	   &	1024.48&3.73  & 0.07&  3.6&  0.6&  \emph{2.6 }&  0.5&\emph{4.4	    }&0.4 &\emph{4.1   }&0.5  \\
     	   &	1111.30&4.3   & 0.3 &  5.4&  0.9&  \emph{3.9 }&  0.5&\emph{5.0	    }&0.2 &\emph{4.1   }&0.4  \\
     	   &	1112.31&3.8   & 0.2 &  3.8&  0.8&  \emph{2.1 }&  0.5&\emph{3.1	    }&0.3 &\emph{2.4   }&0.2  \\
     	   &	1113.30&3.7   & 0.2 &  4.3&  0.7&  \emph{2.1 }&  0.5&\emph{3.2	    }&0.2 &\emph{2.5   }&0.2  \\
     	   &	1114.31&3.6   & 0.2 &  5.4&  0.8&  \emph{3.2 }&  0.5&\emph{3.7	    }&0.2 &\emph{2.9   }&0.2  \\
CQ Tau	   &	1111.74&1.7   & 0.2 &    3&    1&     \emph{10}&    3&\emph{23	   }&2   &\emph{17    }&1    \\
     	   &	1112.76&2.8   & 0.4 &    2&    1&	  0&	4&\emph{23	}&2   &\emph{17    }&2    \\
     	   &	1113.75&2.9   & 0.3 &  2.4&  0.8&	  0&	4&\emph{20 	}&2   &\emph{15    }&1    \\
     	   &	1114.73&2.6   & 0.2 &  2.1&  0.7&      \emph{7}&    4&\emph{20	   }&1   &\emph{15    }&1    \\
     	   &	1209.60&2.3   & 0.1 &  2.6&  0.8&	\emph{8}&    2&\emph{14.5    }&0.7 &\emph{14    }&1    \\
RR Tau	   &	1111.73&3.51  & 0.08&  5.0&  0.7& \emph{1.8}&  0.3&\emph{6.5	    }&0.3 &\emph{5.9   }&0.4  \\
     	   &	1112.77&3.53  & 0.09&  4.8&  0.6& \emph{4.8}&  0.4&\emph{4.4	    }&0.3 &\emph{4.2   }&0.4  \\
     	   &	1113.78&3.4   & 0.1 &  5.8&  0.8& \emph{4.4}&  0.5&\emph{11.1      }&0.5 &\emph{10.2  }&0.6  \\
     	   &	1114.75&3.32  & 0.05&  4.4&  0.7& \emph{7.2}&  0.5&\emph{9.9	    }&0.4 &\emph{8.5   }&0.5  \\
     	   &	1209.61&2.19  & 0.08&  4.0&  0.4& \emph{4.8}&  0.6&\emph{8.1	    }&0.4 &\emph{6.8   }&0.8  \\
RY Tau	   &	1111.62&3.9   & 0.1 &   20&    3& \emph{4} &  1&\emph{8.2}&0.7 &\emph{6.7}&0.8  \\
     	   &	1112.59&3.5   & 0.2 &   20&    3& \emph{5} &  1&\emph{8.2}&0.8 &\emph{6.8}&0.8  \\
     	   &	1113.60&3.8   & 0.2 &   21&    3&	 3&   1&3.1	&0.5 &3.2  &0.5  \\
     	   &	1114.59&4.5   & 0.2 &   19&    3&	 4&   1&5	&1   &6.0  &0.8  \\
     	   &	1209.41&4.10  & 0.07&   21&    3&    2.9&  0.6&5.6	&0.7 &5.2  &0.7  \\
PX Vul	   &	1024.51&1.33  & 0.05&  0.5&  0.2&\emph{    1.5 }&  0.3&\emph{1.5	}&0.2 &\emph{1.4   }&0.1  \\
     	   &	1111.34&1.3   & 0.1 &  1.1&  0.3&\emph{    0.6 }&  0.1&\emph{0.9	}&0.1 &\emph{0.9   }&0.1  \\
     	   &	1112.34&1.2   & 0.1 &  0.6&  0.2&\emph{    0.9 }&  0.2&\emph{0.9	}&0.1 &\emph{0.9   }&0.1  \\
     	   &	1113.34&1.34  & 0.06&  1.0&  0.3&\emph{    1.8 }&  0.6&\emph{1.2	}&0.1 &\emph{1.1   }&0.1  \\
     	   &	1114.35&1.4   & 0.1 &  0.8&  0.3&\emph{    0.3 }&  0.1&\emph{0.8	}&0.1 &\emph{0.8   }&0.1  \\
WW Vul	   &	0949.53&2.88  & 0.07&  1.0&  0.1&\emph{    10.5}&  0.6&\emph{8.2	}&0.4 &\emph{5.0   }&0.3  \\
     	   &	0950.54&2.59  & 0.08&  1.1&  0.1&\emph{    6.1 }&  0.8&\emph{5.8	}&0.4 &\emph{4.0   }&0.3  \\
     	   &	1111.32&2.8   & 0.1 &  1.6&  0.5&\emph{    14.3}&  0.7&\emph{18.2	}&0.5 &\emph{15.5  }&0.5  \\
     	   &	1112.33&2.5   & 0.1 &  1.6&  0.3&\emph{    7.6 }&  0.7&\emph{11.9	}&0.8 &\emph{10.0  }&0.8  \\
     	   &	1113.32&2.65  & 0.07&  1.5&  0.2&\emph{    9.9 }&  0.4&\emph{8.2	}&0.2 &\emph{7.5   }&0.3  \\
     	   &	1114.33&2.44  & 0.04&  1.7&  0.3&\emph{    11.8}&  0.4&\emph{9.9	}&0.3 &\emph{7.5   }&0.2  \\
LkHa 234   &	1111.39&2.33  & 0.04&  2.2&  0.3&\emph{    1.7 }&  0.2&0.4	&0.2 &0.5  &0.2  \\
   	   &	1112.44&2.42  & 0.08&  1.8&  0.3&\emph{    1.5 }&  0.2&0.2	&0.1 &0.3  &0.1  \\
   	   &	1113.44&2.23  & 0.06&  1.7&  0.3&\emph{    1.4 }&  0.1&0.9	&0.2 &0.4  &0.2  \\
   	   &	1114.42&2.41  & 0.07&  1.9&  0.2&\emph{    1.1 }&  0.1&1.0	&0.2 &0.5  &0.1  \\

\end{longtable}
\normalsize
\begin{minipage}{18cm}

  \underline{Notes to Table \ref{Table:fluxes}}: Line fluxes and their
  uncertainties (F and $\delta$F, respectively, in erg cm$^{-2}$
  s$^{-1}$) derived from the equivalent widths in Table
  \ref{Table:extract} and the simultaneous $V$ and $R$ magnitudes from
  \citet{Oudmaijer01}. Flux uncertainties come from the
propagation of the individual errors in the EWs and the
magnitudes. Italic numbers refer to lines seen in absorption. ``...'' indicates that the line is not detected
  in any of the spectra of the corresponding star for which the fluxes
  can be derived. When a line is measured only in several spectra of a given object, upper limits are shown for the non-detections.
\end{minipage}
\end{appendix}
\newpage
\begin{appendix}
\section{Mean spectra and relative variability distributions}
\label{appendixonline2}
\onecolumn
\begin{table}
\centering
\renewcommand\arraystretch{10}
\begin{tabular}{cc}
\includegraphics[height=47mm,clip=true]{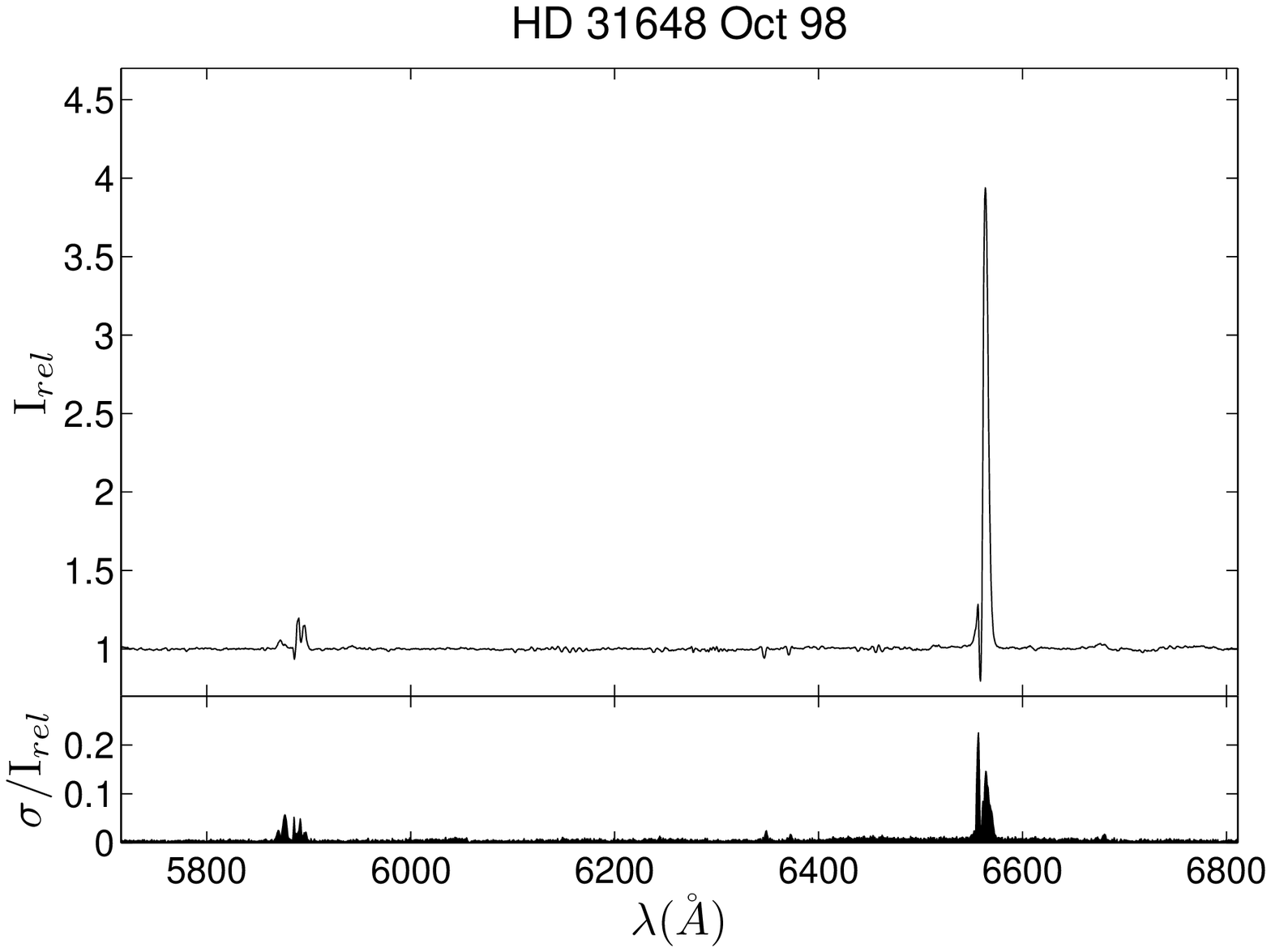}& 
\includegraphics[bb=4 77 763 470,height=45mm,clip=true]{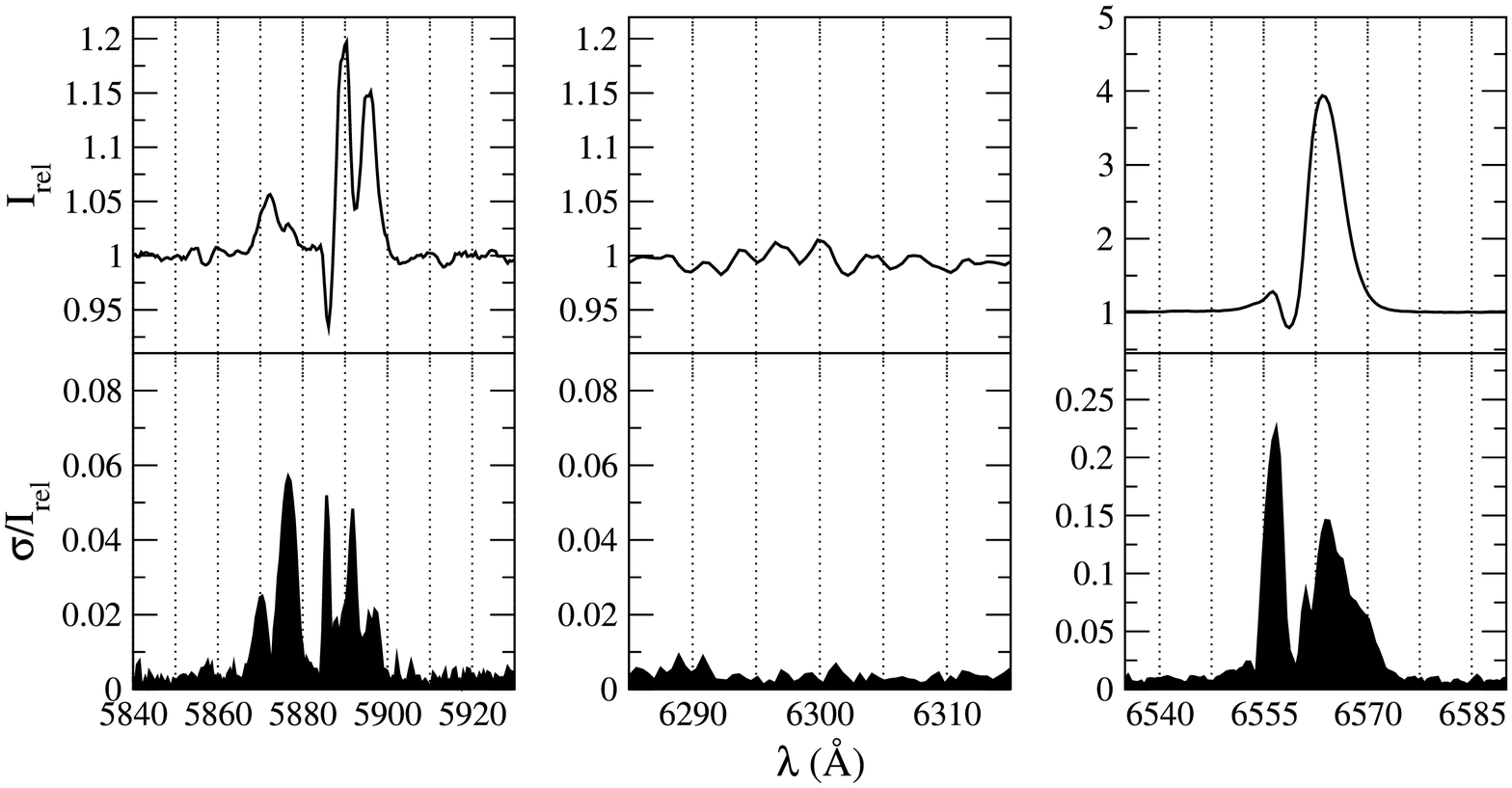} \\
\includegraphics[height=47mm,clip=true]{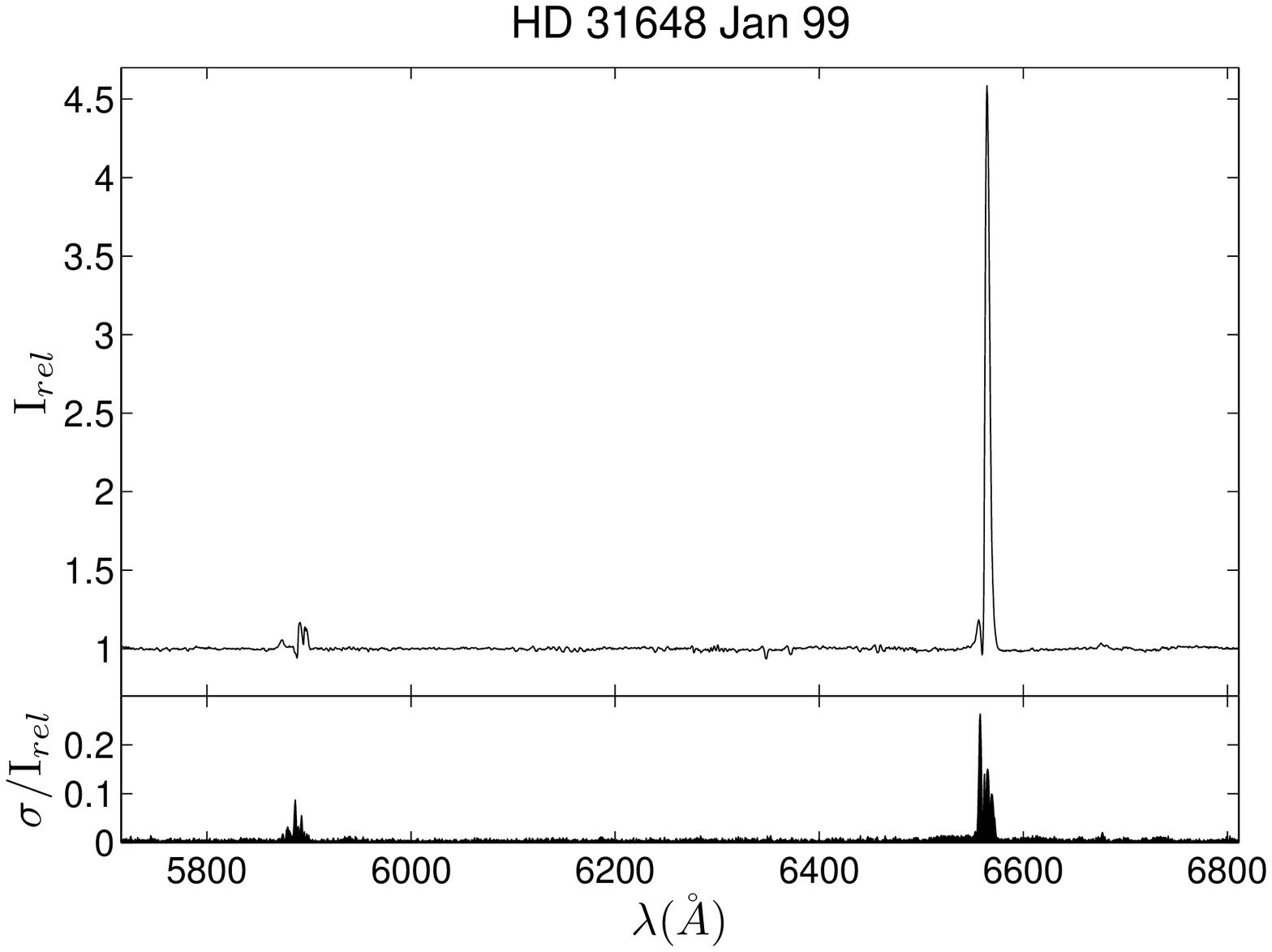}& 
\includegraphics[bb=4 77 763 470,height=45mm,clip=true]{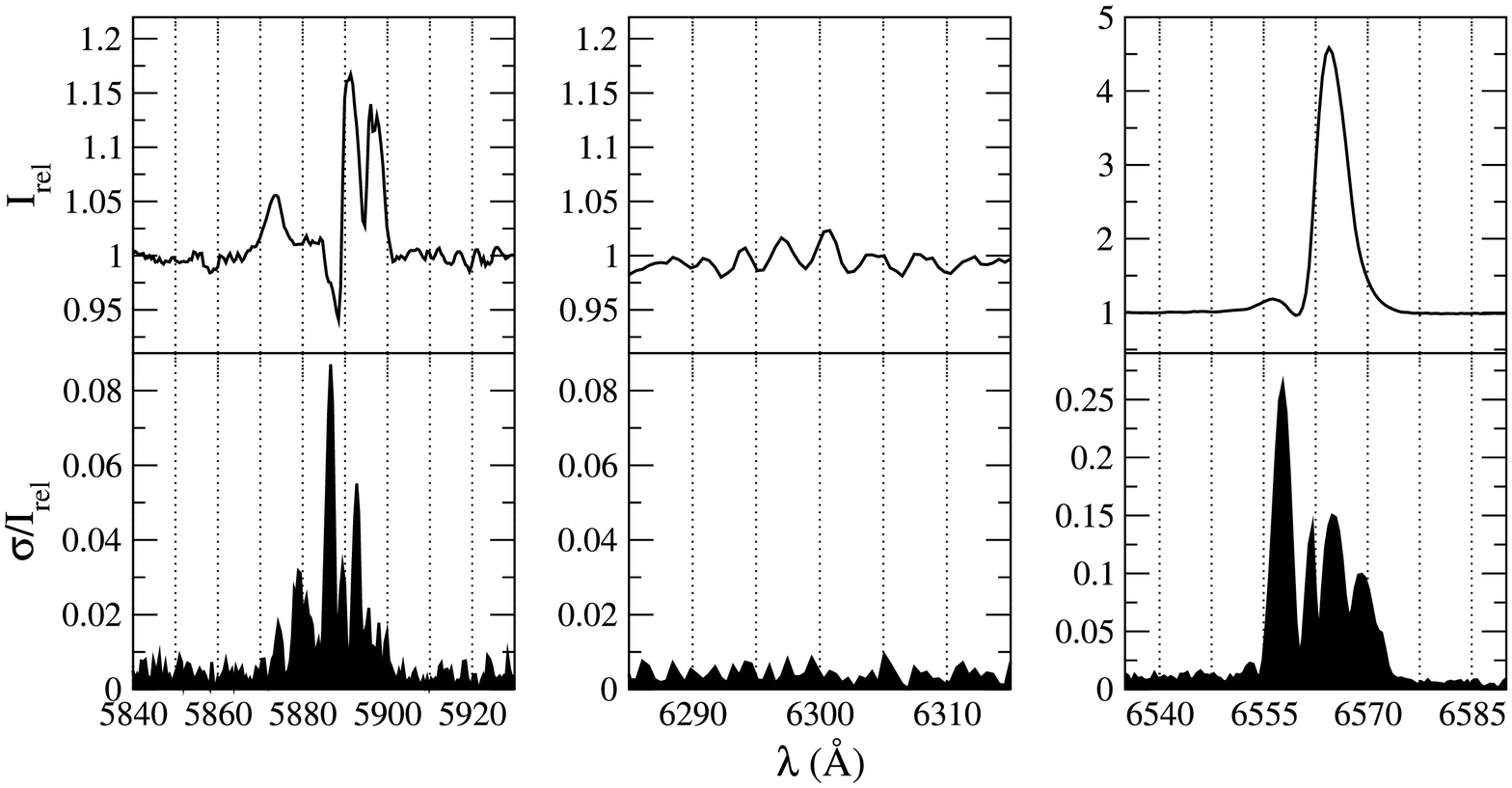} \\
\includegraphics[height=47mm,clip=true]{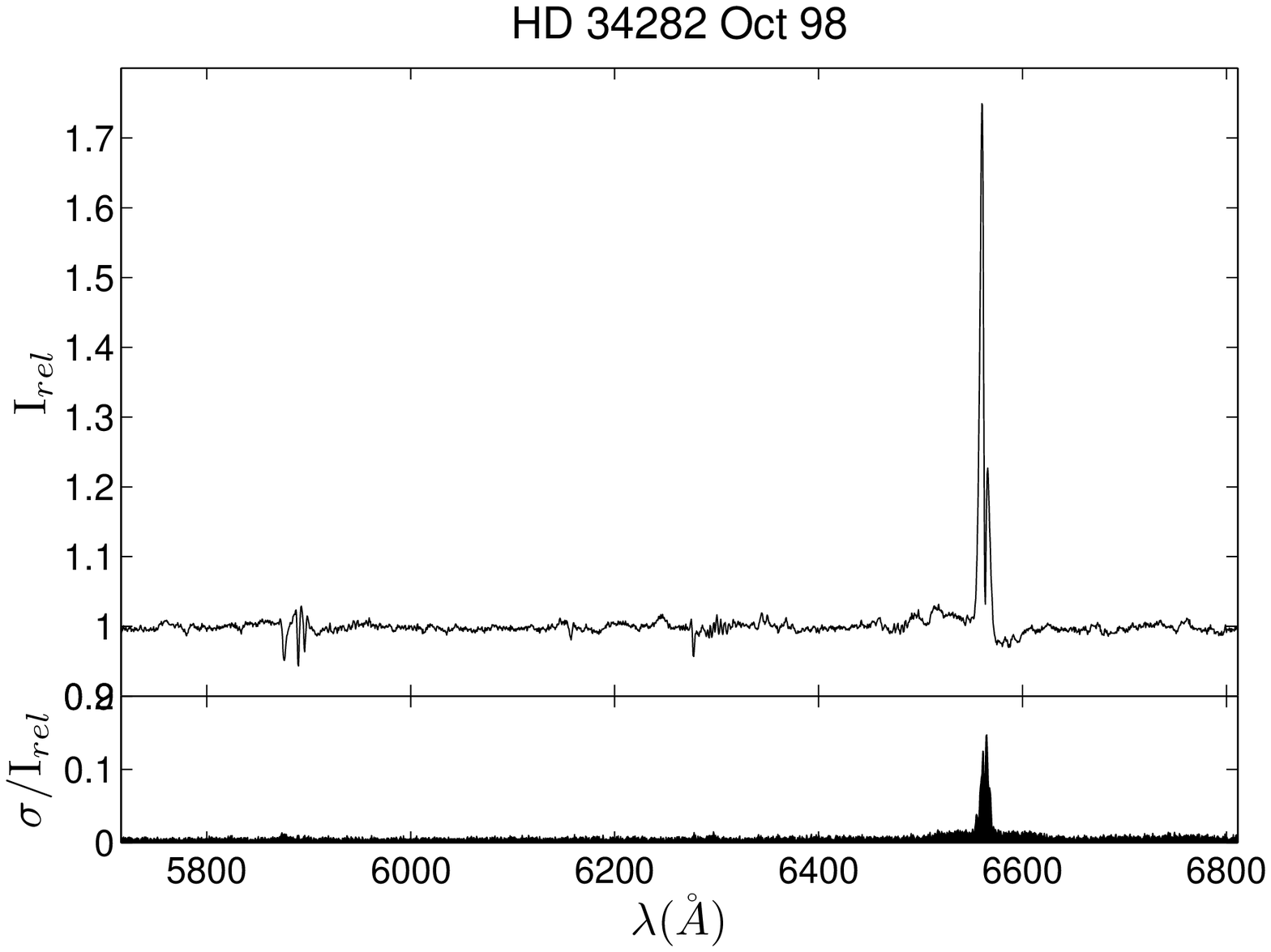}& 
\includegraphics[bb=4 77 763 470,height=45mm,clip=true]{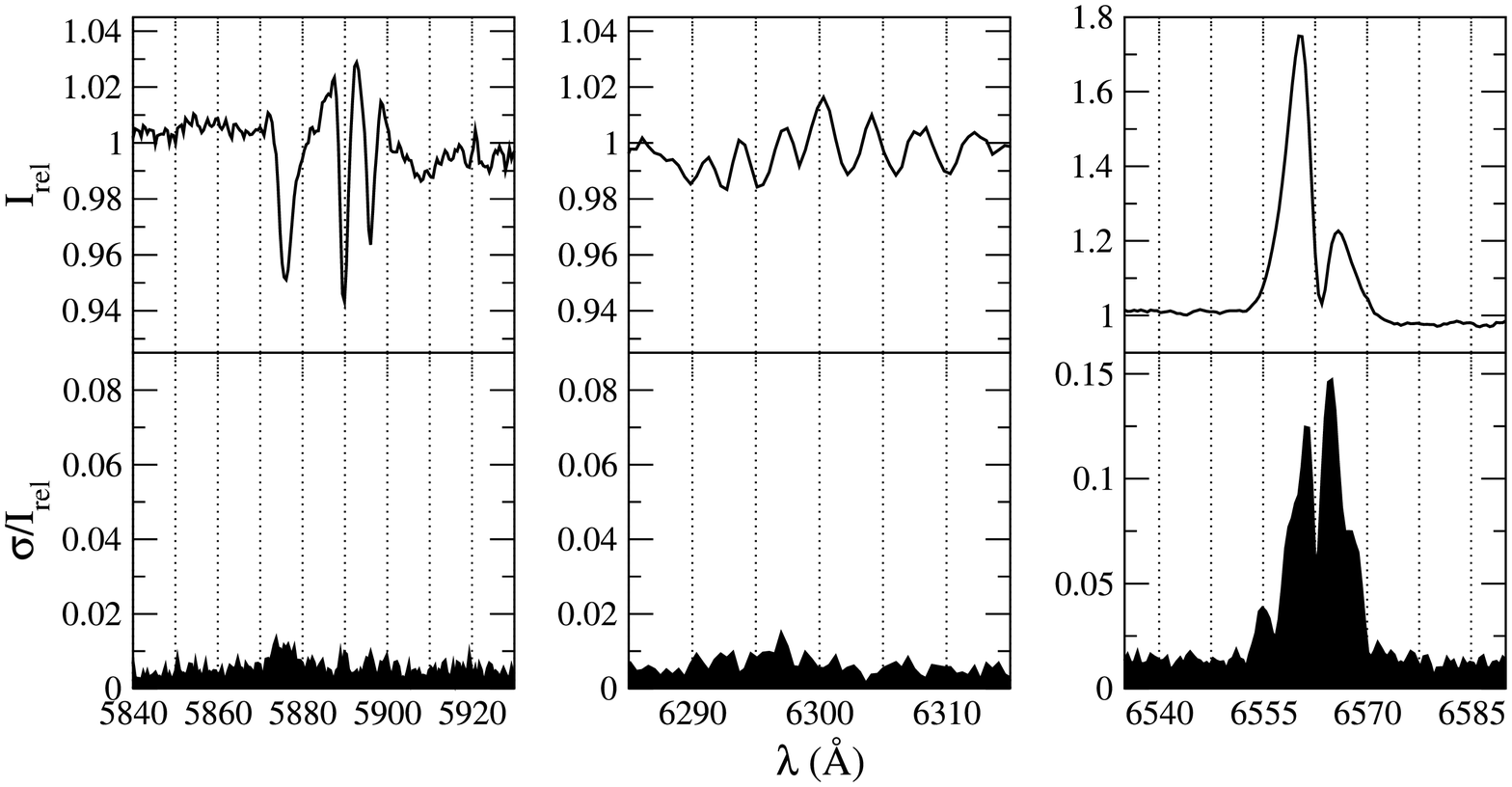} \\
\includegraphics[height=47mm,clip=true]{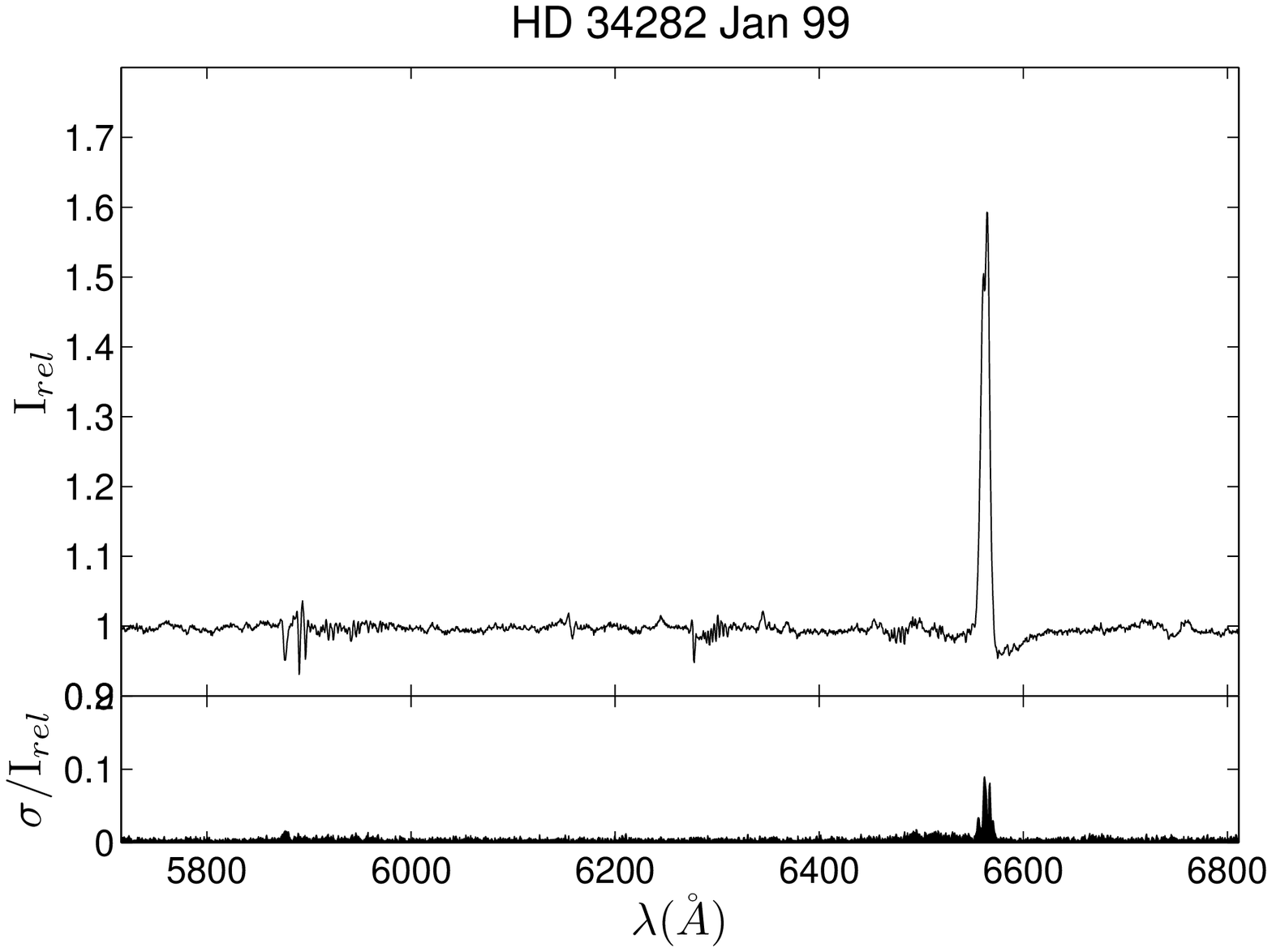}& 
\includegraphics[bb=4 77 763 470,height=45mm,clip=true]{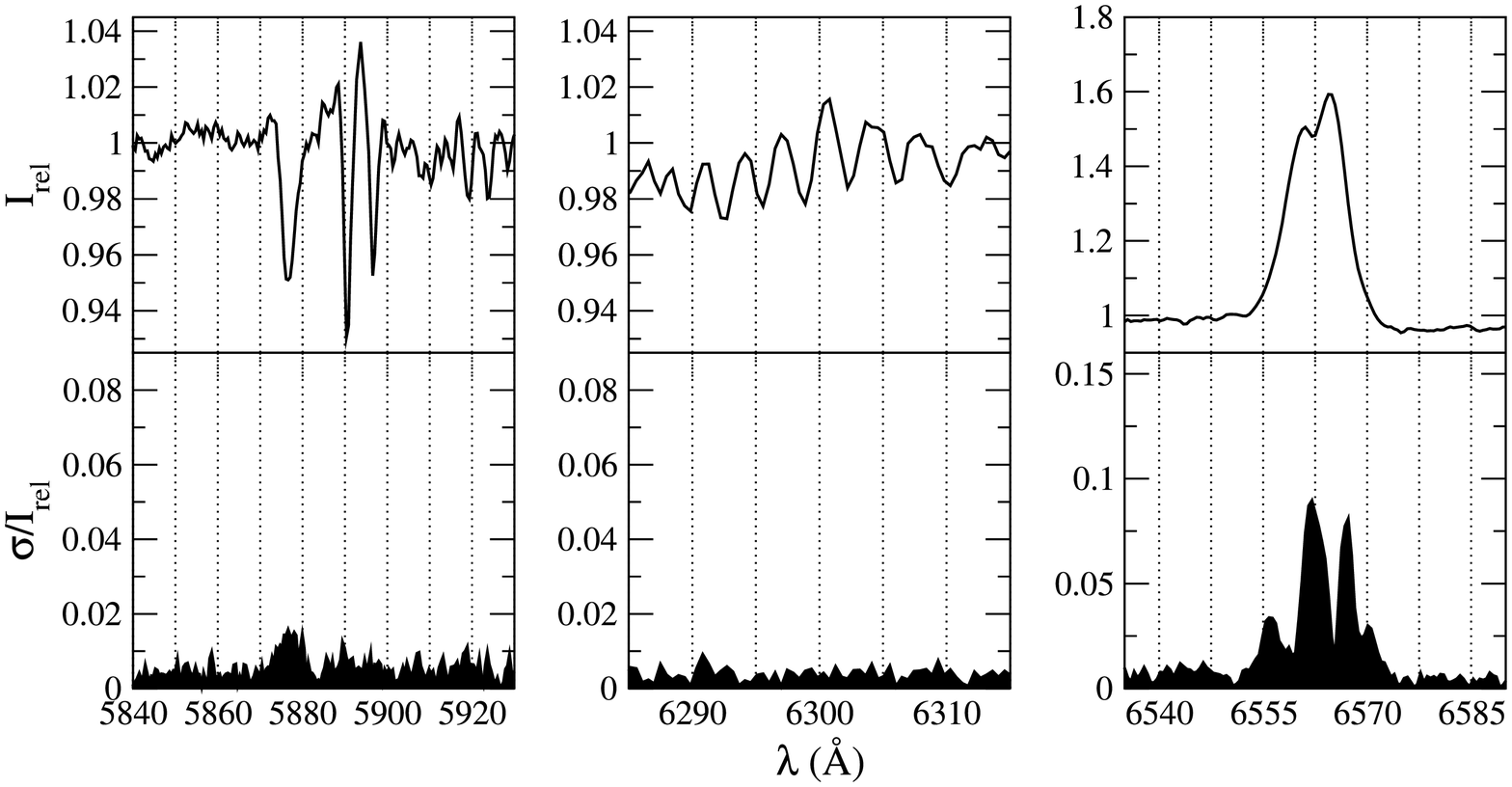} \\
\end{tabular}
\end{table}
\clearpage
\begin{table}
\centering
\renewcommand\arraystretch{10}
\begin{tabular}{cc}
\includegraphics[height=47mm,clip=true]{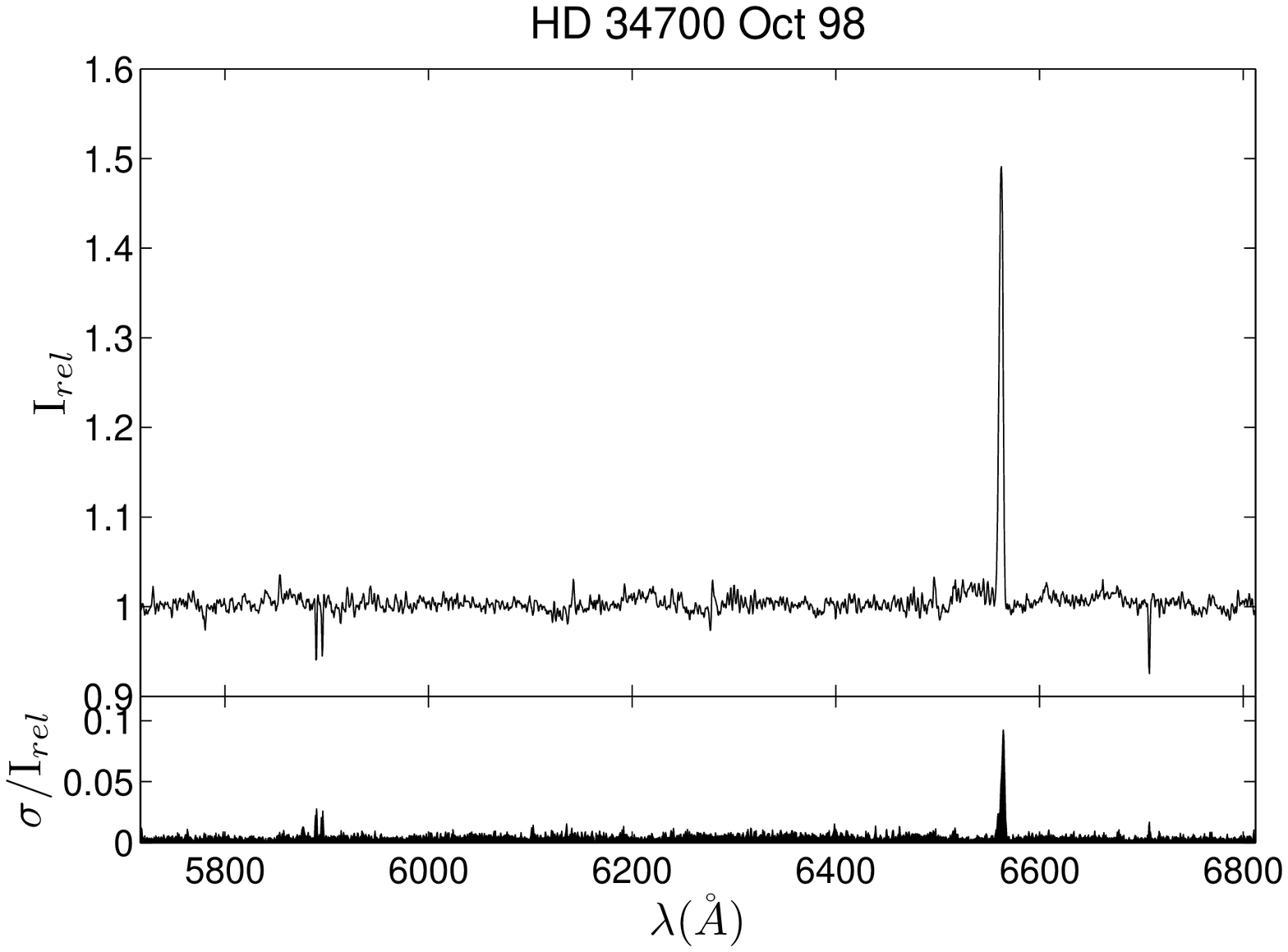}& 
\includegraphics[bb=4 77 763 470,height=45mm,clip=true]{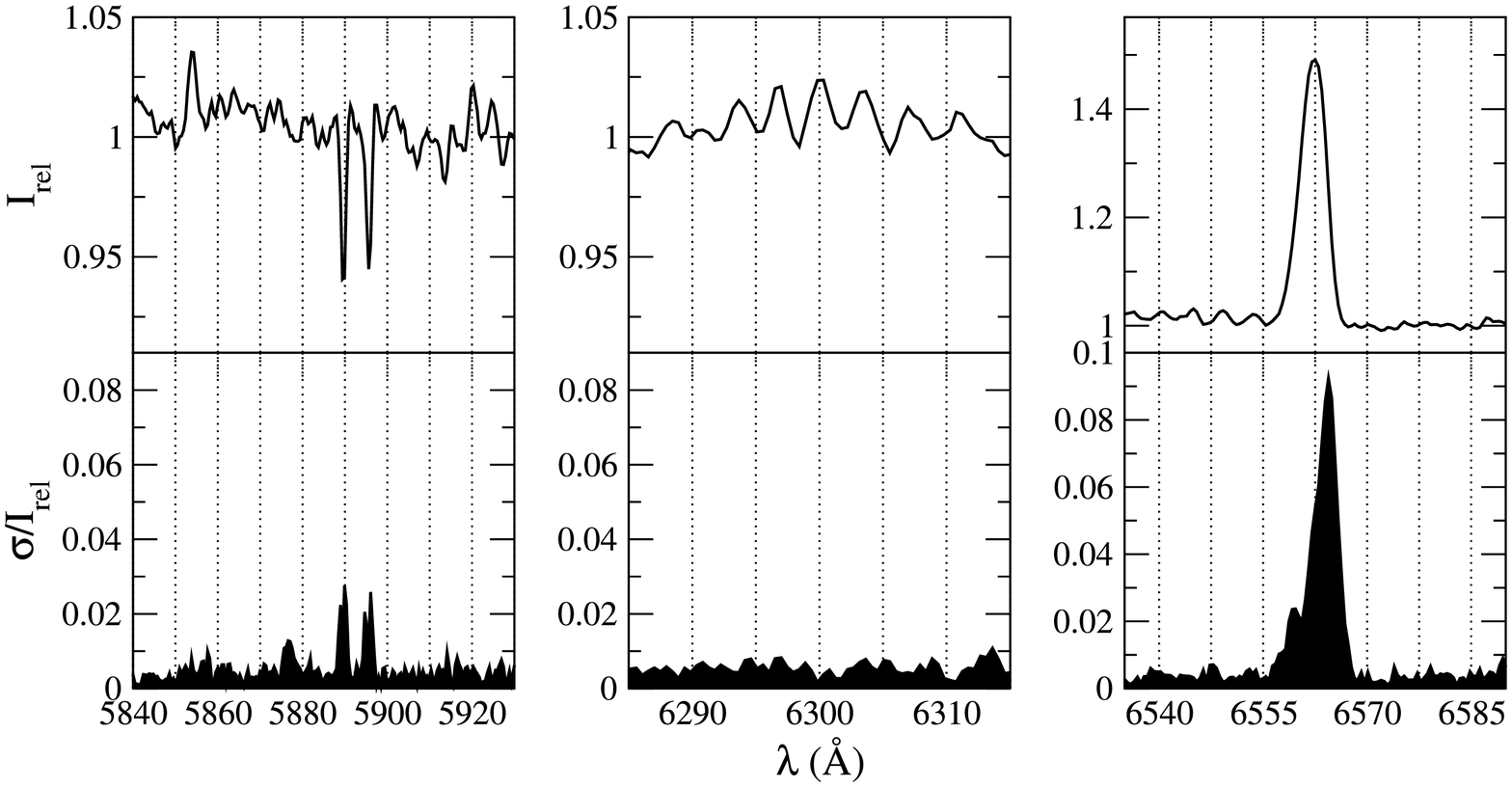} \\
\includegraphics[height=47mm,clip=true]{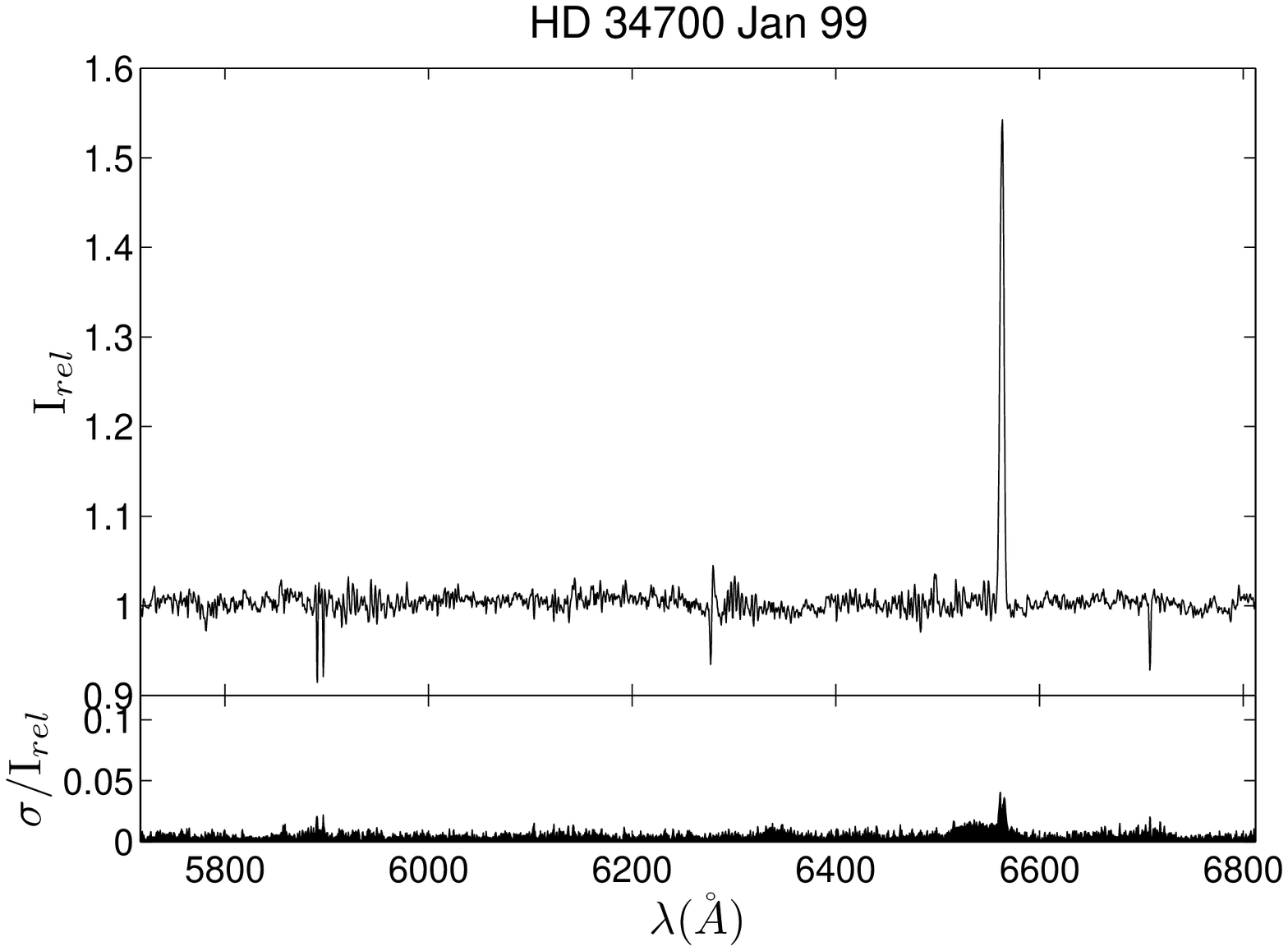}&
\includegraphics[bb=4 77 763 470,height=45mm,clip=true]{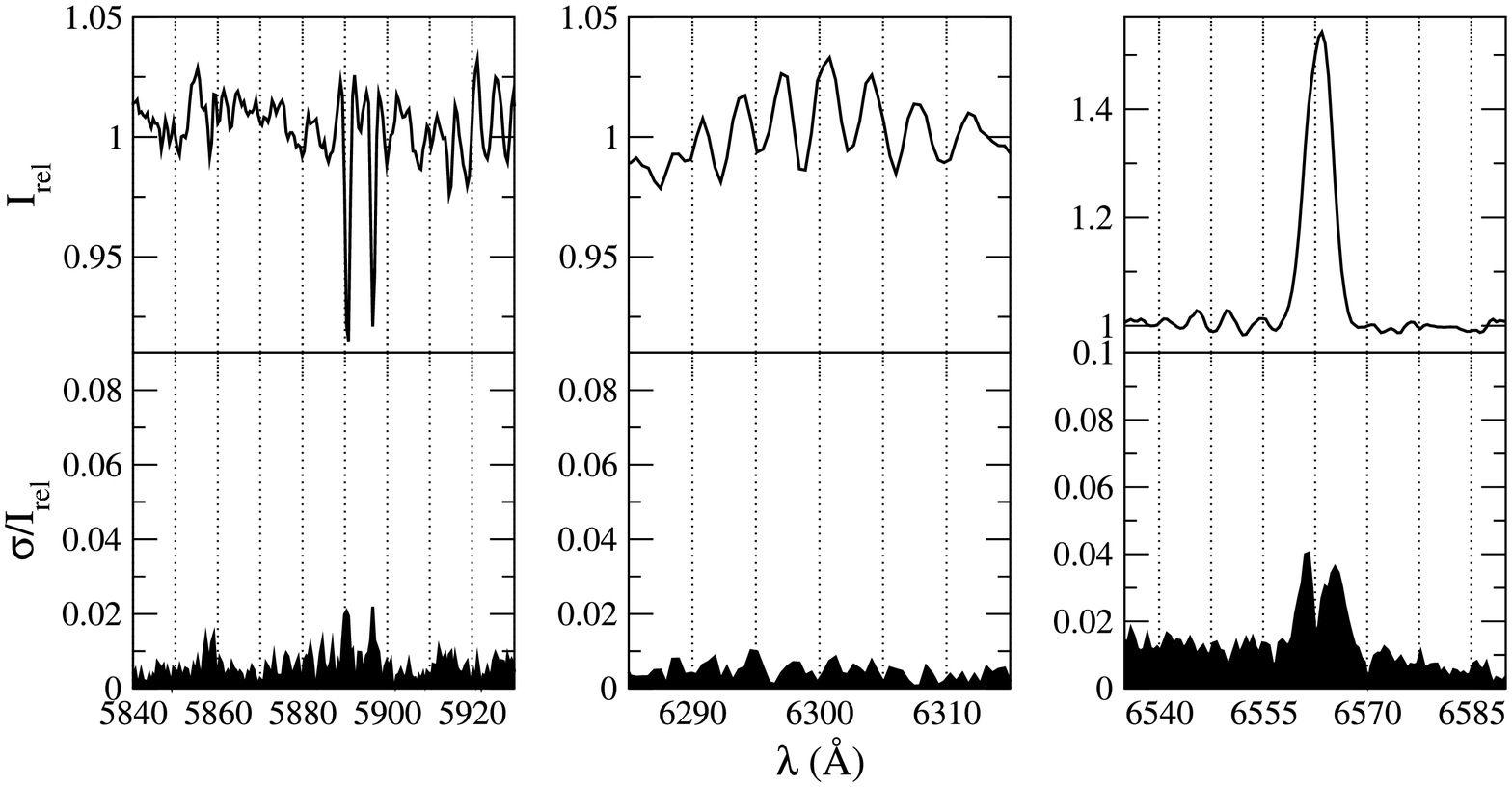} \\
\includegraphics[height=47mm,clip=true]{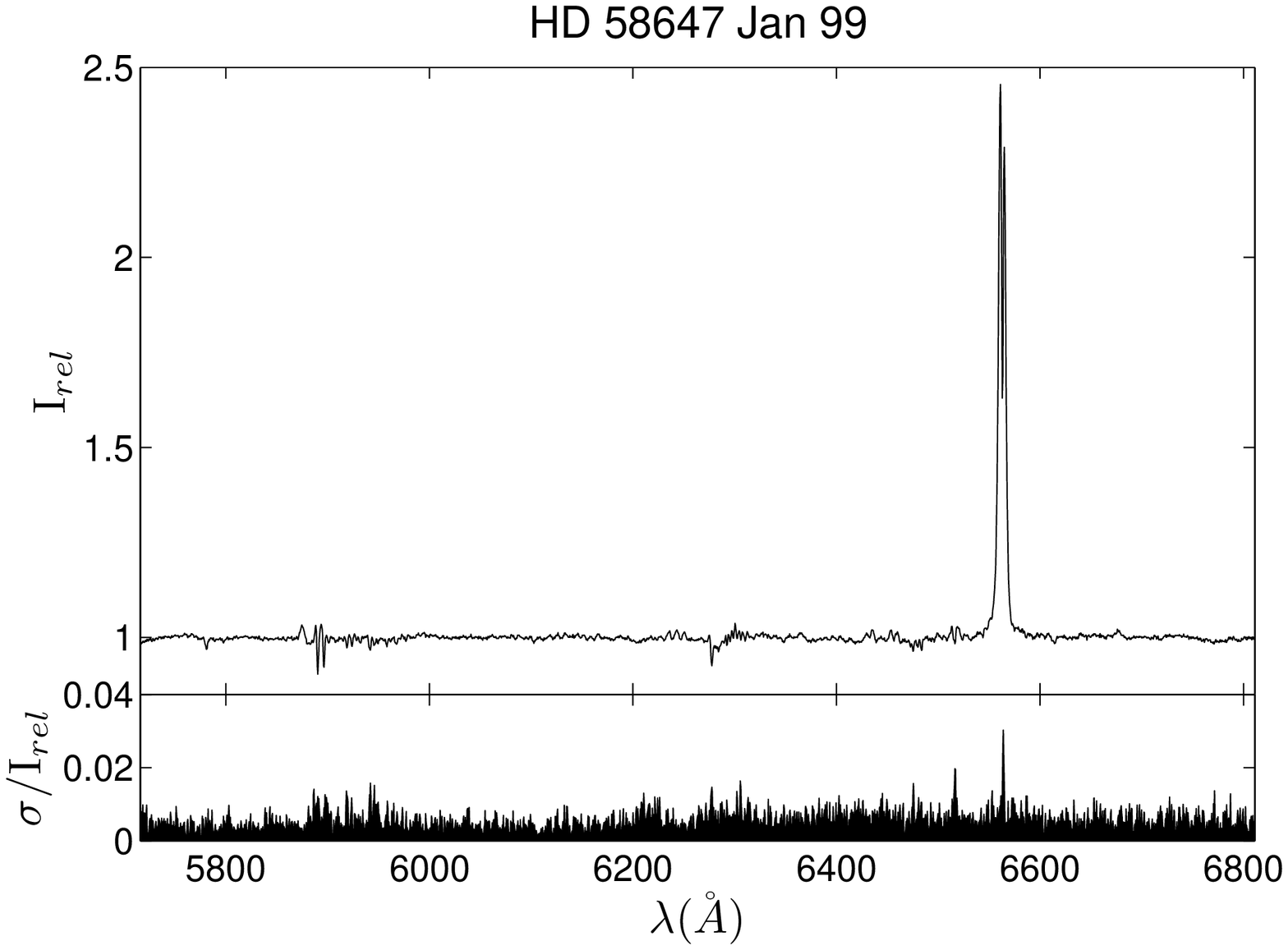}&
\includegraphics[bb=4 77 763 470,height=45mm,clip=true]{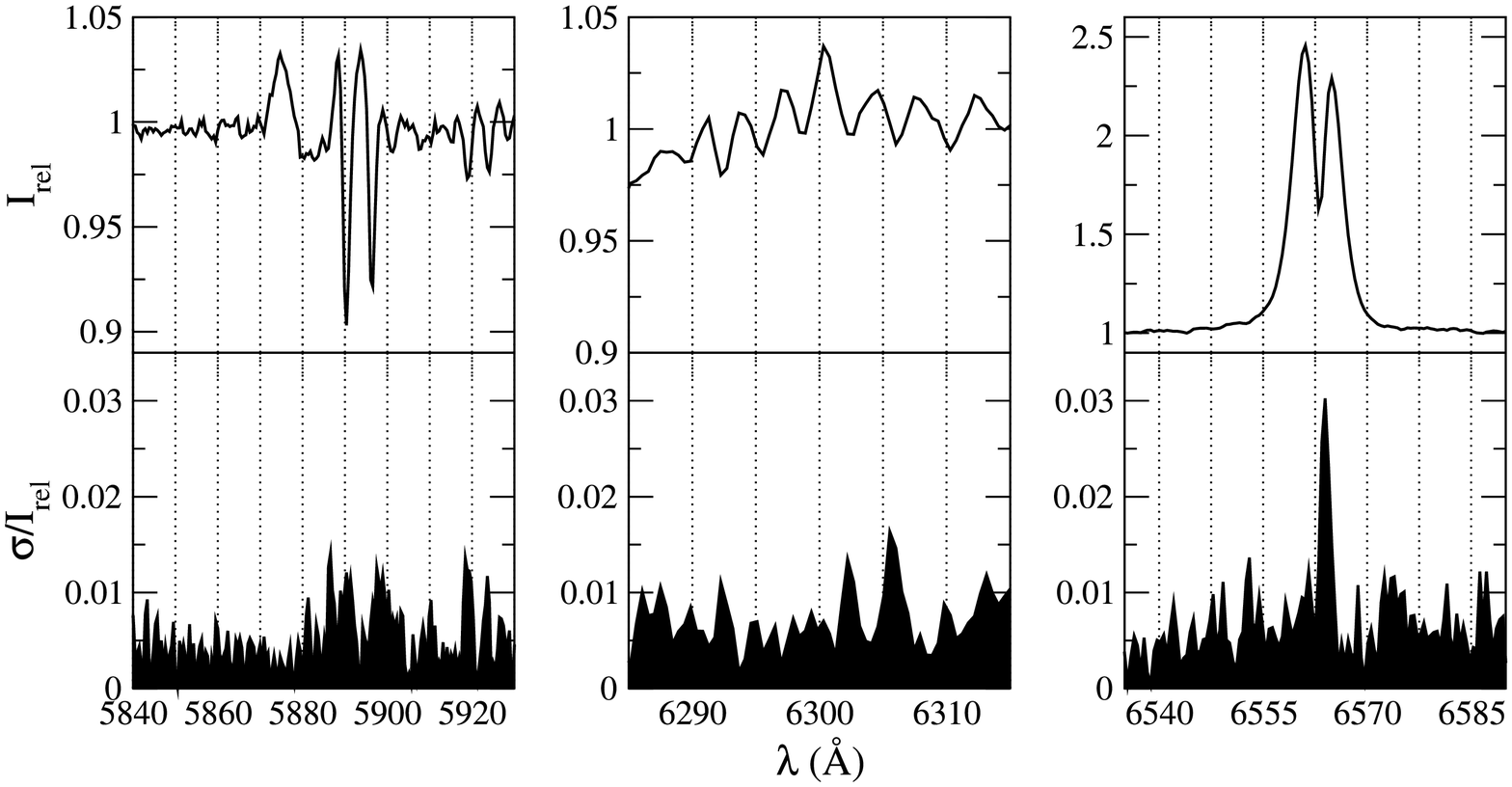} \\
\includegraphics[height=47mm,clip=true]{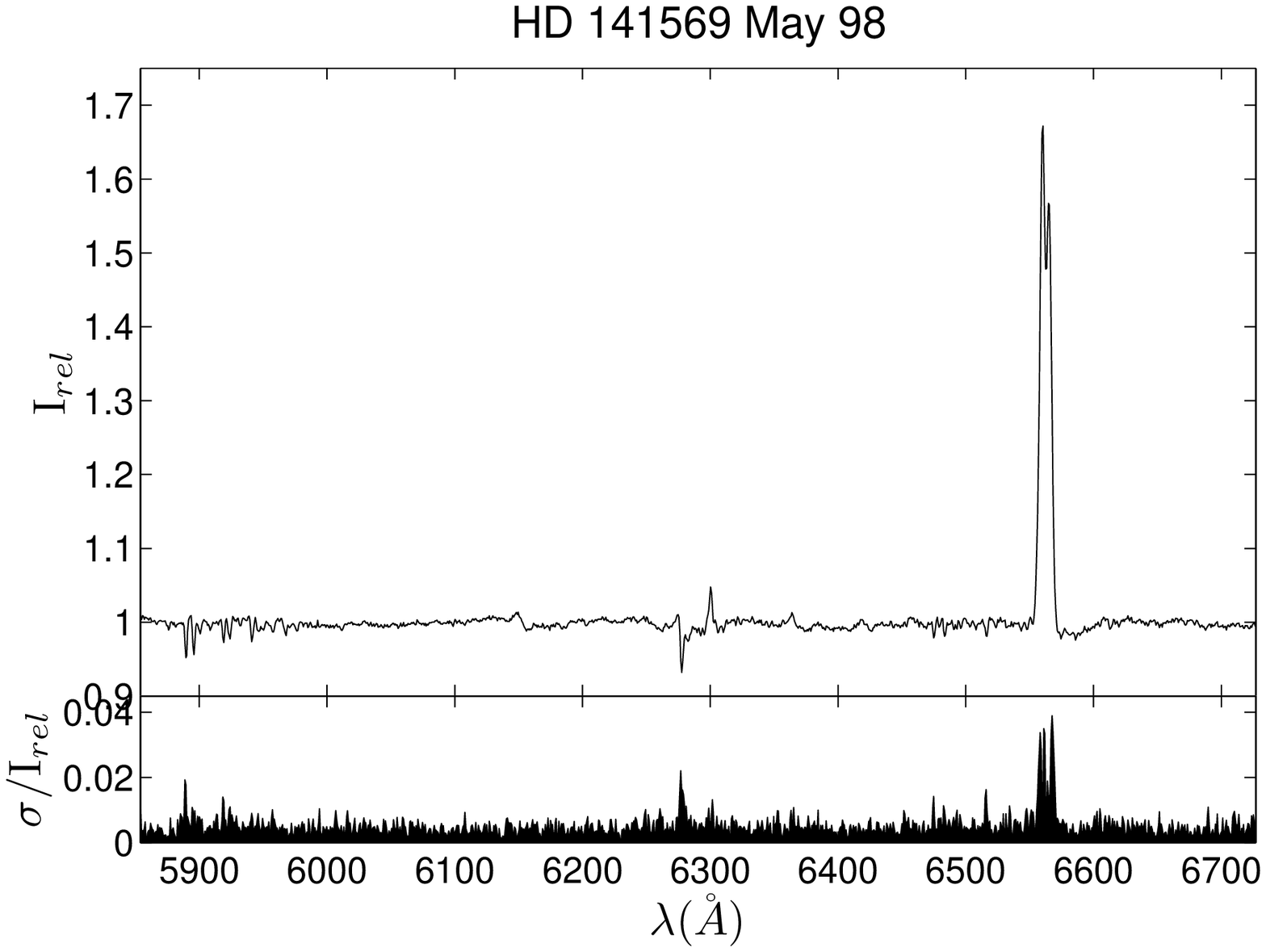}&
\includegraphics[bb=4 77 763 470,height=45mm,clip=true]{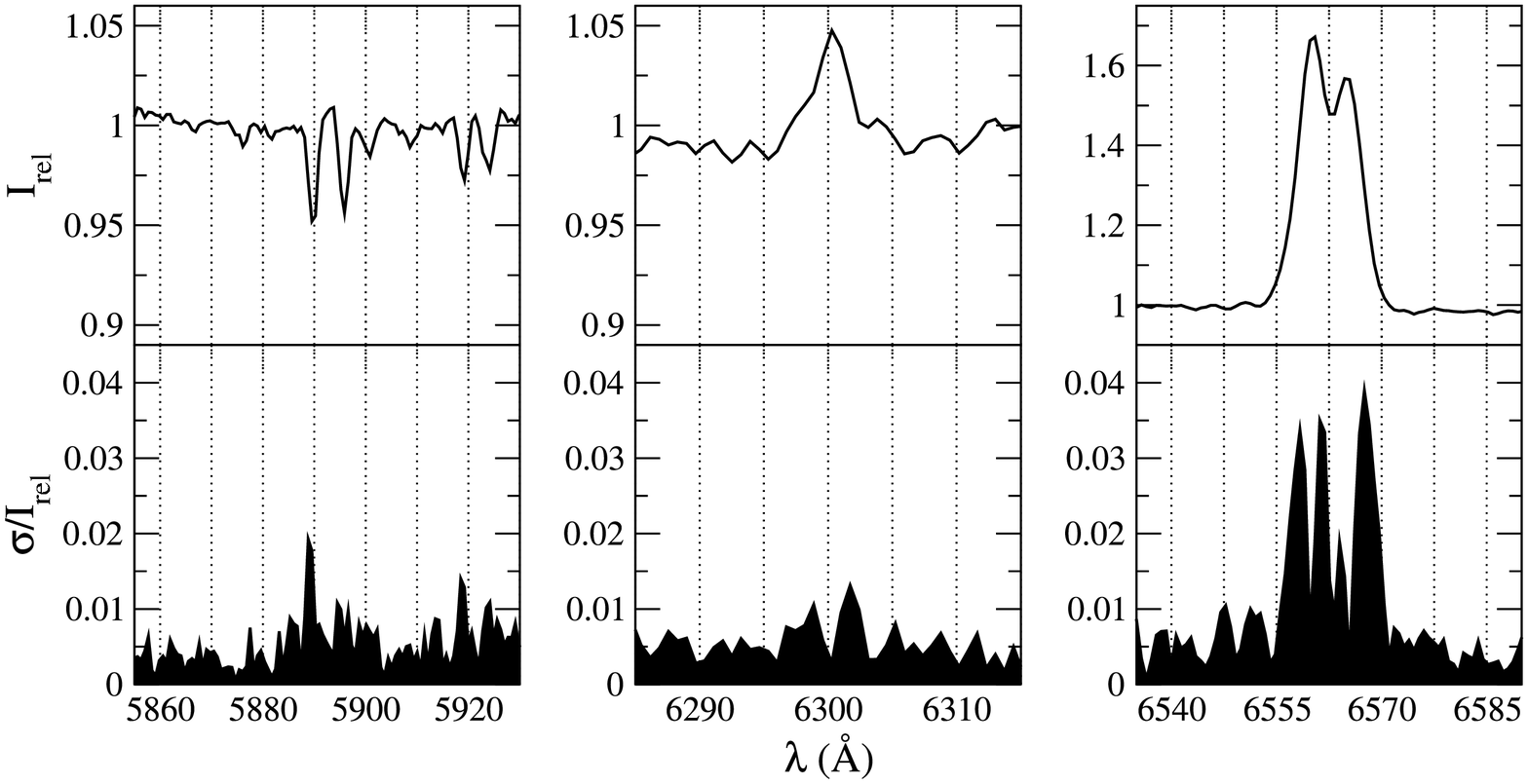} \\
\end{tabular}
\end{table}
\clearpage
\begin{table}
\centering
\renewcommand\arraystretch{10}
\begin{tabular}{cc}
\includegraphics[height=47mm,clip=true]{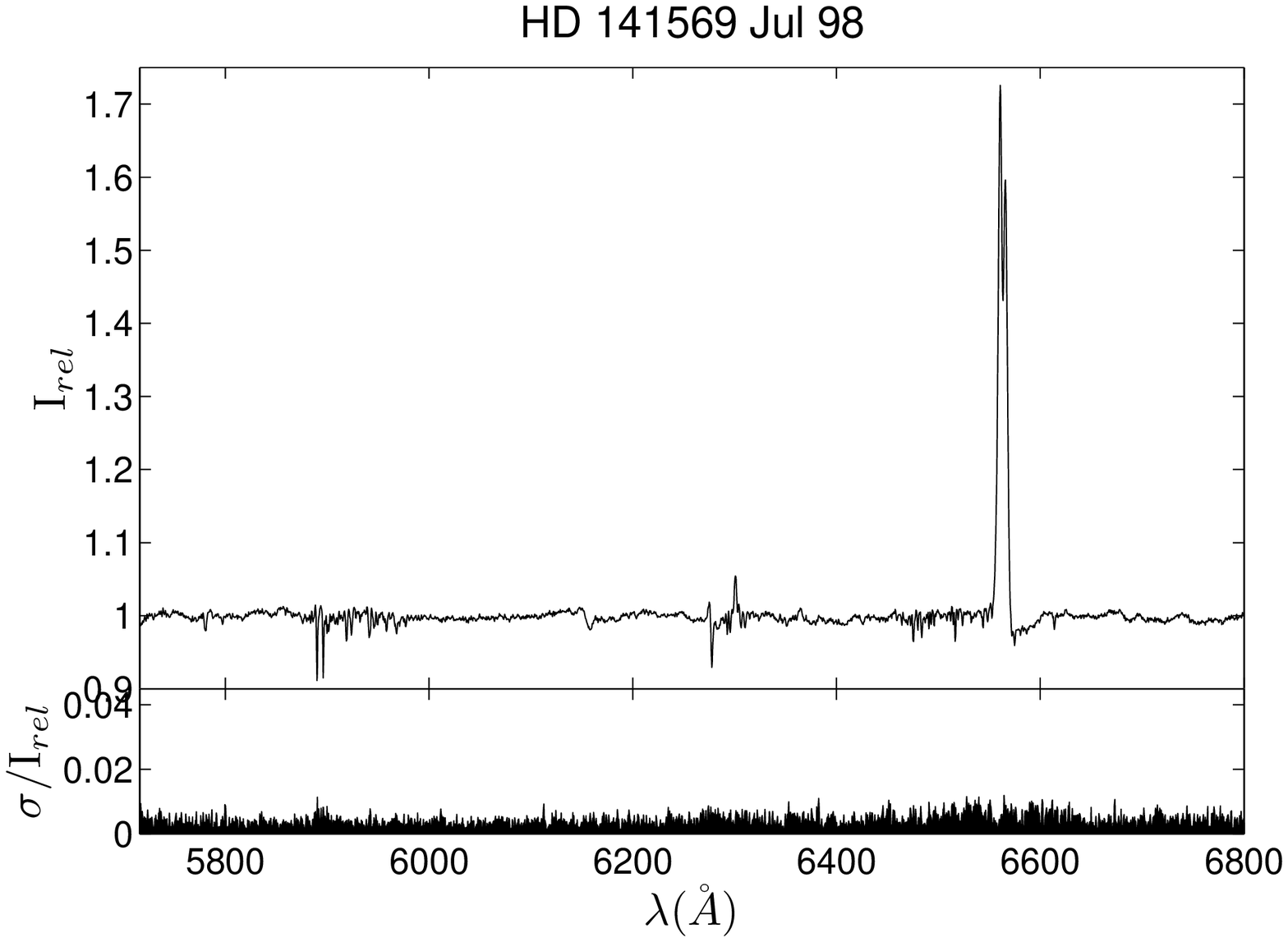}&
\includegraphics[bb=4 77 763 470,height=45mm,clip=true]{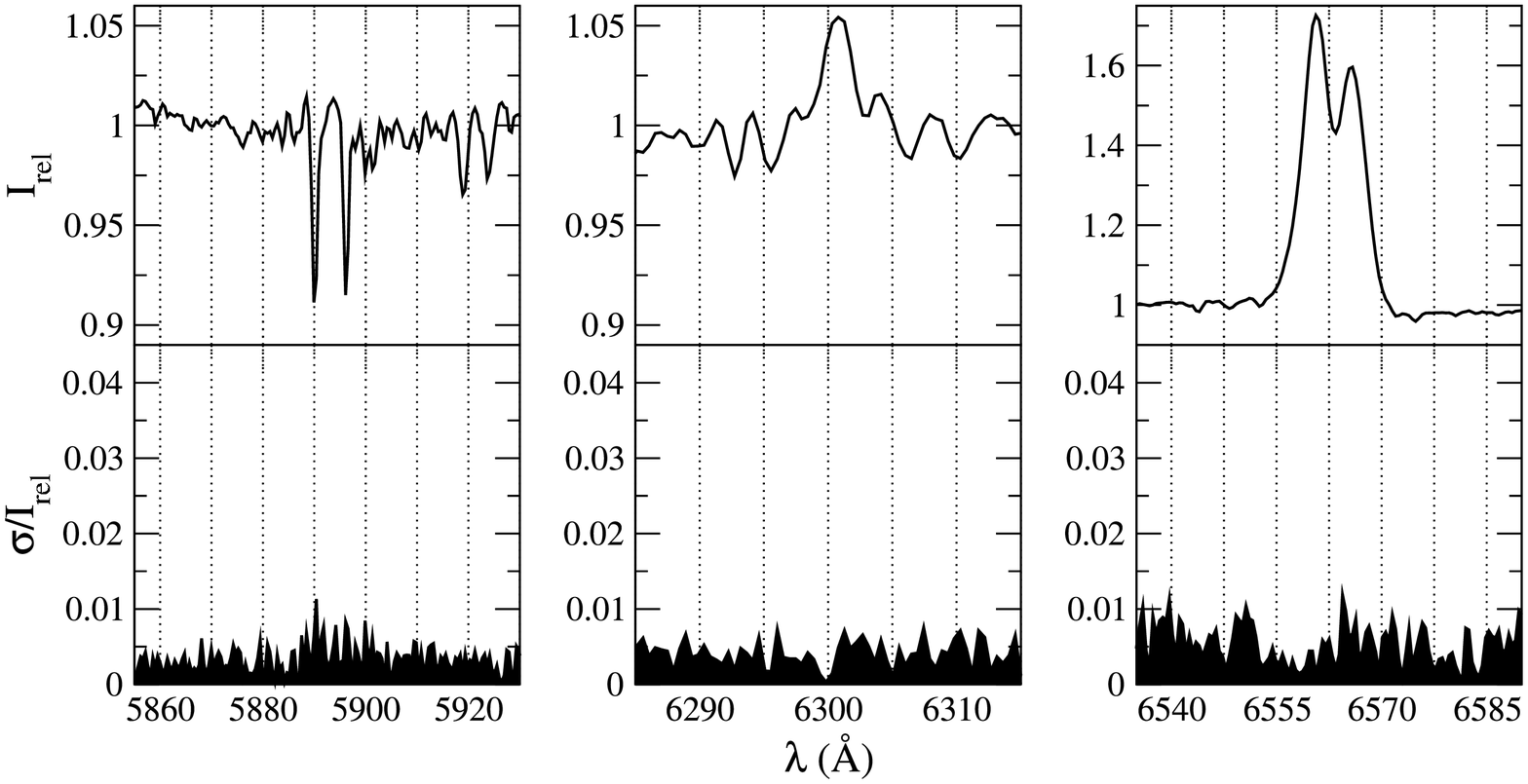} \\
\includegraphics[height=47mm,clip=true]{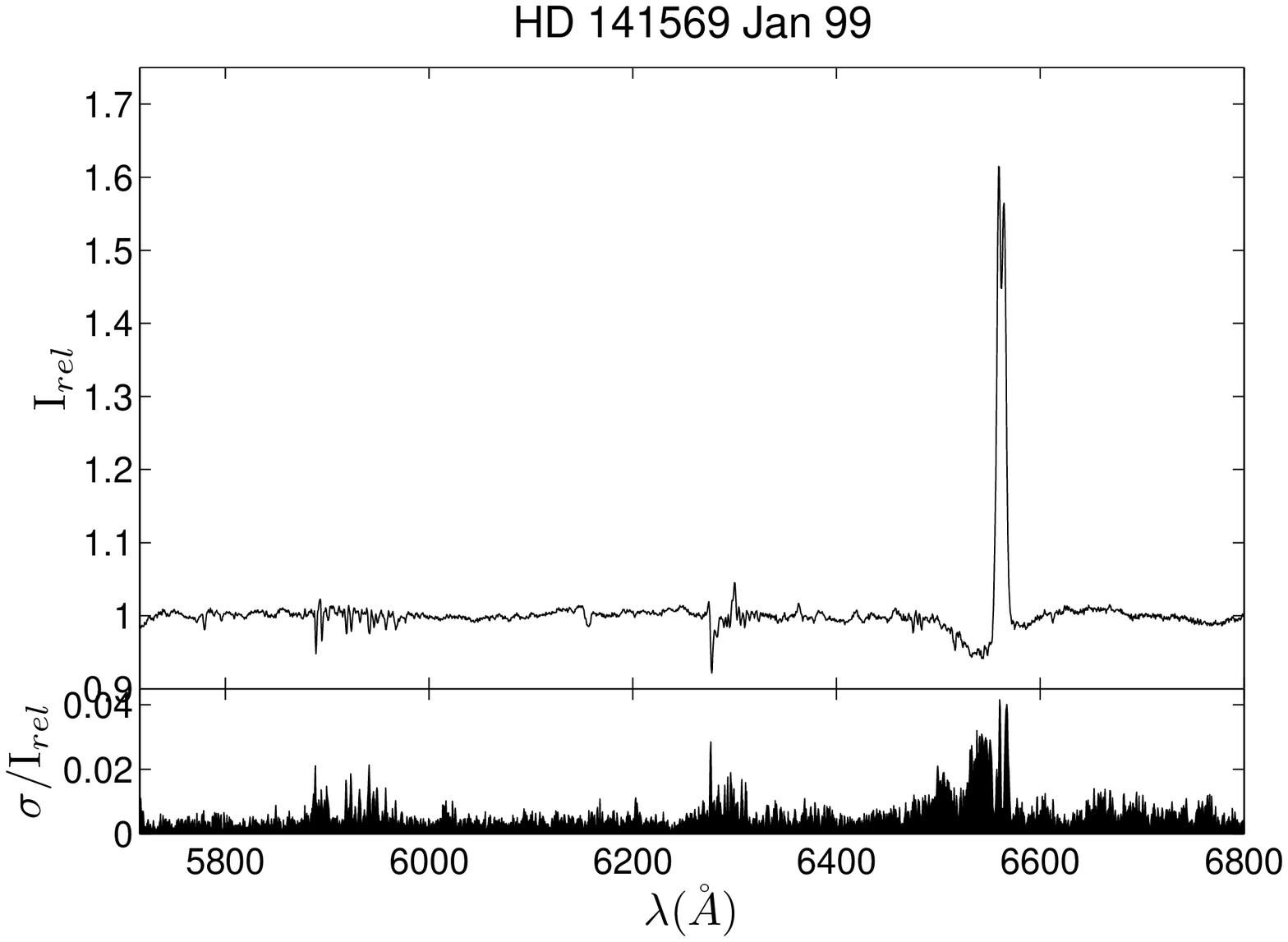}&
\includegraphics[bb=4 77 763 470,height=45mm,clip=true]{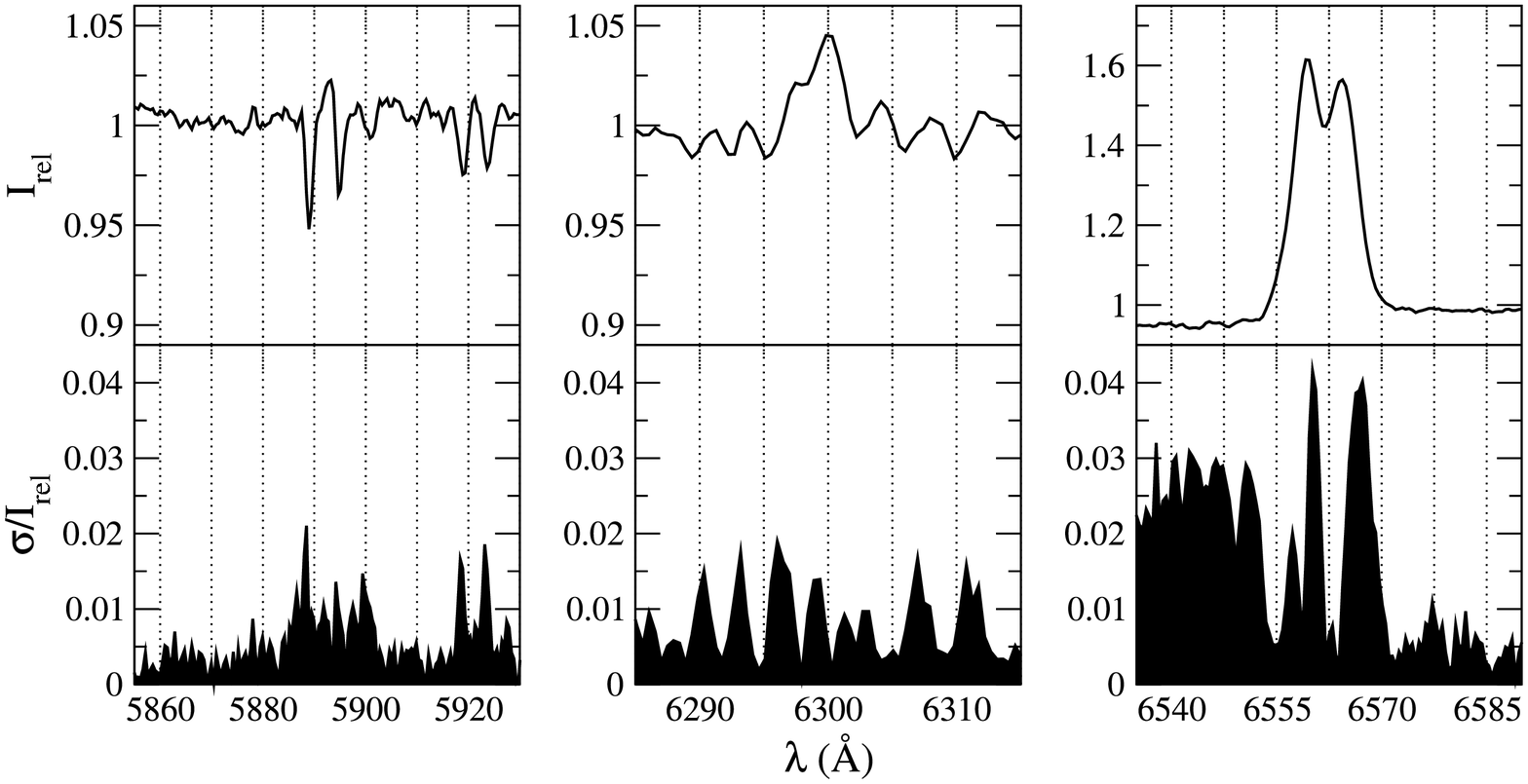} \\
\includegraphics[height=47mm,clip=true]{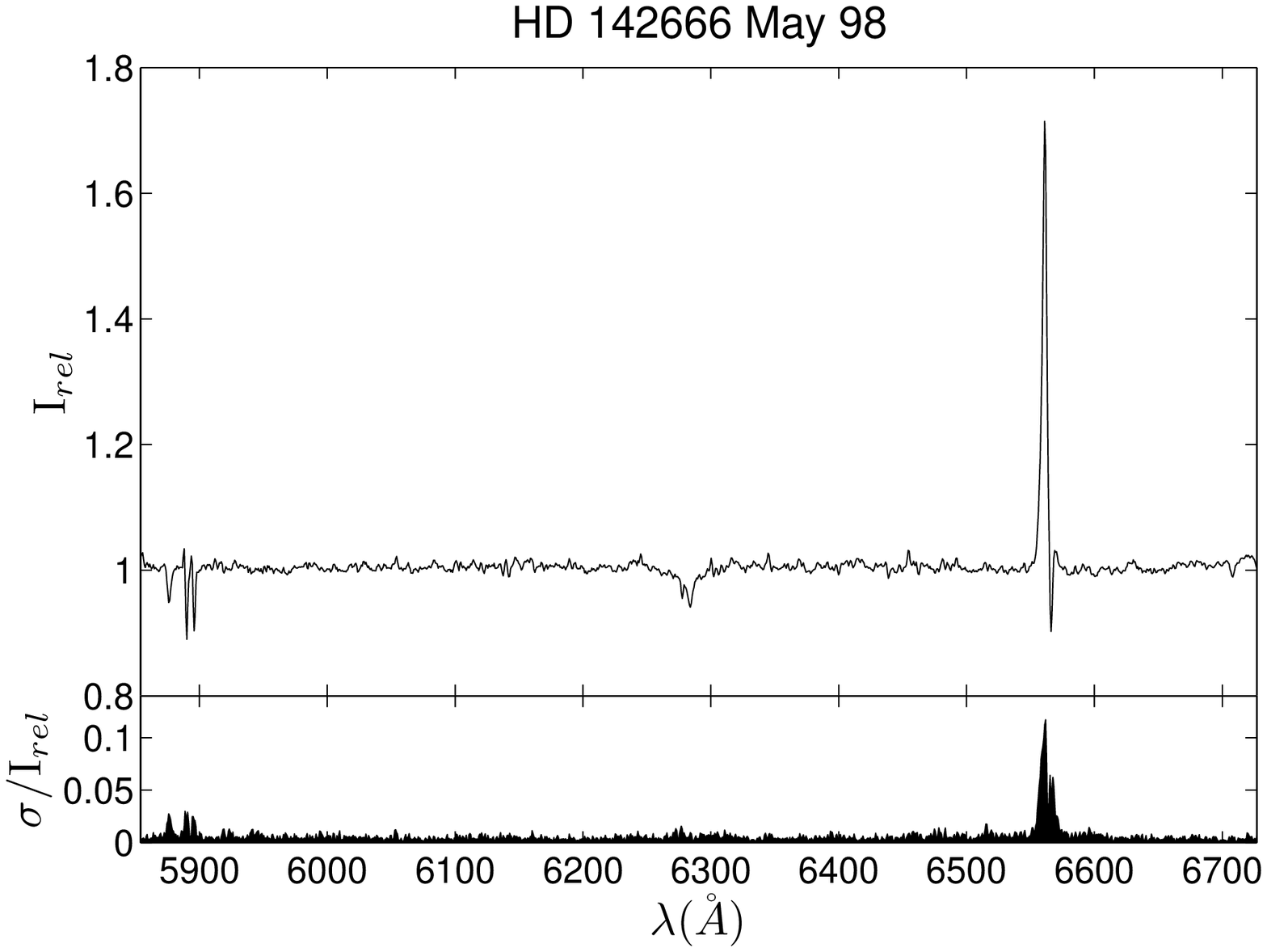}&
\includegraphics[bb=4 77 763 470,height=45mm,clip=true]{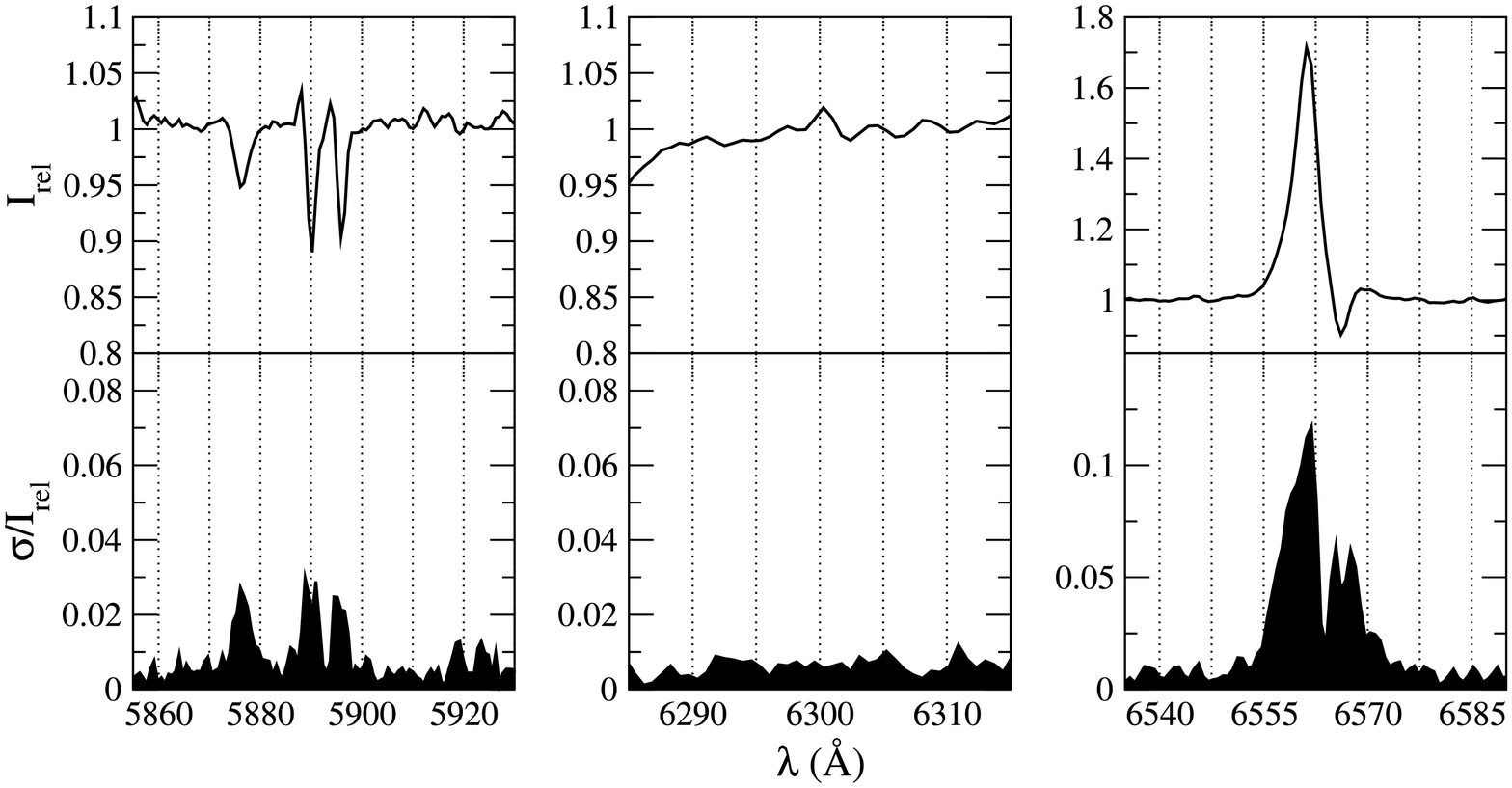} \\
\includegraphics[height=47mm,clip=true]{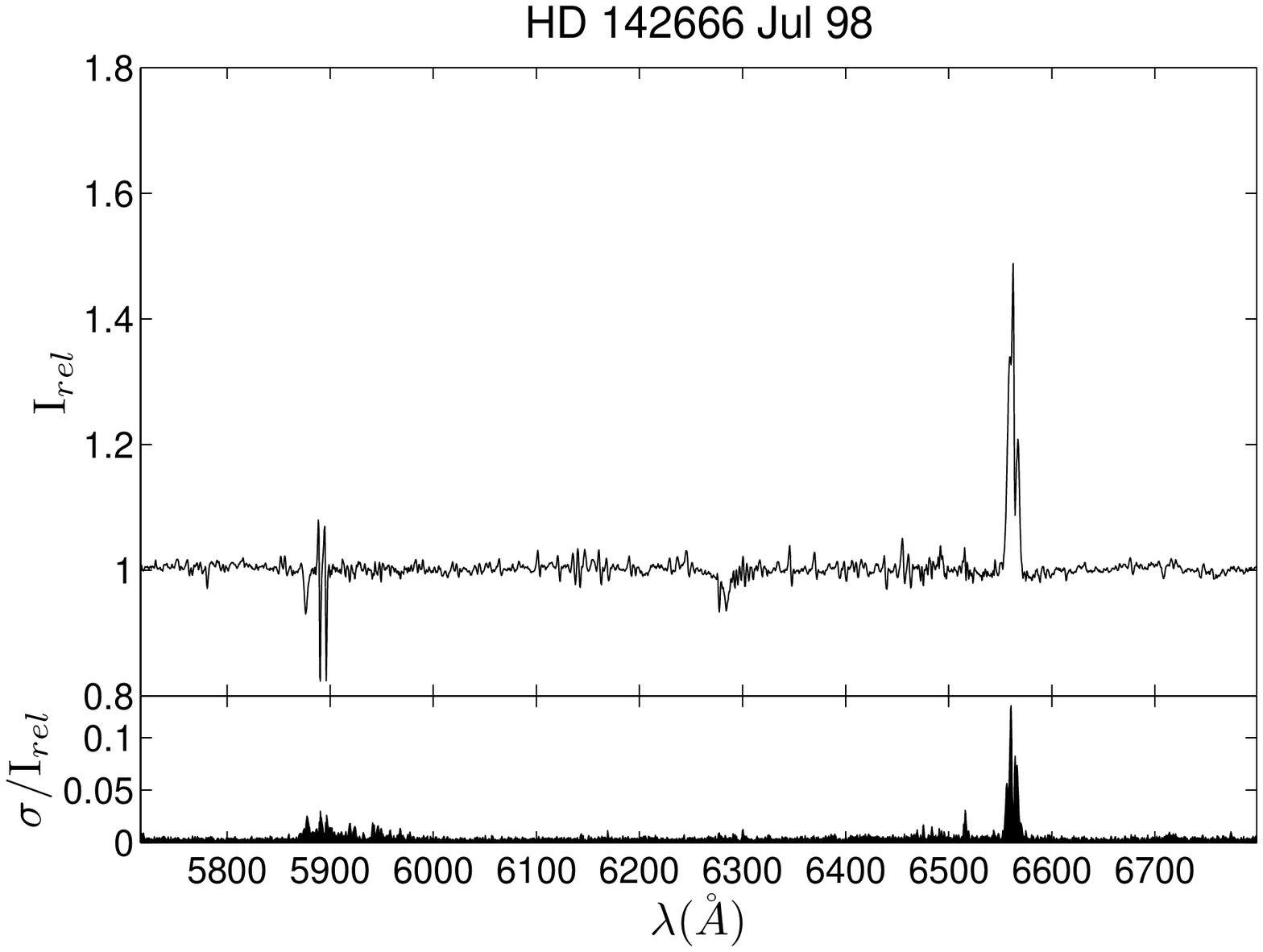}&
\includegraphics[bb=4 77 763 470,height=45mm,clip=true]{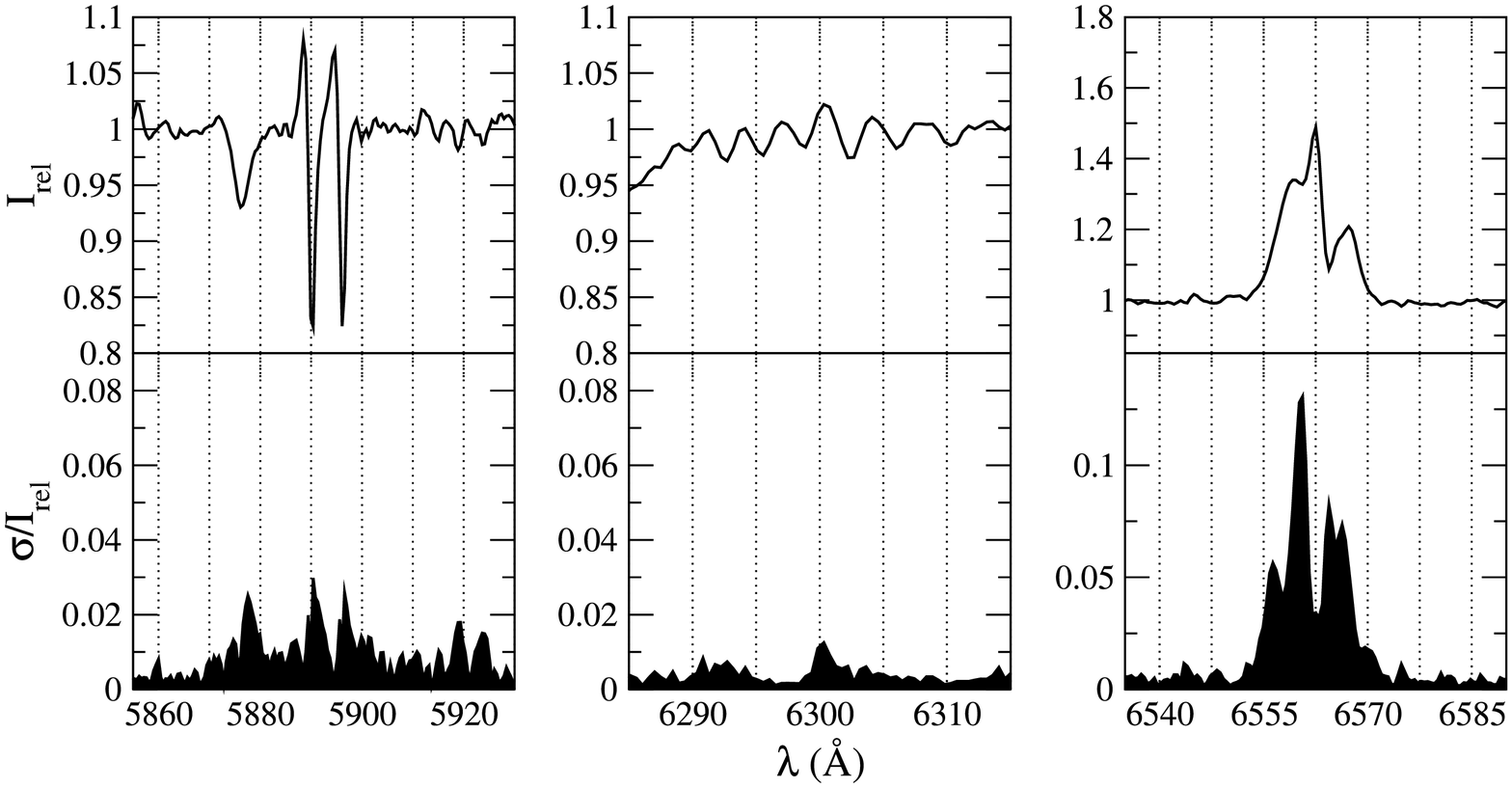} \\
\end{tabular}
\end{table}
\clearpage
\begin{table}
\centering
\renewcommand\arraystretch{10}
\begin{tabular}{cc}
\includegraphics[height=47mm,clip=true]{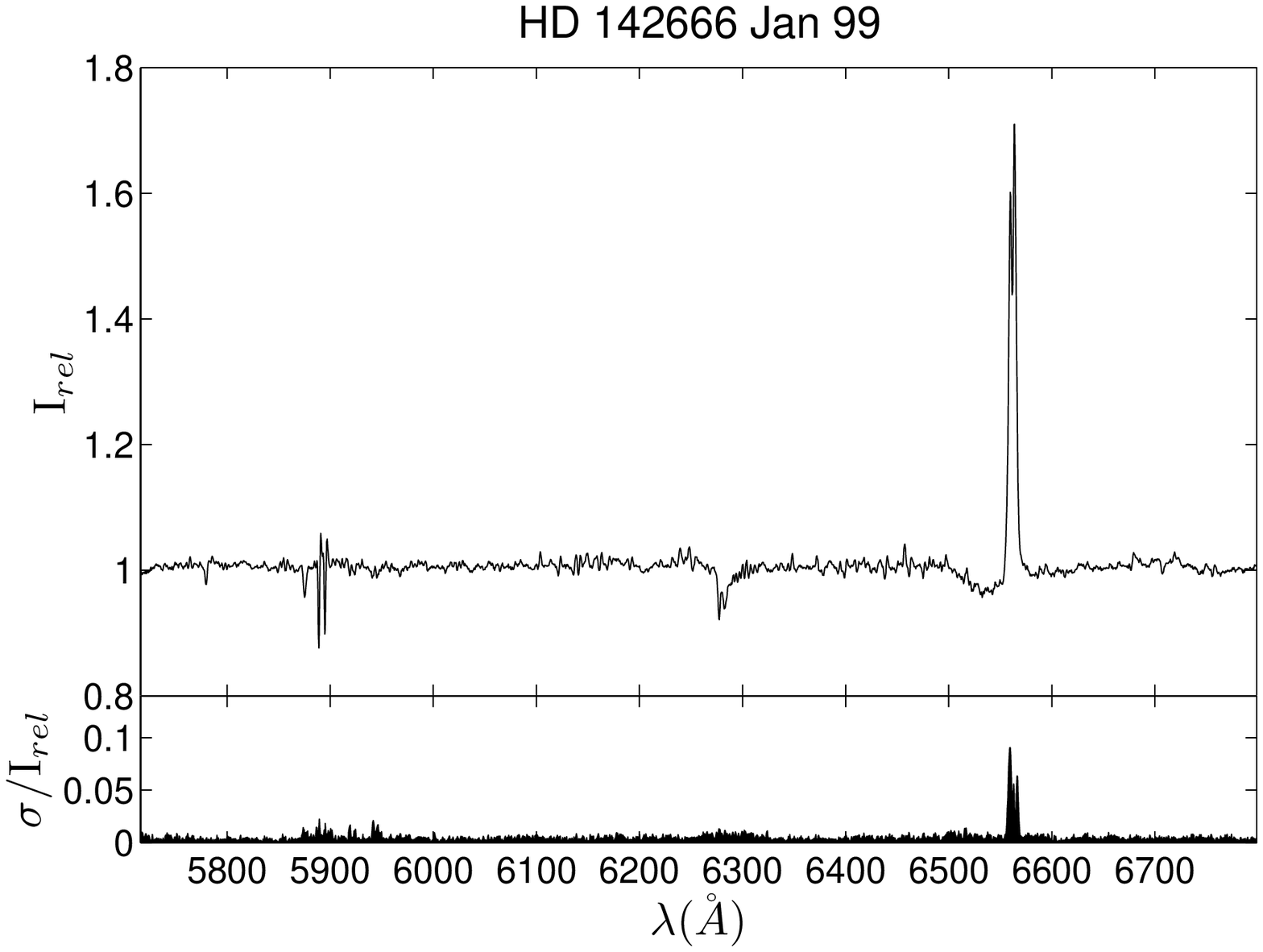}&
\includegraphics[bb=4 77 763 470,height=45mm,clip=true]{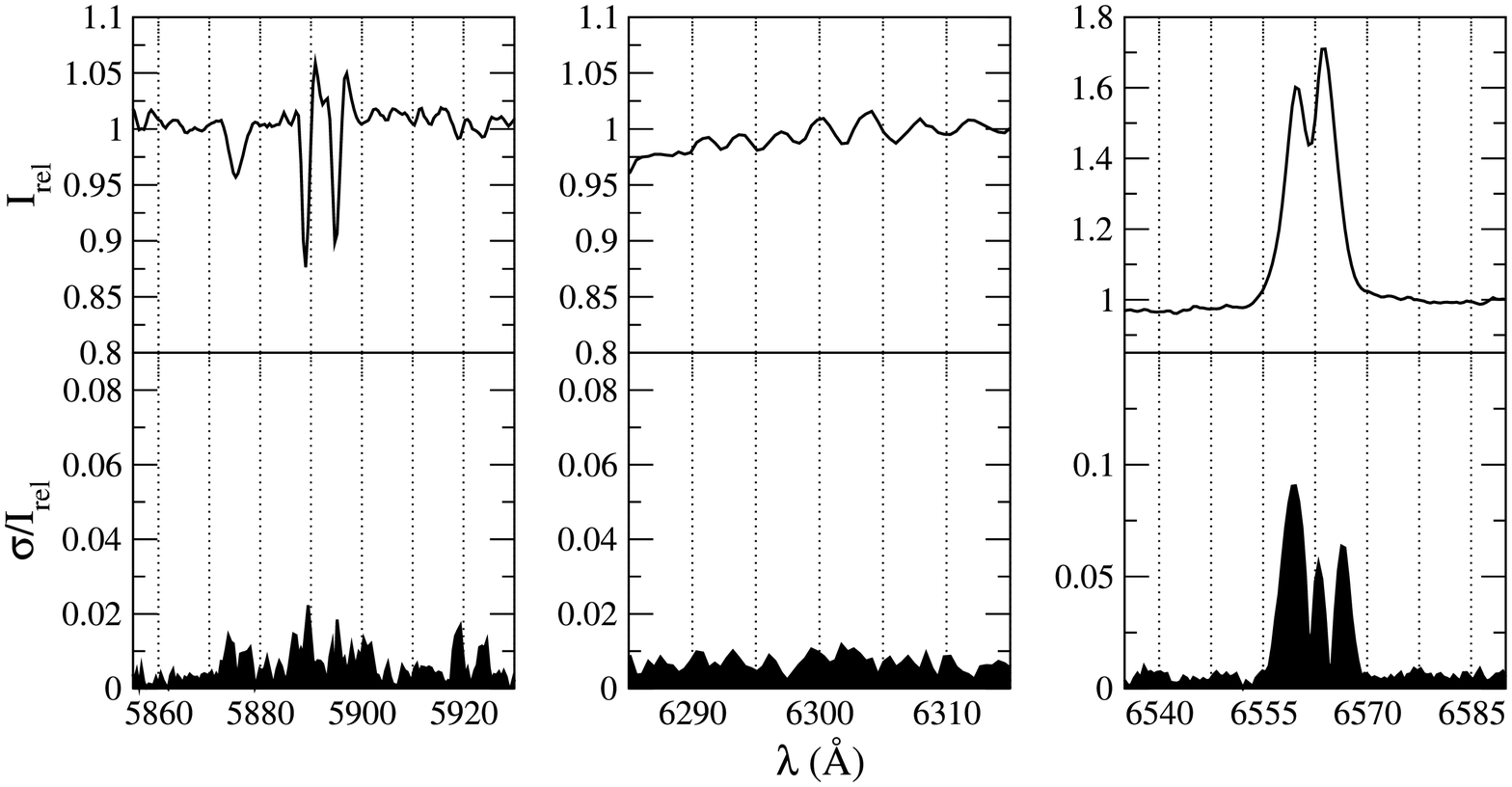} \\ 
\includegraphics[height=47mm,clip=true]{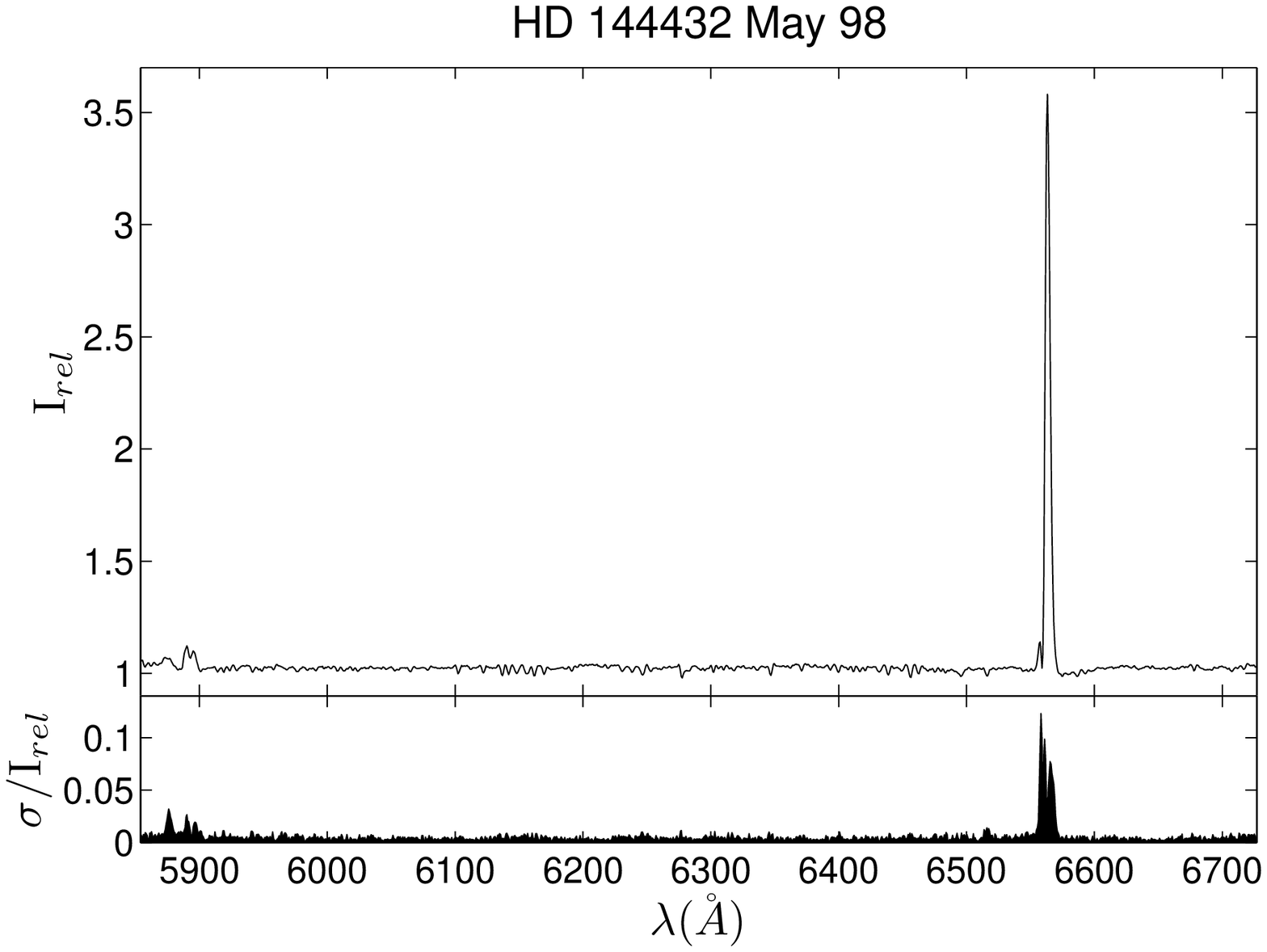}&
\includegraphics[bb=4 77 763 470,height=45mm,clip=true]{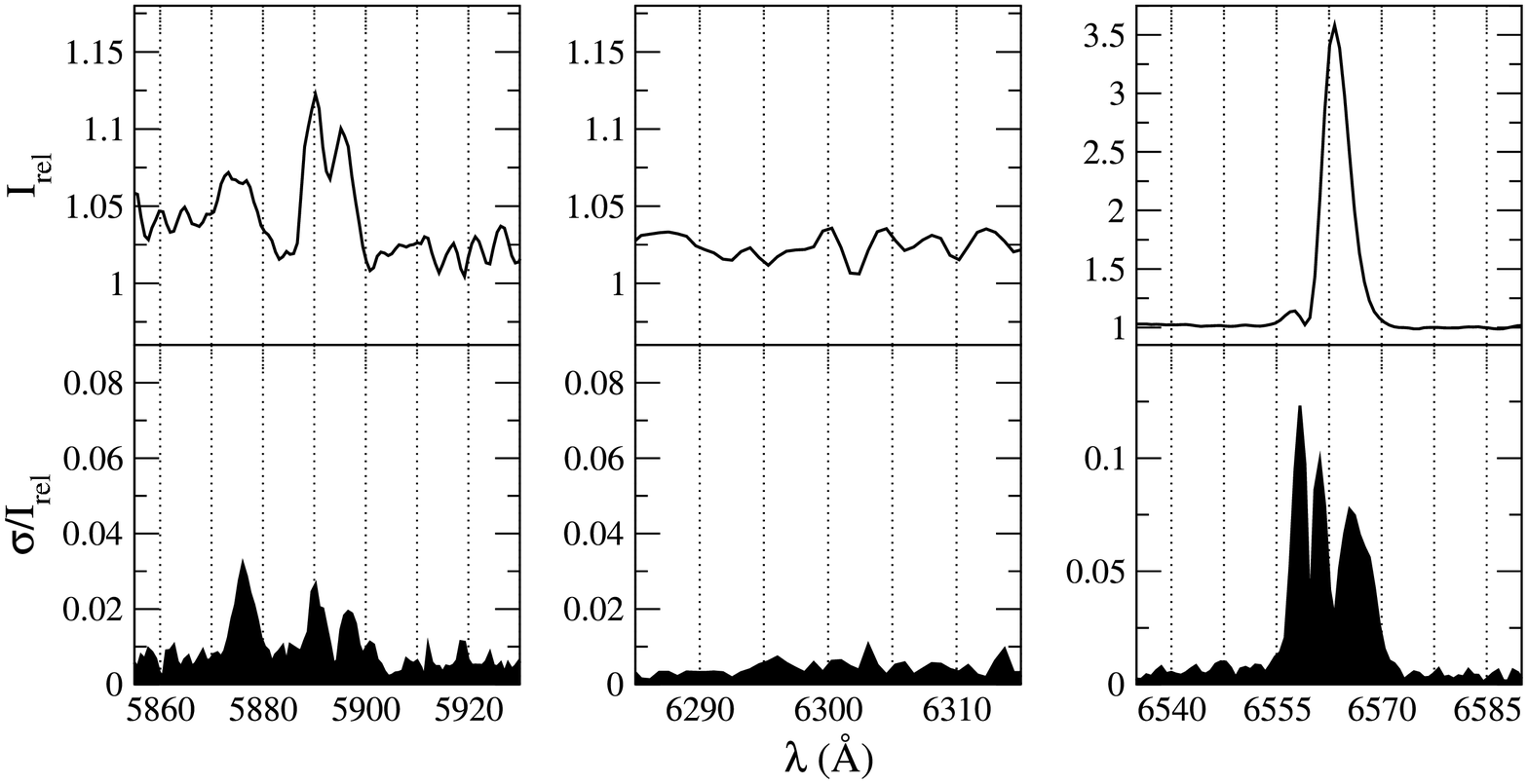} \\
\includegraphics[height=47mm,clip=true]{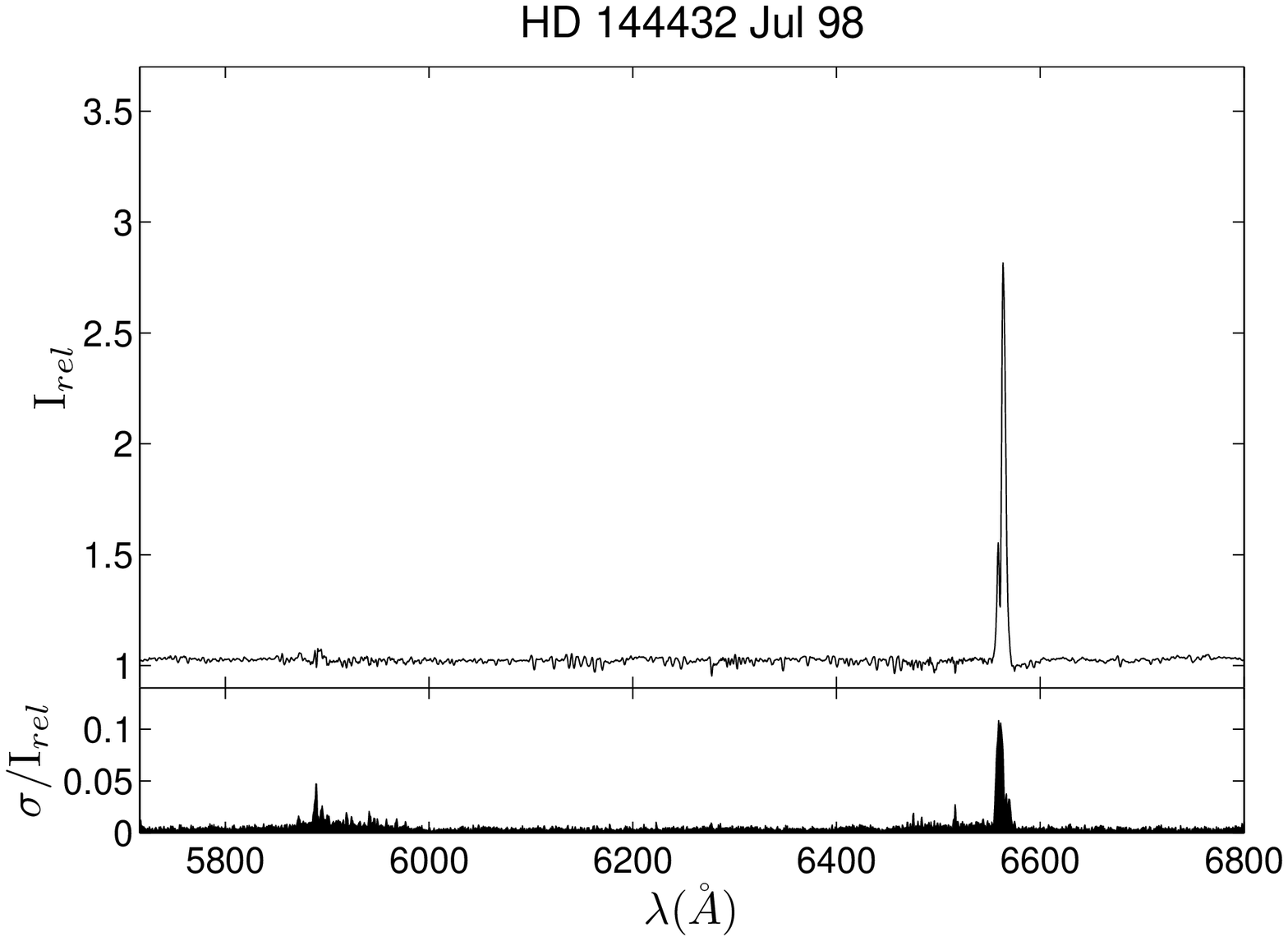}&
\includegraphics[bb=4 77 763 470,height=45mm,clip=true]{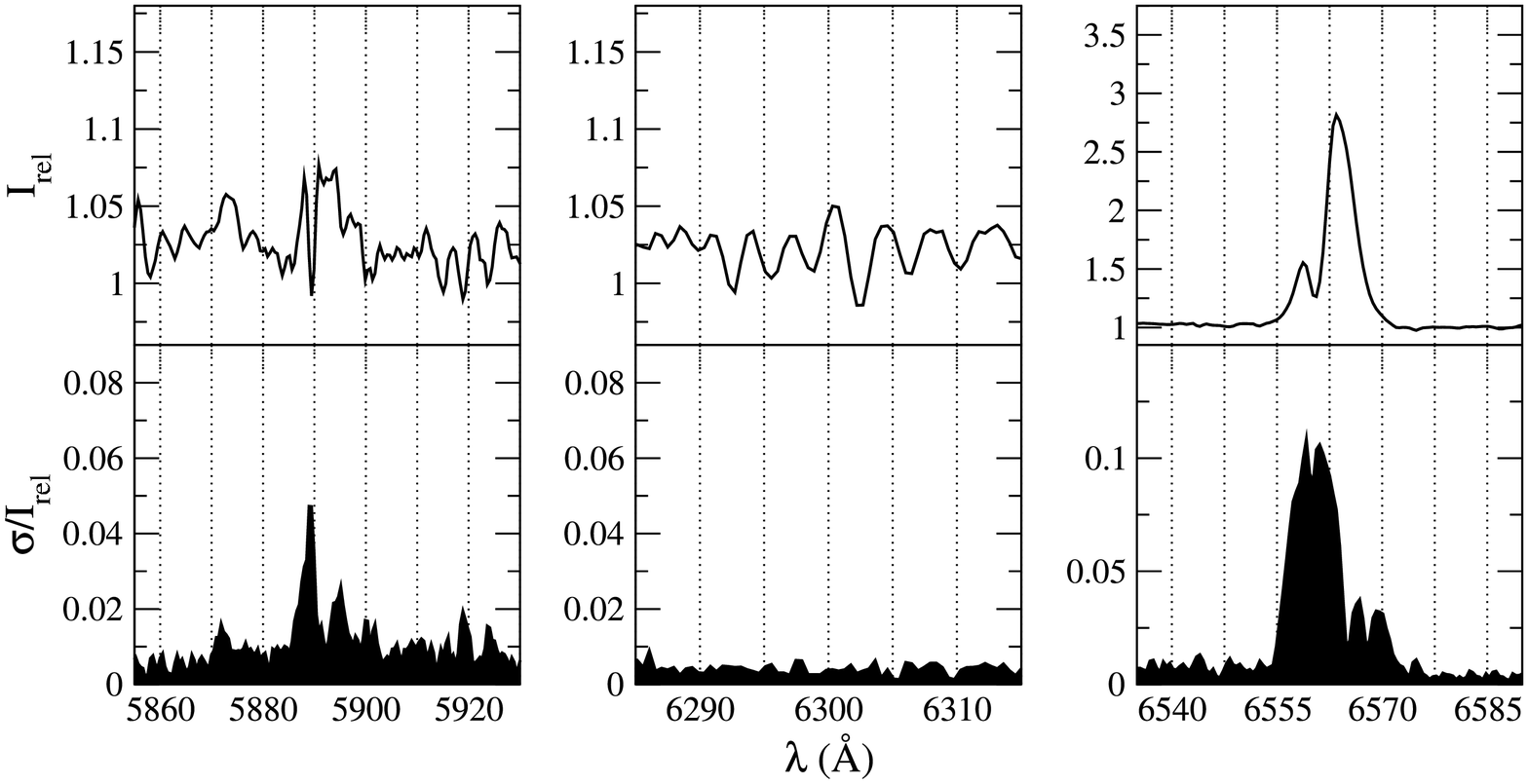} \\
\includegraphics[height=47mm,clip=true]{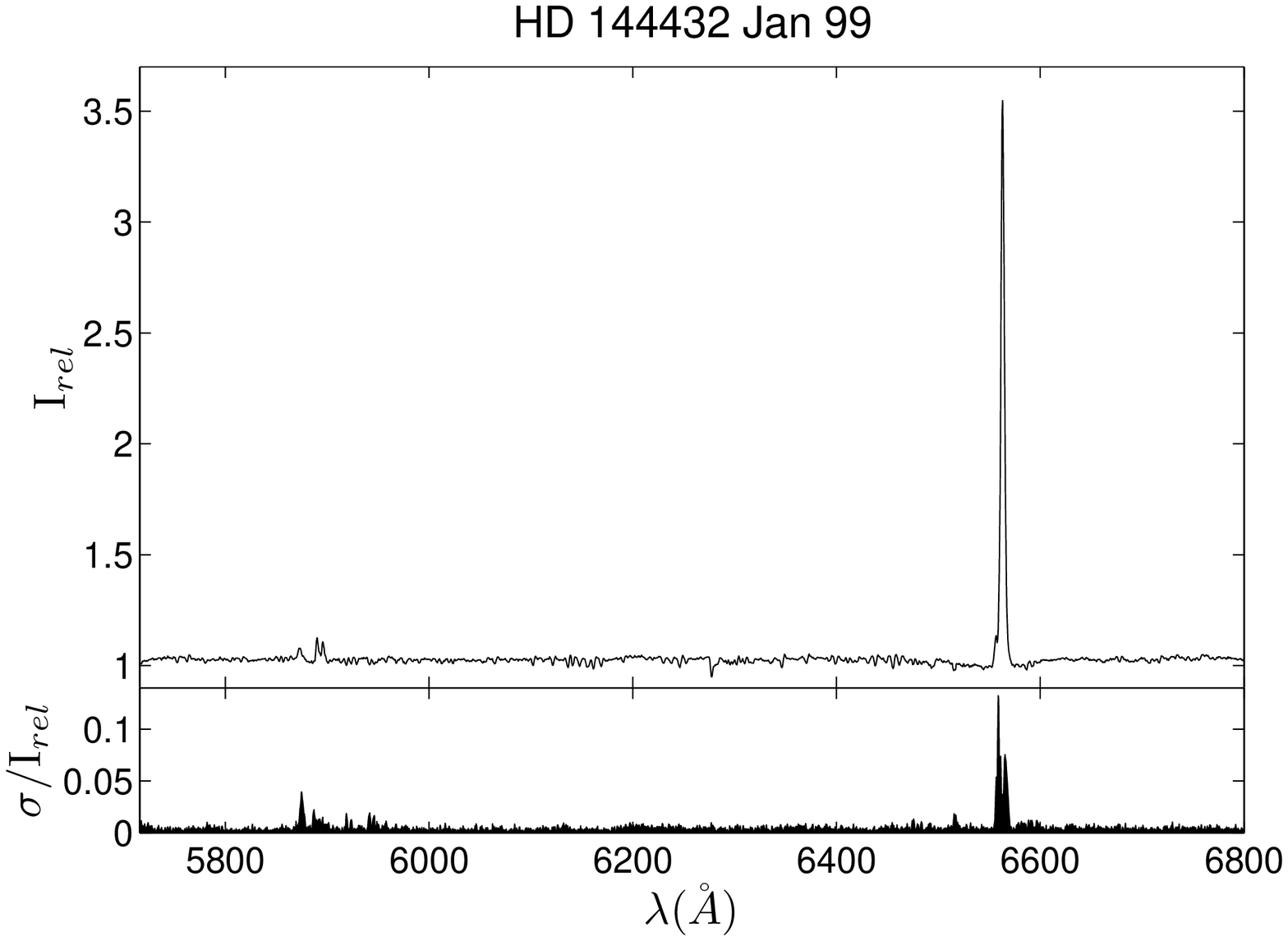}&
\includegraphics[bb=4 77 763 470,height=45mm,clip=true]{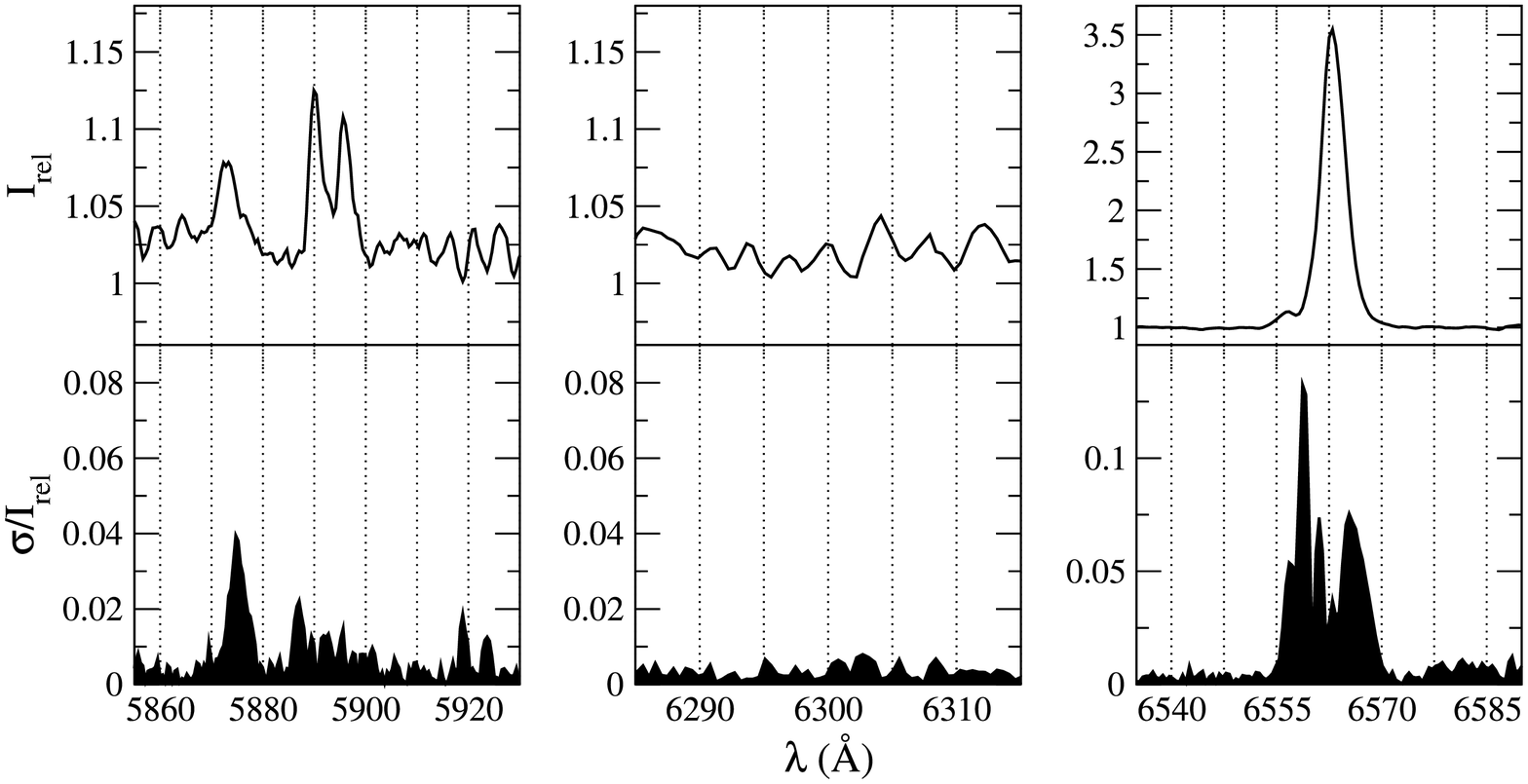} \\ 
\end{tabular}
\end{table}
\clearpage
\begin{table}
\centering
\renewcommand\arraystretch{10}
\begin{tabular}{cc}
\includegraphics[height=47mm,clip=true]{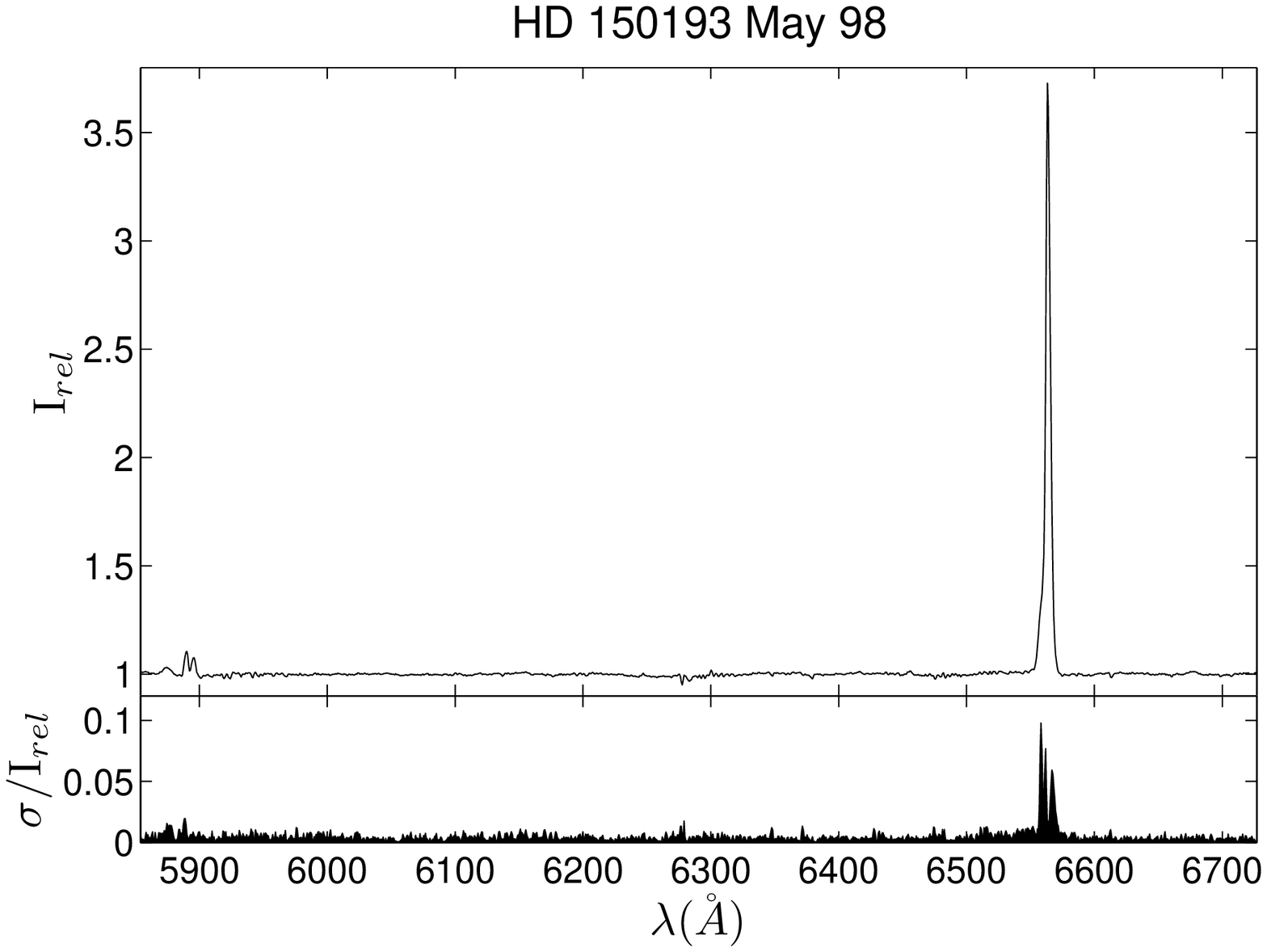}&
\includegraphics[bb=4 77 763 470,height=45mm,clip=true]{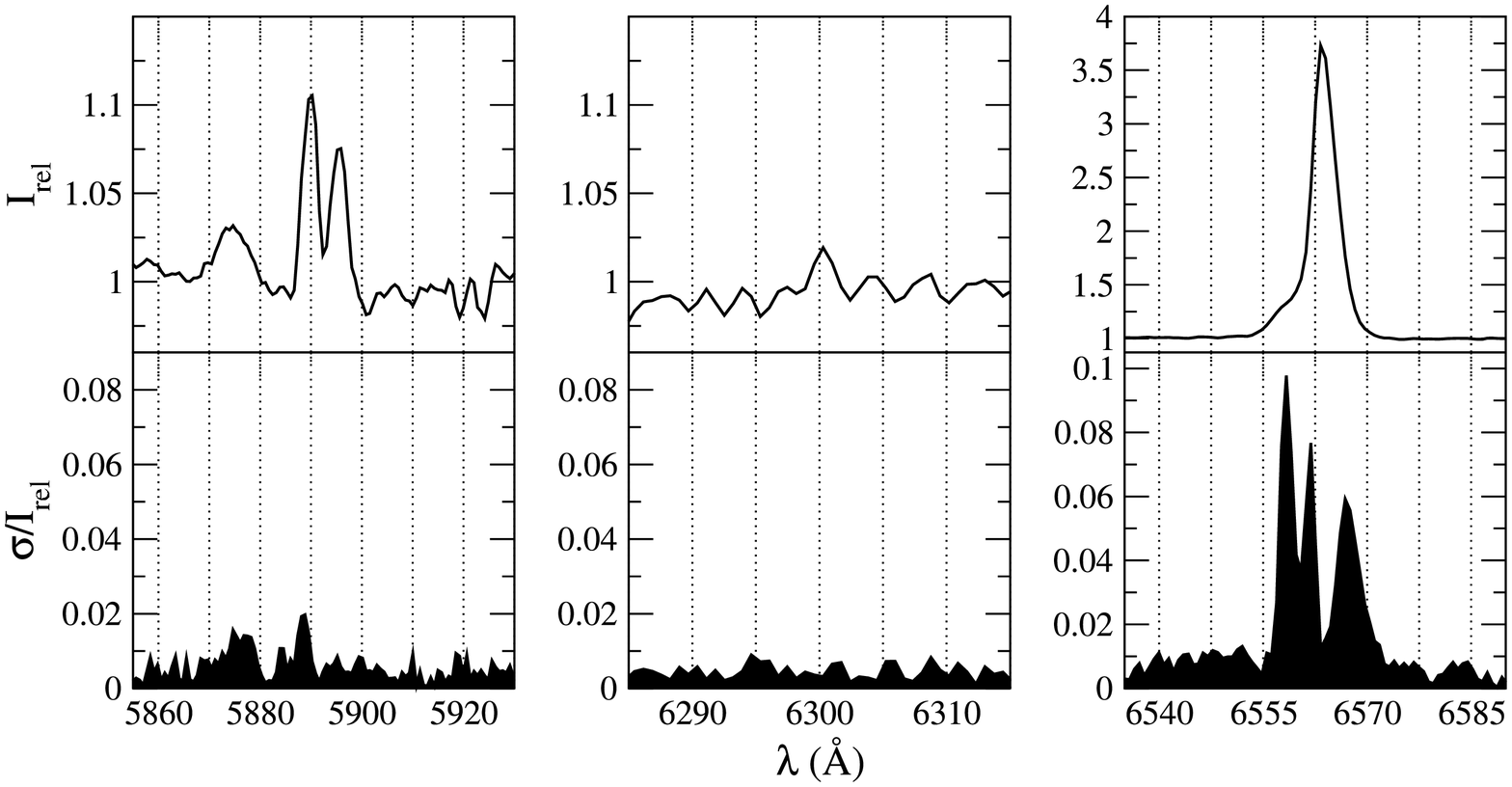} \\
\includegraphics[height=47mm,clip=true]{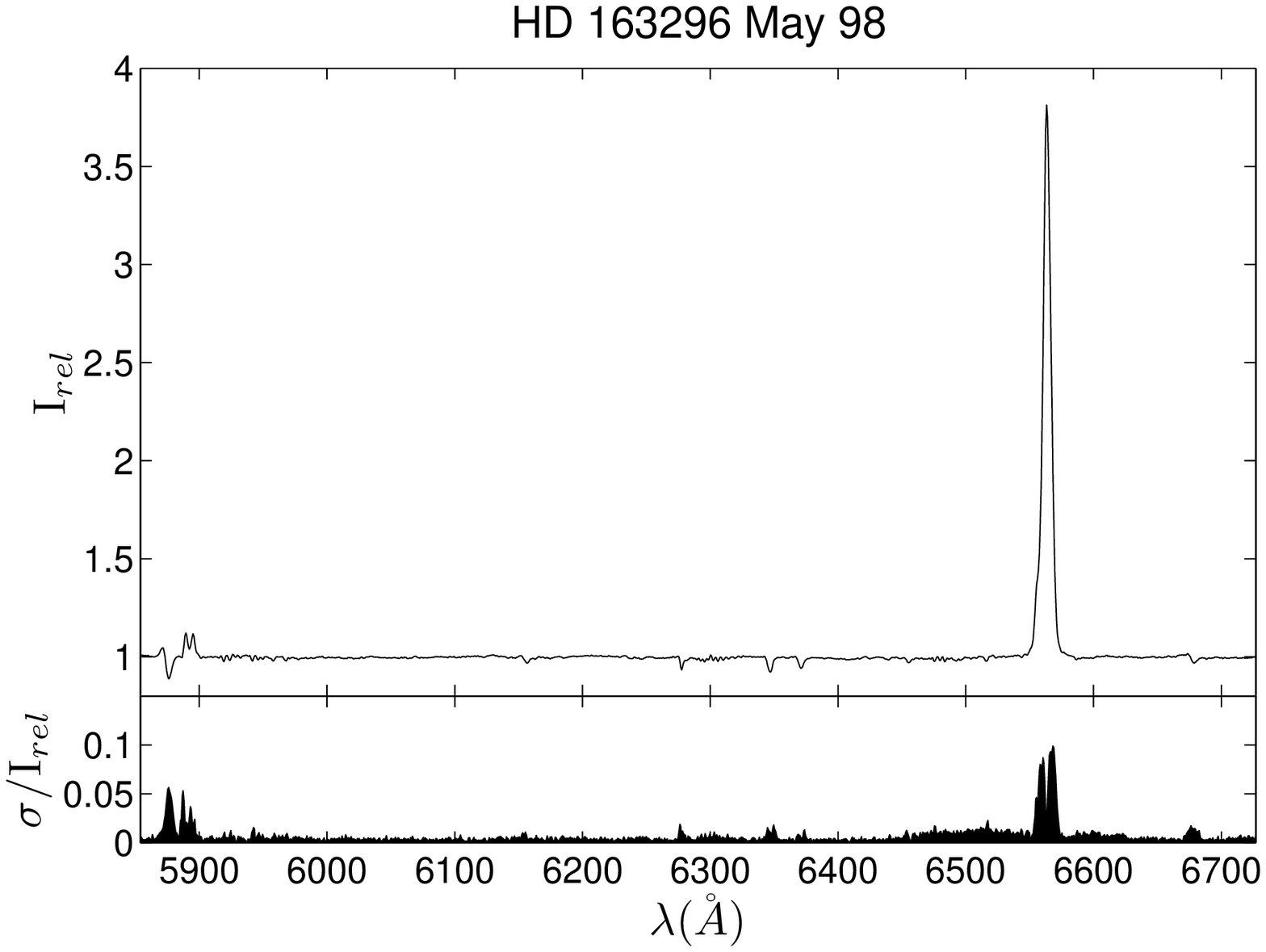}&
\includegraphics[bb=4 77 763 470,height=45mm,clip=true]{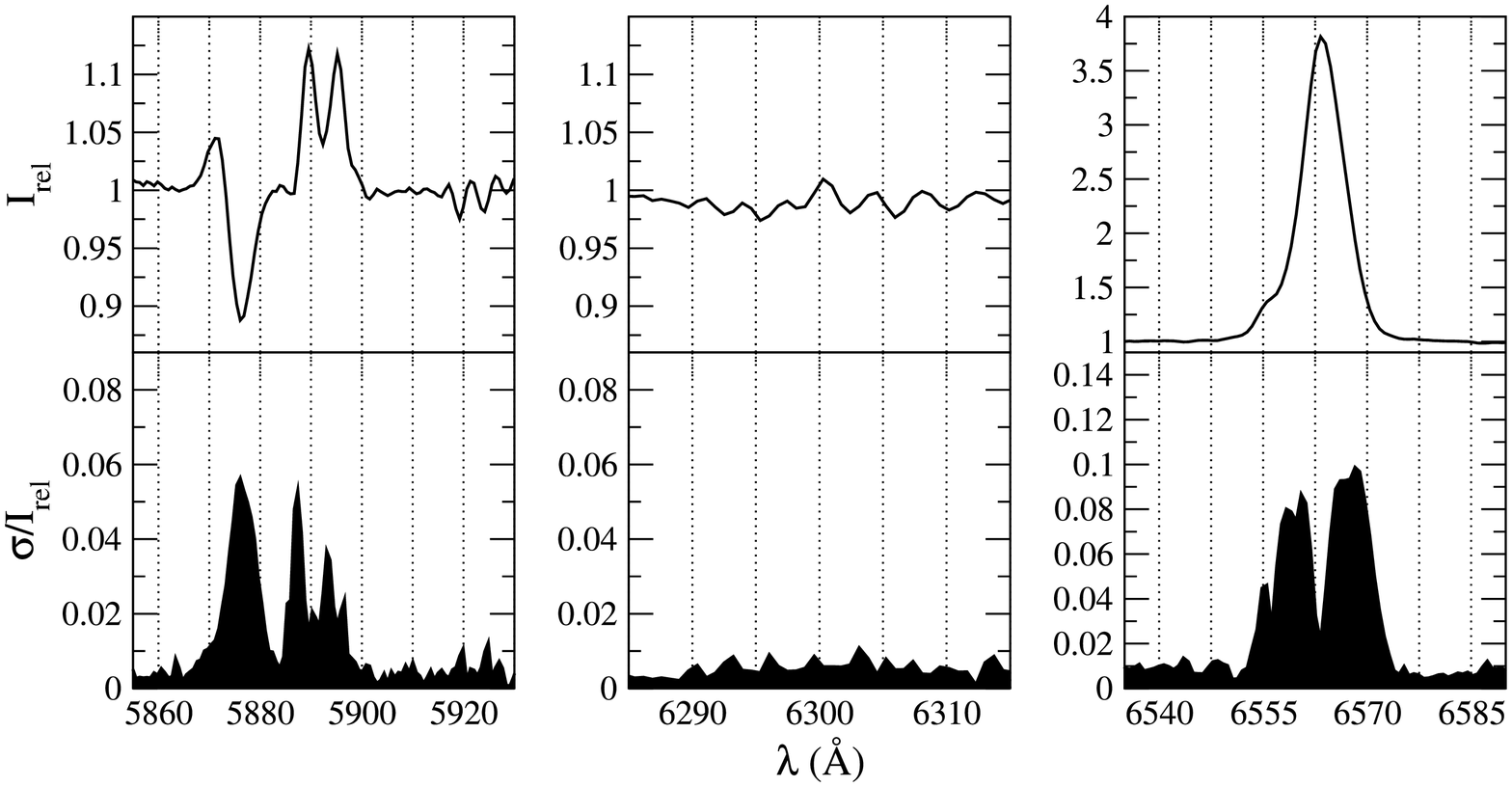} \\
\includegraphics[height=47mm,clip=true]{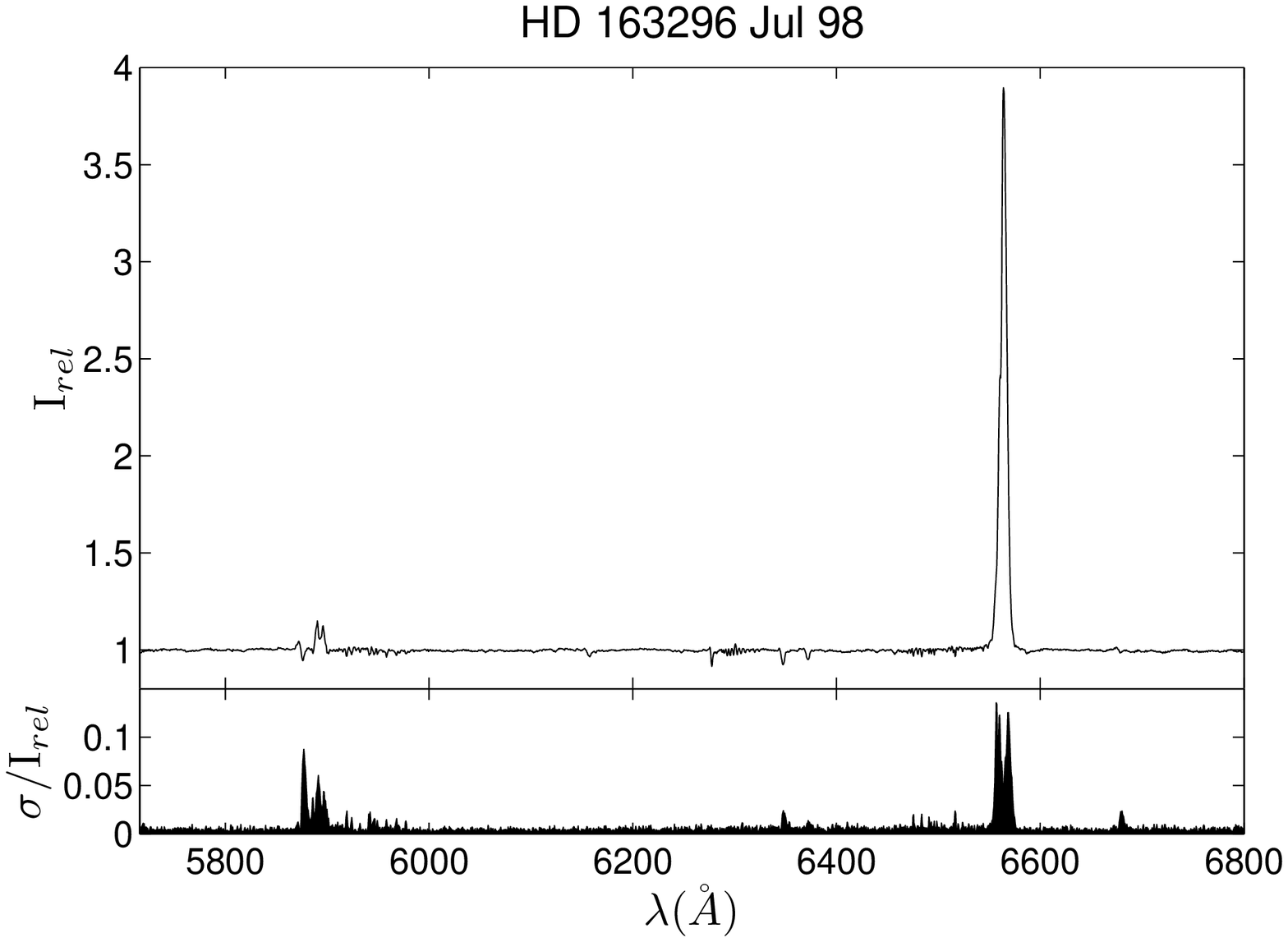}&
\includegraphics[bb=4 77 763 470,height=45mm,clip=true]{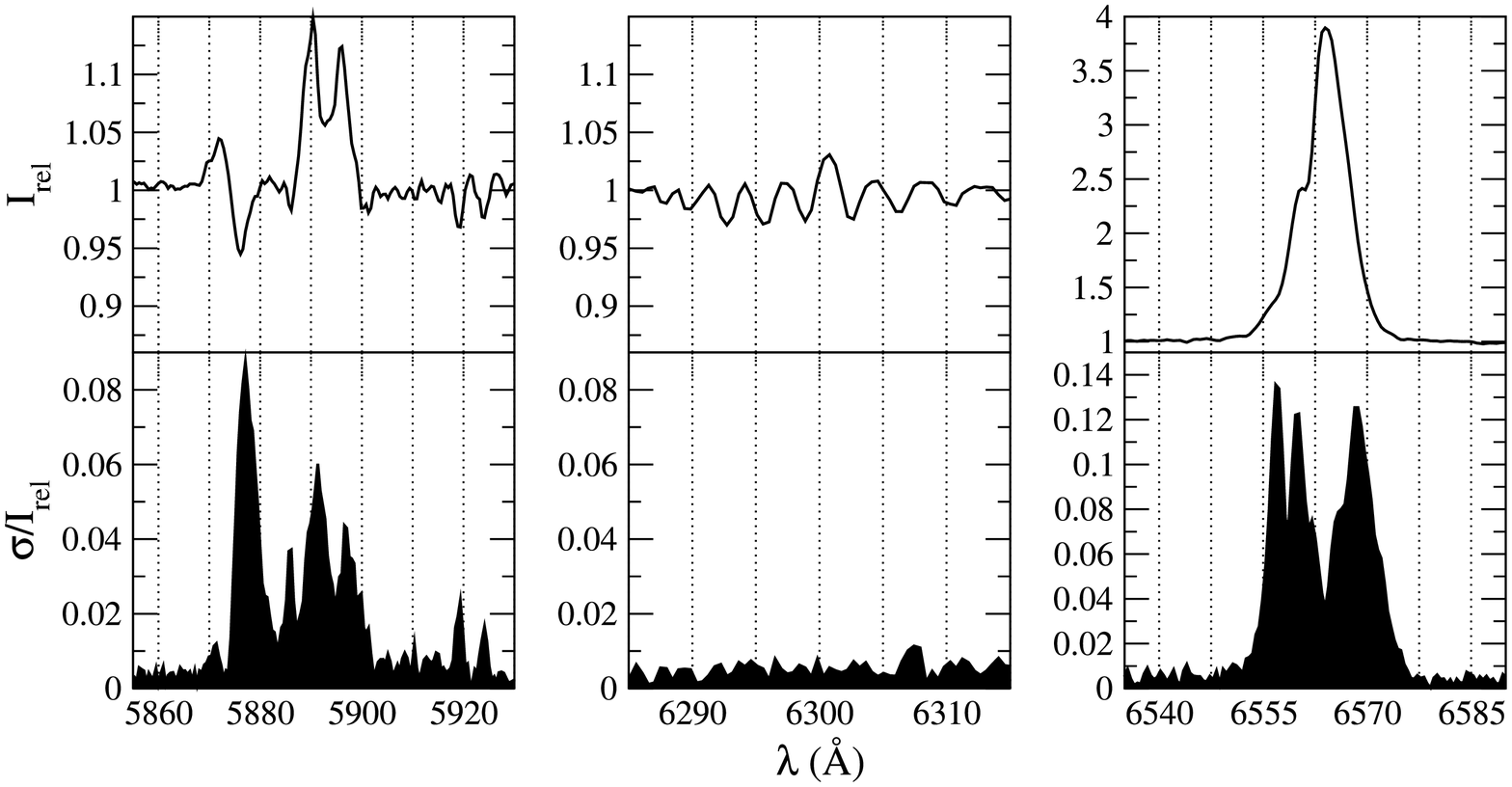} \\
\includegraphics[height=47mm,clip=true]{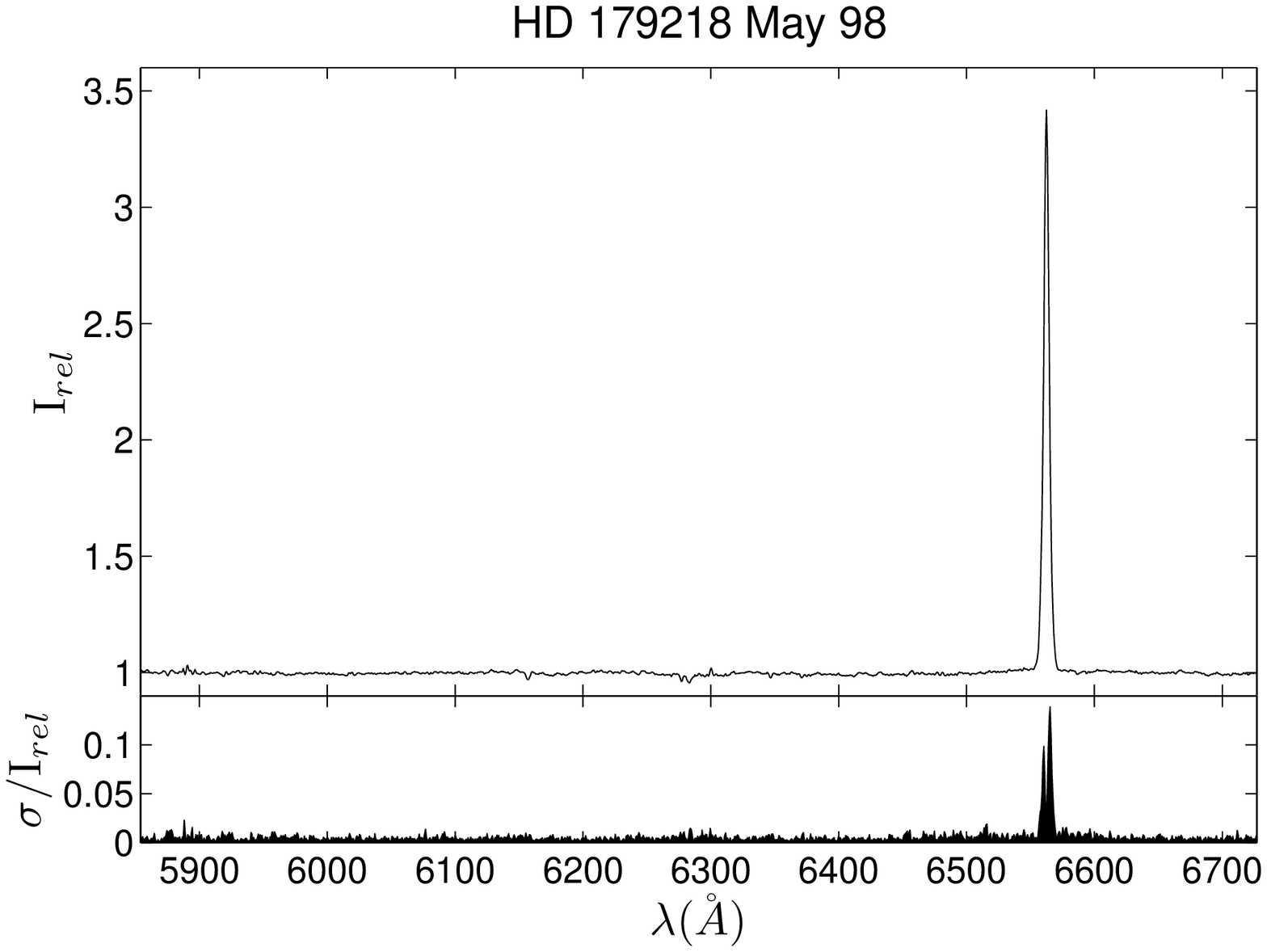}&
\includegraphics[bb=4 77 763 470,height=45mm,clip=true]{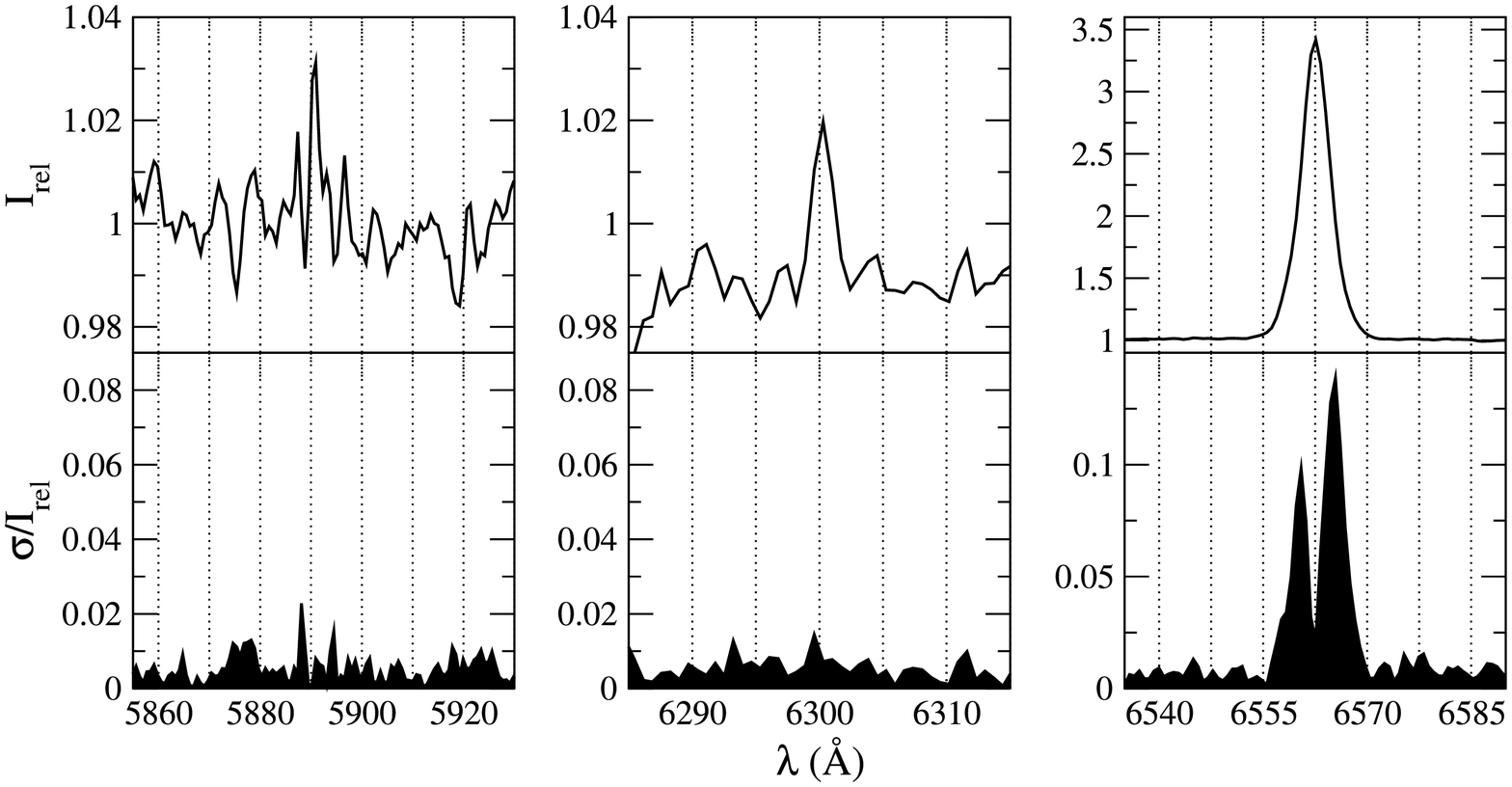} \\
\end{tabular}
\end{table}
\clearpage
\begin{table}
\centering
\renewcommand\arraystretch{10}
\begin{tabular}{cc}
\includegraphics[height=47mm,clip=true]{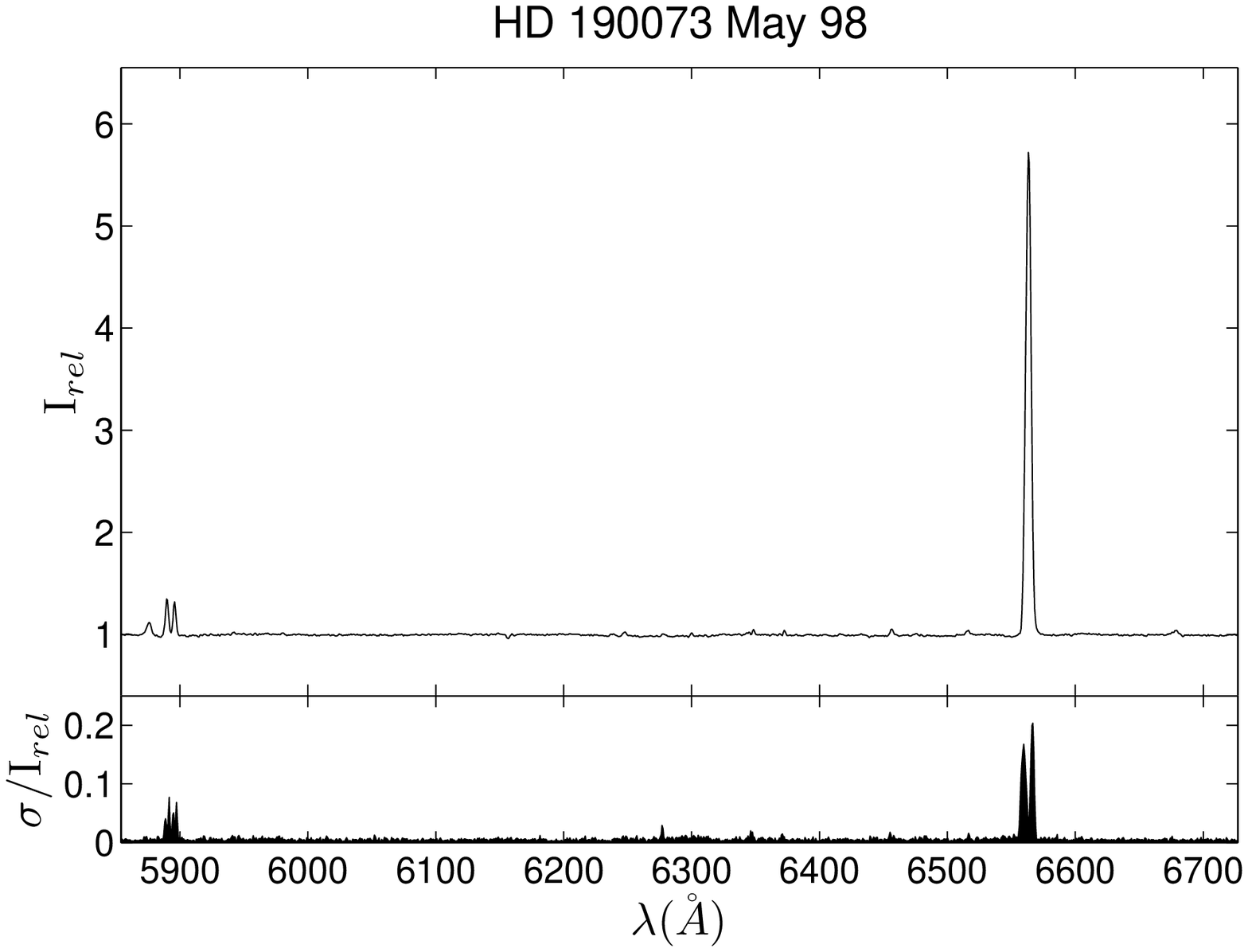}&
\includegraphics[bb=4 77 763 470,height=45mm,clip=true]{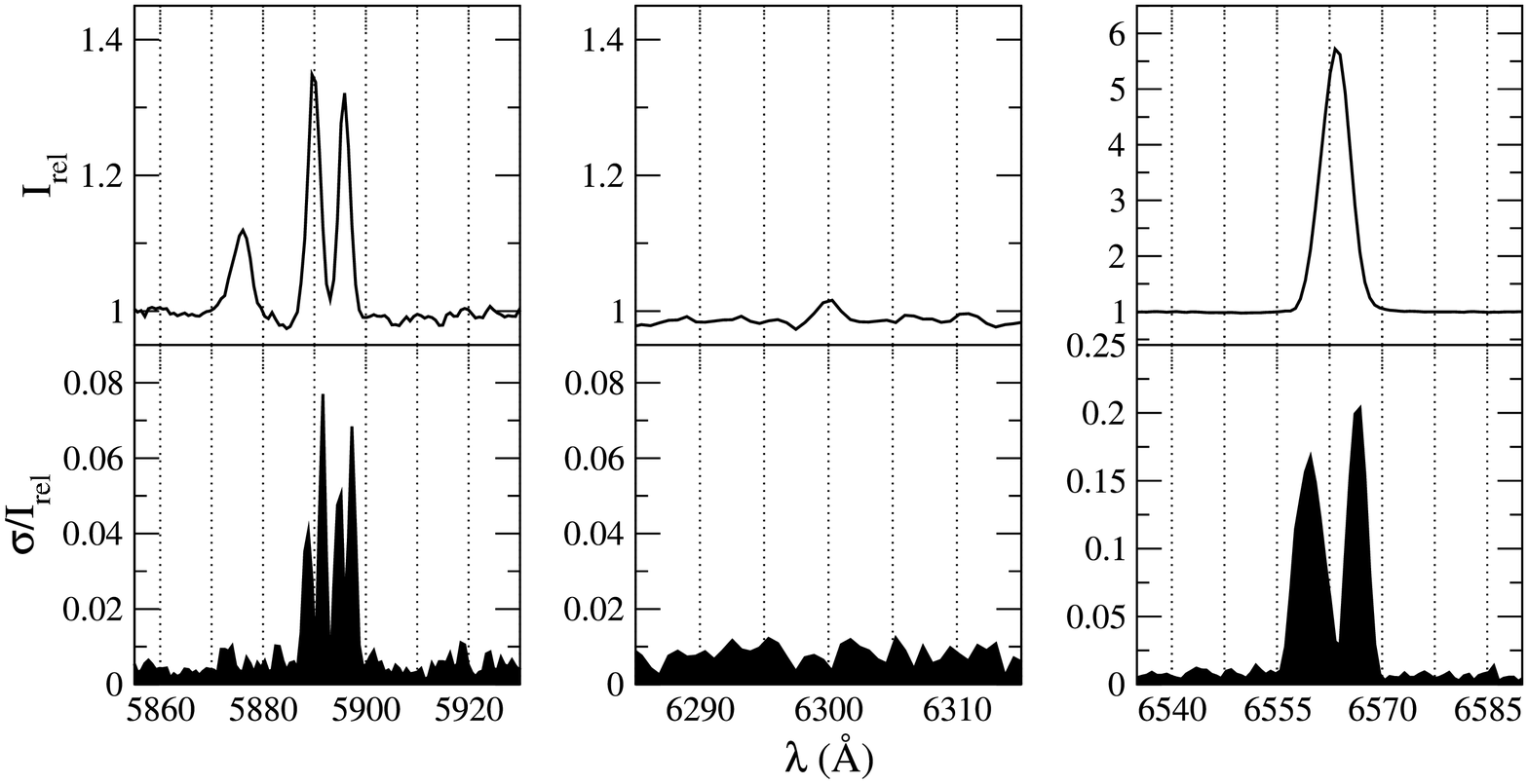} \\
\includegraphics[height=47mm,clip=true]{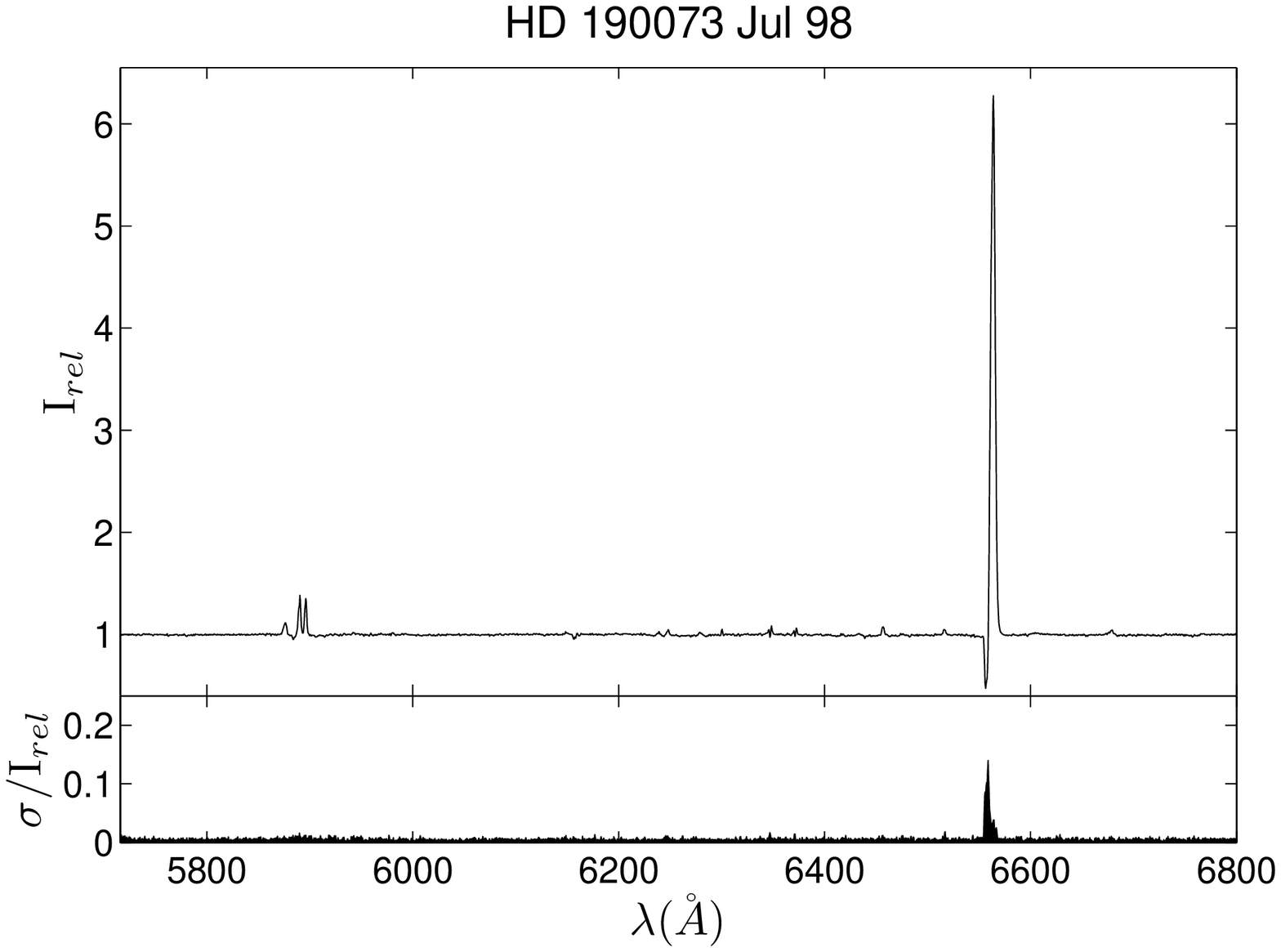}&
\includegraphics[bb=4 77 763 470,height=45mm,clip=true]{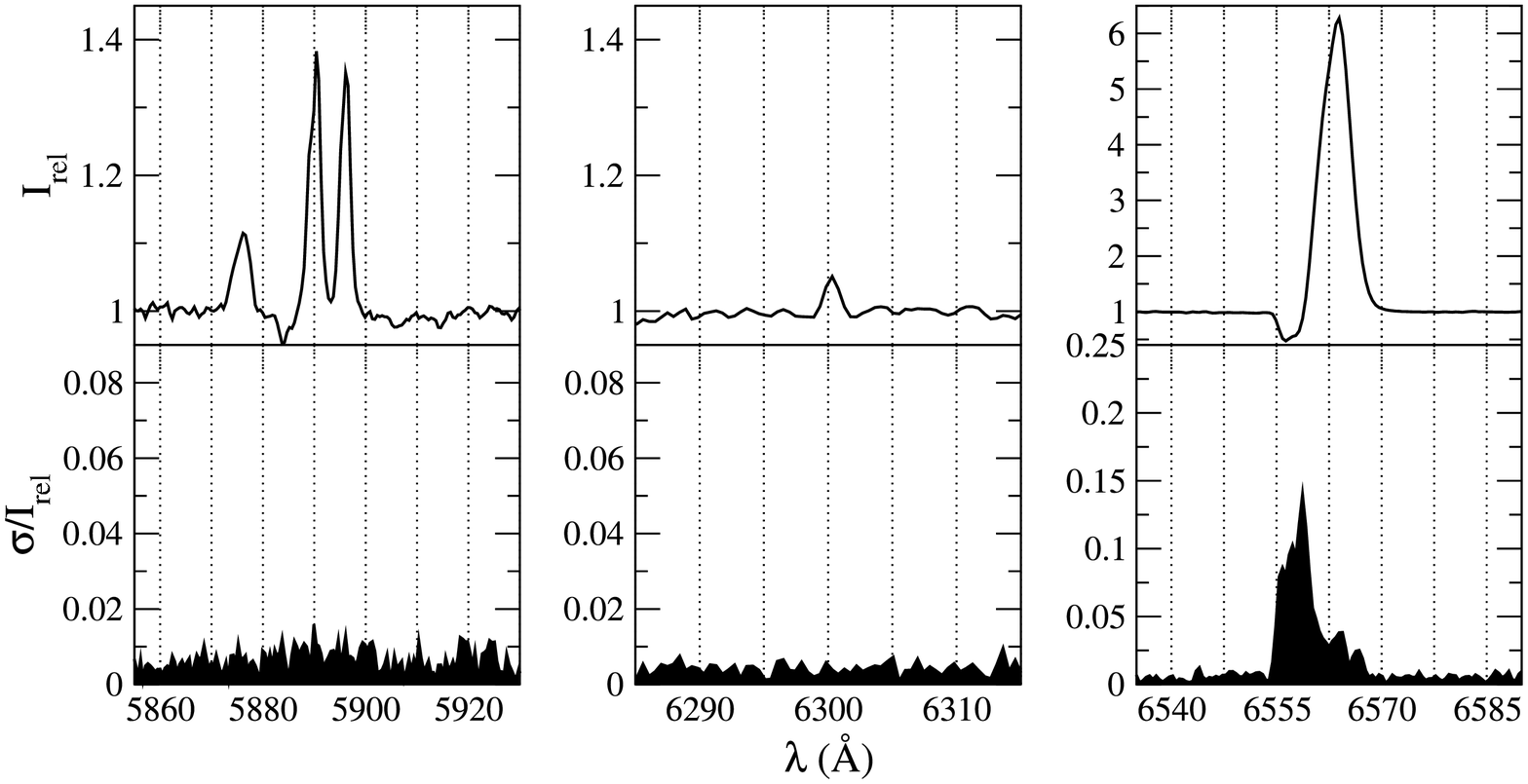} \\
\includegraphics[height=47mm,clip=true]{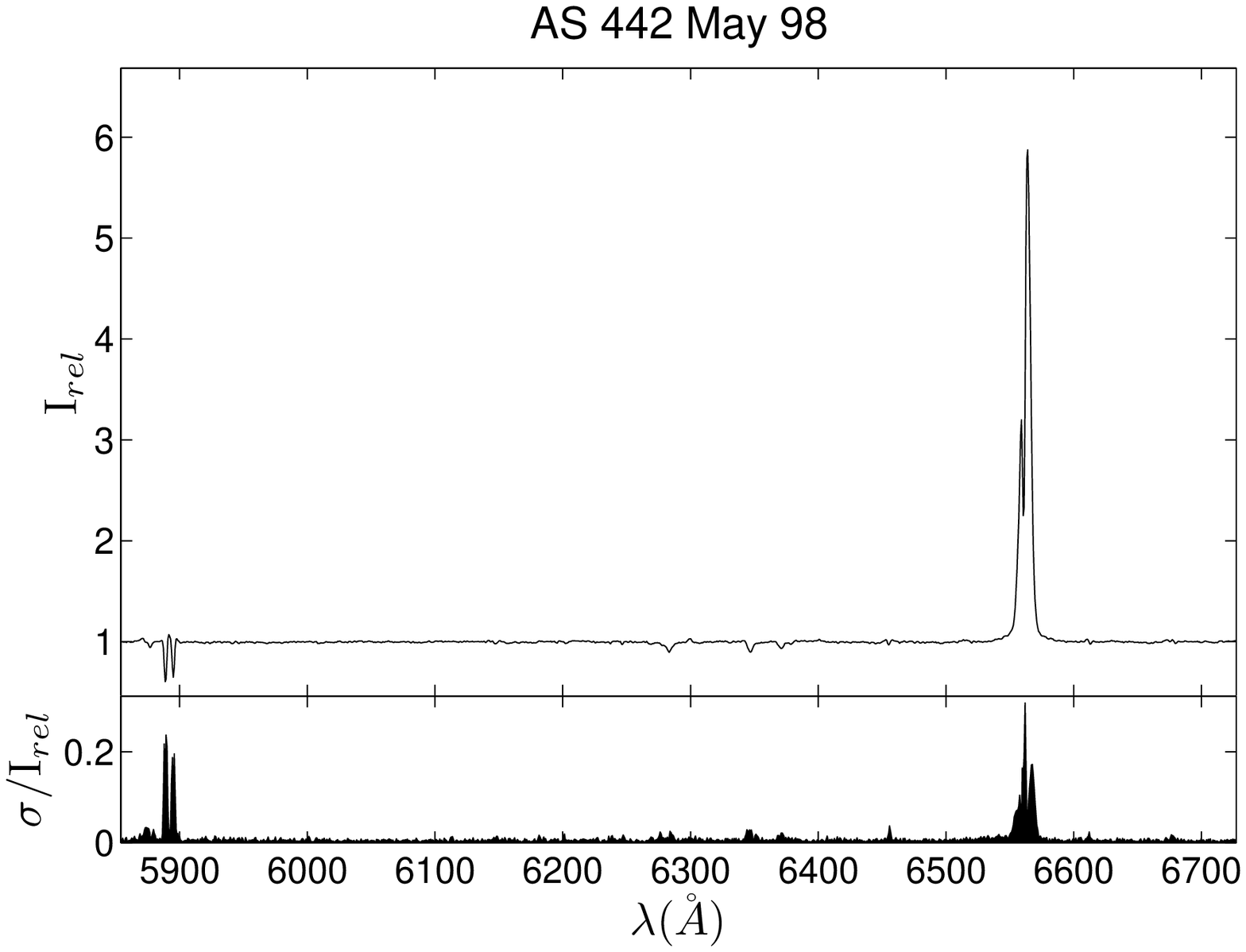}&
\includegraphics[bb=4 77 763 470,height=45mm,clip=true]{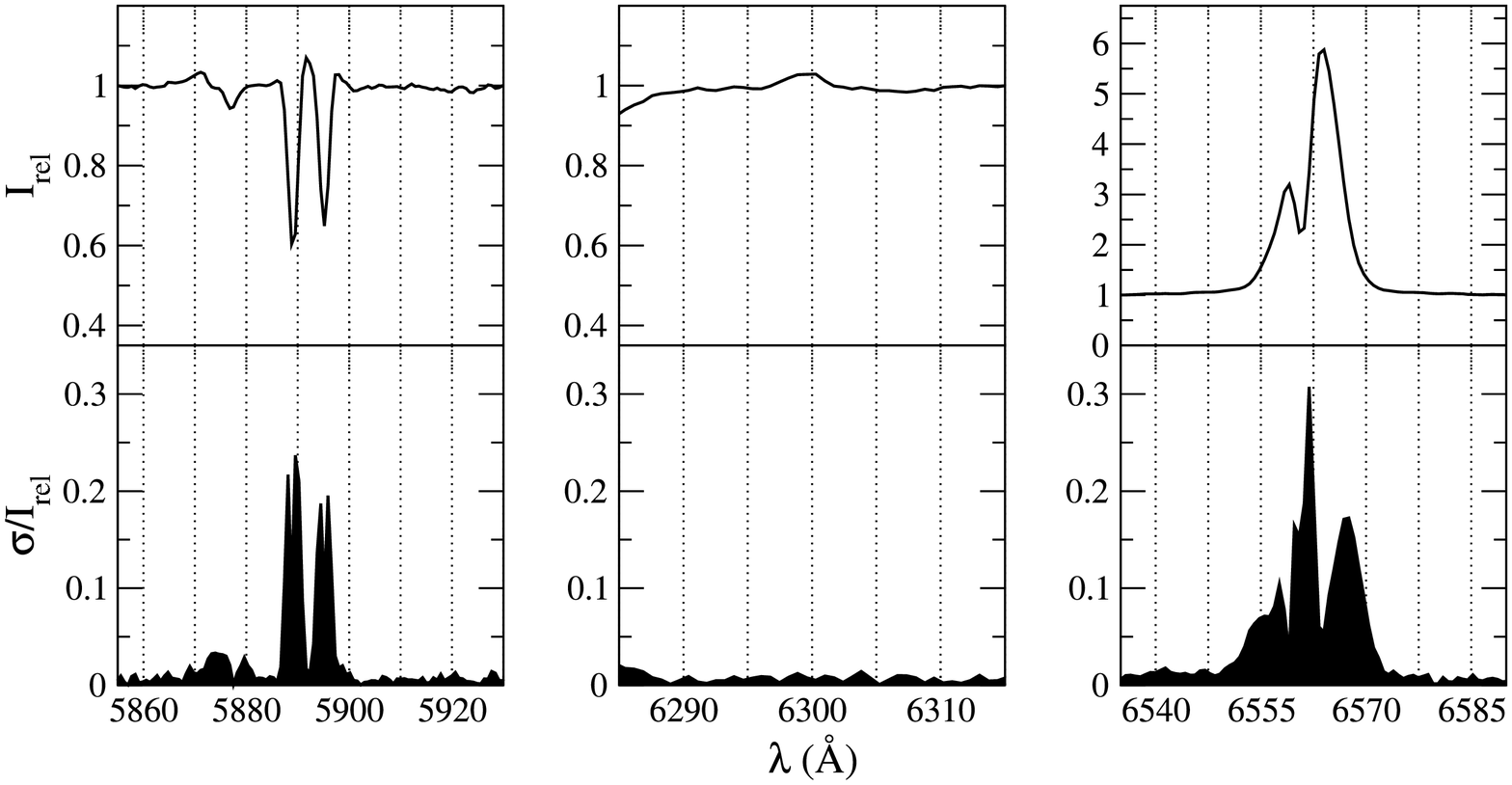} \\
\includegraphics[height=47mm,clip=true]{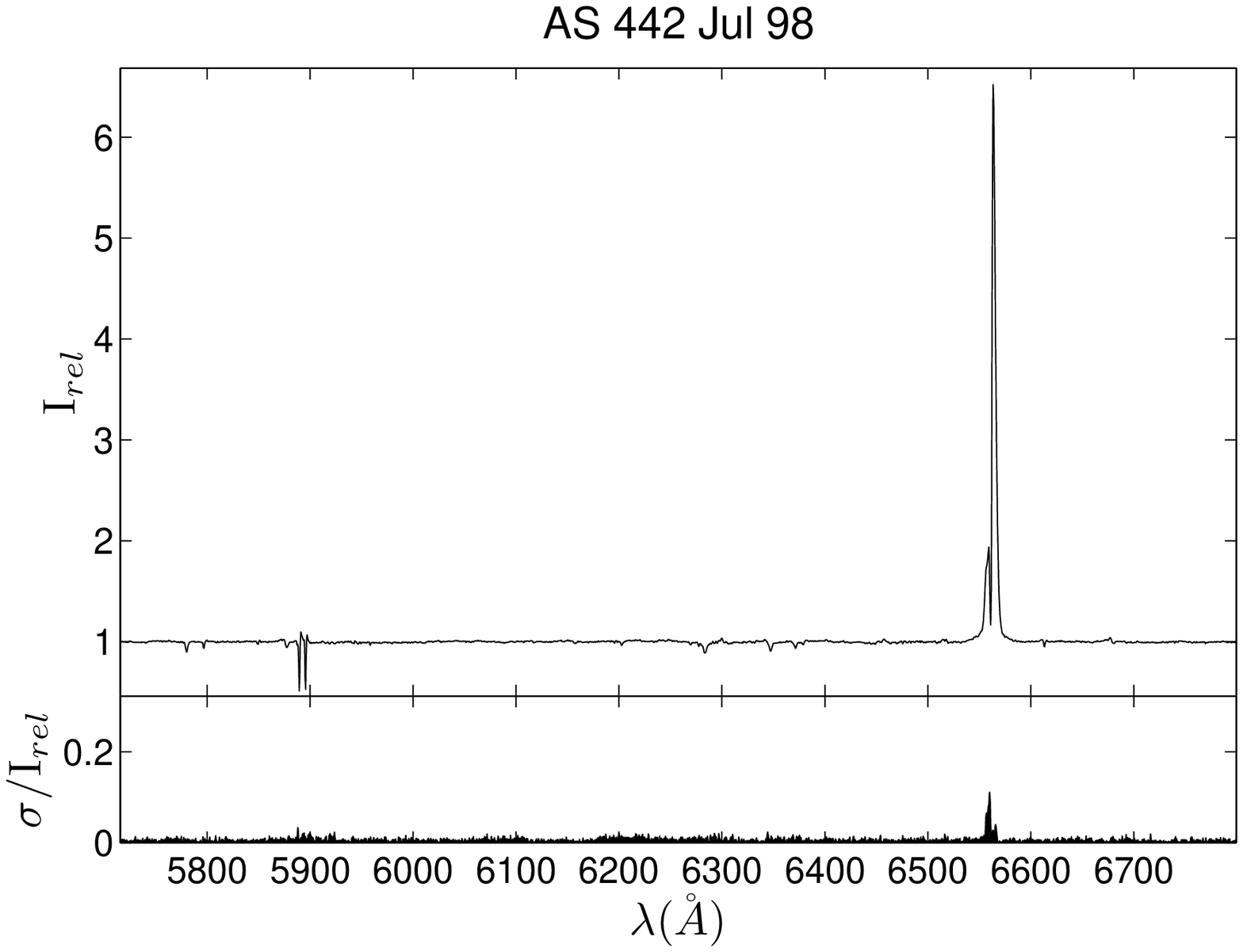}&
\includegraphics[bb=4 77 763 470,height=45mm,clip=true]{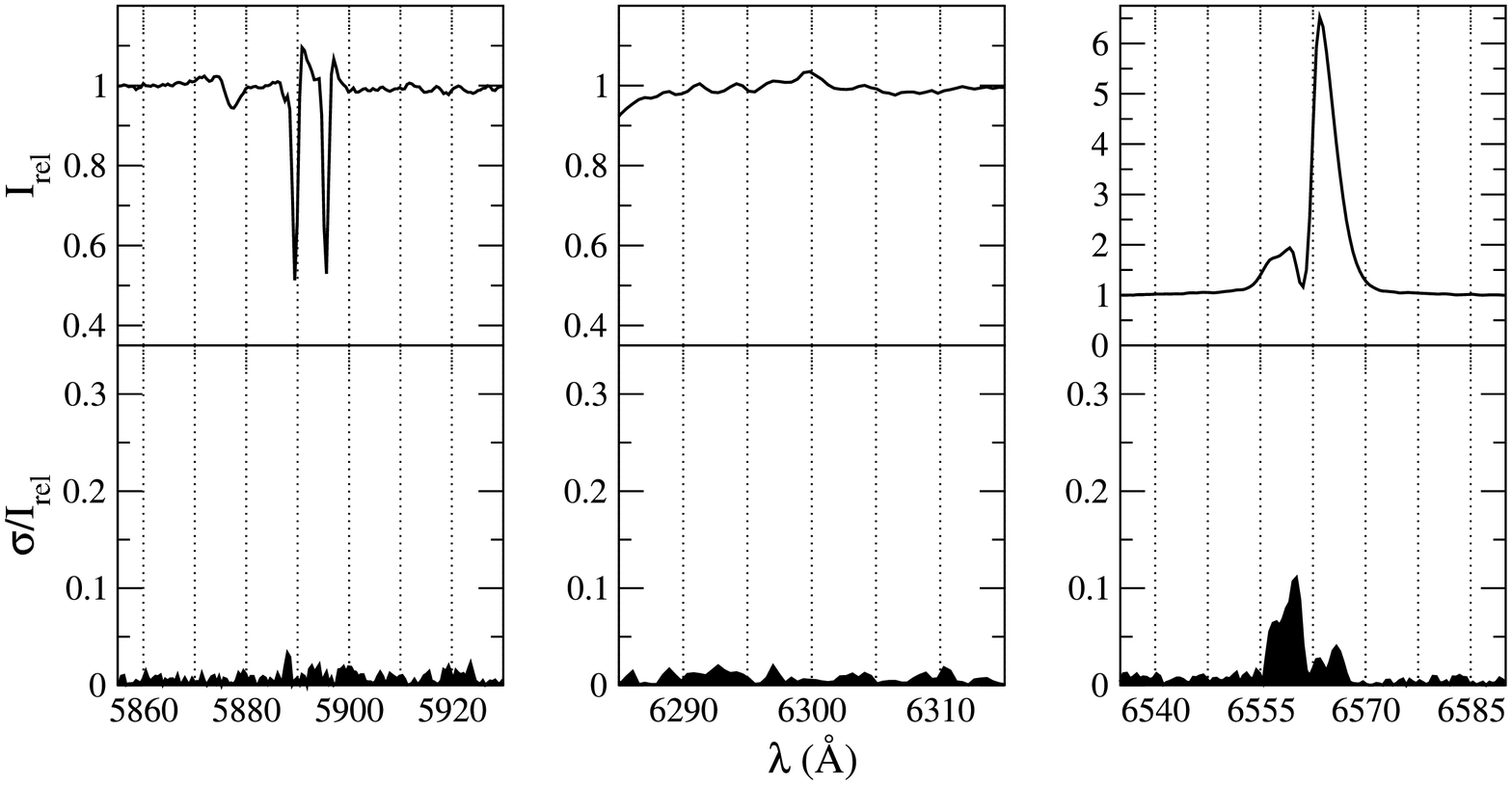} \\
\end{tabular}
\end{table}
\clearpage
\begin{table}
\centering
\renewcommand\arraystretch{10}
\begin{tabular}{cc}
\includegraphics[height=47mm,clip=true]{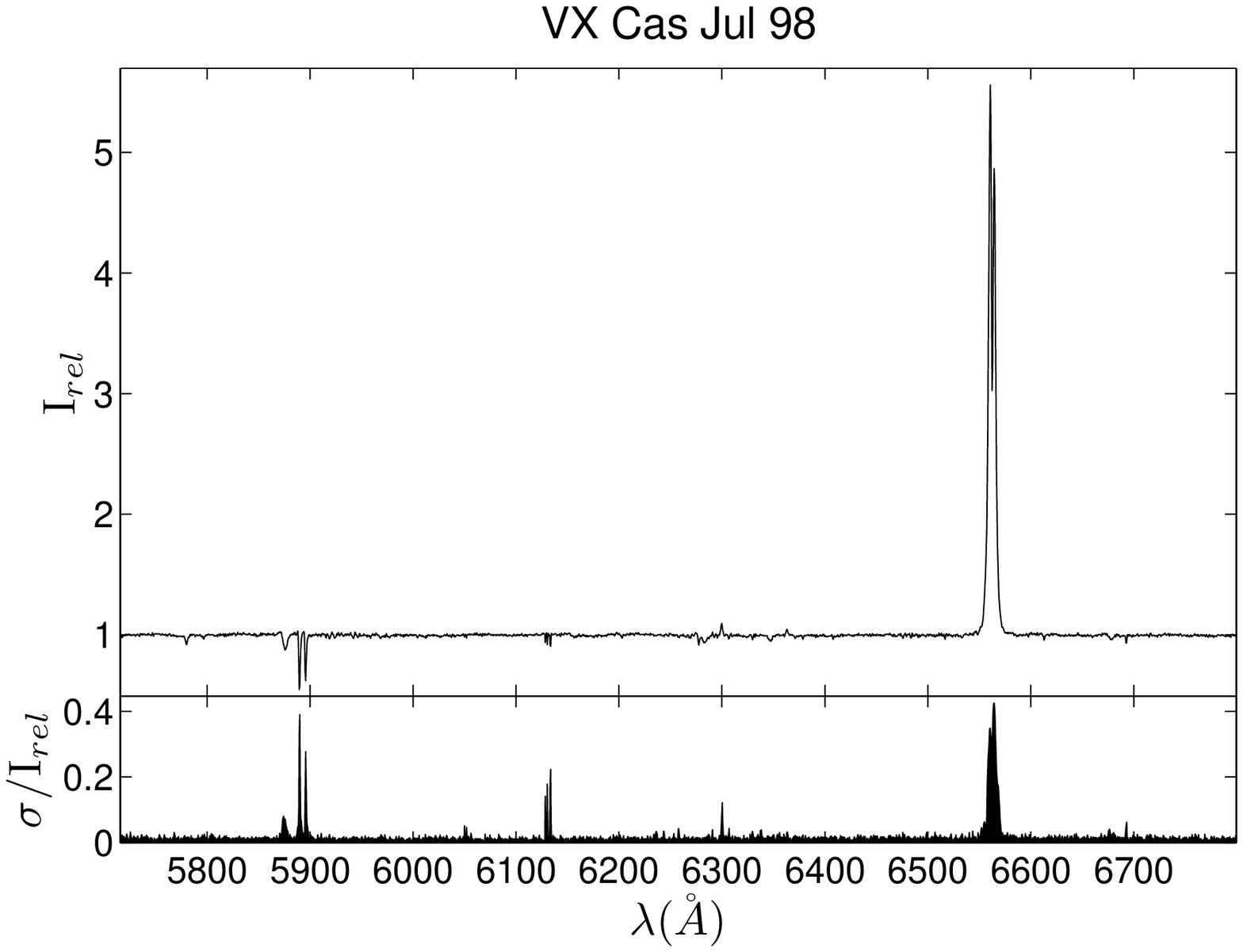}&
\includegraphics[bb=4 77 763 470,height=45mm,clip=true]{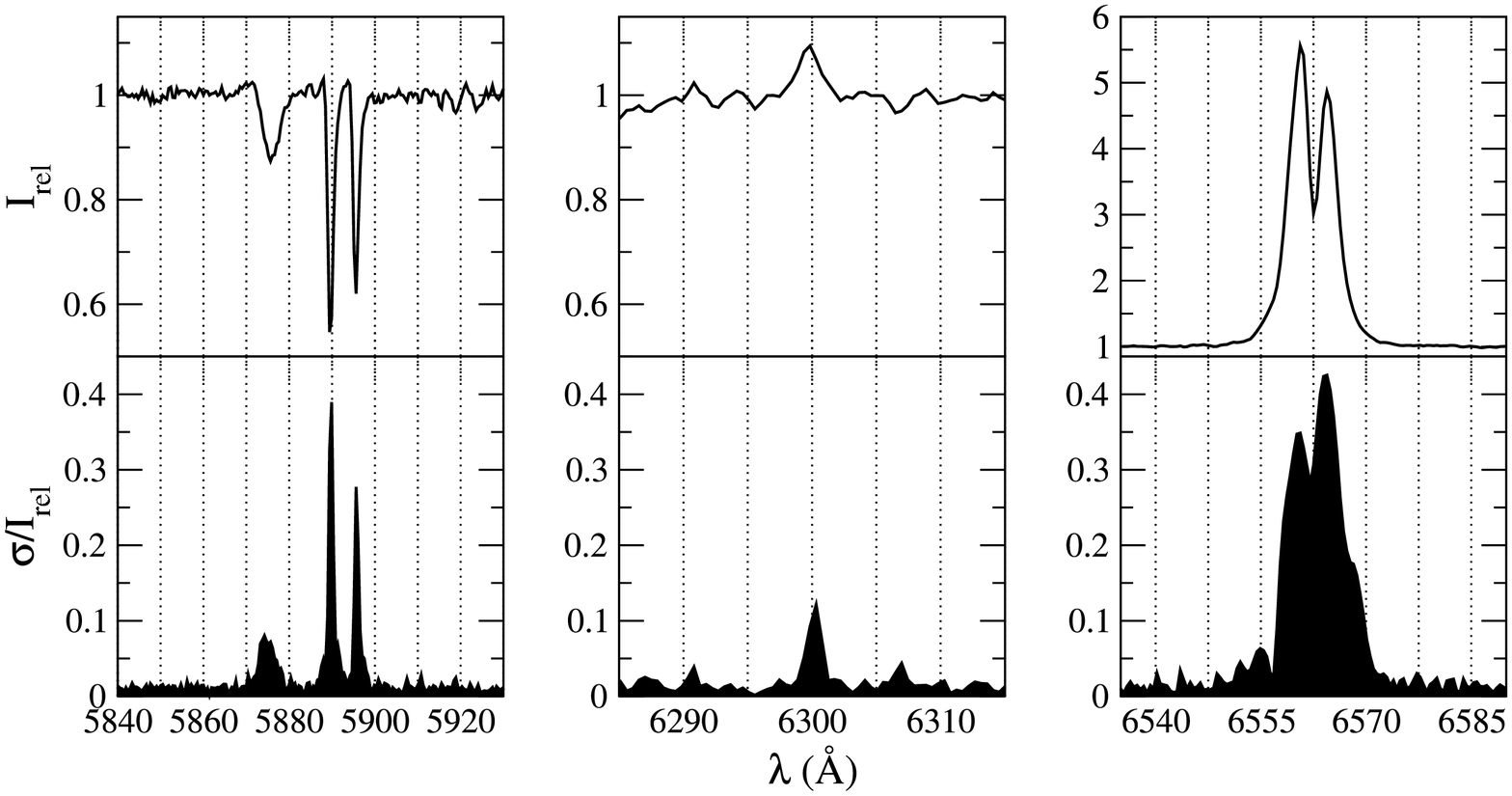} \\ 
\includegraphics[height=47mm,clip=true]{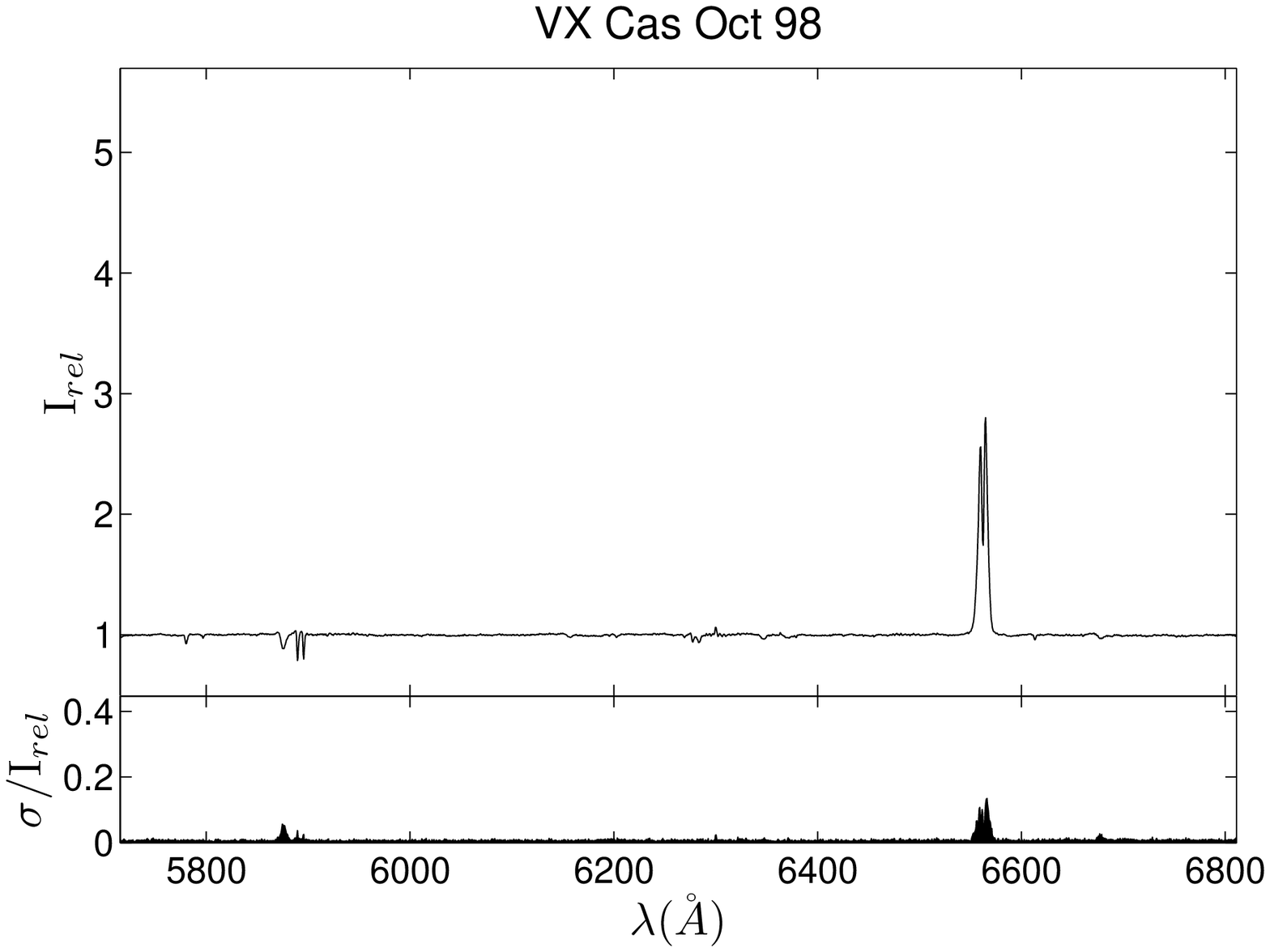}&
\includegraphics[bb=4 77 763 470,height=45mm,clip=true]{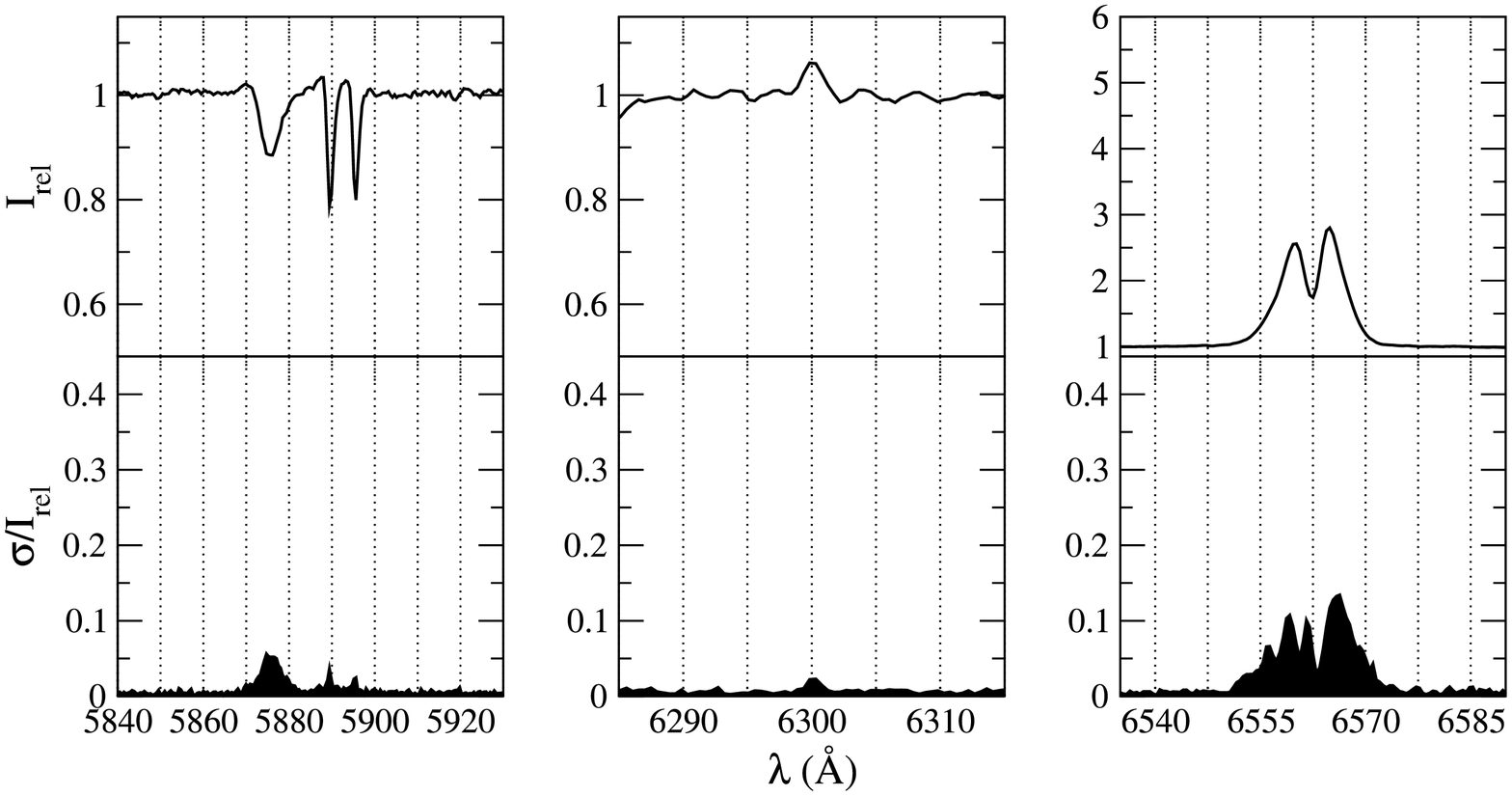} \\ 
\includegraphics[height=47mm,clip=true]{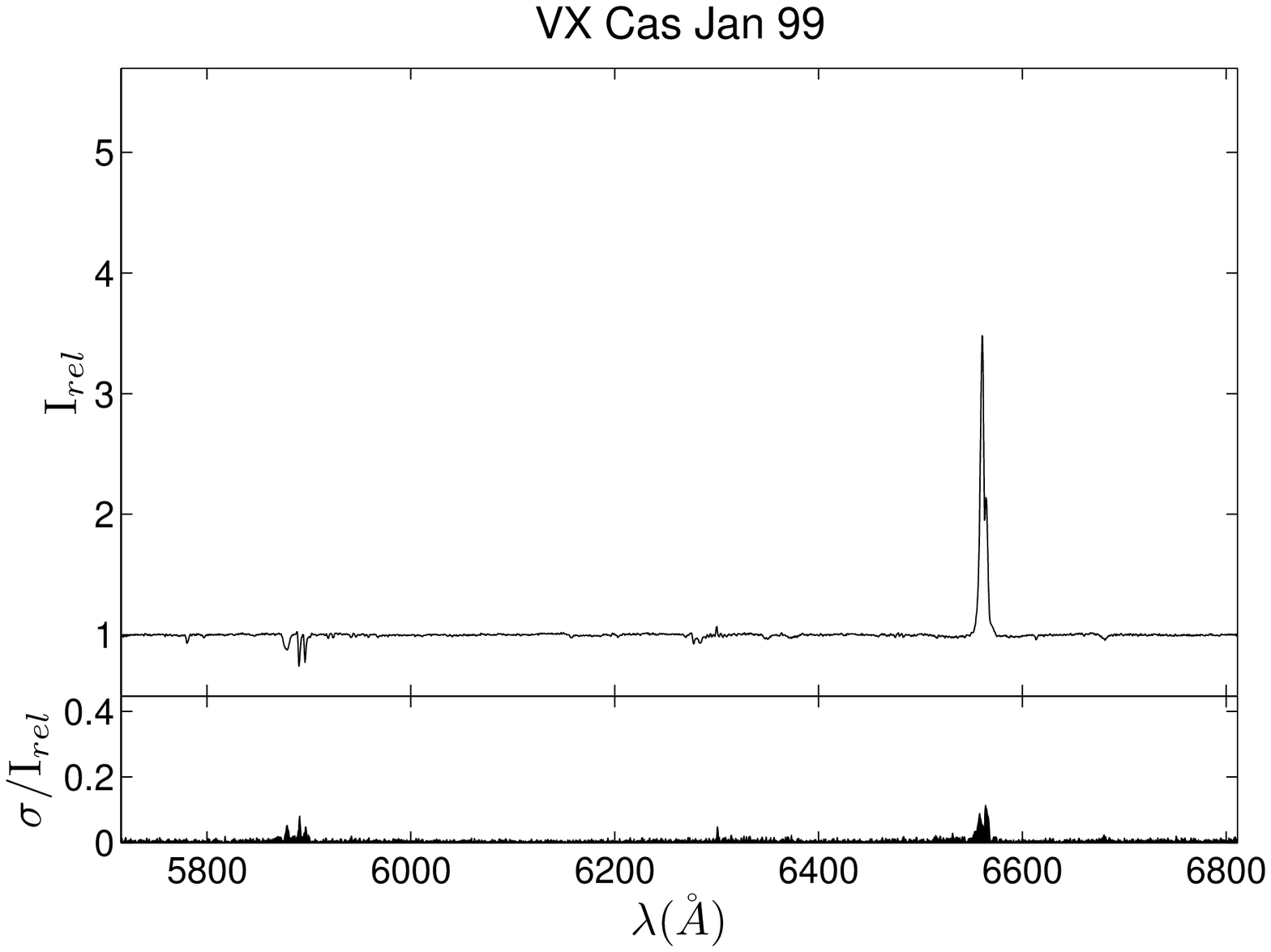}&
\includegraphics[bb=4 77 763 470,height=45mm,clip=true]{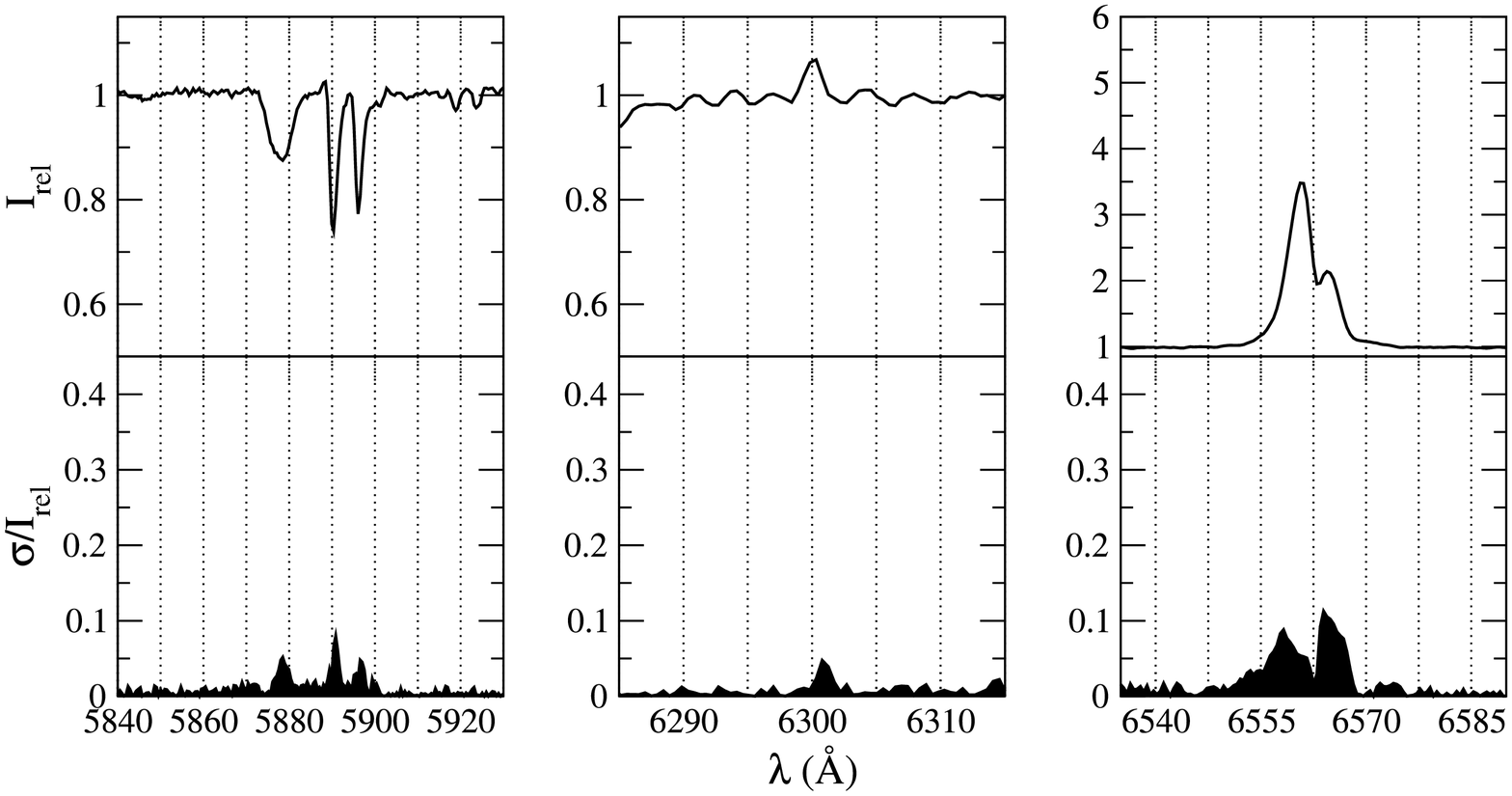} \\ 
\includegraphics[height=47mm,clip=true]{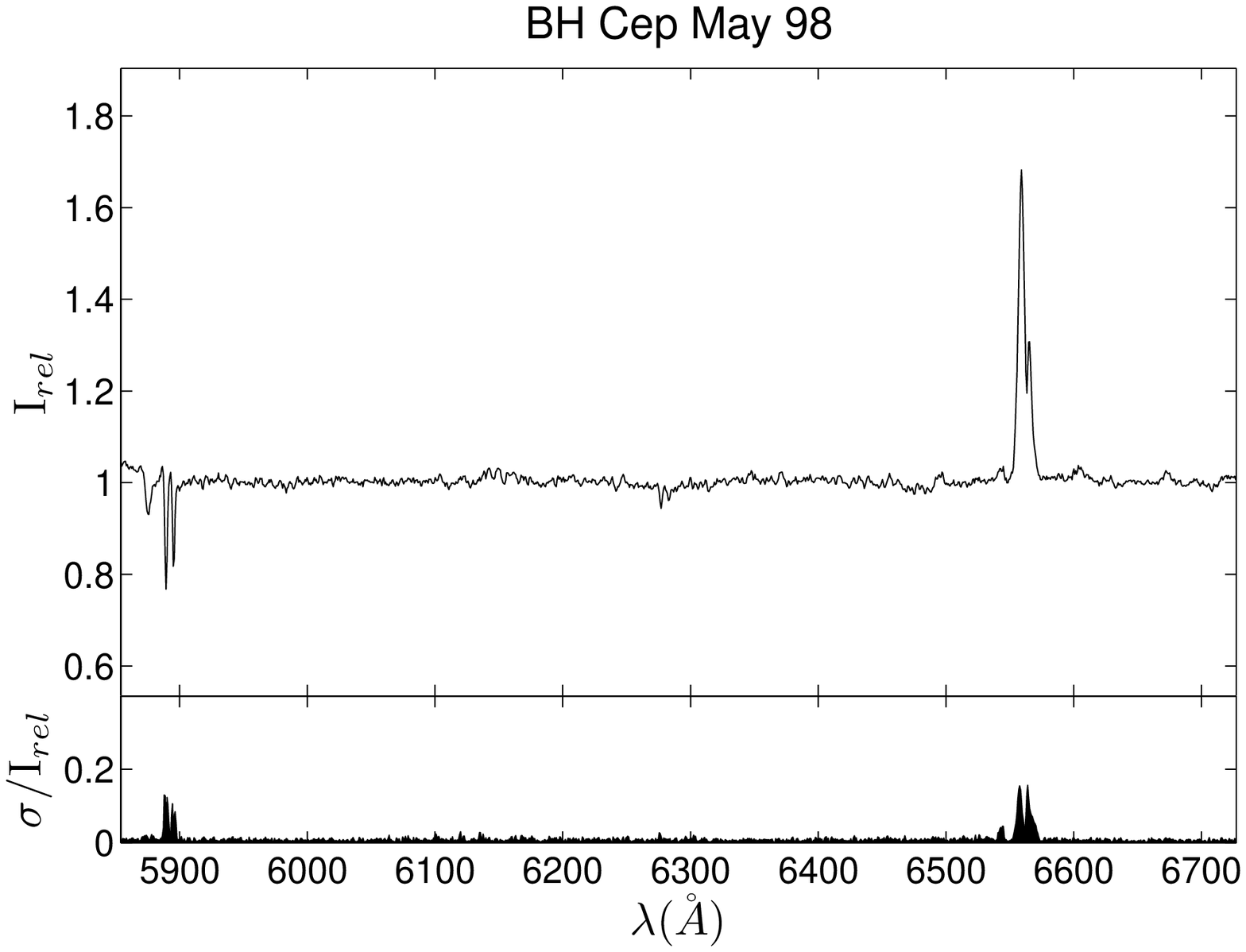}&
\includegraphics[bb=4 77 763 470,height=45mm,clip=true]{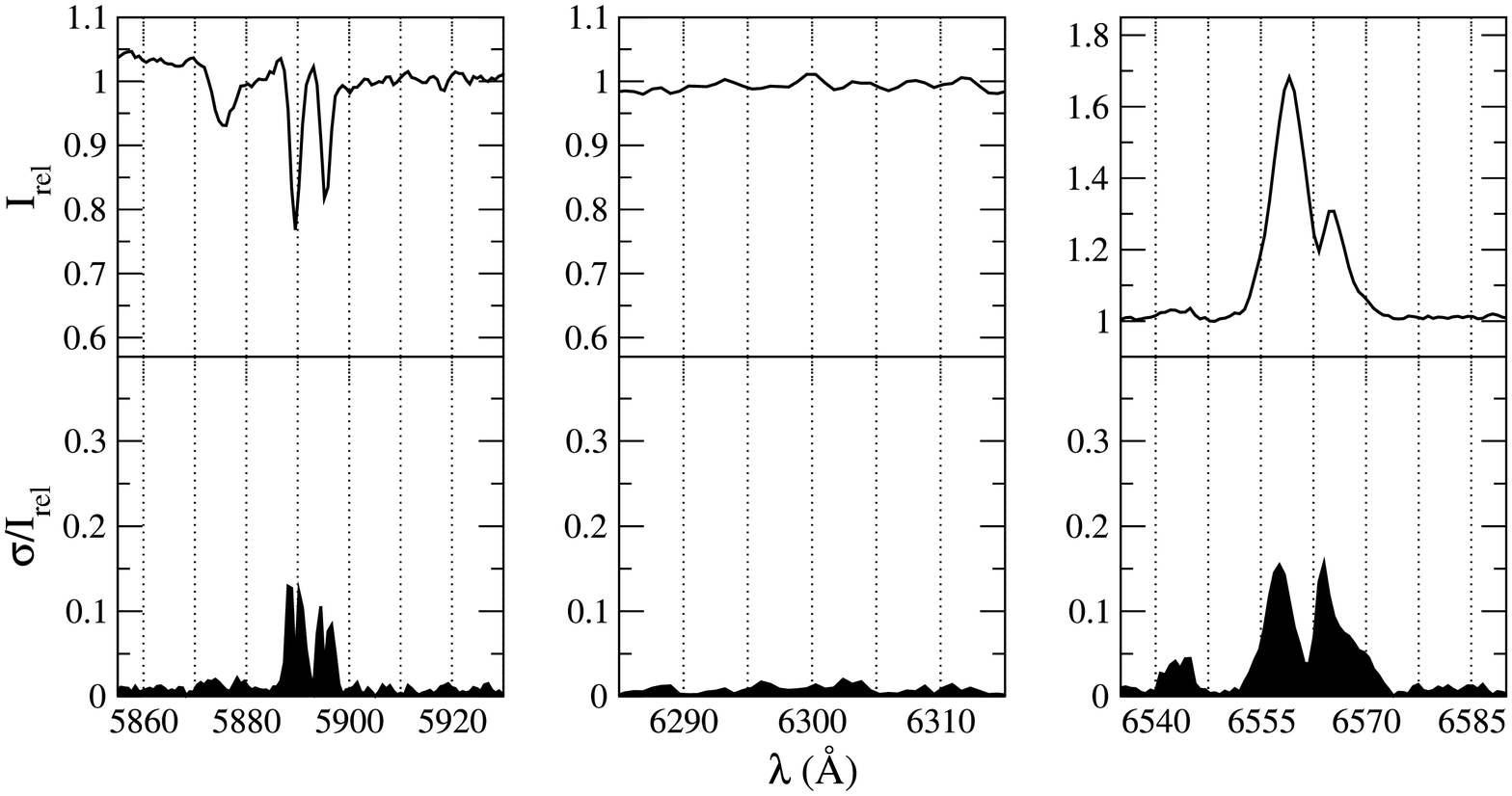} \\ 
\end{tabular}
\end{table}
\clearpage
\begin{table}
\centering
\renewcommand\arraystretch{10}
\begin{tabular}{cc}
\includegraphics[height=47mm,clip=true]{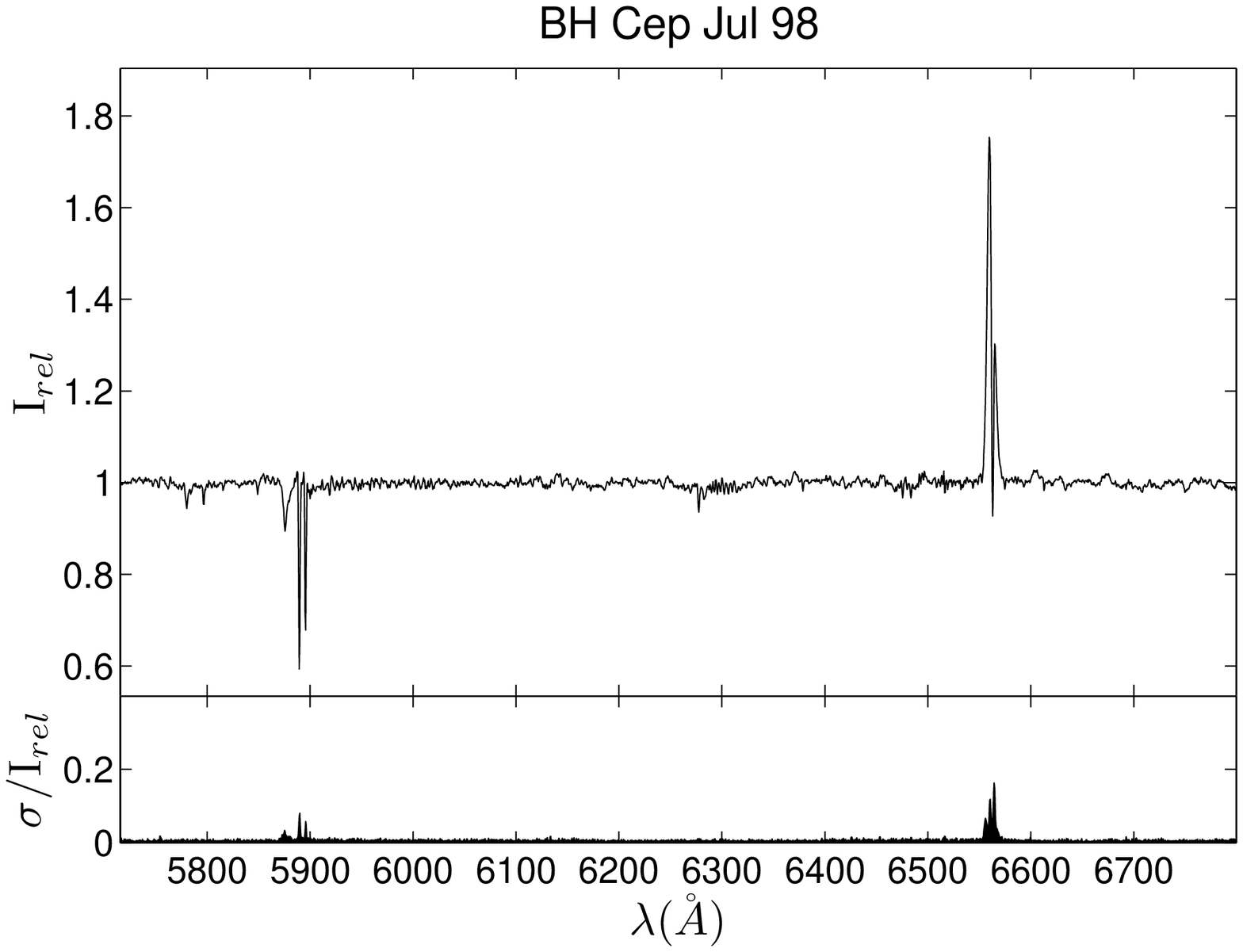}&
\includegraphics[bb=4 77 763 470,height=45mm,clip=true]{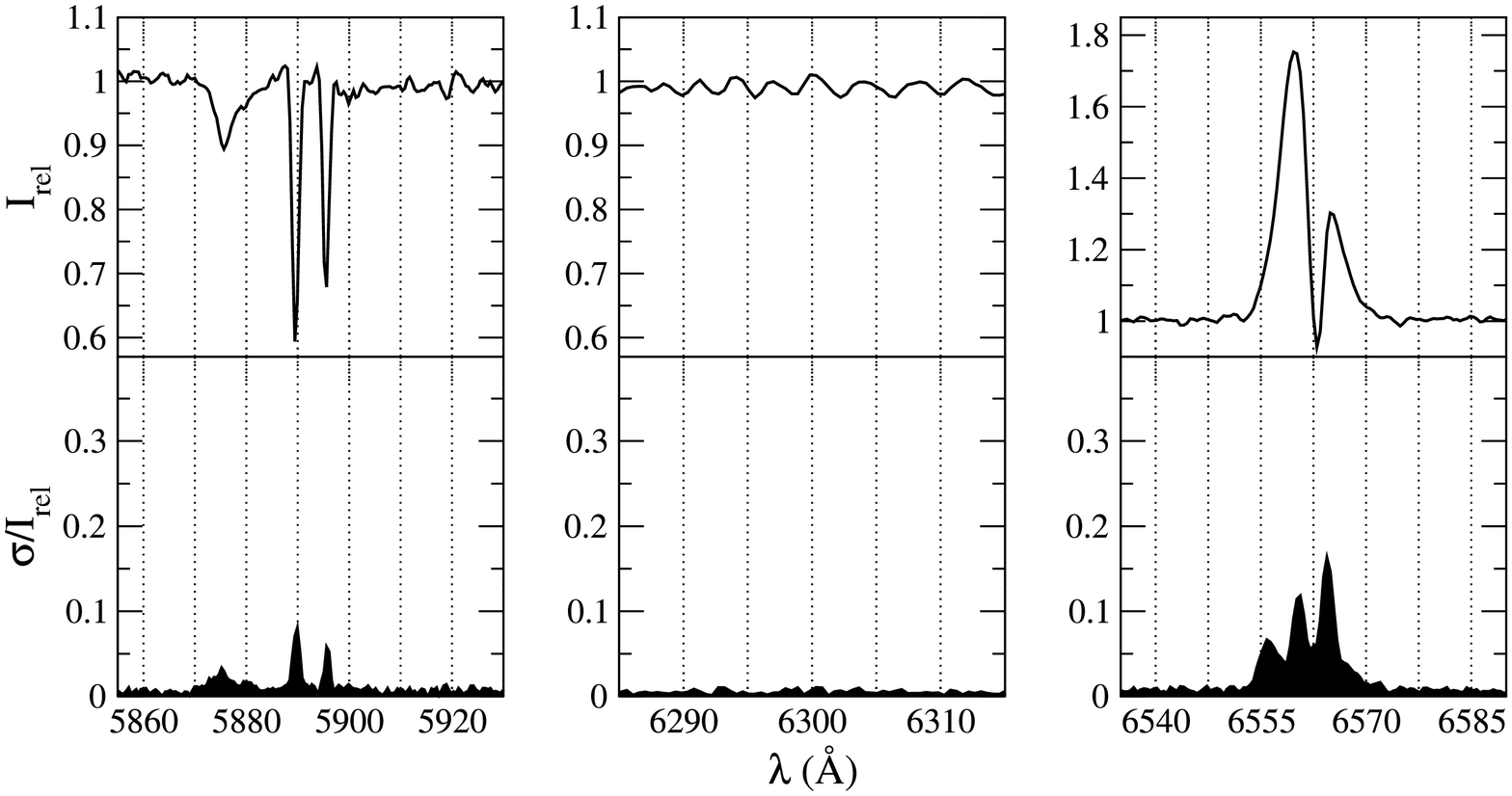} \\ 
\includegraphics[height=47mm,clip=true]{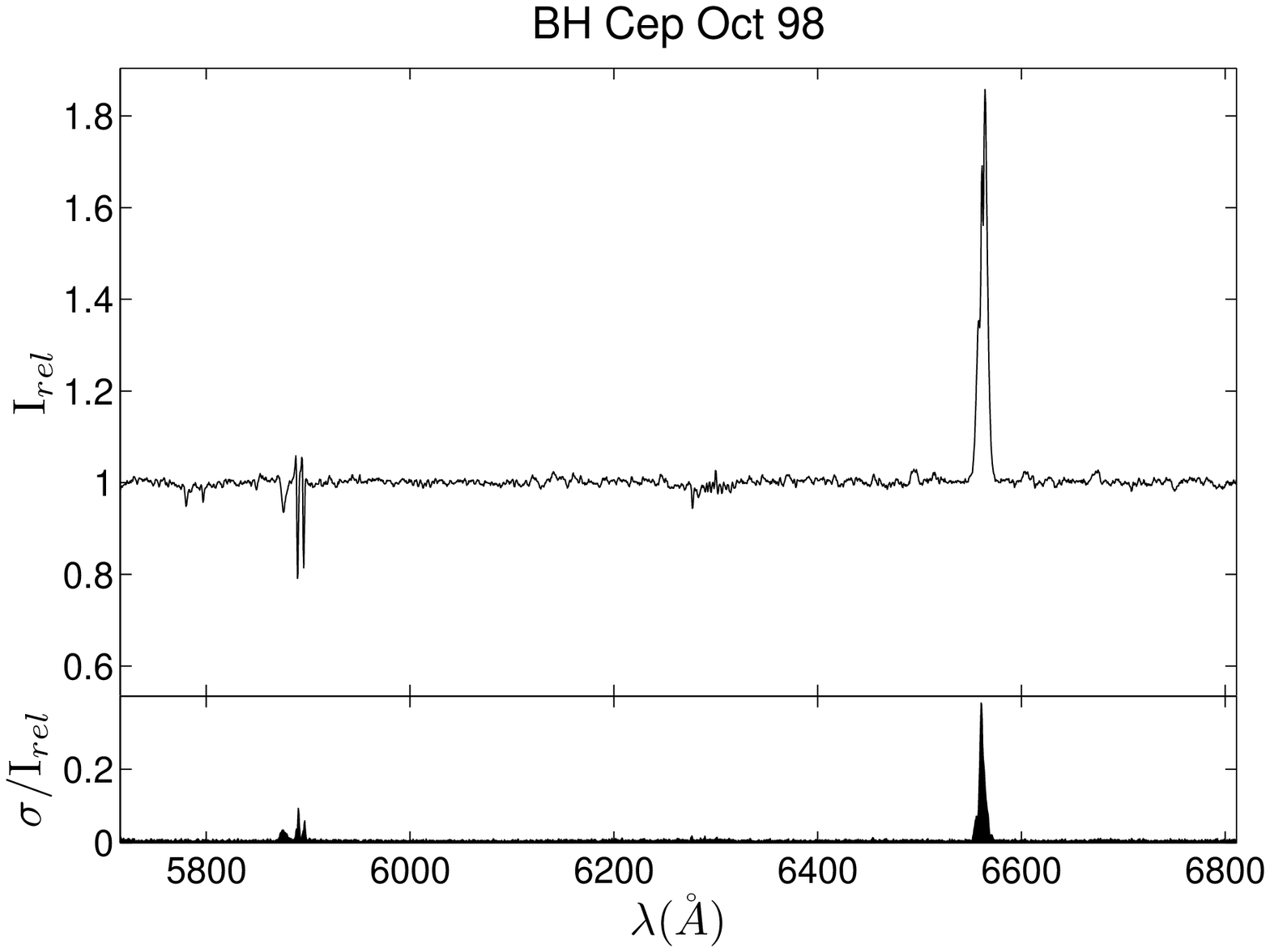}&
\includegraphics[bb=4 77 763 470,height=45mm,clip=true]{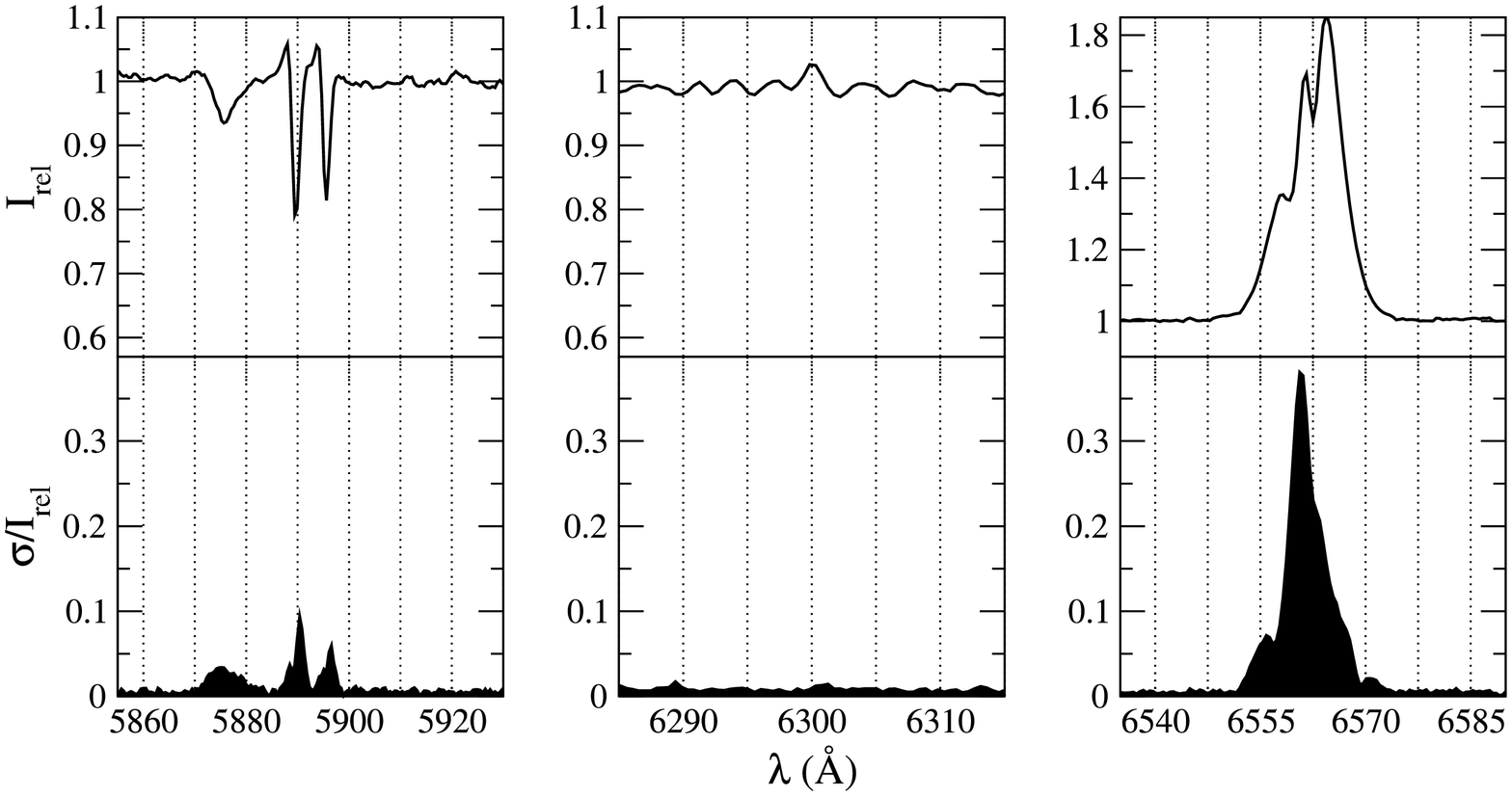} \\ 
\includegraphics[height=47mm,clip=true]{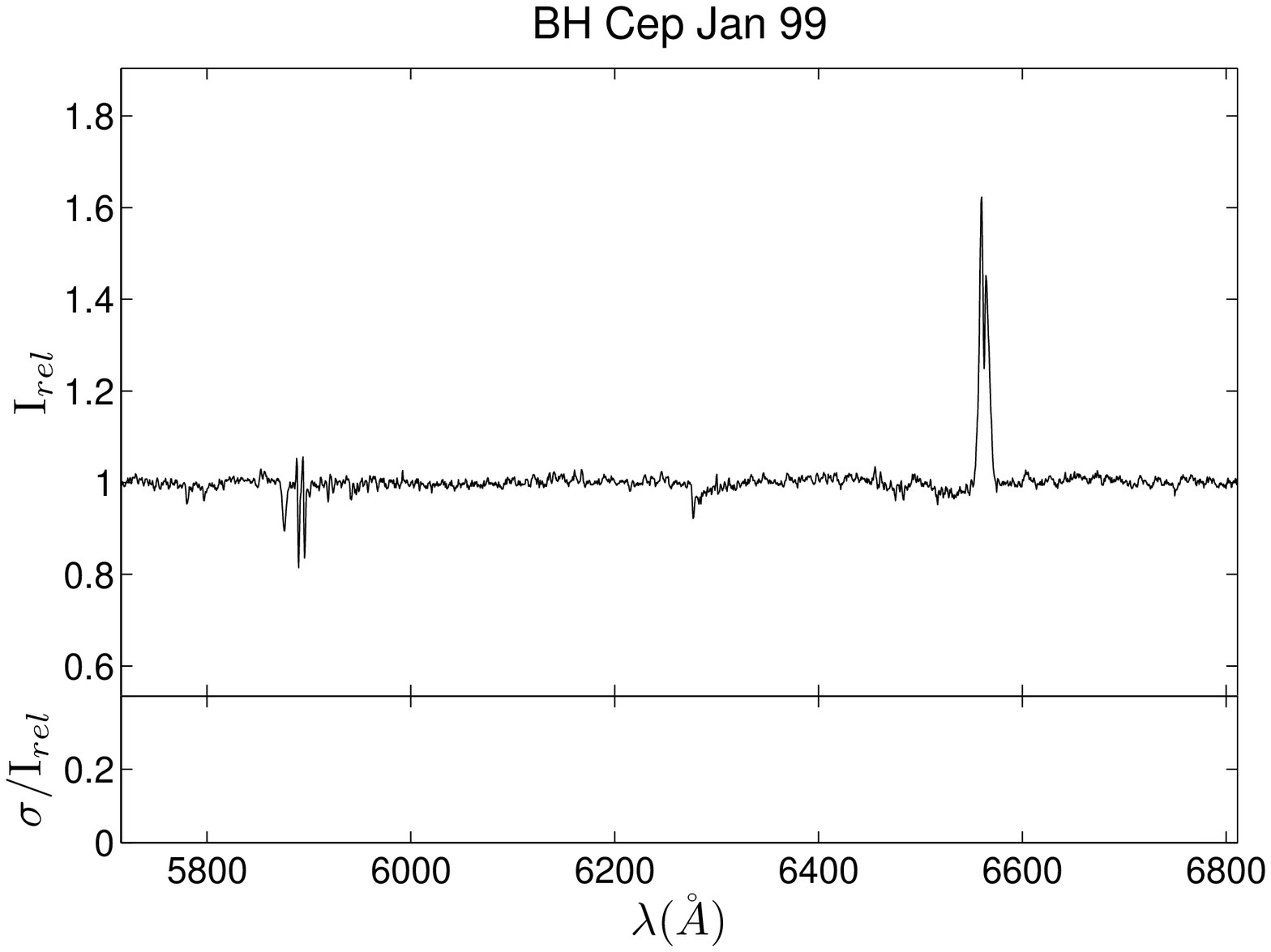}&
\includegraphics[bb=4 77 763 470,height=45mm,clip=true]{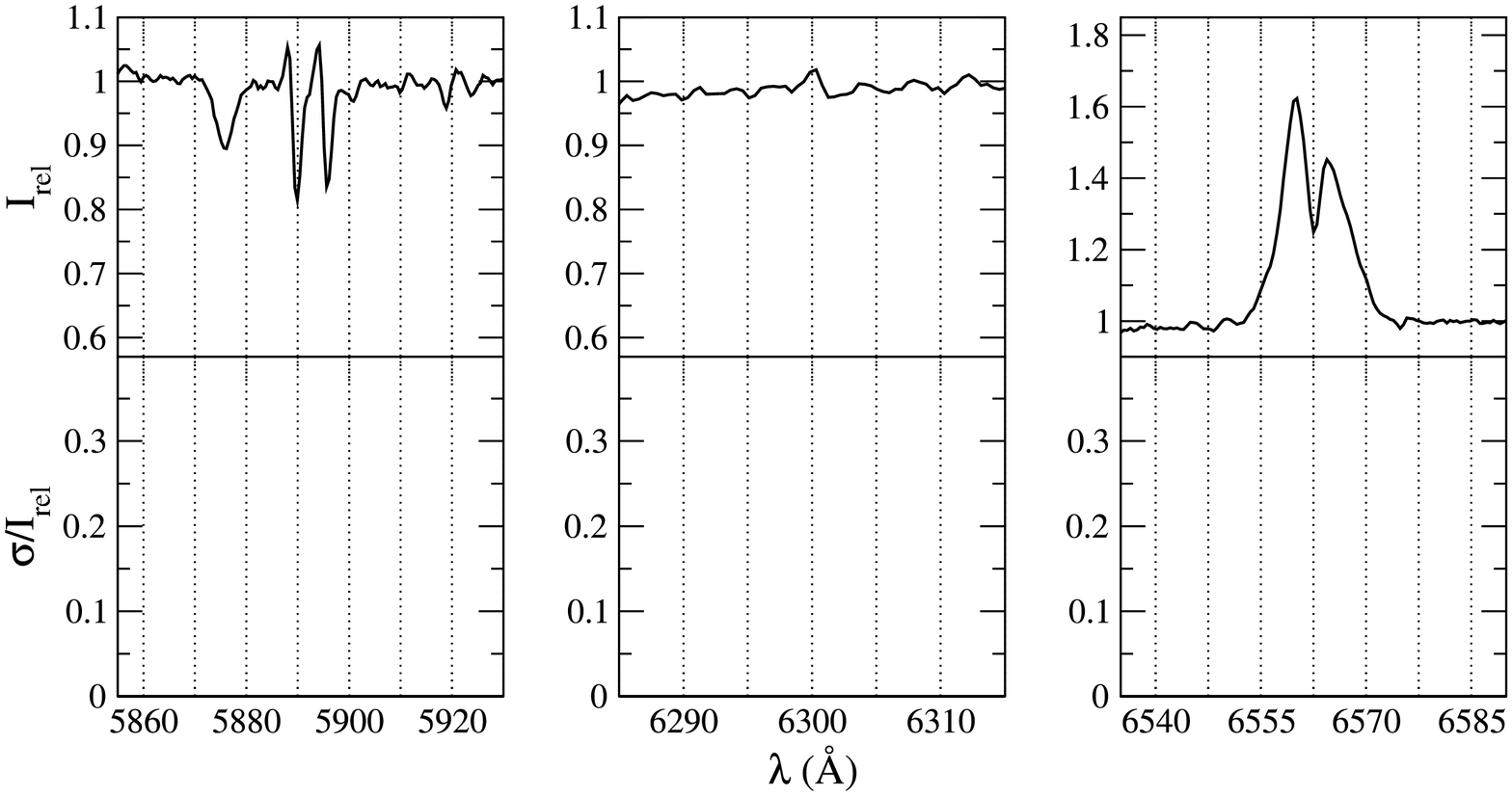} \\ 
\includegraphics[height=47mm,clip=true]{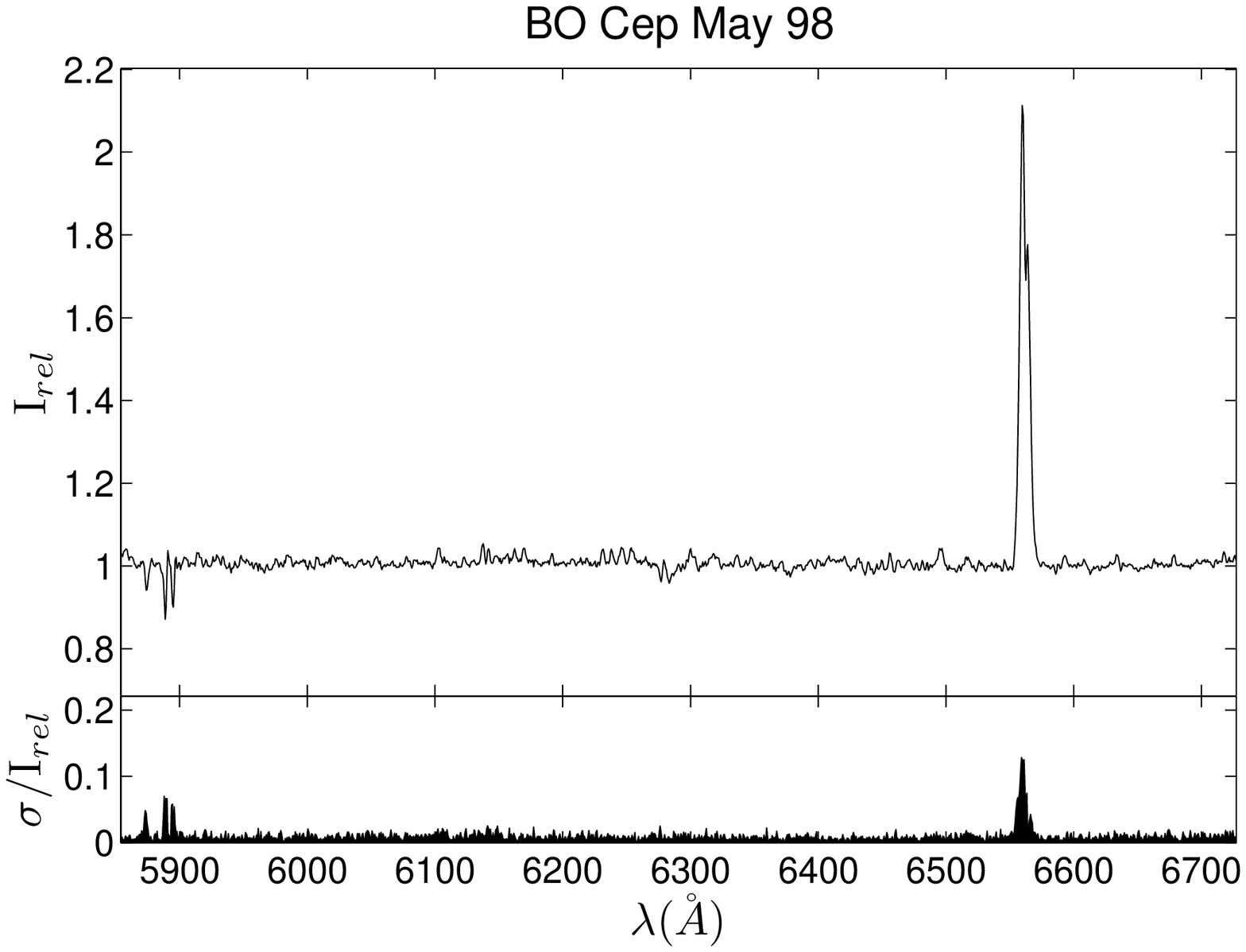}&
\includegraphics[bb=4 77 763 470,height=45mm,clip=true]{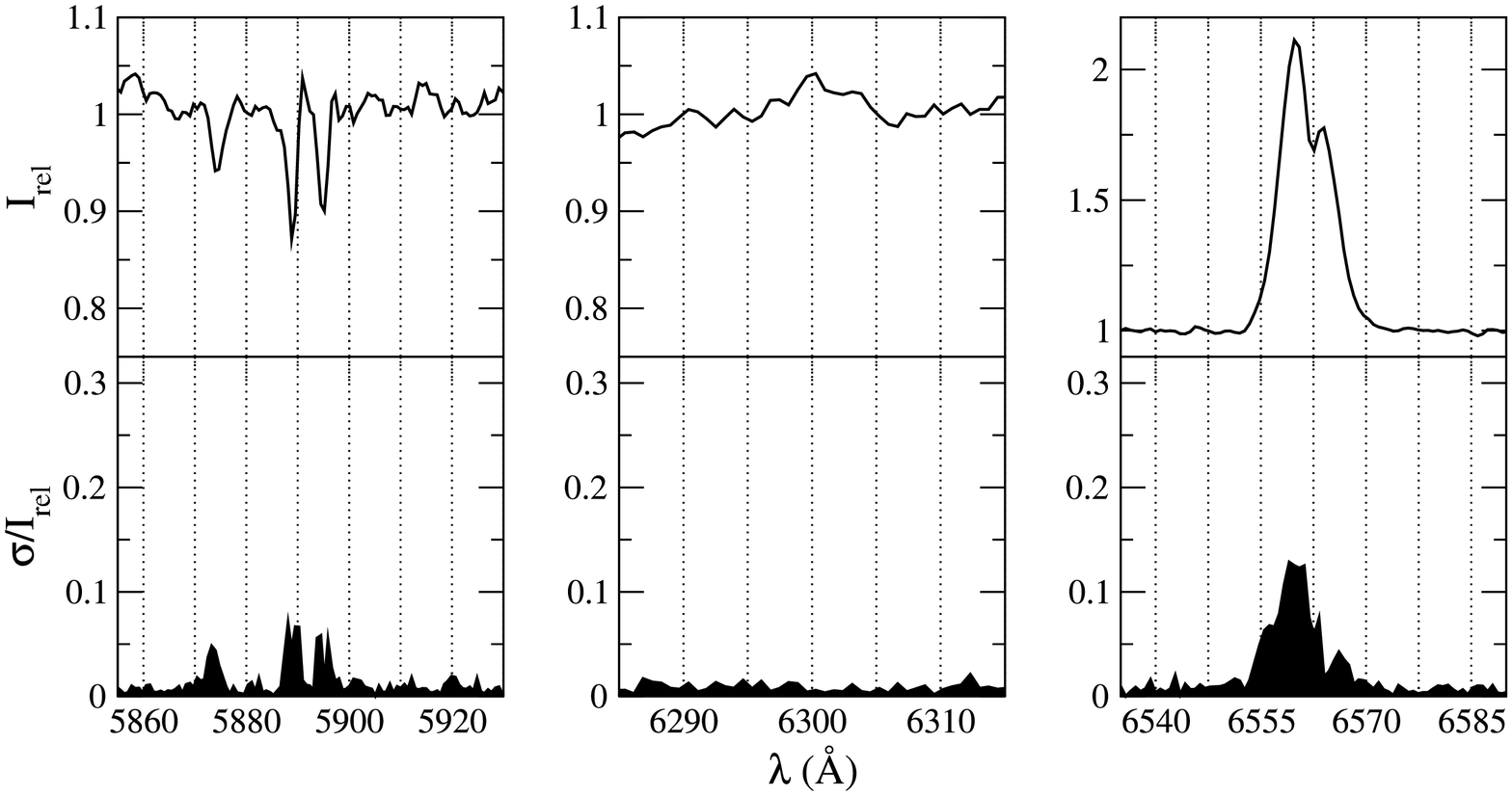} \\ 
\end{tabular}
\end{table}
\clearpage
\begin{table}
\centering
\renewcommand\arraystretch{10}
\begin{tabular}{cc}
\includegraphics[height=47mm,clip=true]{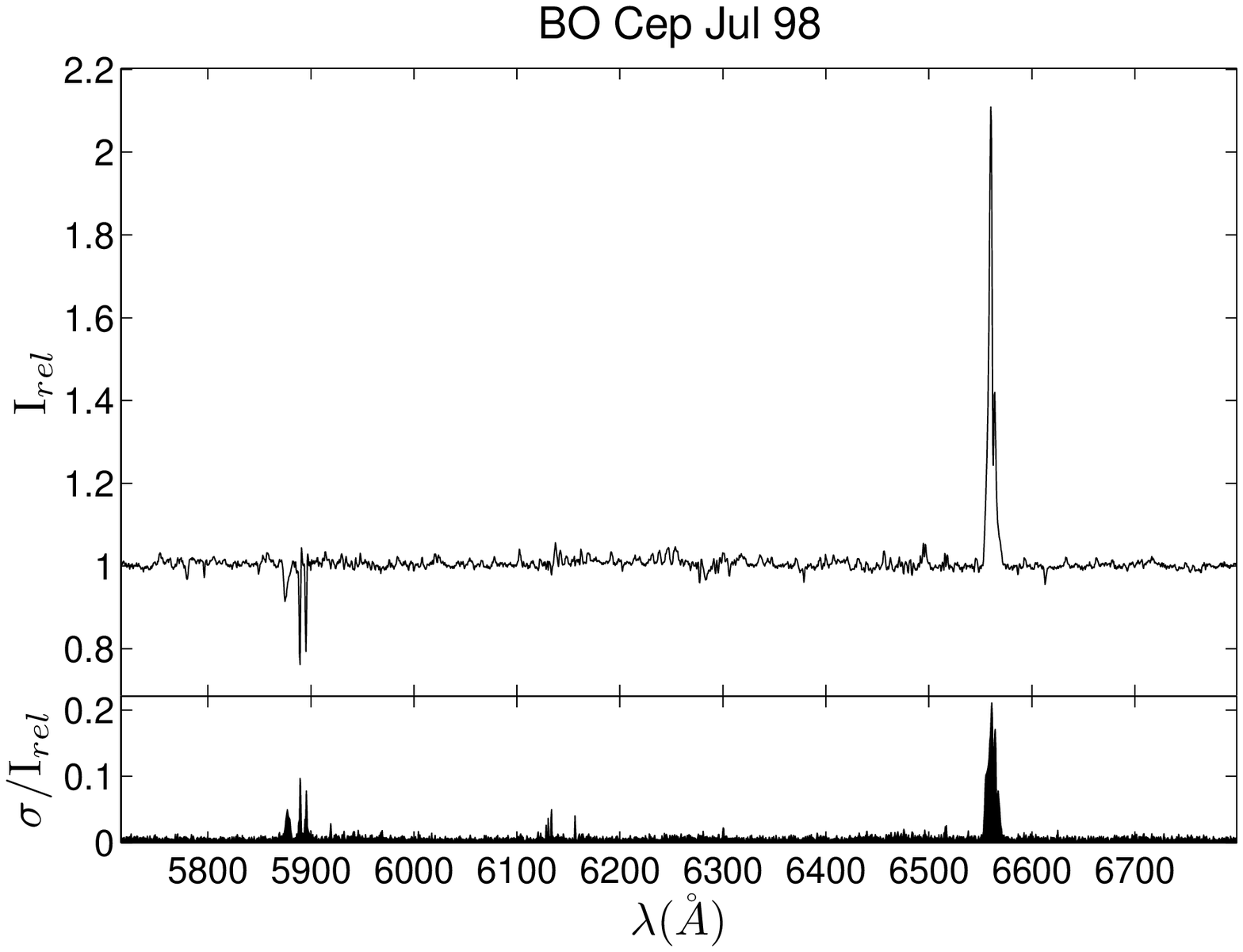}&
\includegraphics[bb=4 77 763 470,height=45mm,clip=true]{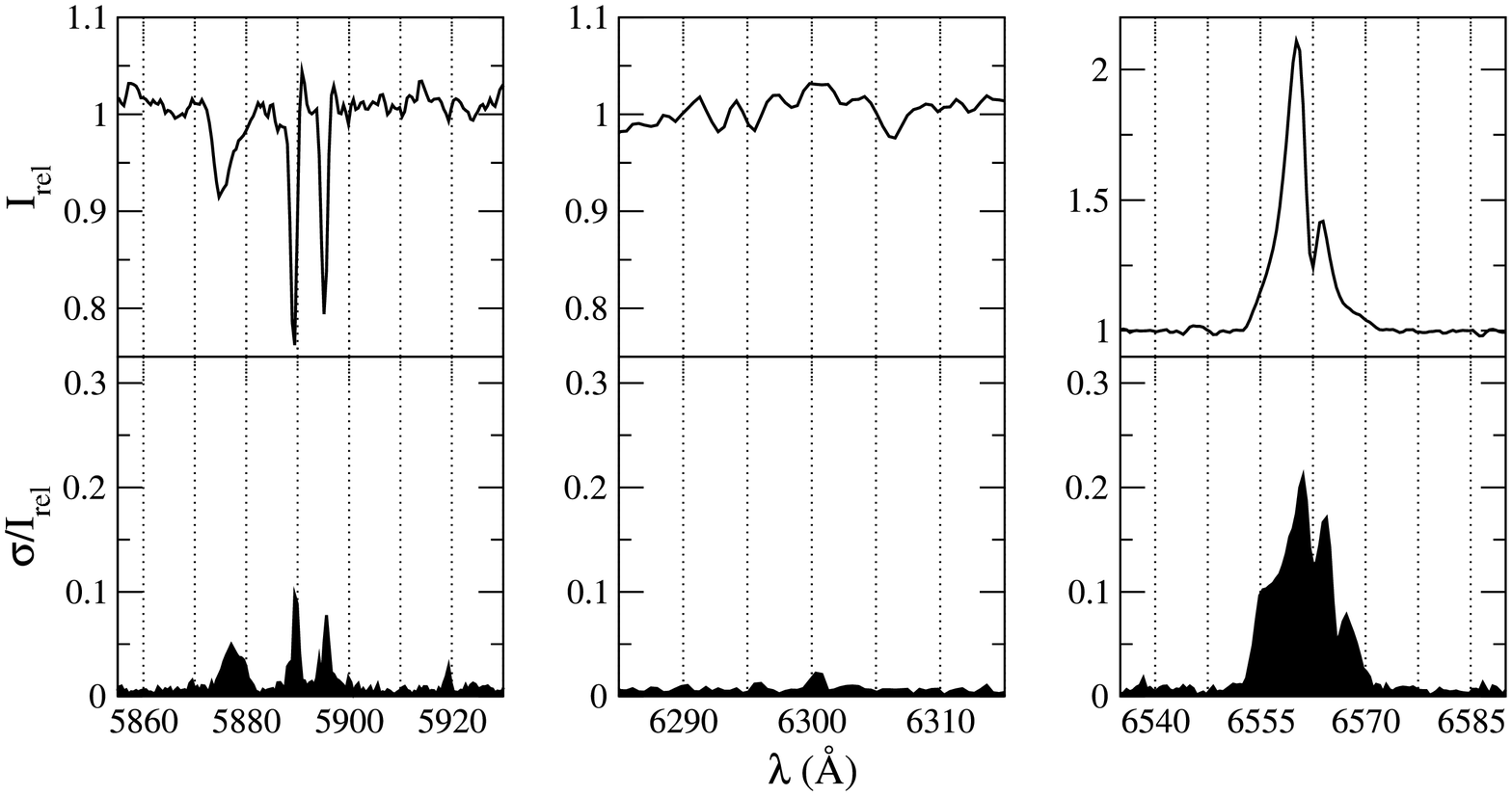} \\ 
\includegraphics[height=47mm,clip=true]{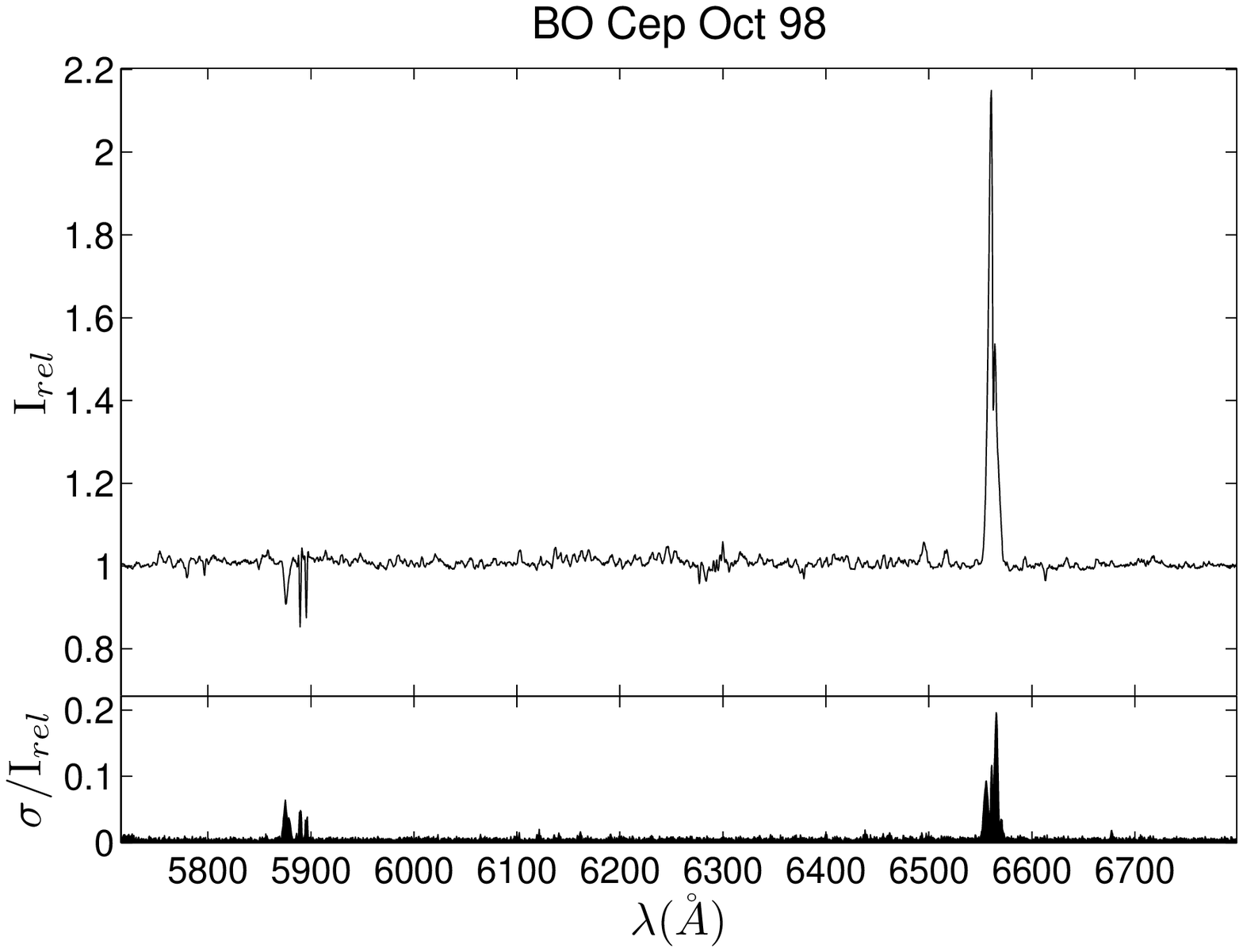}&
\includegraphics[bb=4 77 763 470,height=45mm,clip=true]{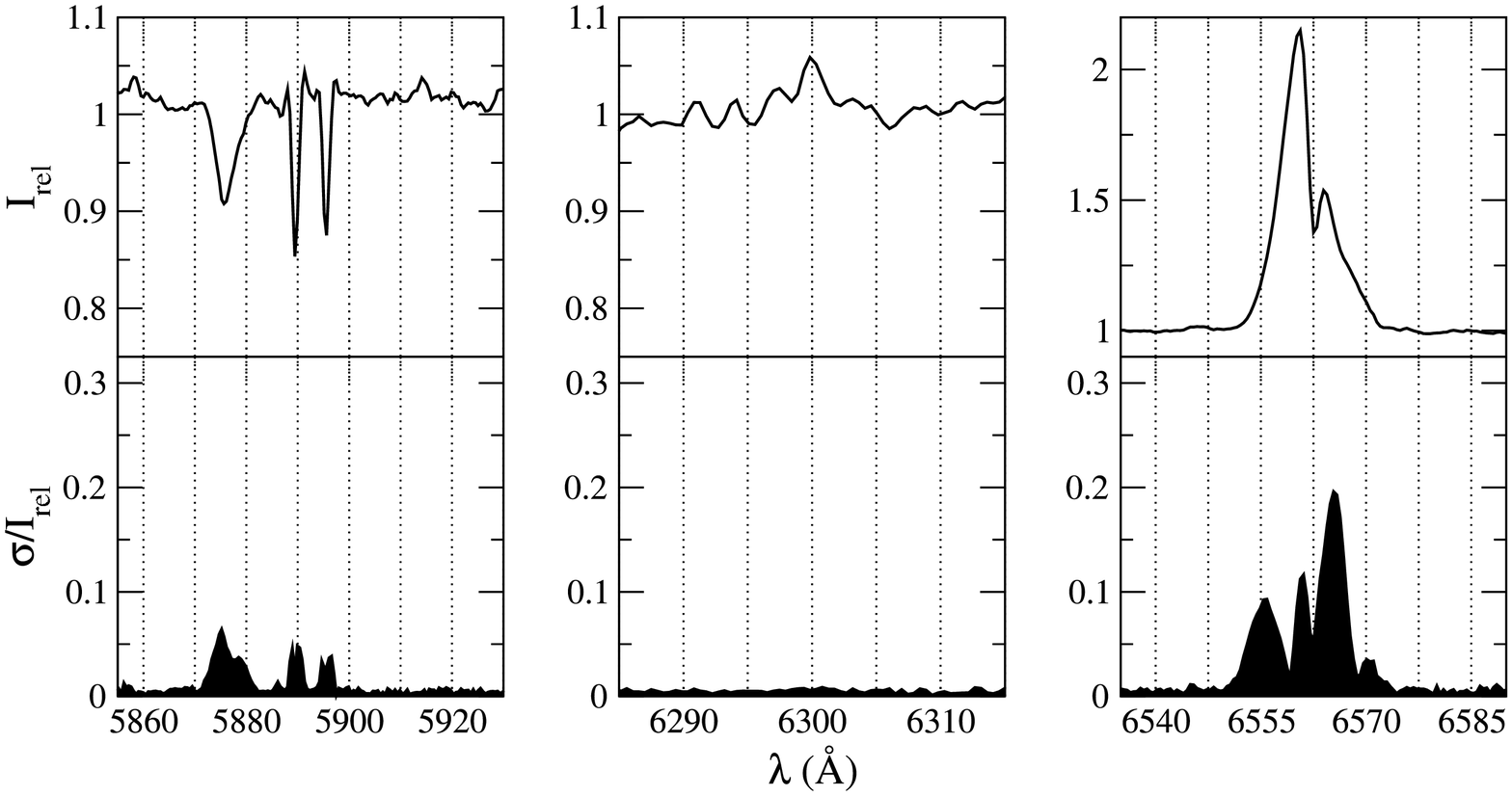} \\ 
\includegraphics[height=47mm,clip=true]{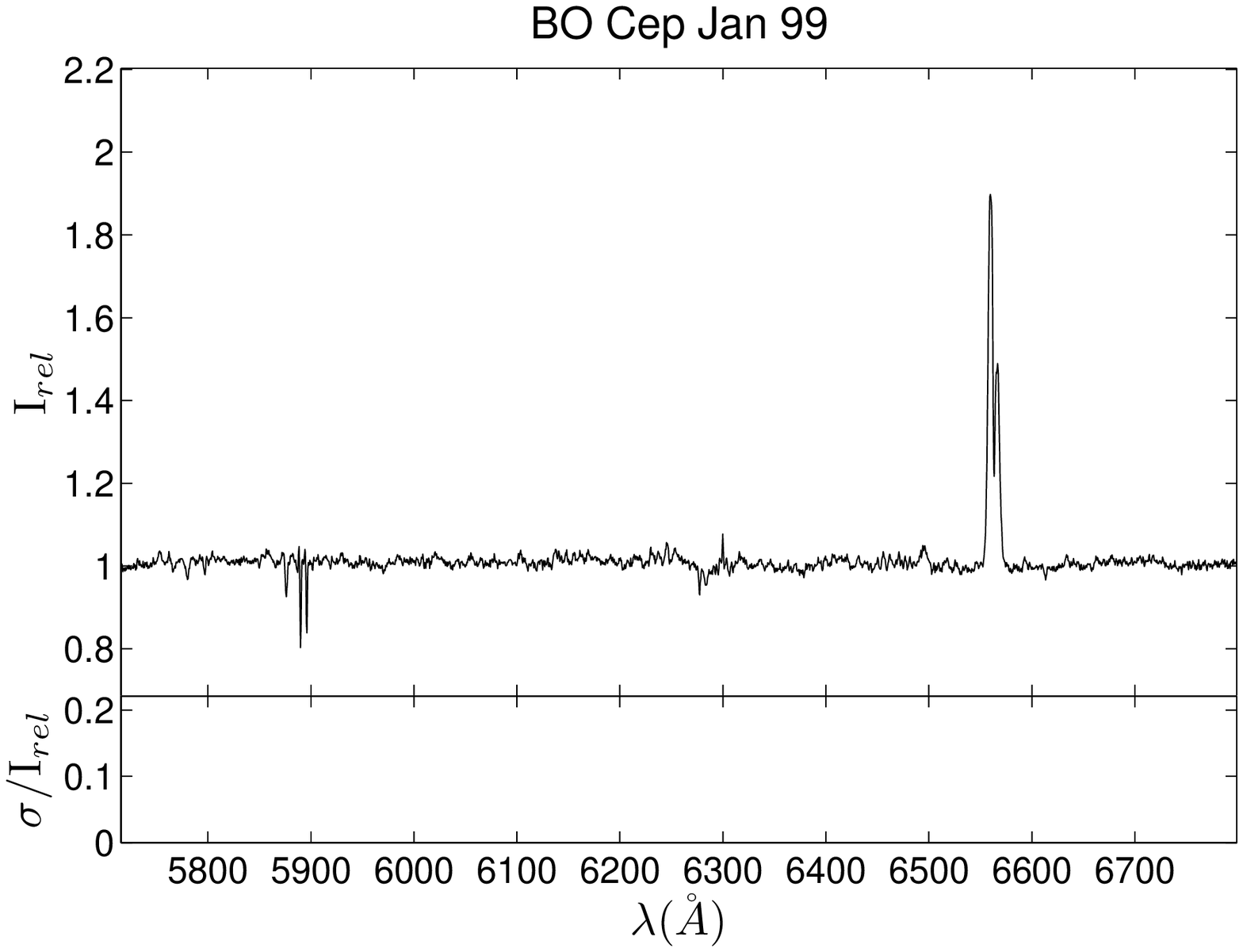}&
\includegraphics[bb=4 77 763 470,height=45mm,clip=true]{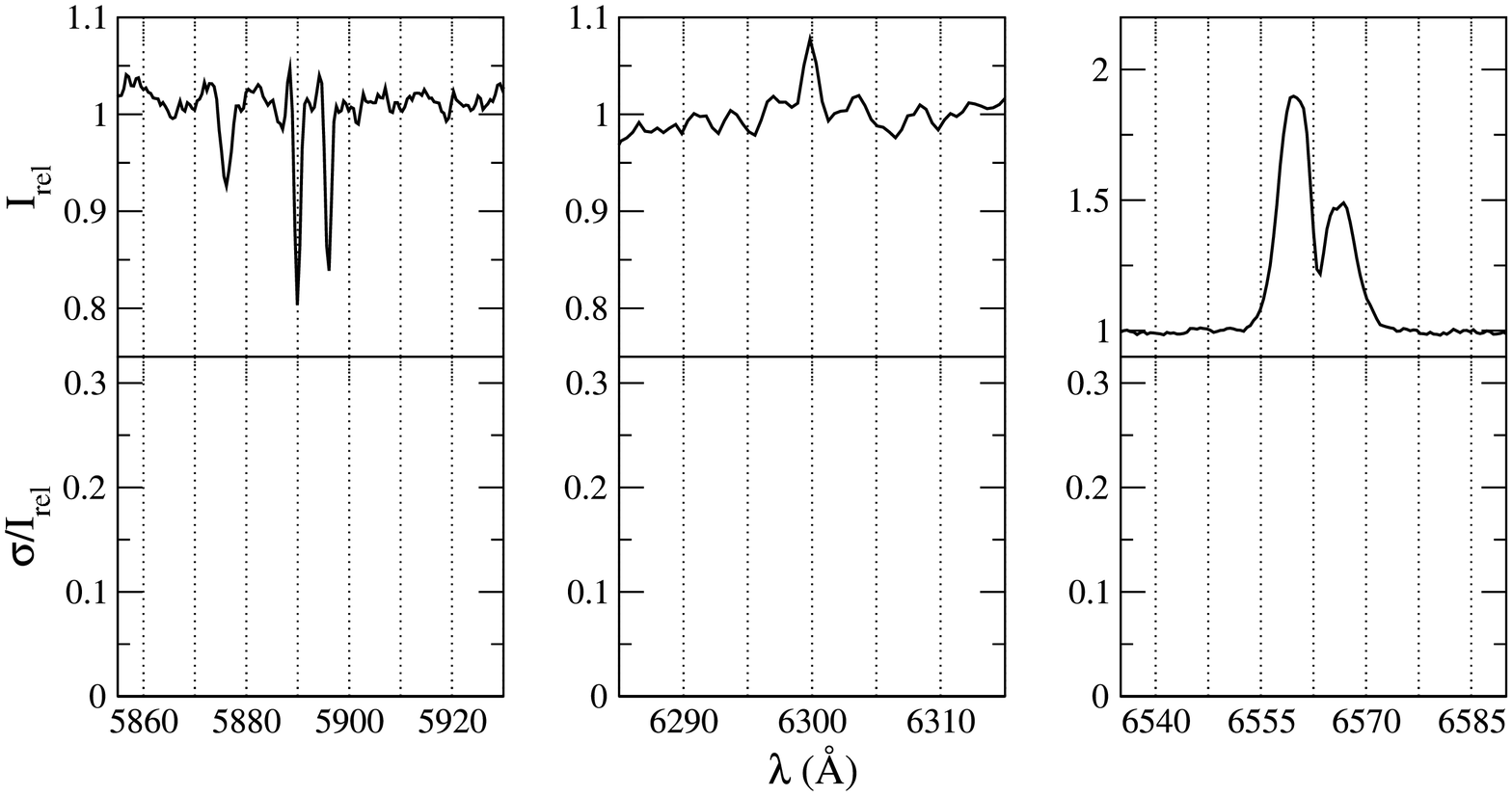} \\
\includegraphics[height=47mm,clip=true]{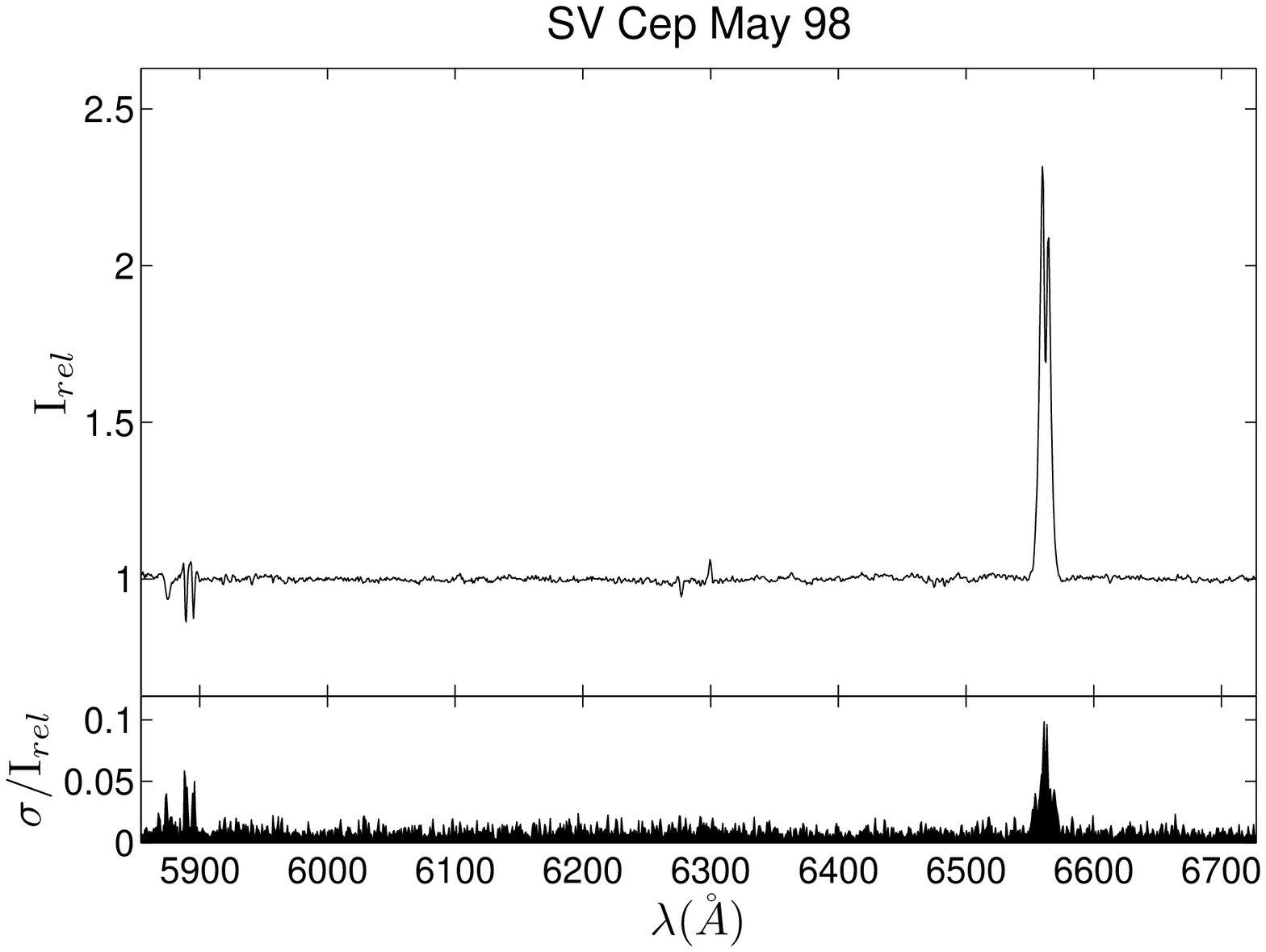}&
\includegraphics[bb=4 77 763 470,height=45mm,clip=true]{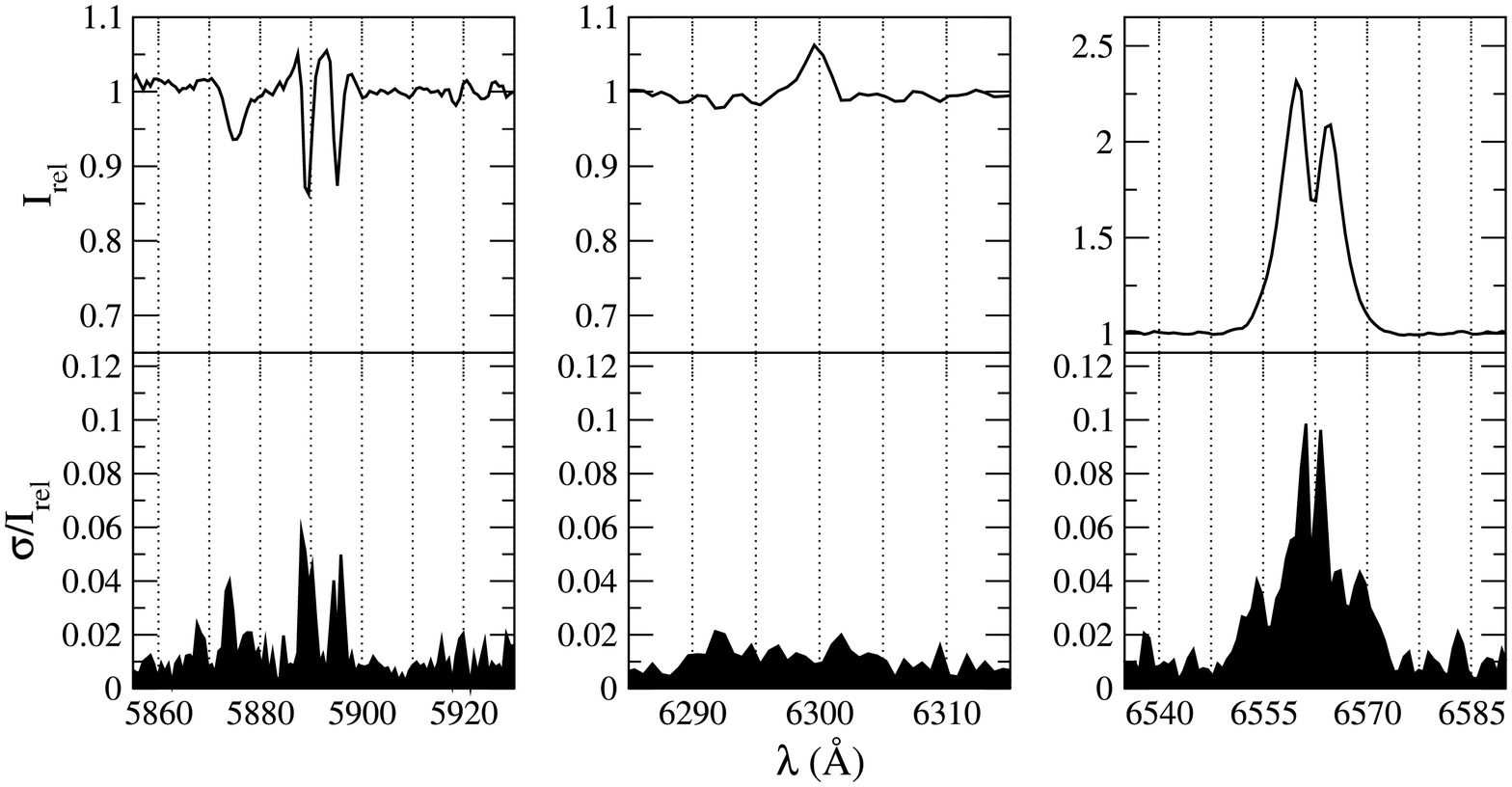} \\  
\end{tabular}
\end{table}
\clearpage
\begin{table}
\centering
\renewcommand\arraystretch{10}
\begin{tabular}{cc}
\includegraphics[height=47mm,clip=true]{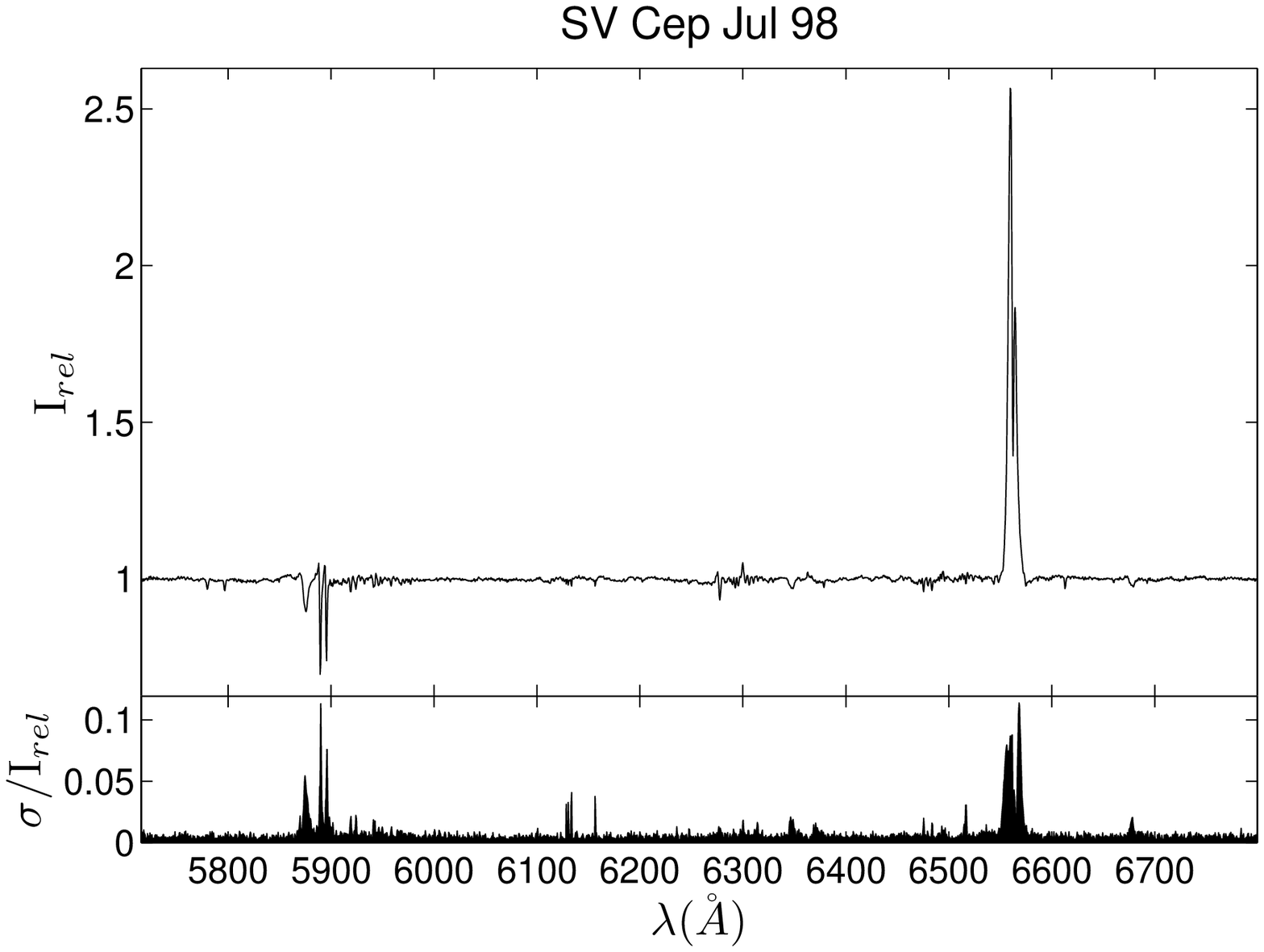}&
\includegraphics[bb=4 77 763 470,height=45mm,clip=true]{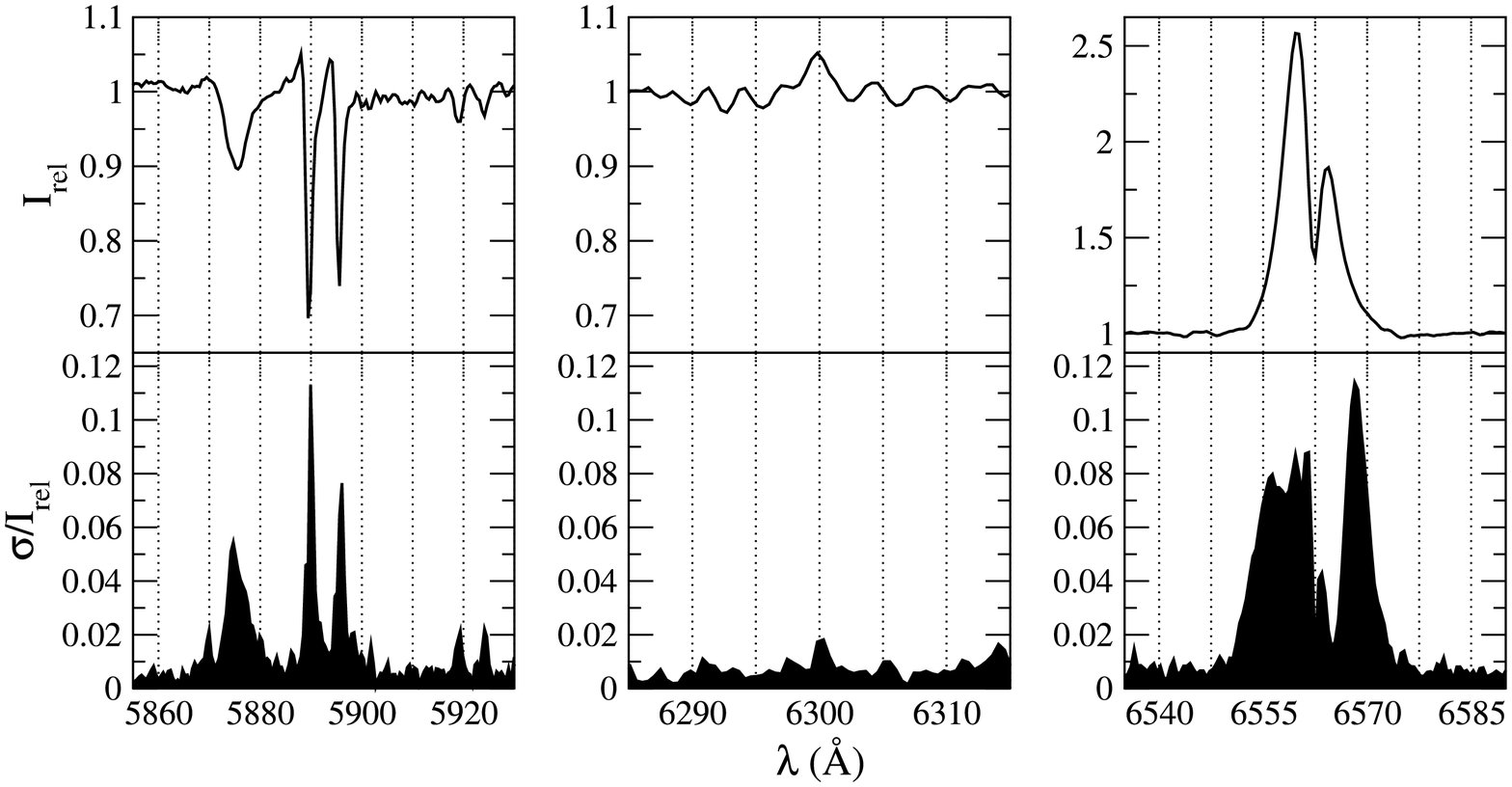} \\ 
\includegraphics[height=47mm,clip=true]{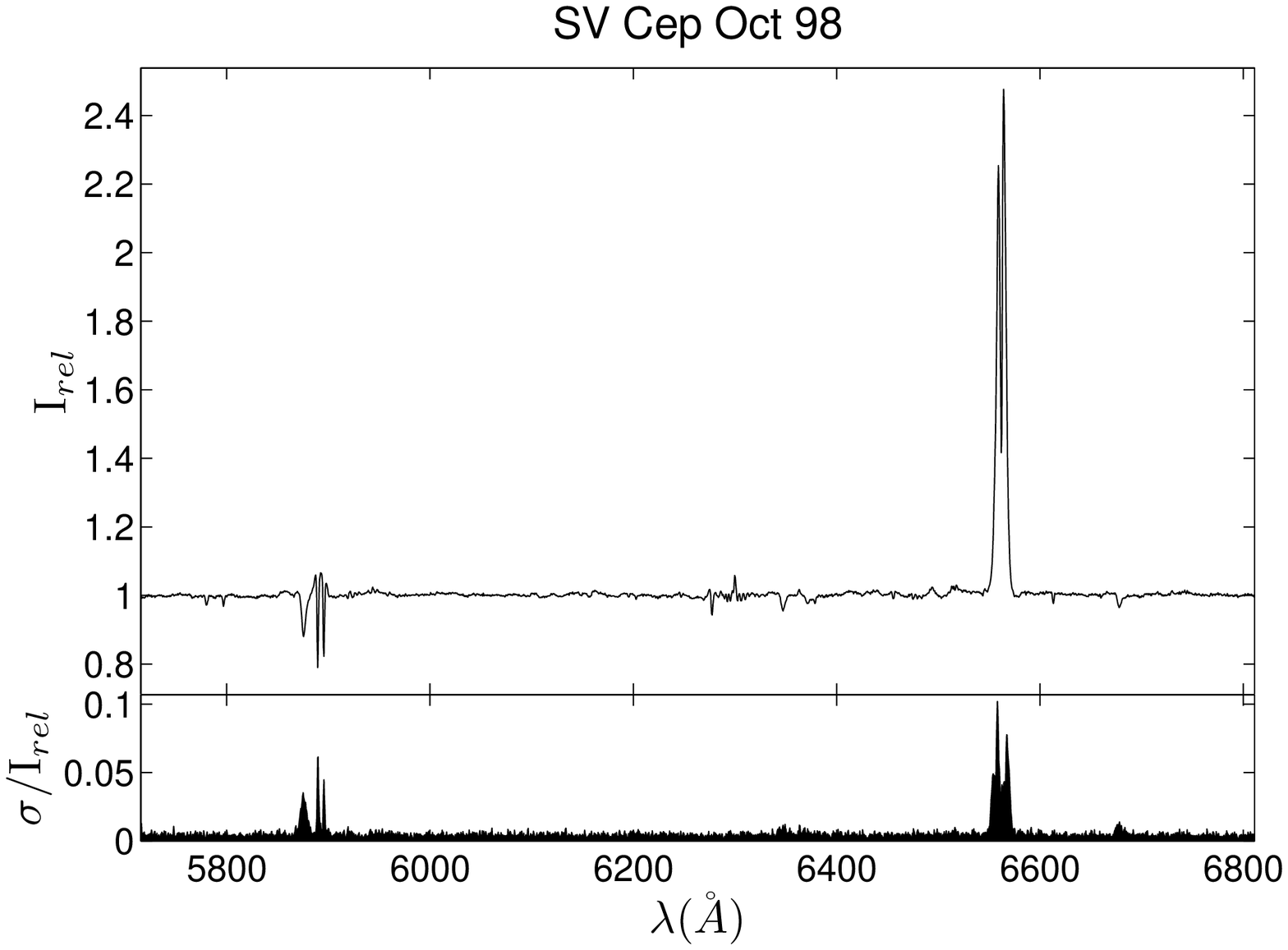}&
\includegraphics[bb=4 77 763 470,height=45mm,clip=true]{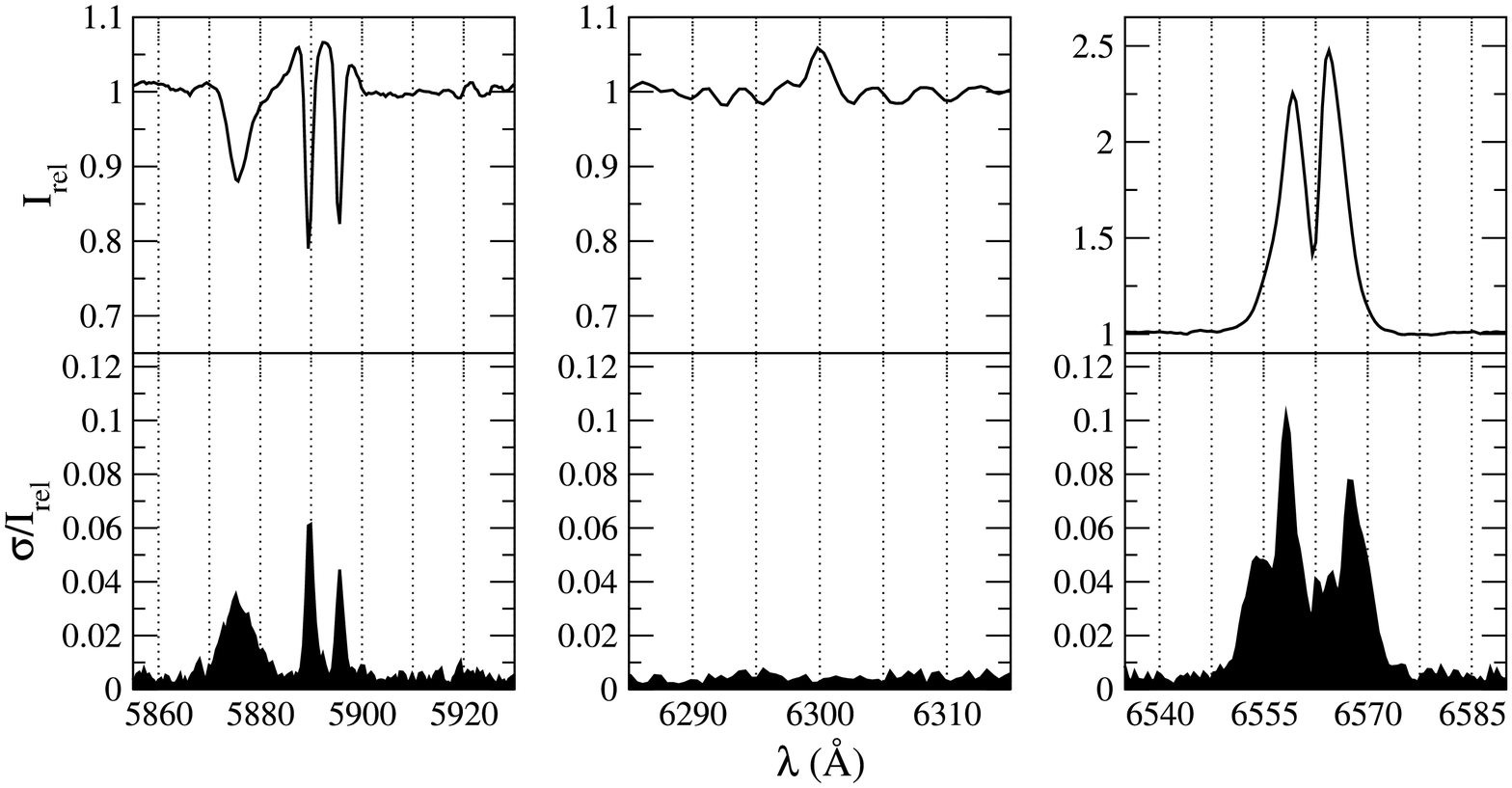} \\ 
\includegraphics[height=47mm,clip=true]{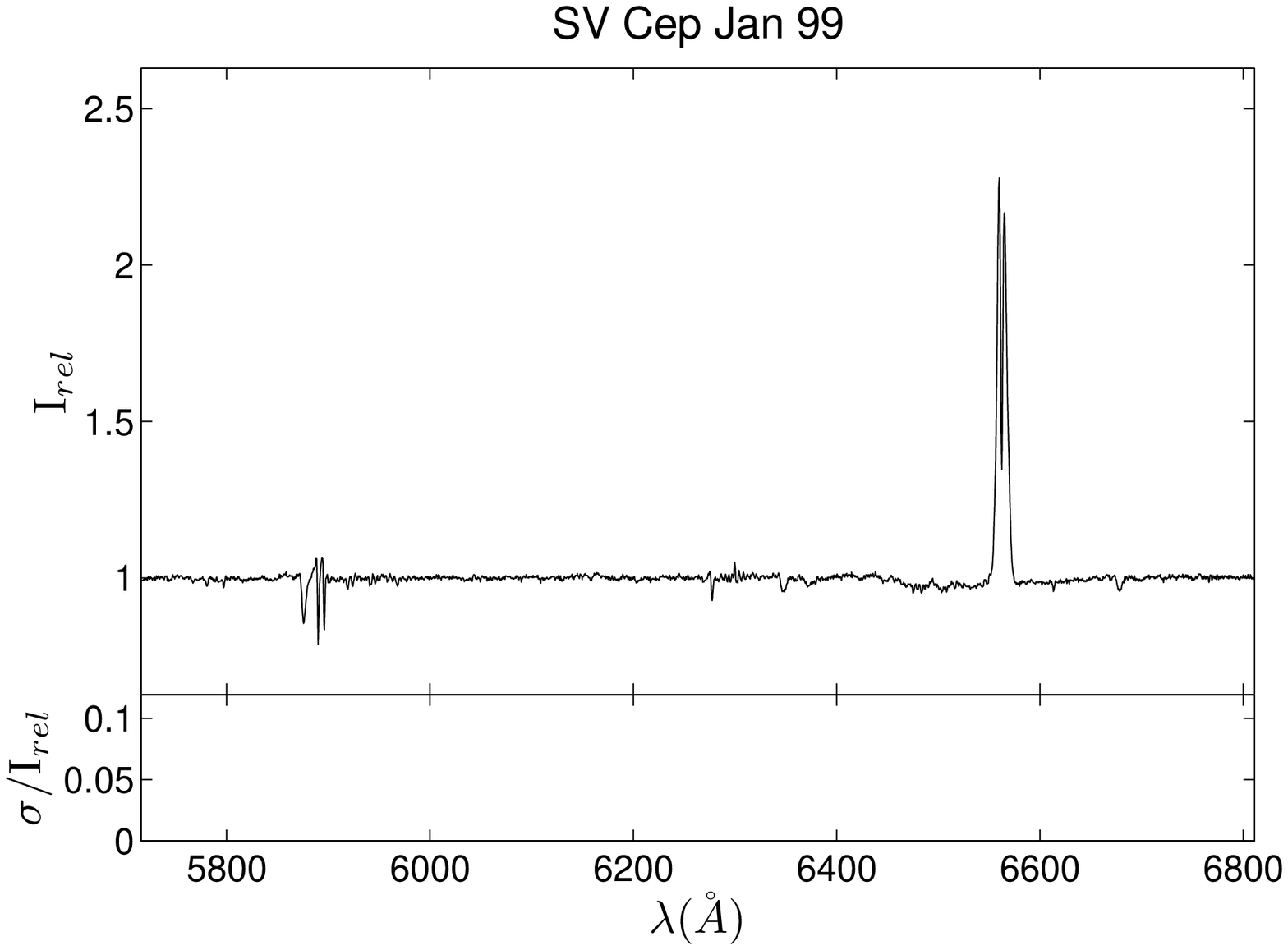}&
\includegraphics[bb=4 77 763 470,height=45mm,clip=true]{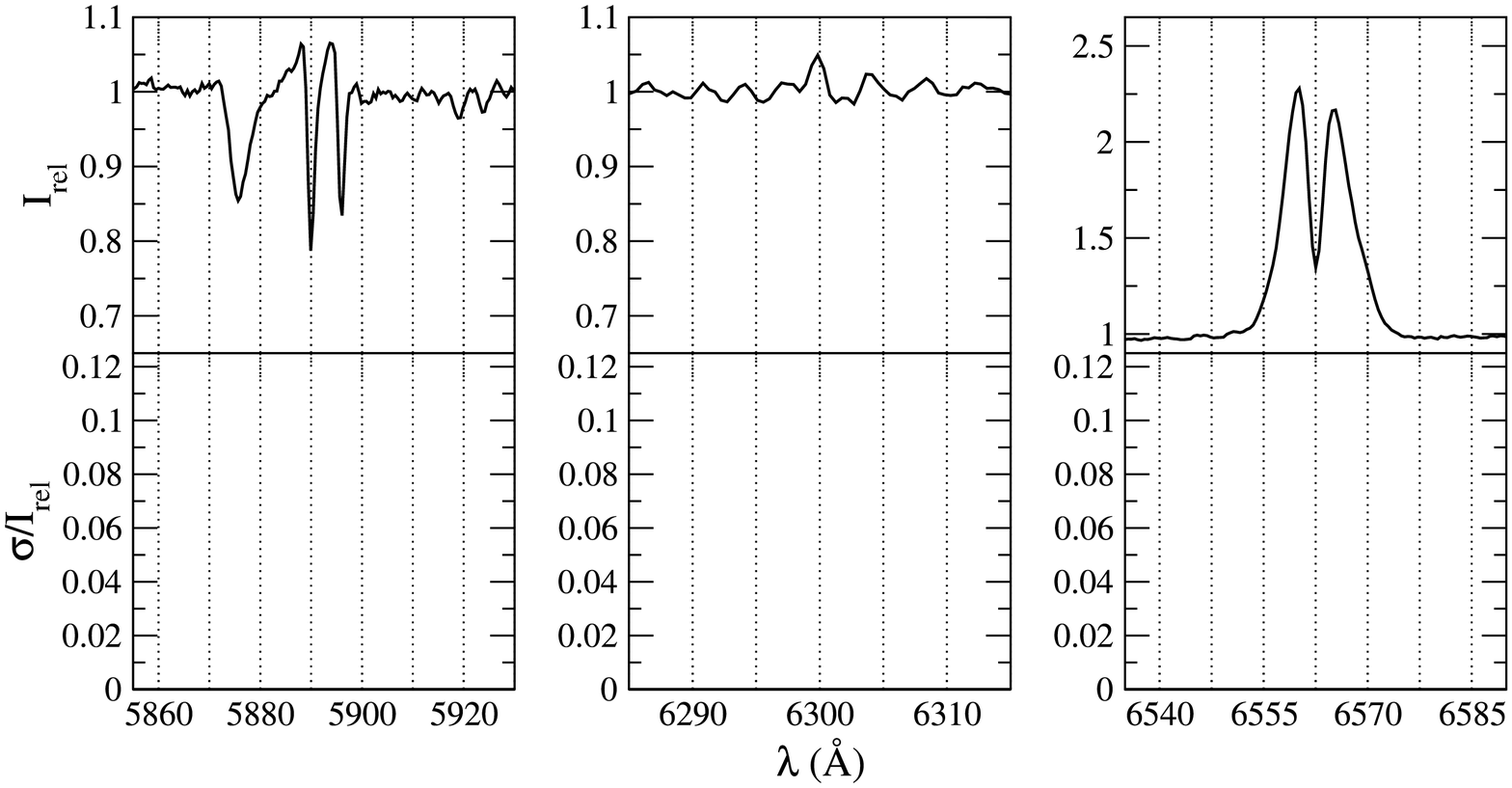} \\  
\includegraphics[height=47mm,clip=true]{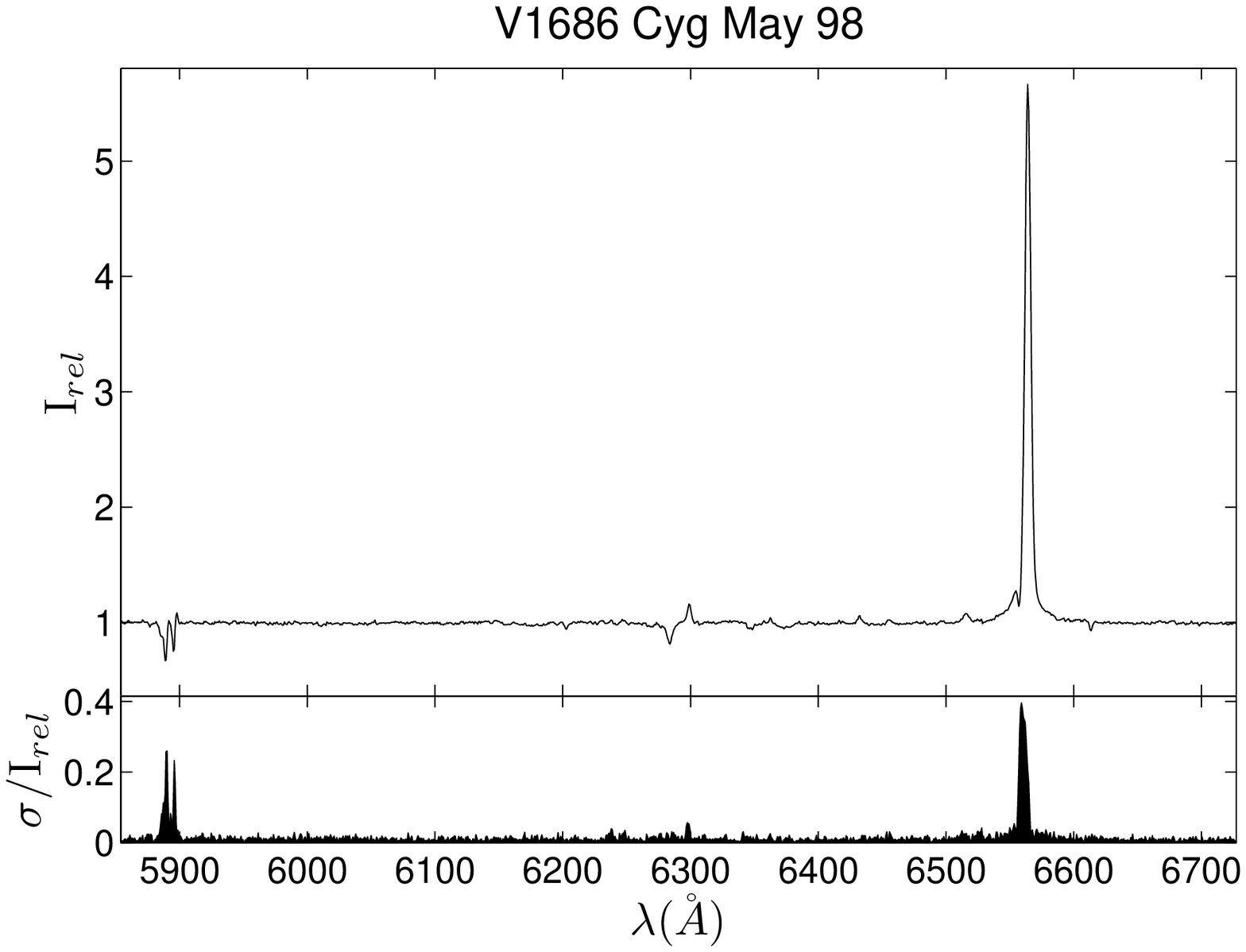}&
\includegraphics[bb=4 77 763 470,height=45mm,clip=true]{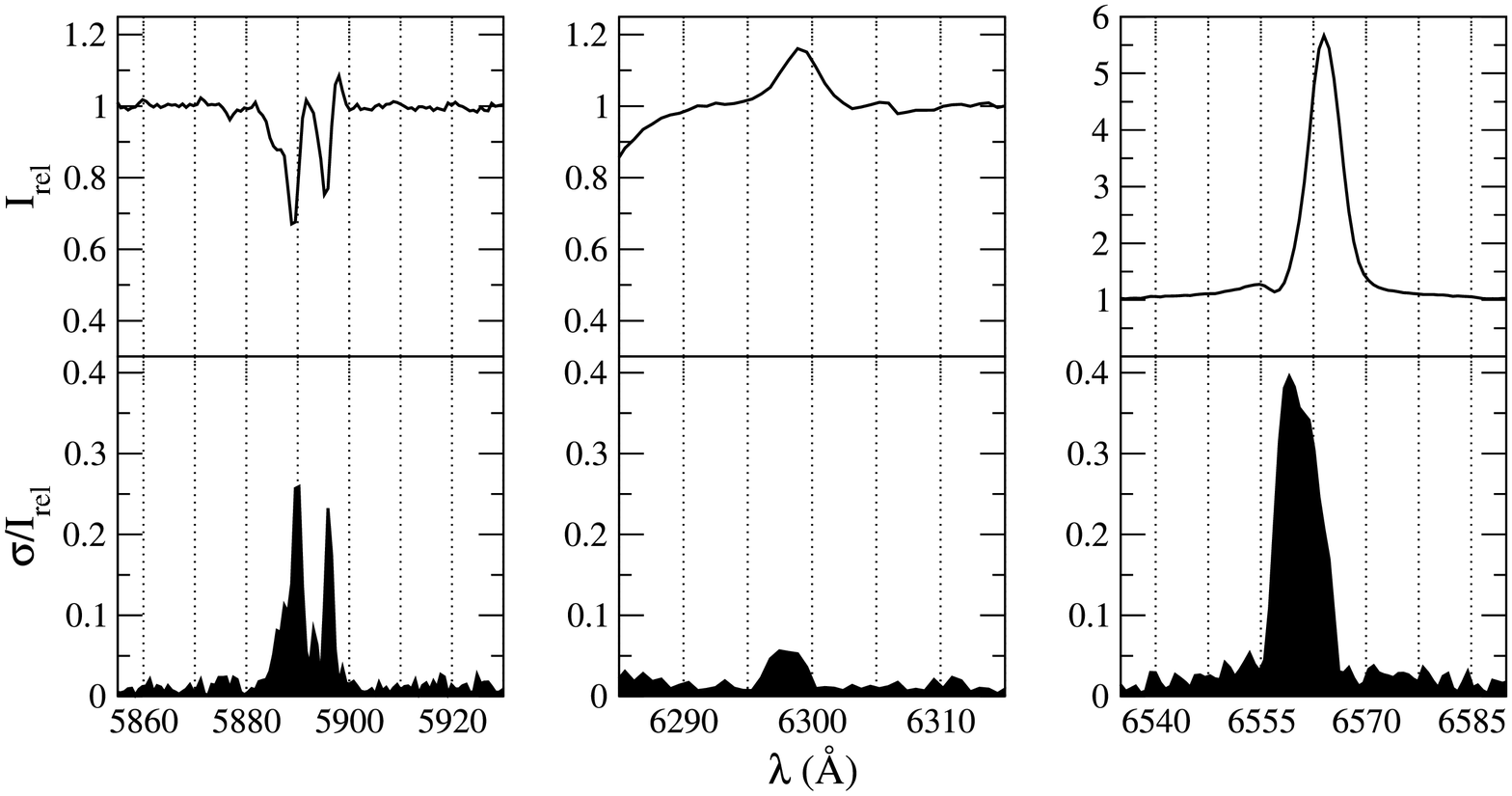} \\  
\end{tabular}
\end{table}
\clearpage
\begin{table}
\centering
\renewcommand\arraystretch{10}
\begin{tabular}{cc}
\includegraphics[height=47mm,clip=true]{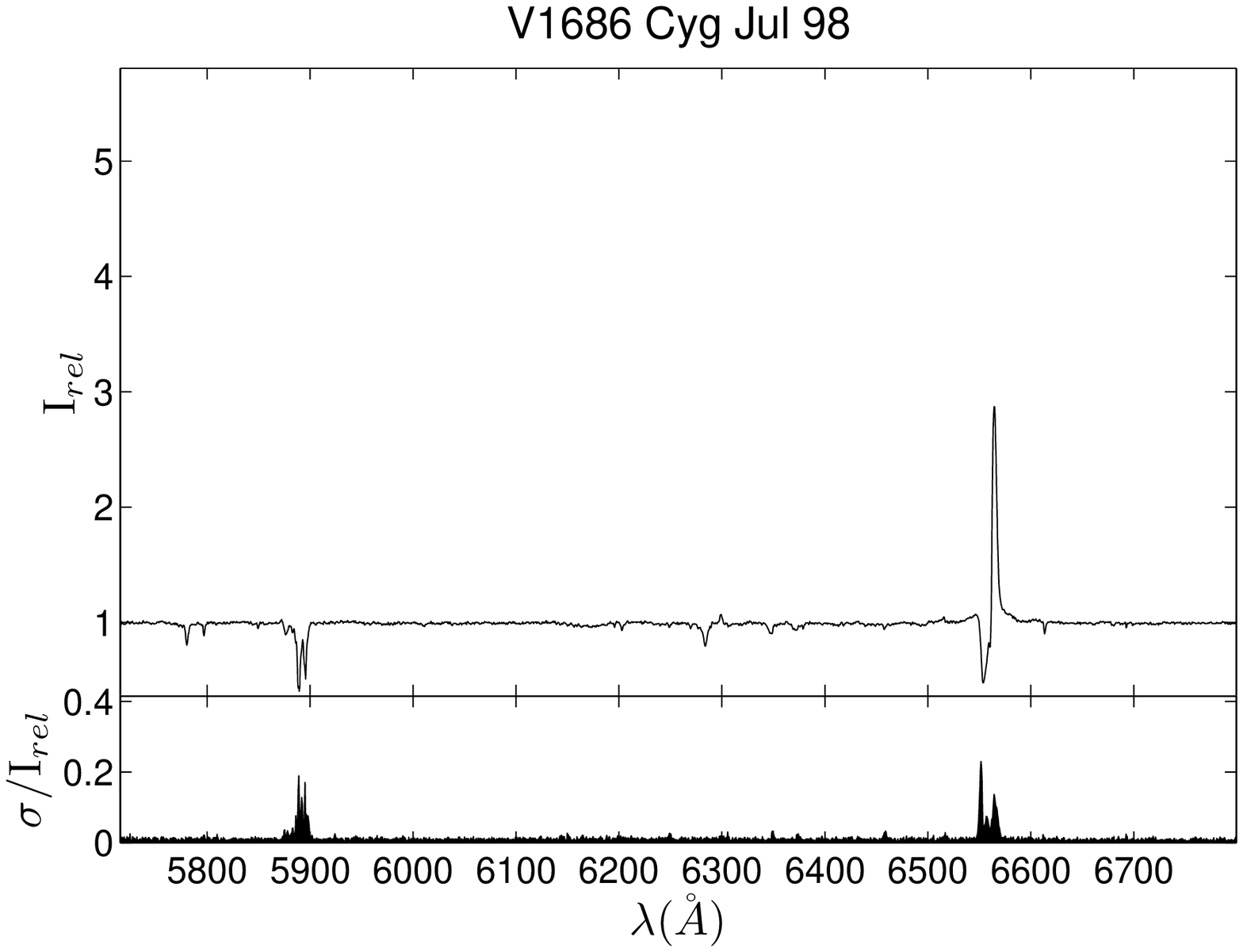}&
\includegraphics[bb=4 77 763 470,height=45mm,clip=true]{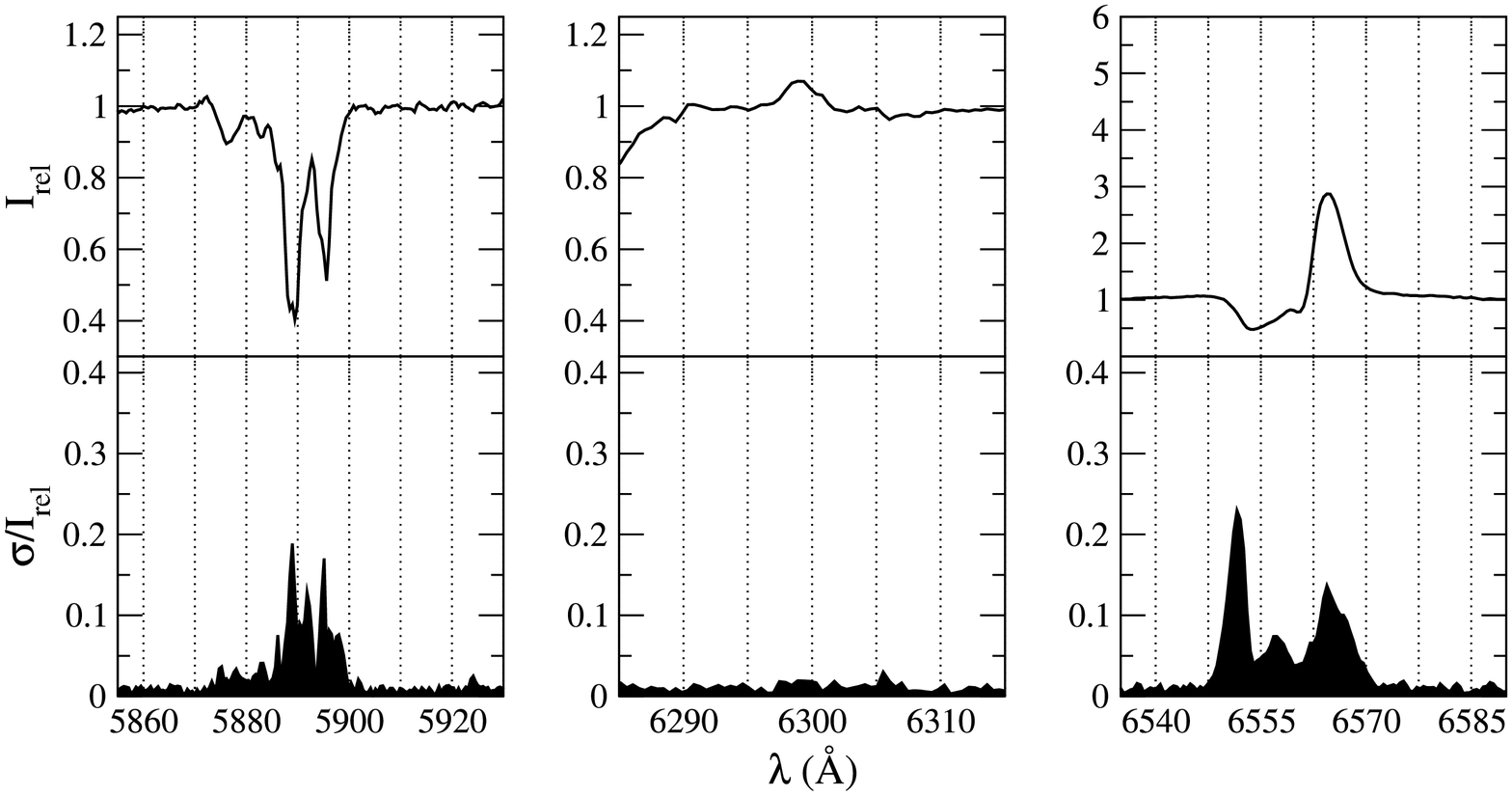} \\  
\includegraphics[height=47mm,clip=true]{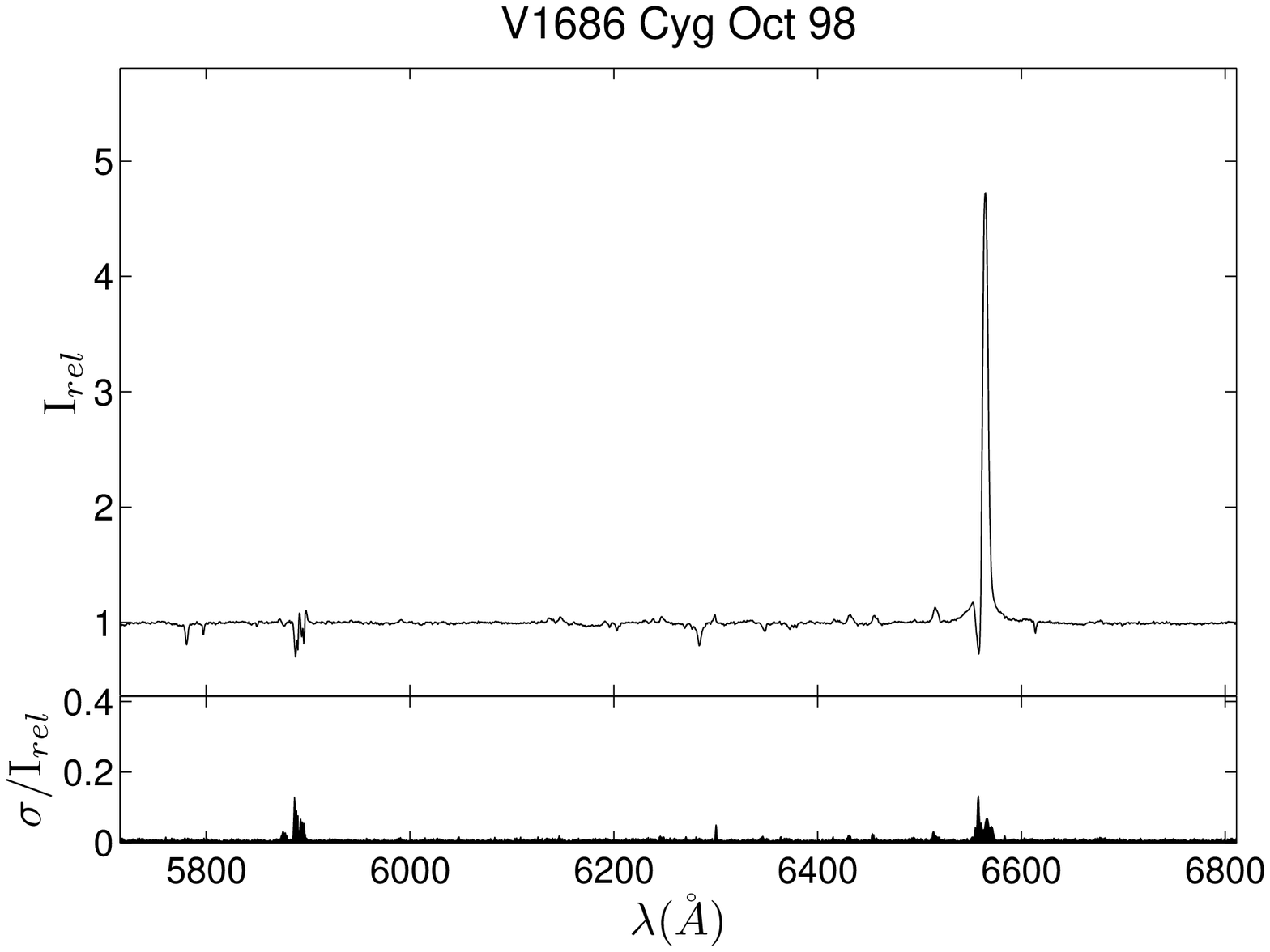}&
\includegraphics[bb=4 77 763 470,height=45mm,clip=true]{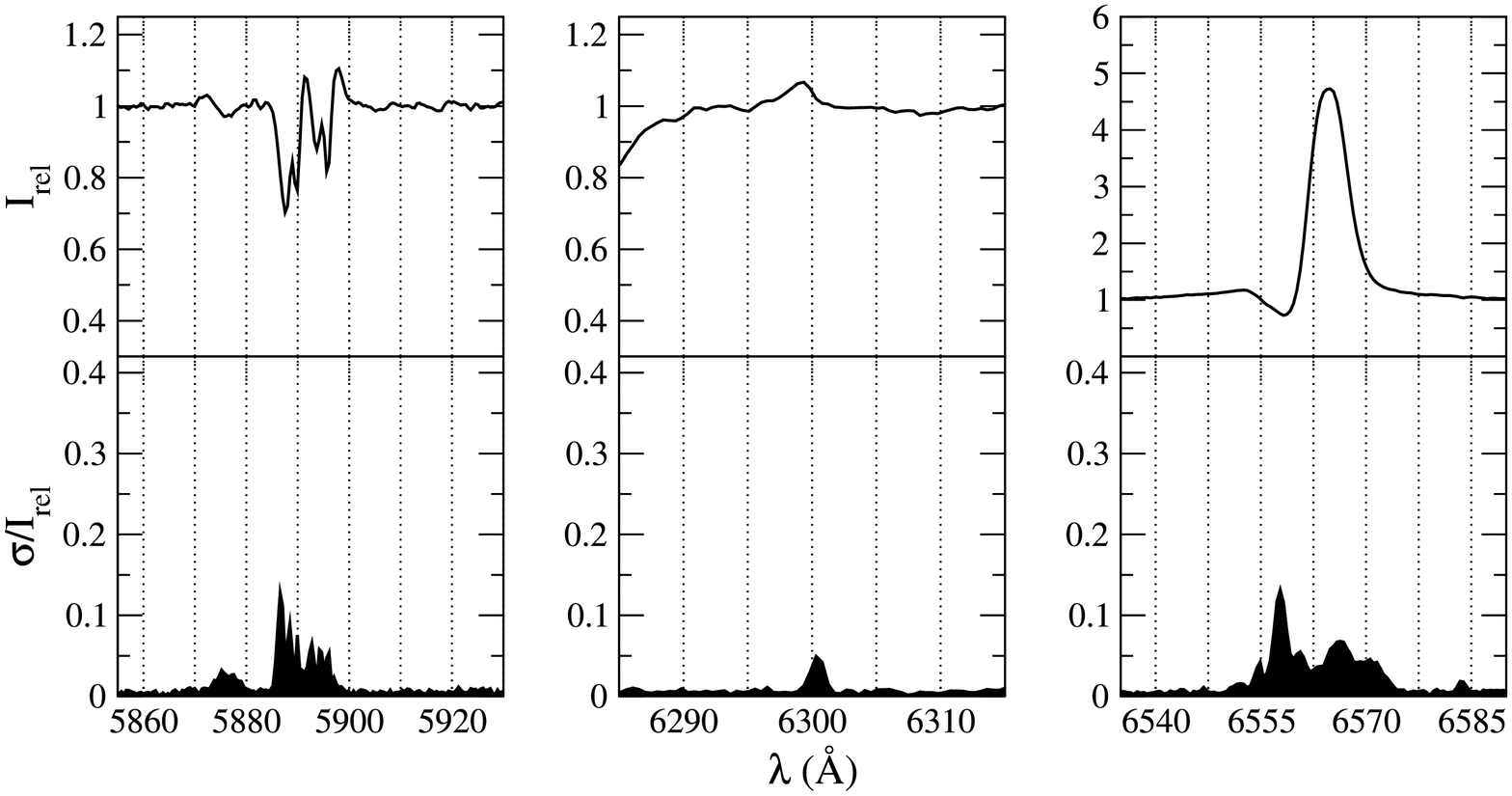} \\  
\includegraphics[height=47mm,clip=true]{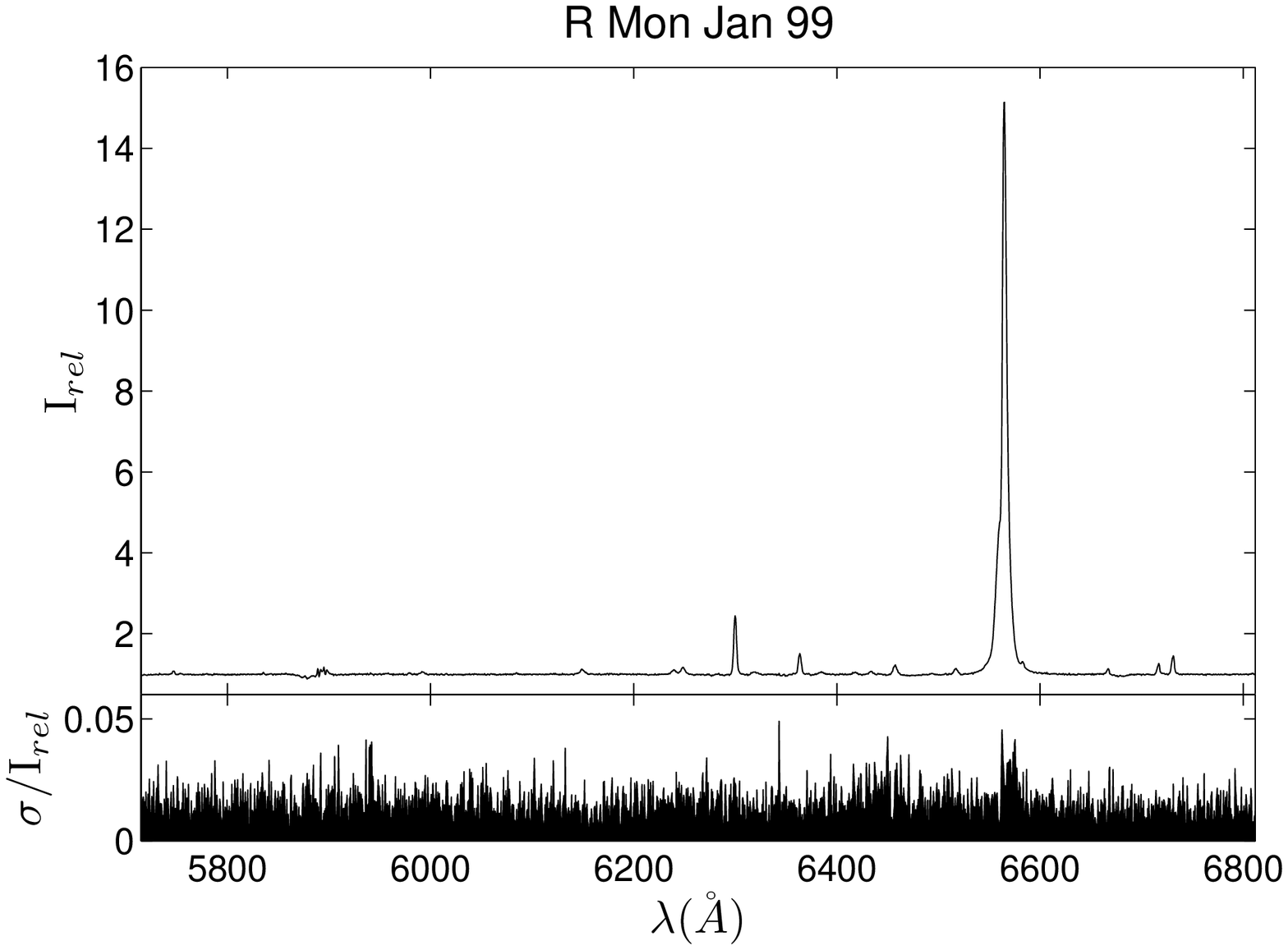}&
\includegraphics[bb=4 77 763 470,height=45mm,clip=true]{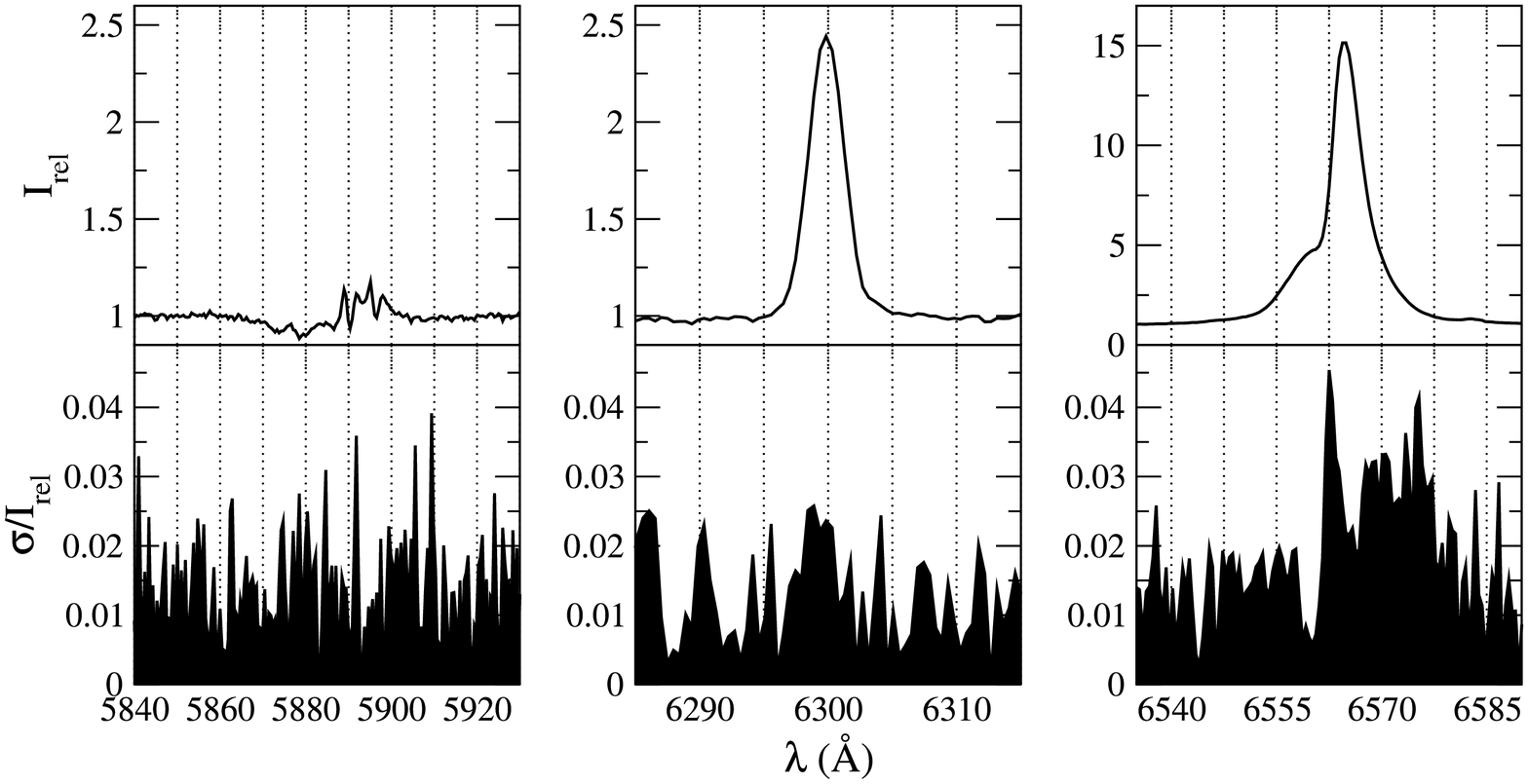} \\ 
\includegraphics[height=47mm,clip=true]{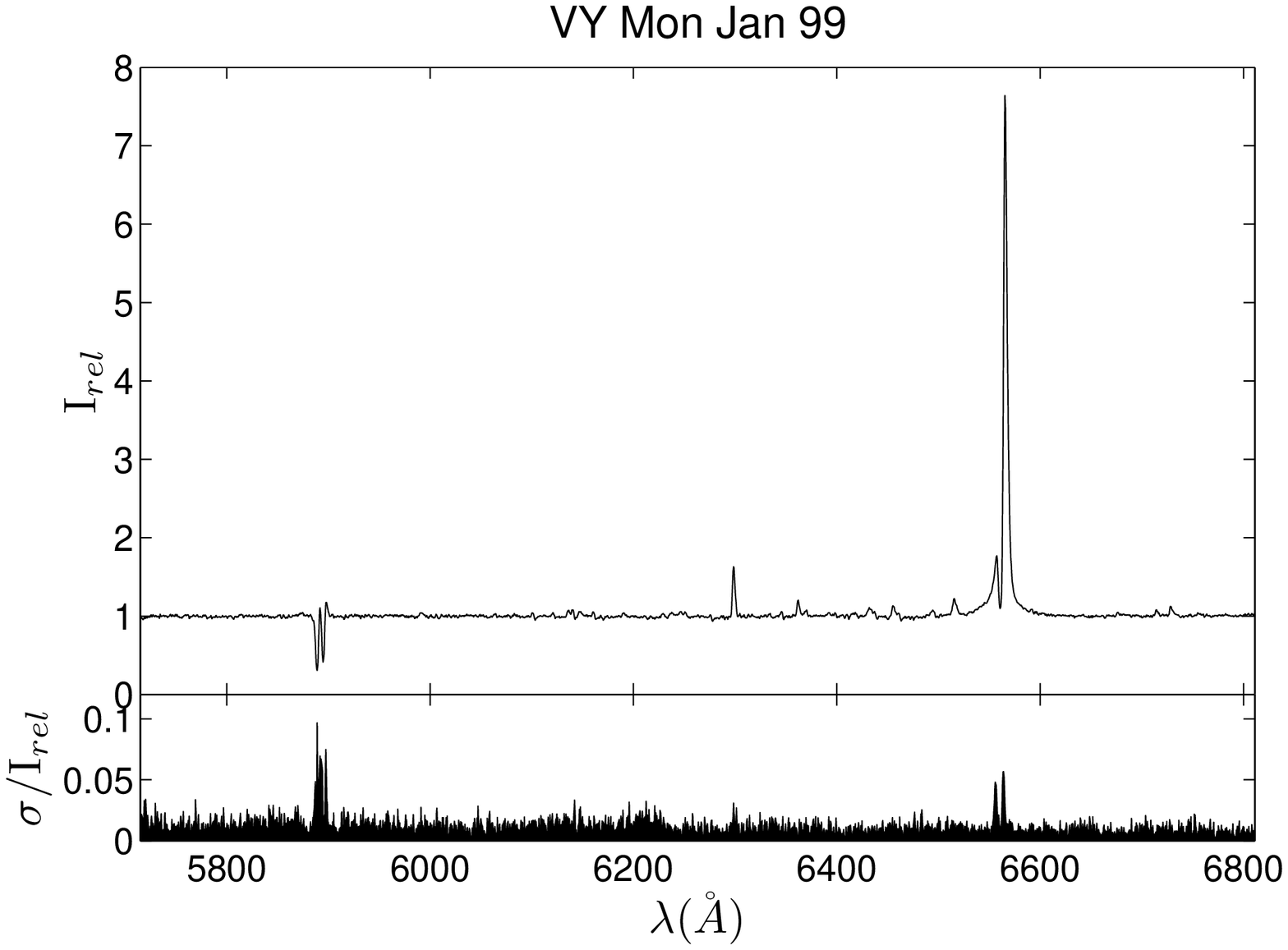}&
\includegraphics[bb=4 77 763 470,height=45mm,clip=true]{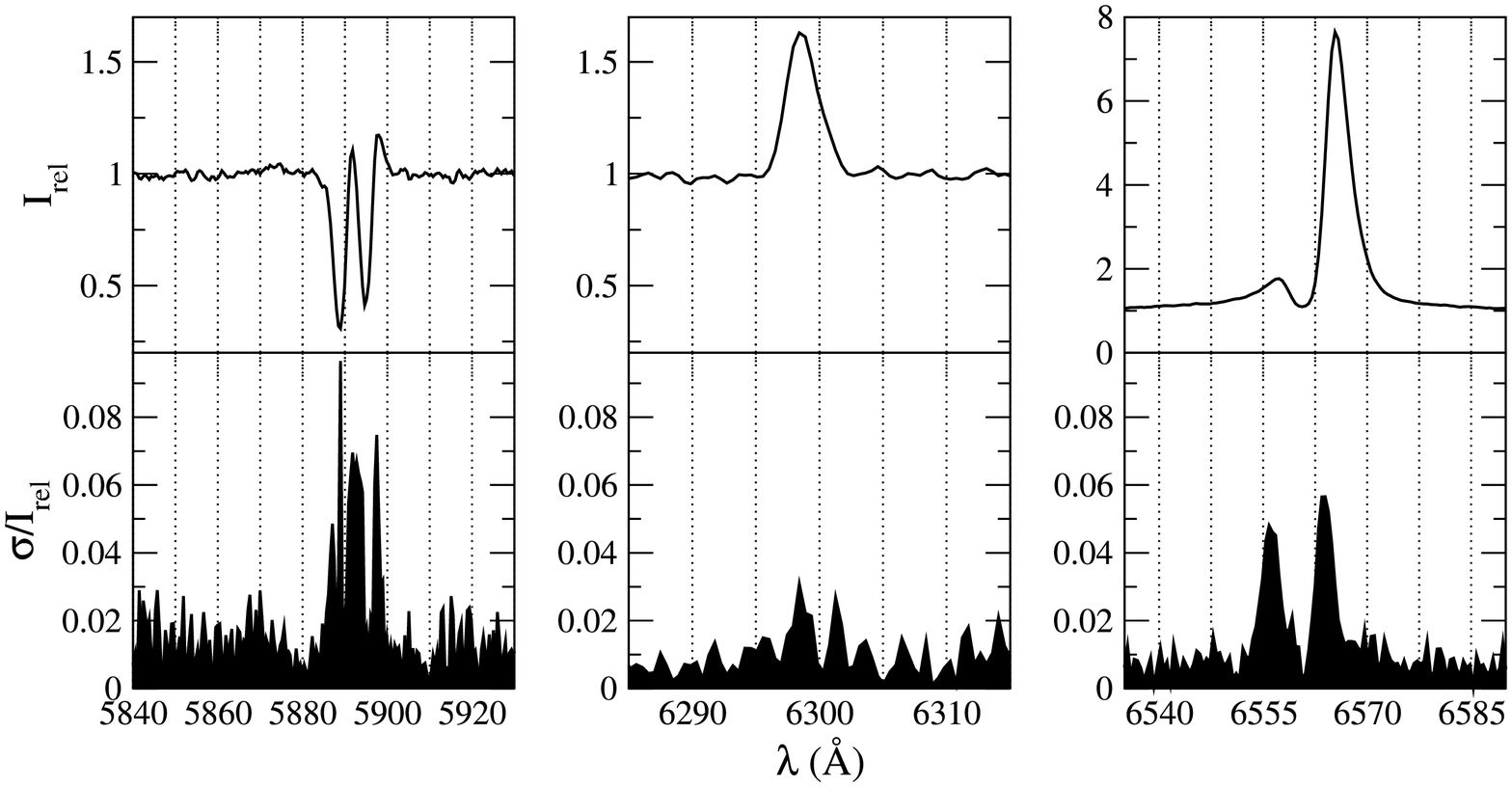} \\   
\end{tabular}
\end{table}
\clearpage
\begin{table}
\centering
\renewcommand\arraystretch{10}
\begin{tabular}{cc}
\includegraphics[height=47mm,clip=true]{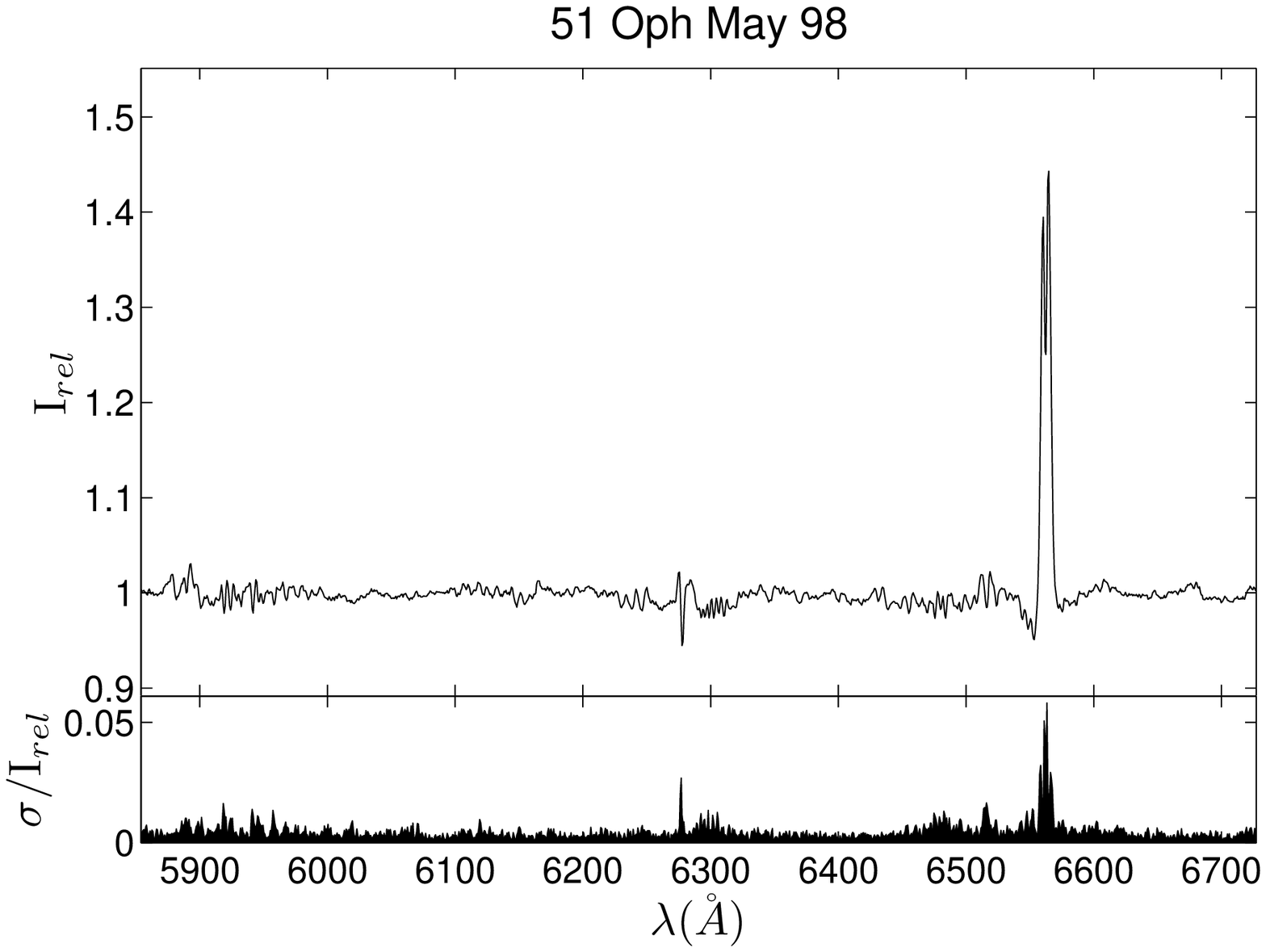}&
\includegraphics[bb=4 77 763 470,height=45mm,clip=true]{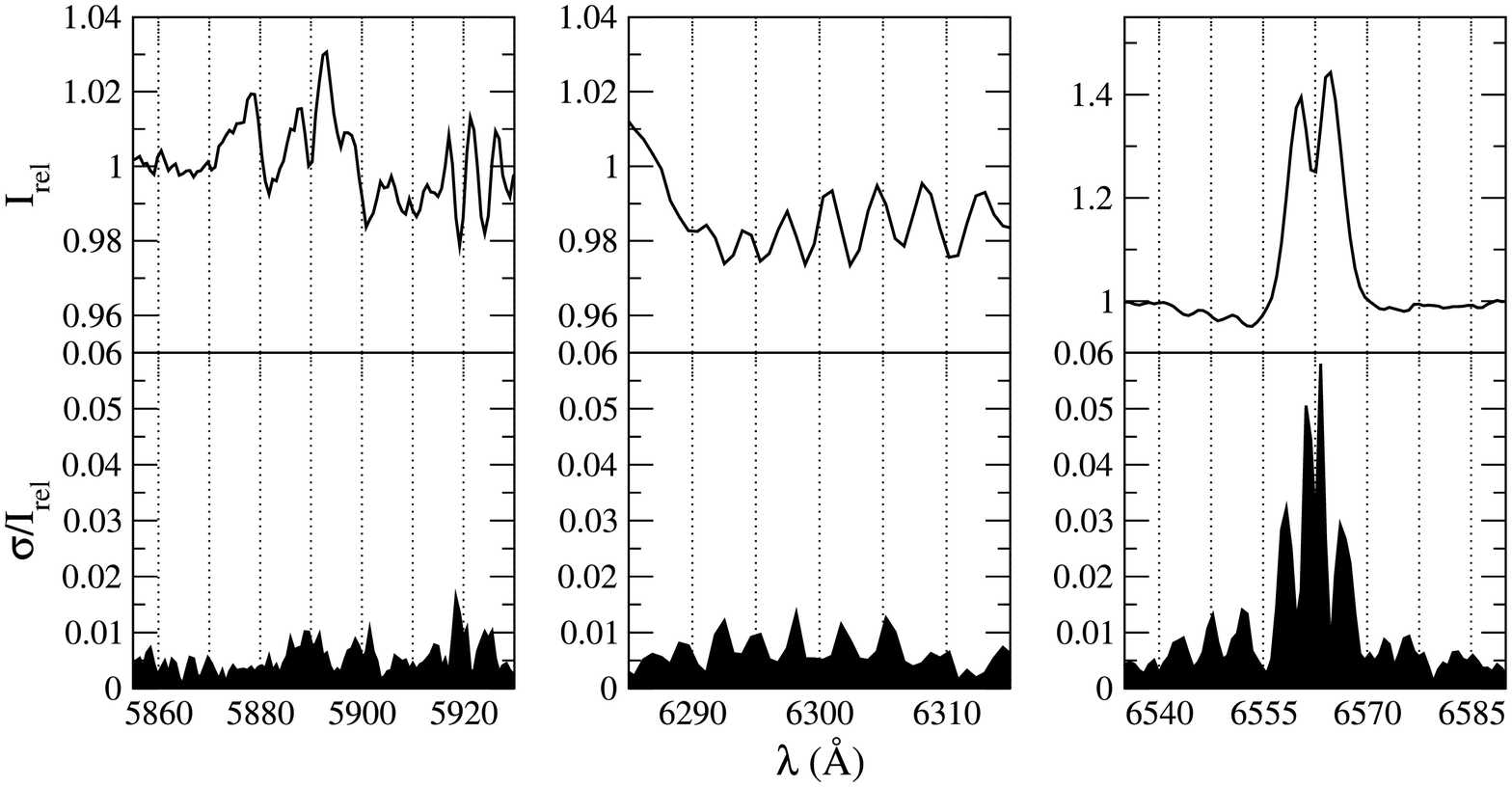} \\
\includegraphics[height=47mm,clip=true]{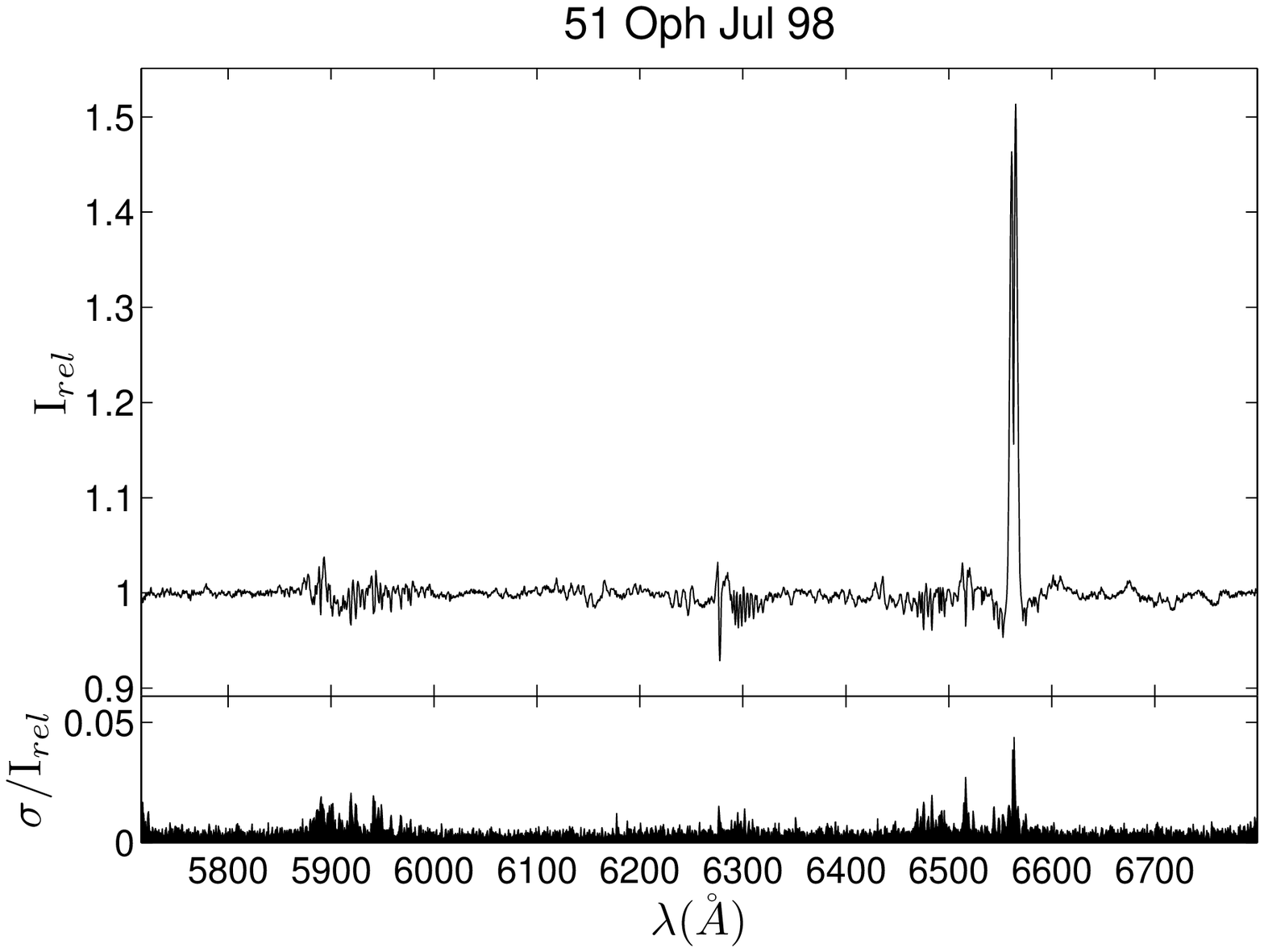}&
\includegraphics[bb=4 77 763 470,height=45mm,clip=true]{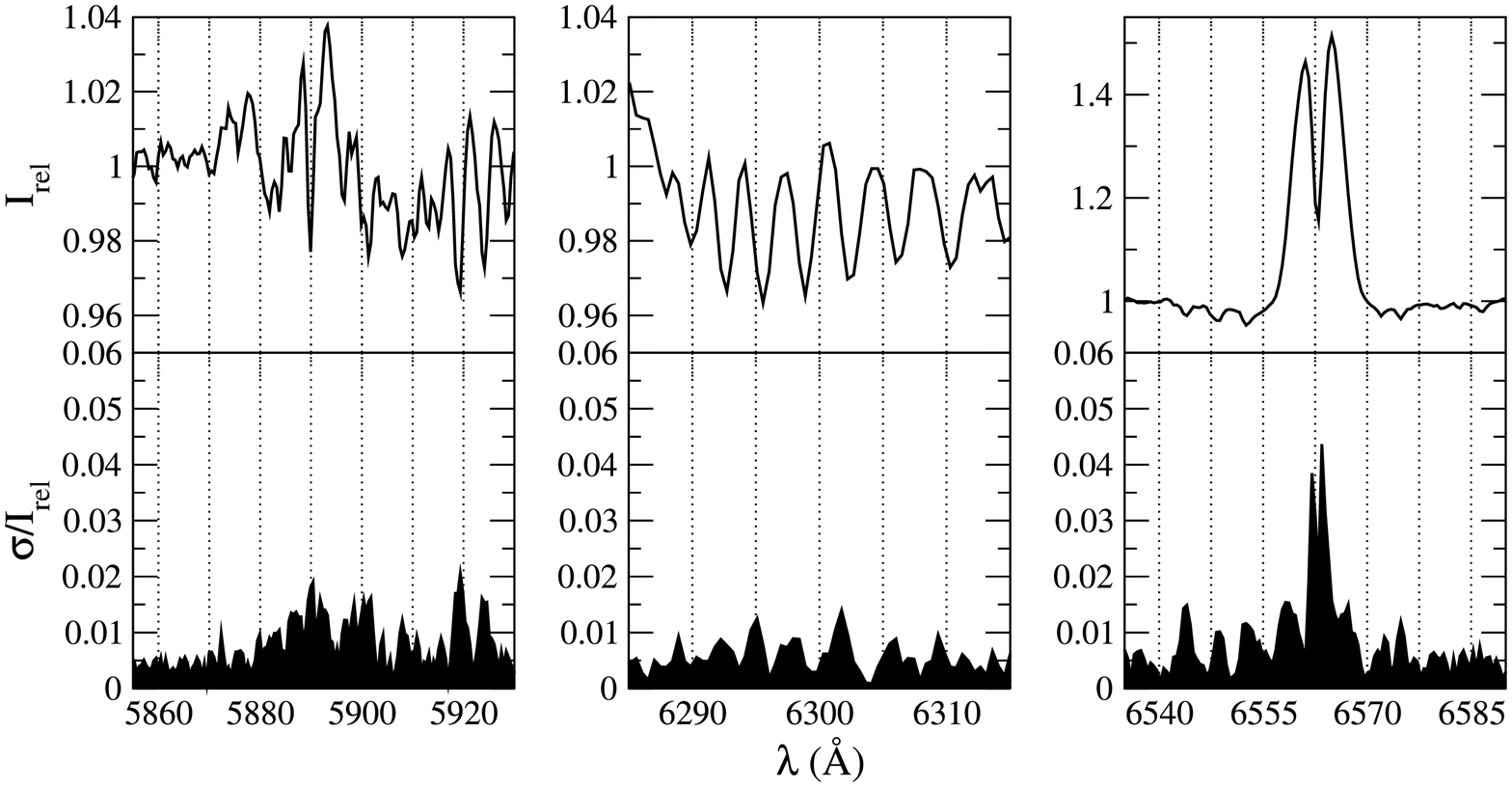} \\ 
\includegraphics[height=47mm,clip=true]{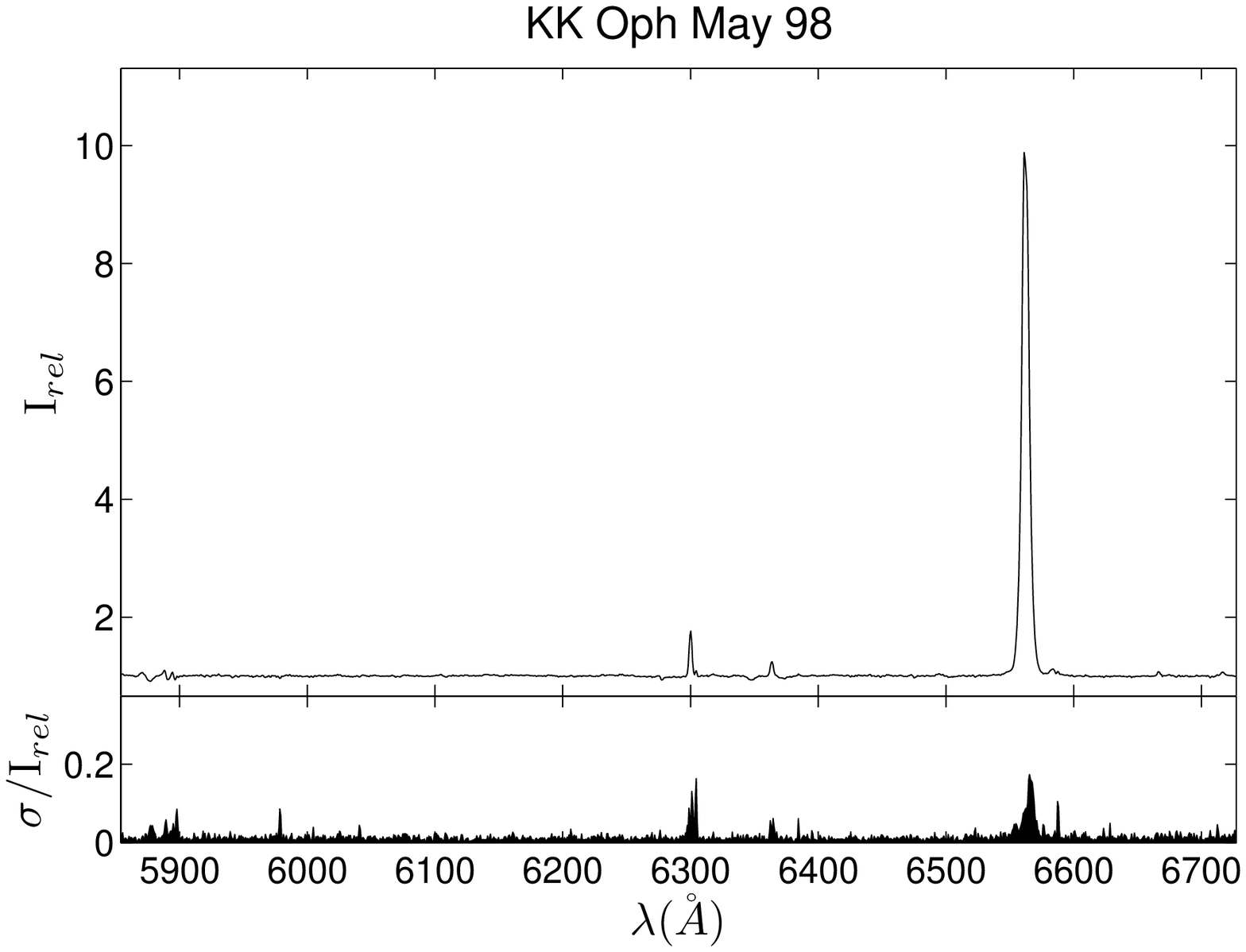}&
\includegraphics[bb=4 77 763 470,height=45mm,clip=true]{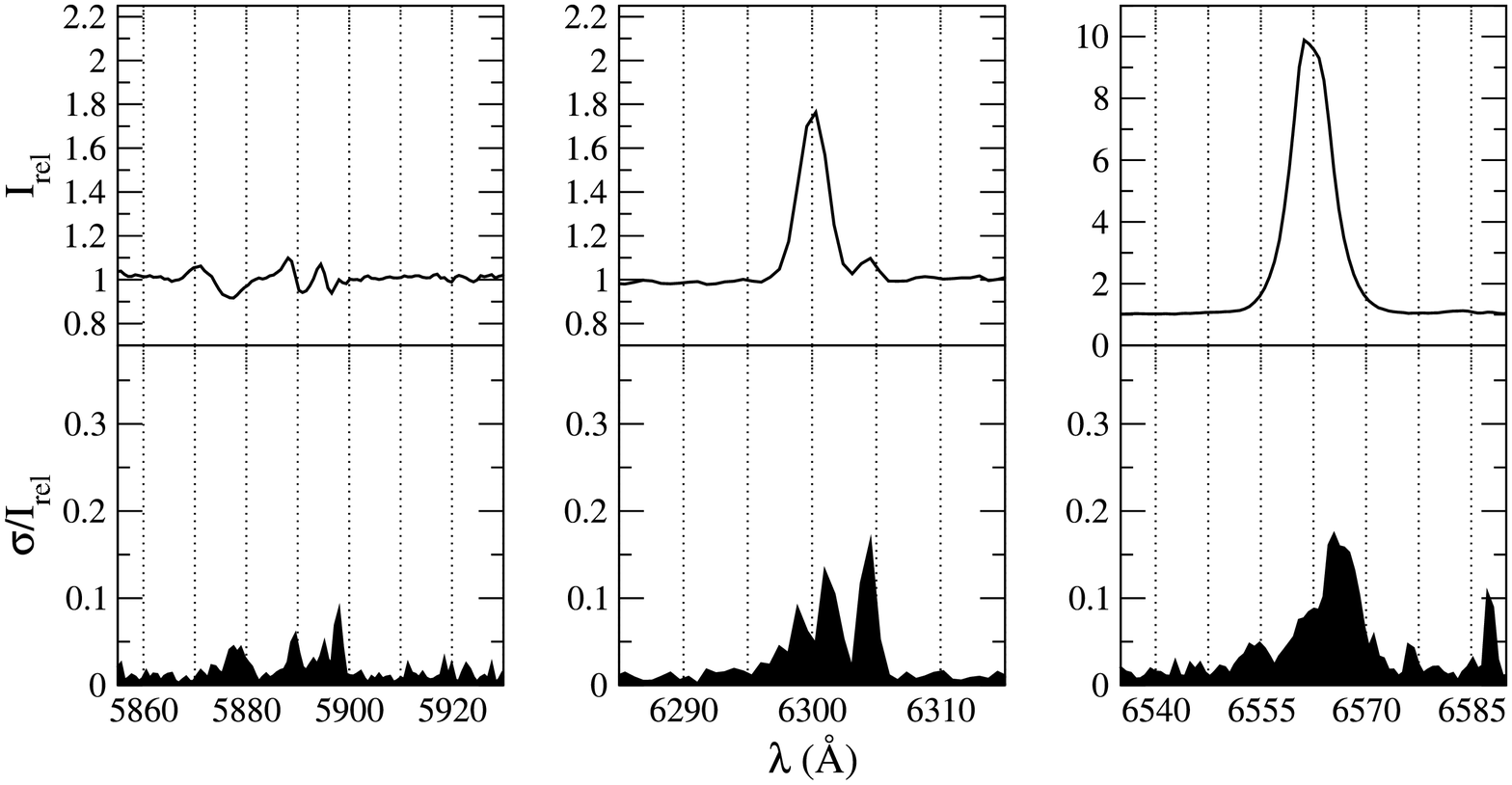} \\
\includegraphics[height=47mm,clip=true]{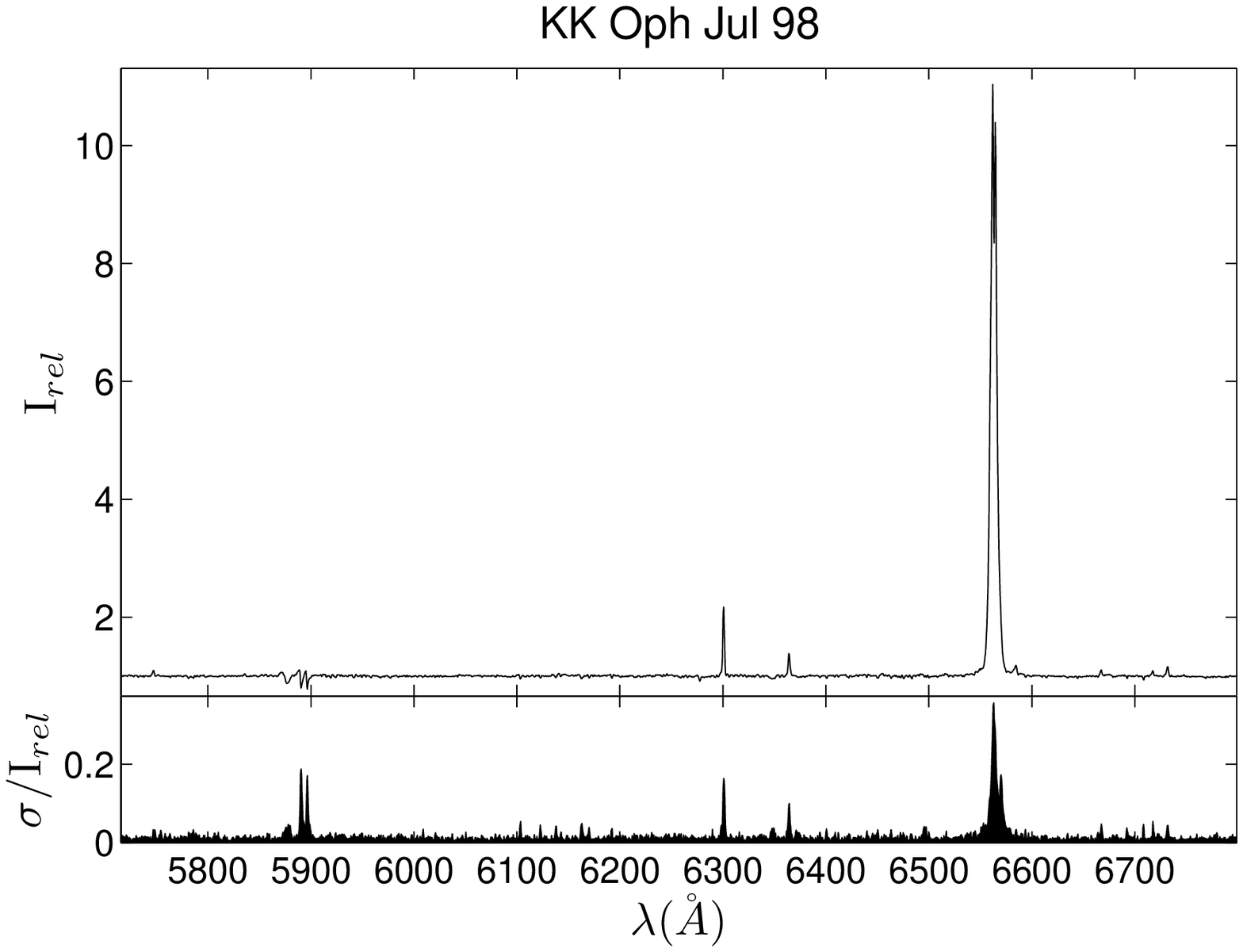}&
\includegraphics[bb=4 77 763 470,height=45mm,clip=true]{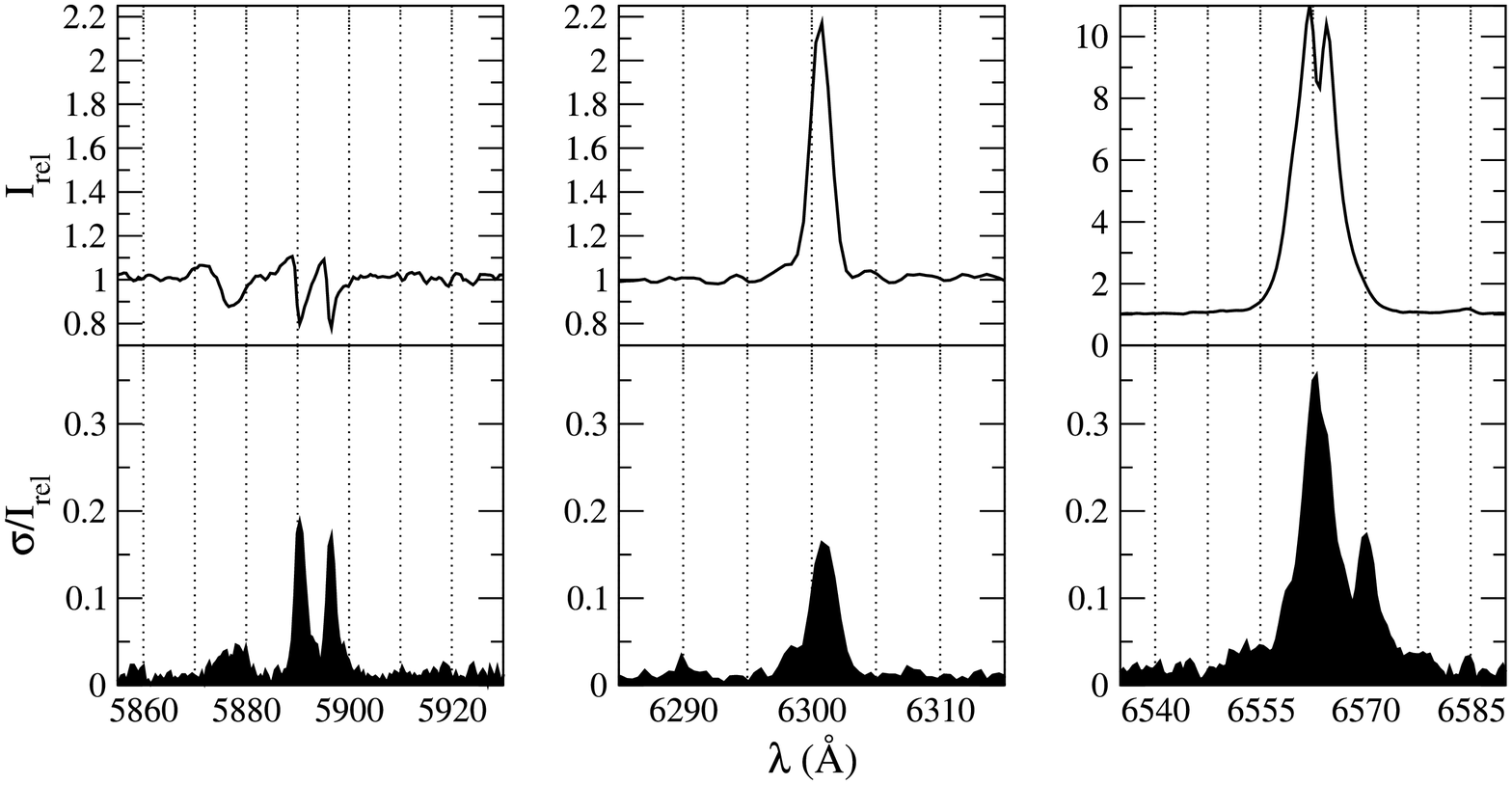} \\ 
\end{tabular}
\end{table}
\clearpage
\begin{table}
\centering
\renewcommand\arraystretch{10}
\begin{tabular}{cc}
\includegraphics[height=47mm,clip=true]{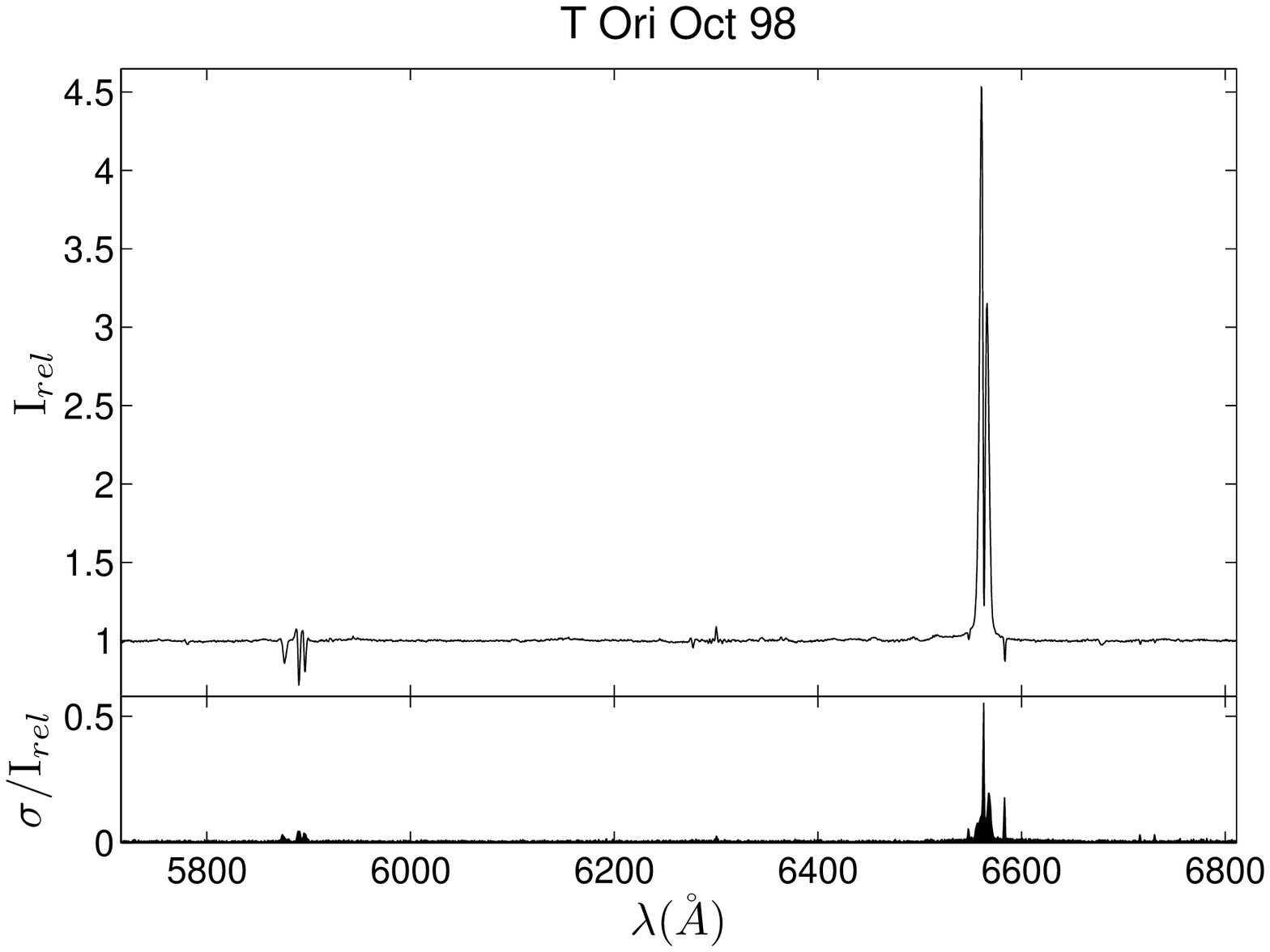}&
\includegraphics[bb=4 77 763 470,height=45mm,clip=true]{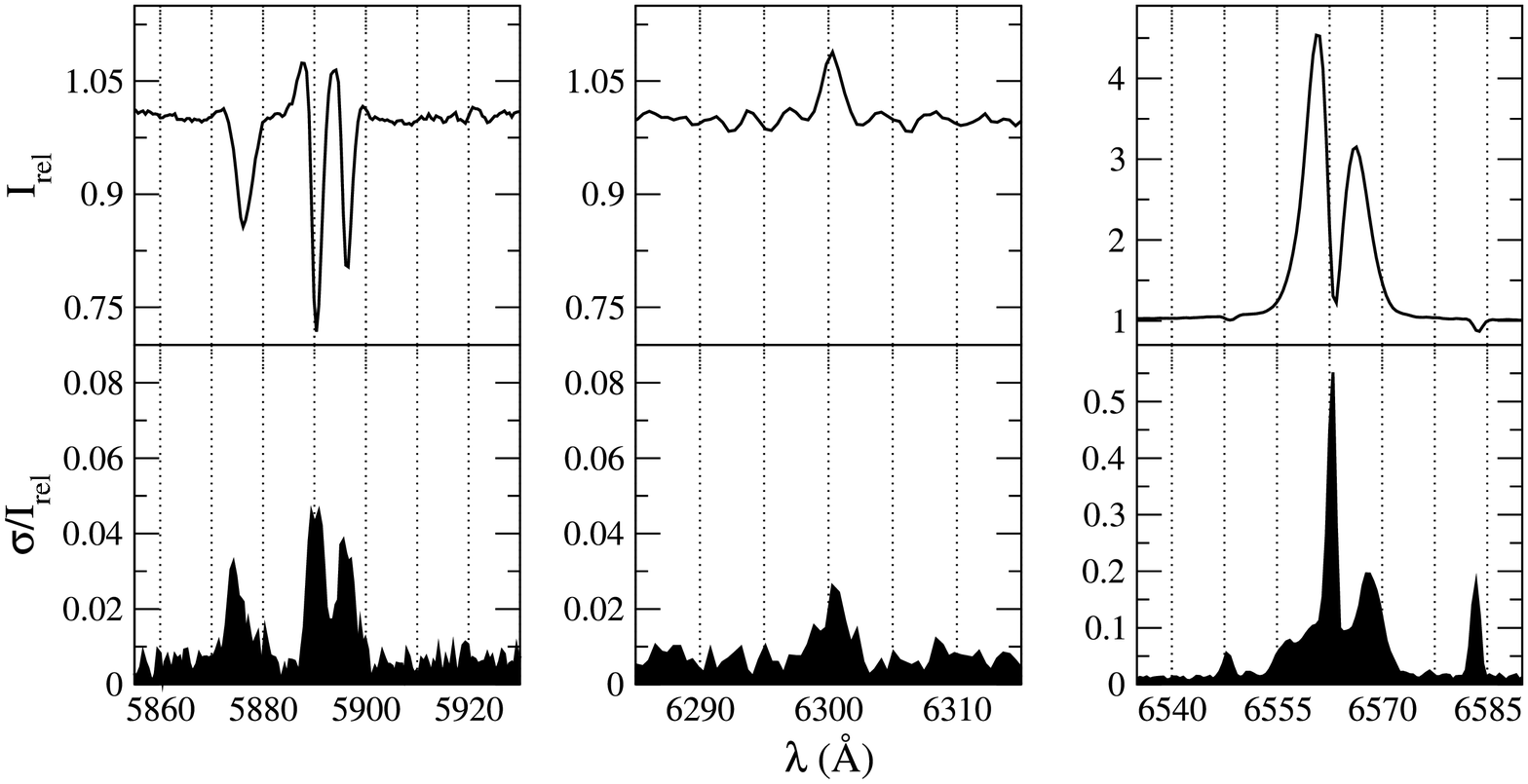} \\ 
\includegraphics[height=47mm,clip=true]{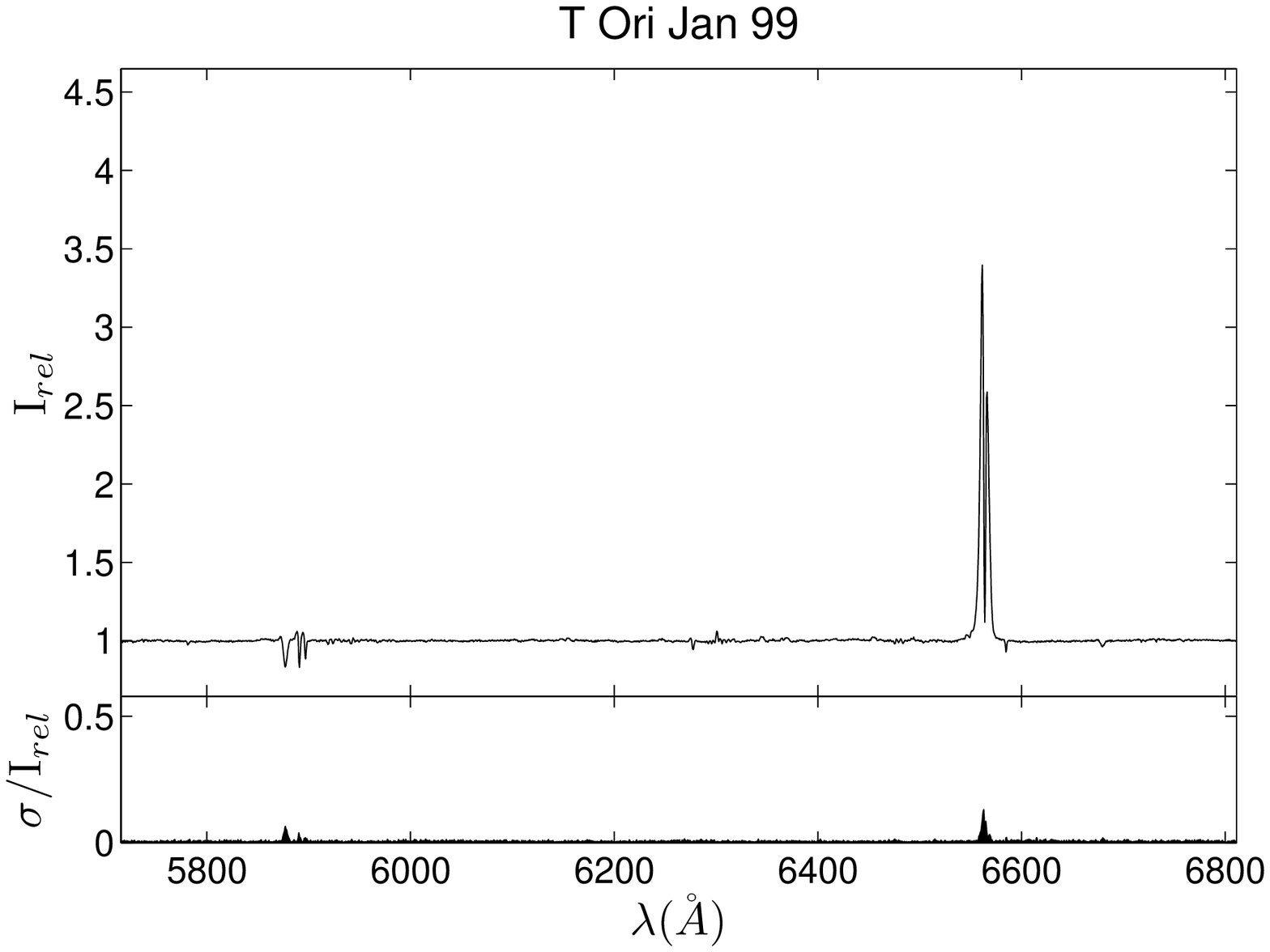}&
\includegraphics[bb=4 77 763 470,height=45mm,clip=true]{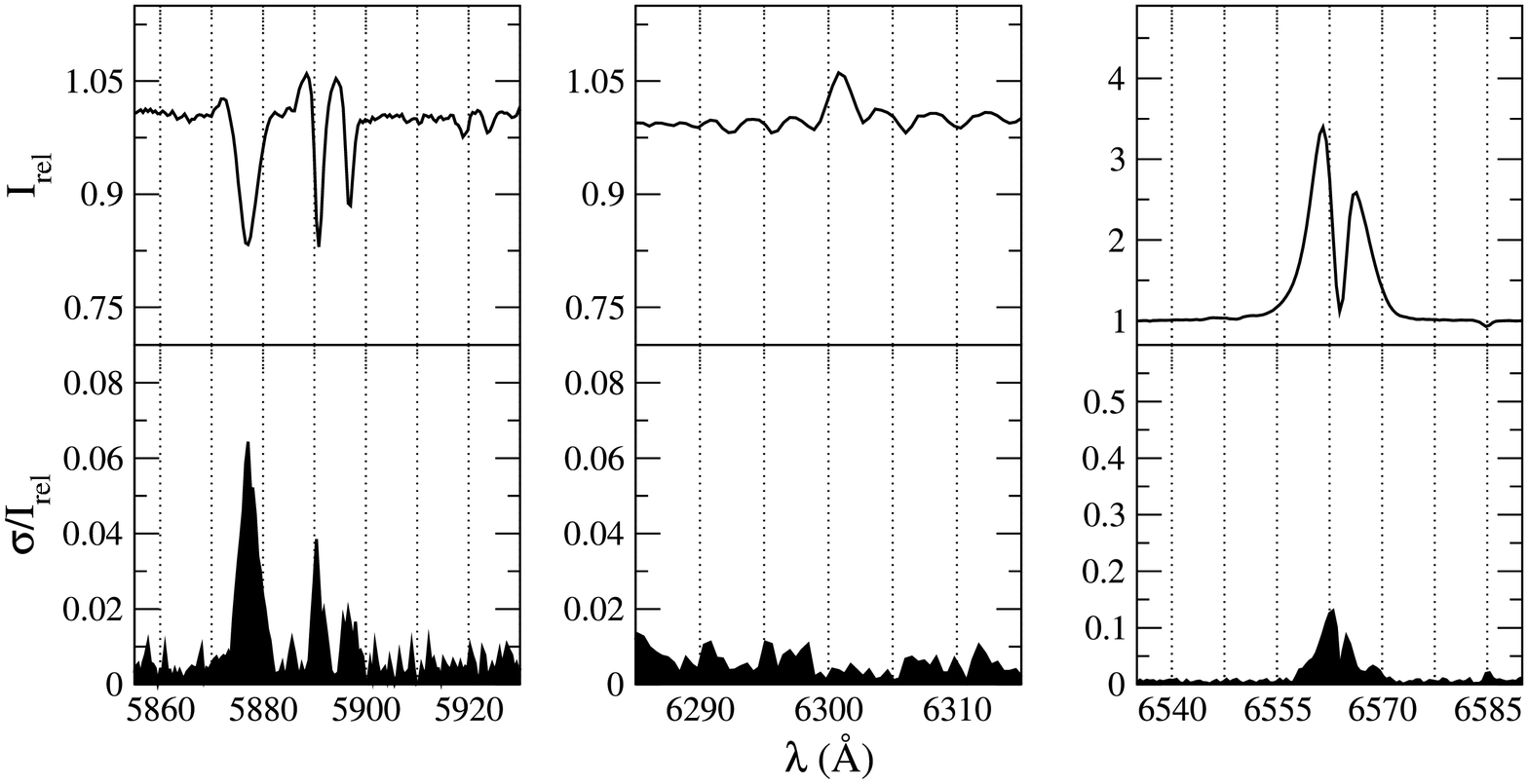} \\ 
\includegraphics[height=47mm,clip=true]{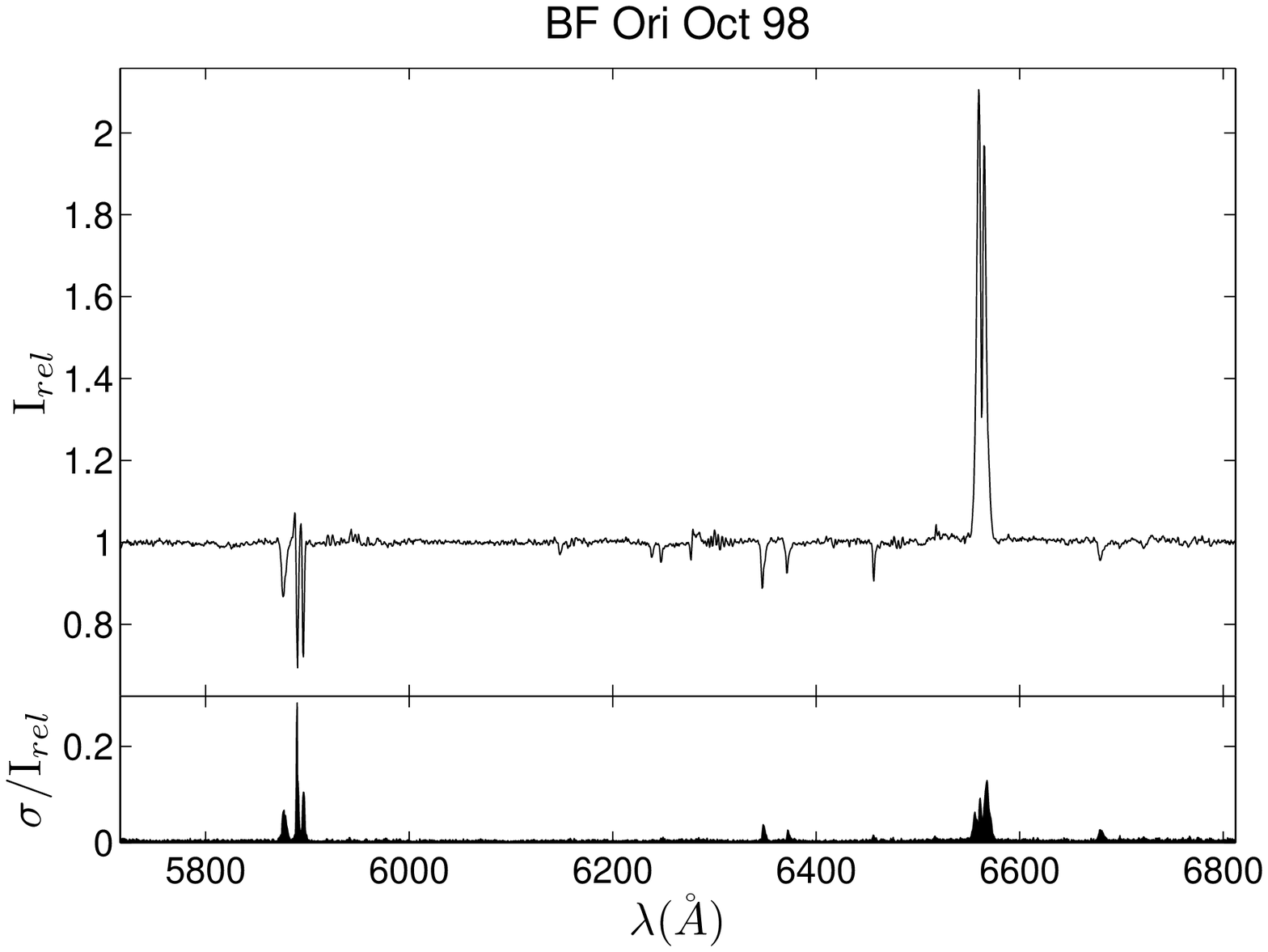}&
\includegraphics[bb=4 77 763 470,height=45mm,clip=true]{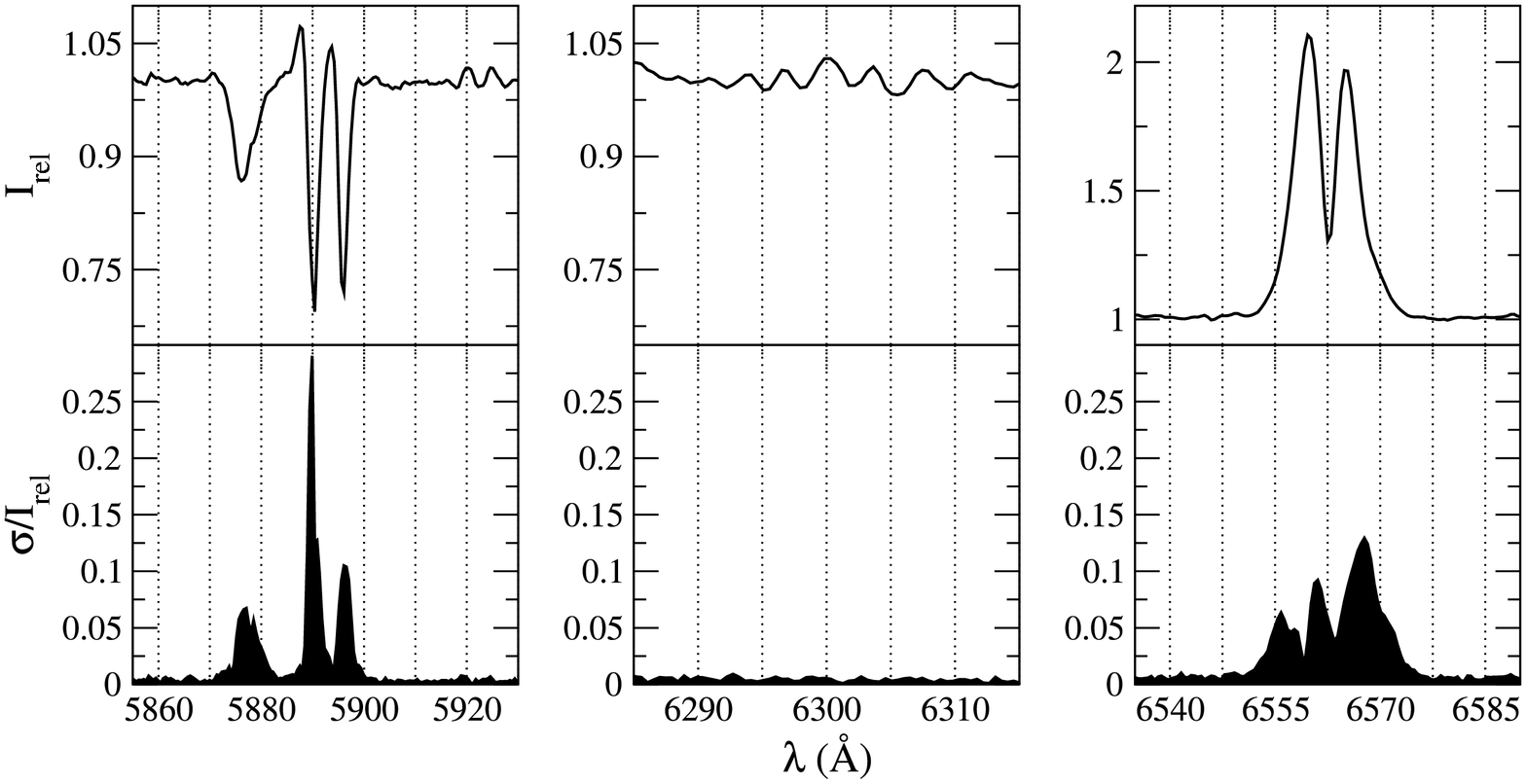} \\ 
\includegraphics[height=47mm,clip=true]{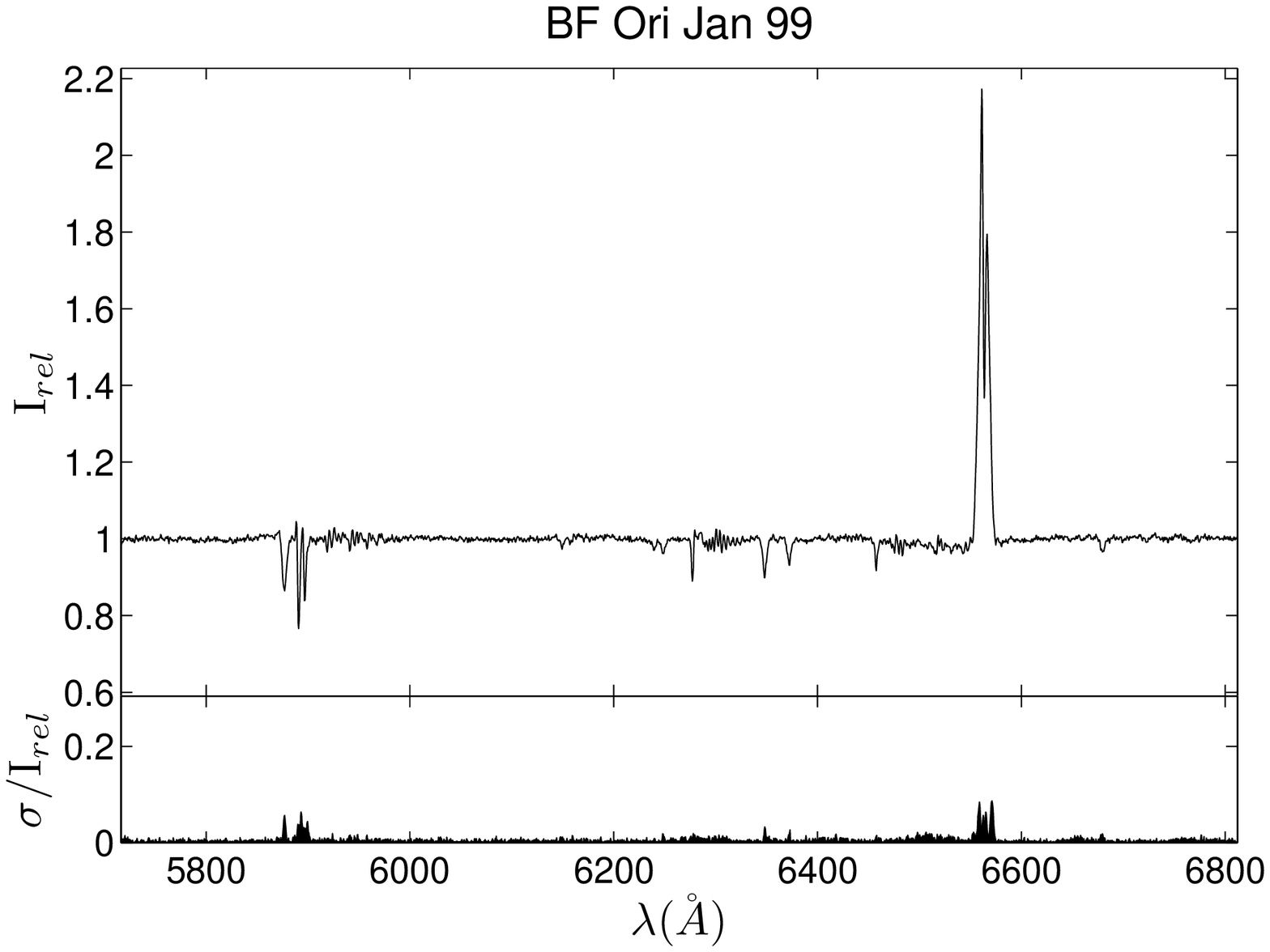}&
\includegraphics[bb=4 77 763 470,height=45mm,clip=true]{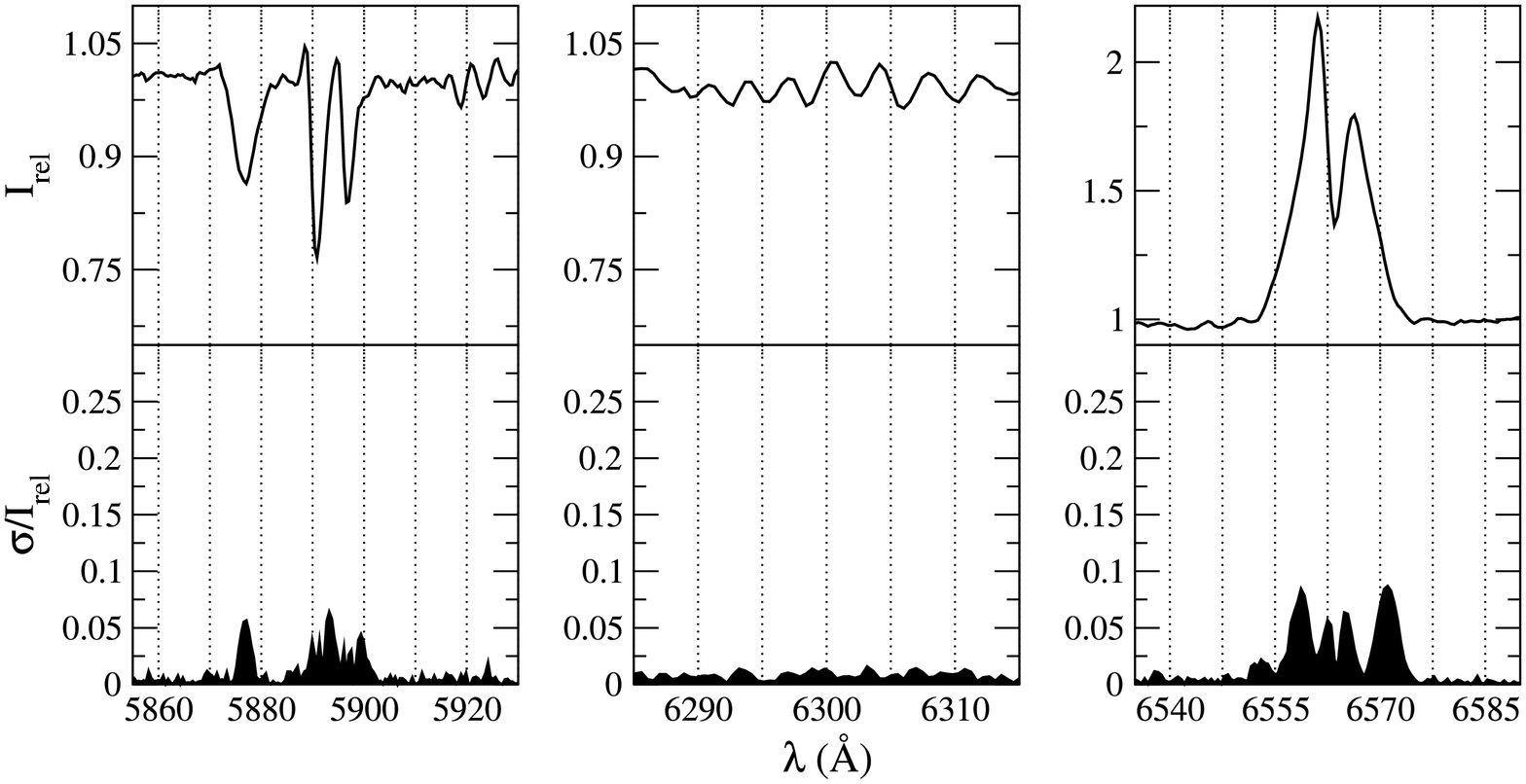} \\ 
\end{tabular}
\end{table}
\clearpage
\begin{table}
\centering
\renewcommand\arraystretch{10}
\begin{tabular}{cc}
\includegraphics[height=47mm,clip=true]{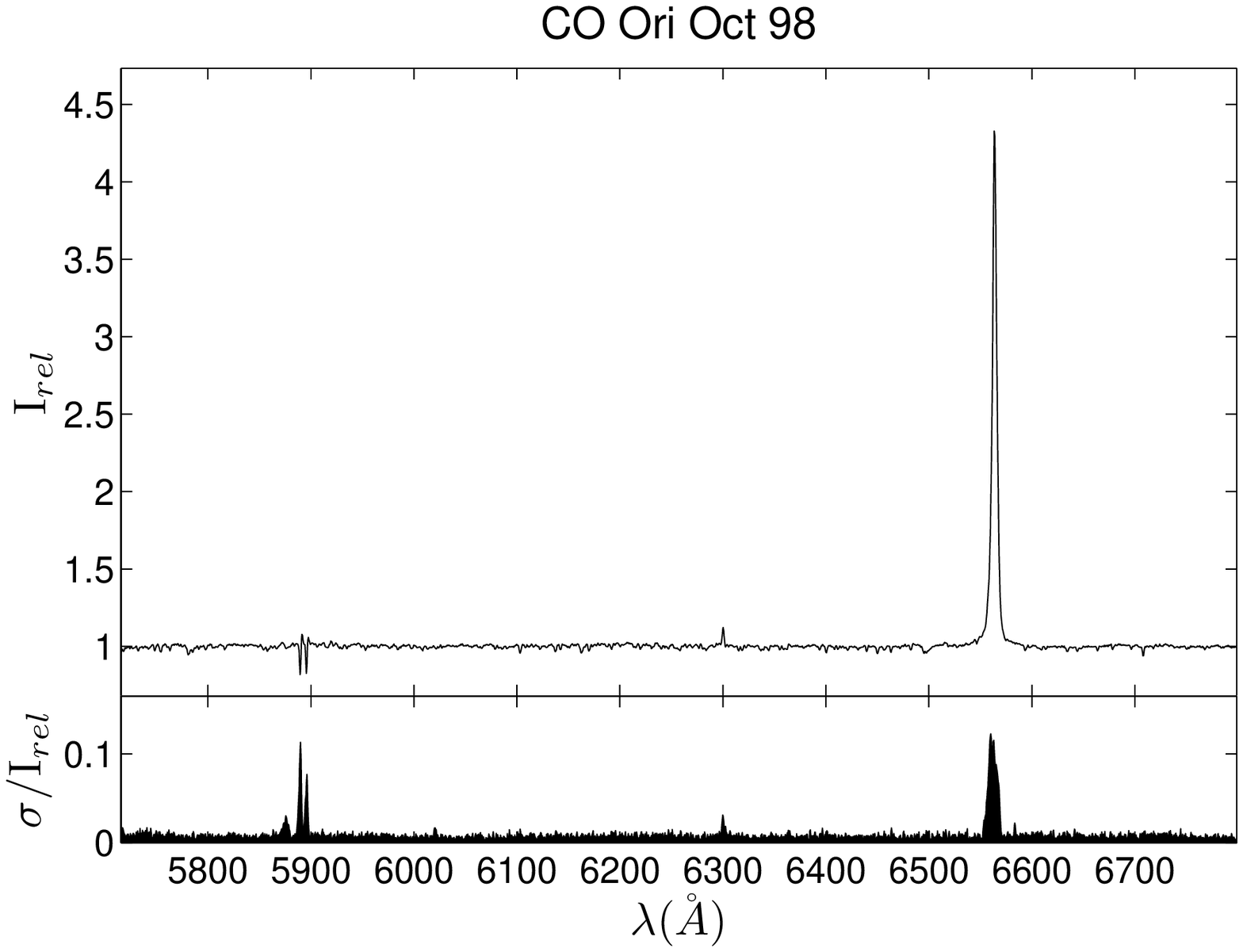}&
\includegraphics[bb=4 77 763 470,height=45mm,clip=true]{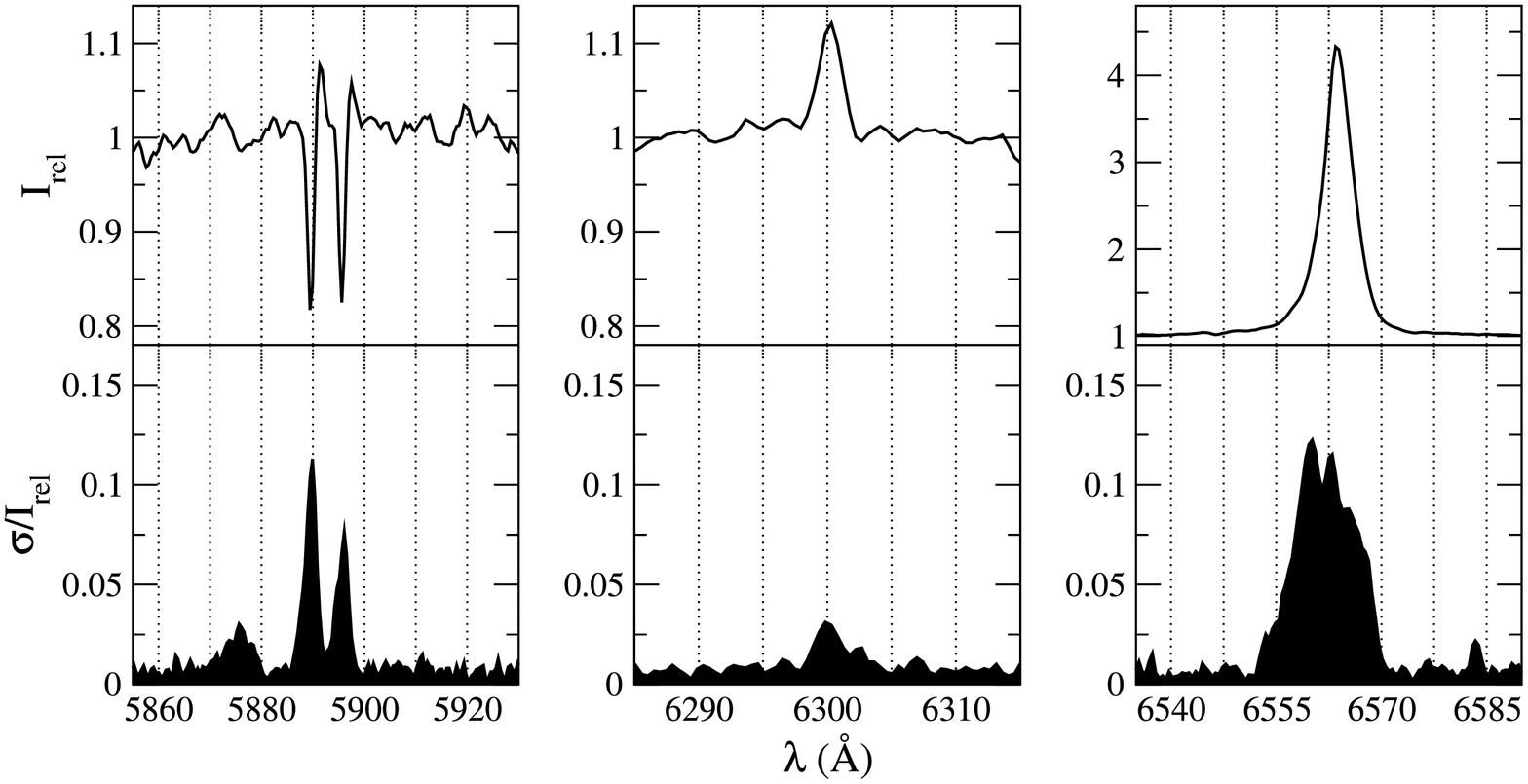} \\ 
\includegraphics[height=47mm,clip=true]{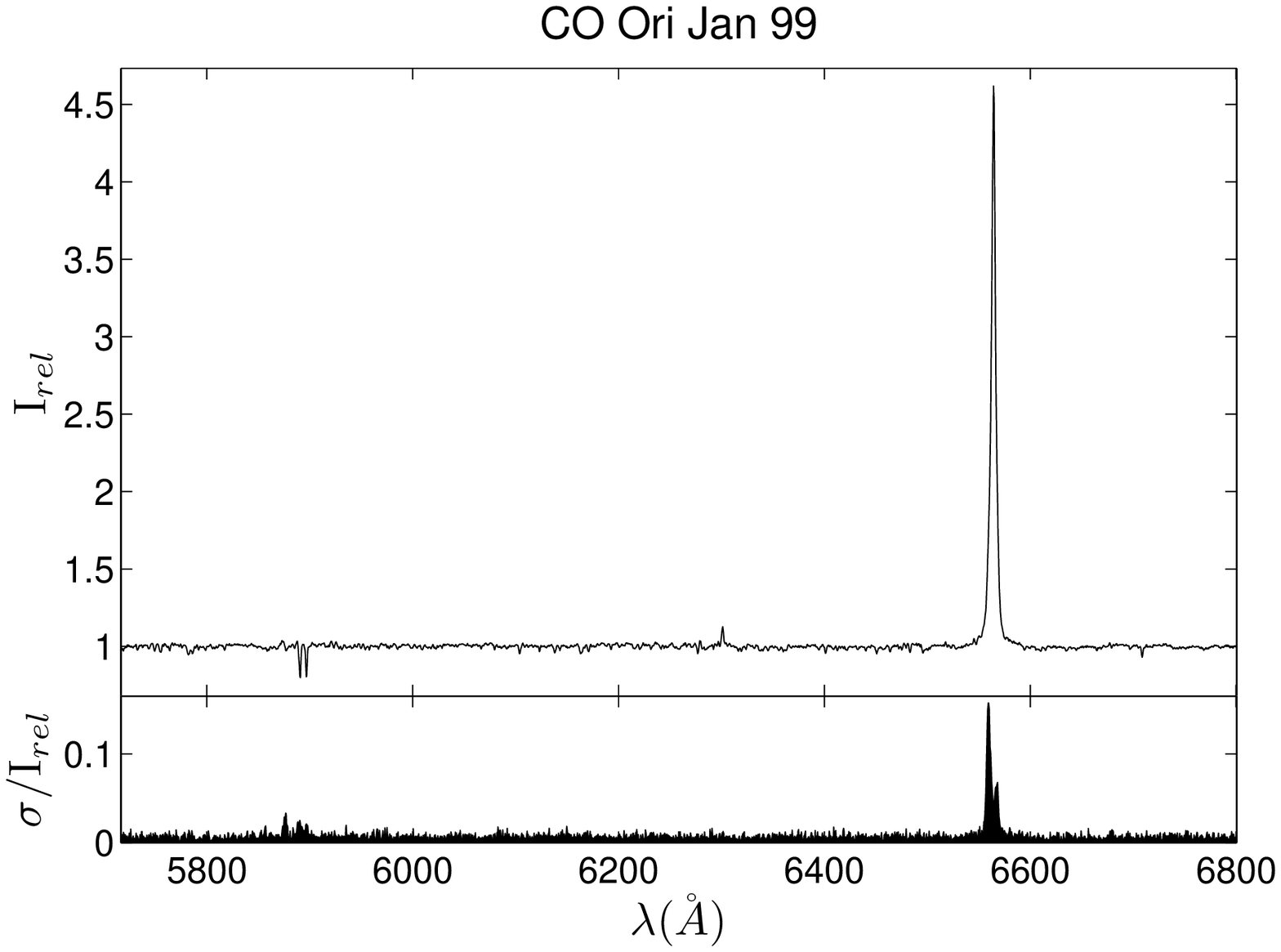}&
\includegraphics[bb=4 77 763 470,height=45mm,clip=true]{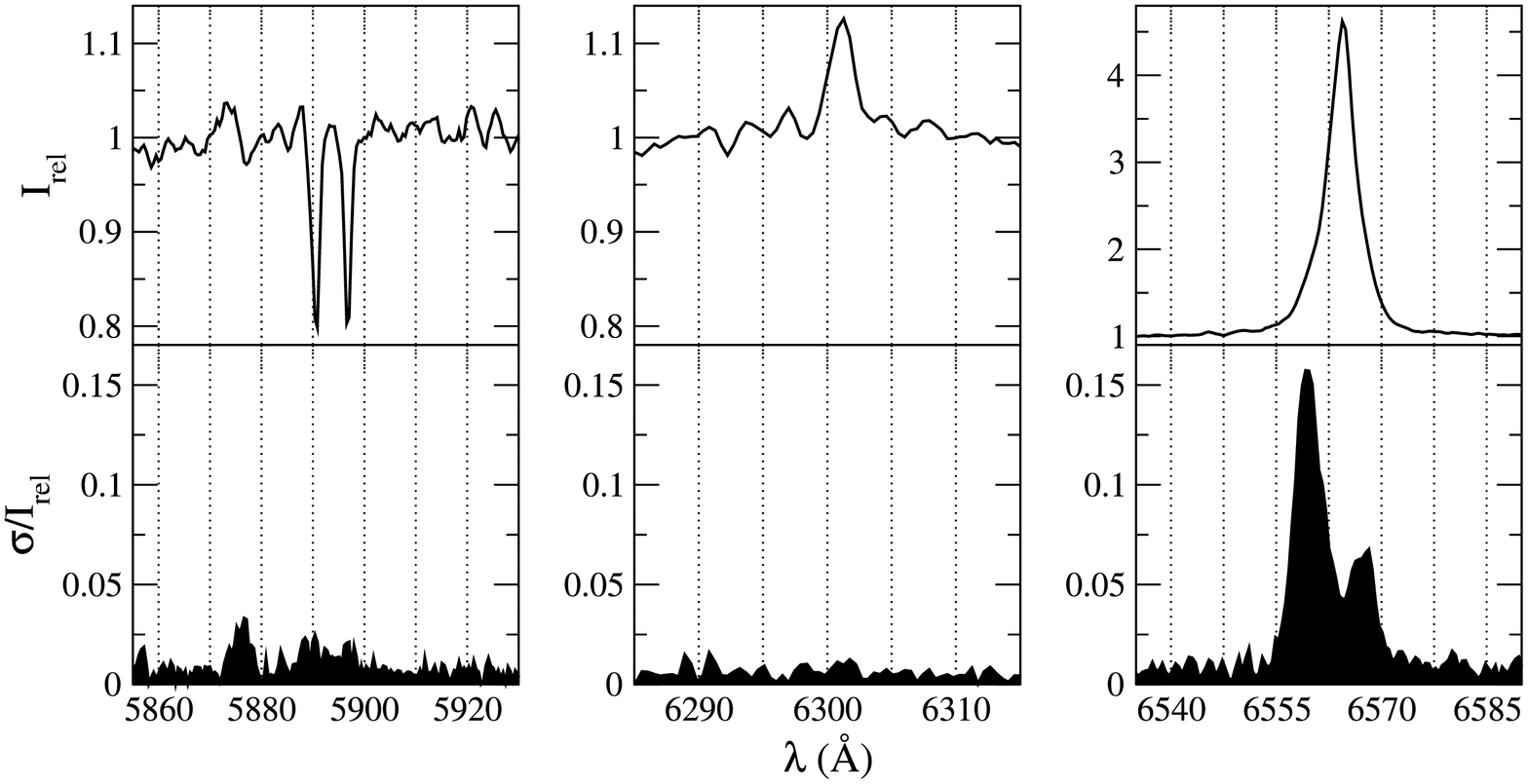} \\ 
\includegraphics[height=47mm,clip=true]{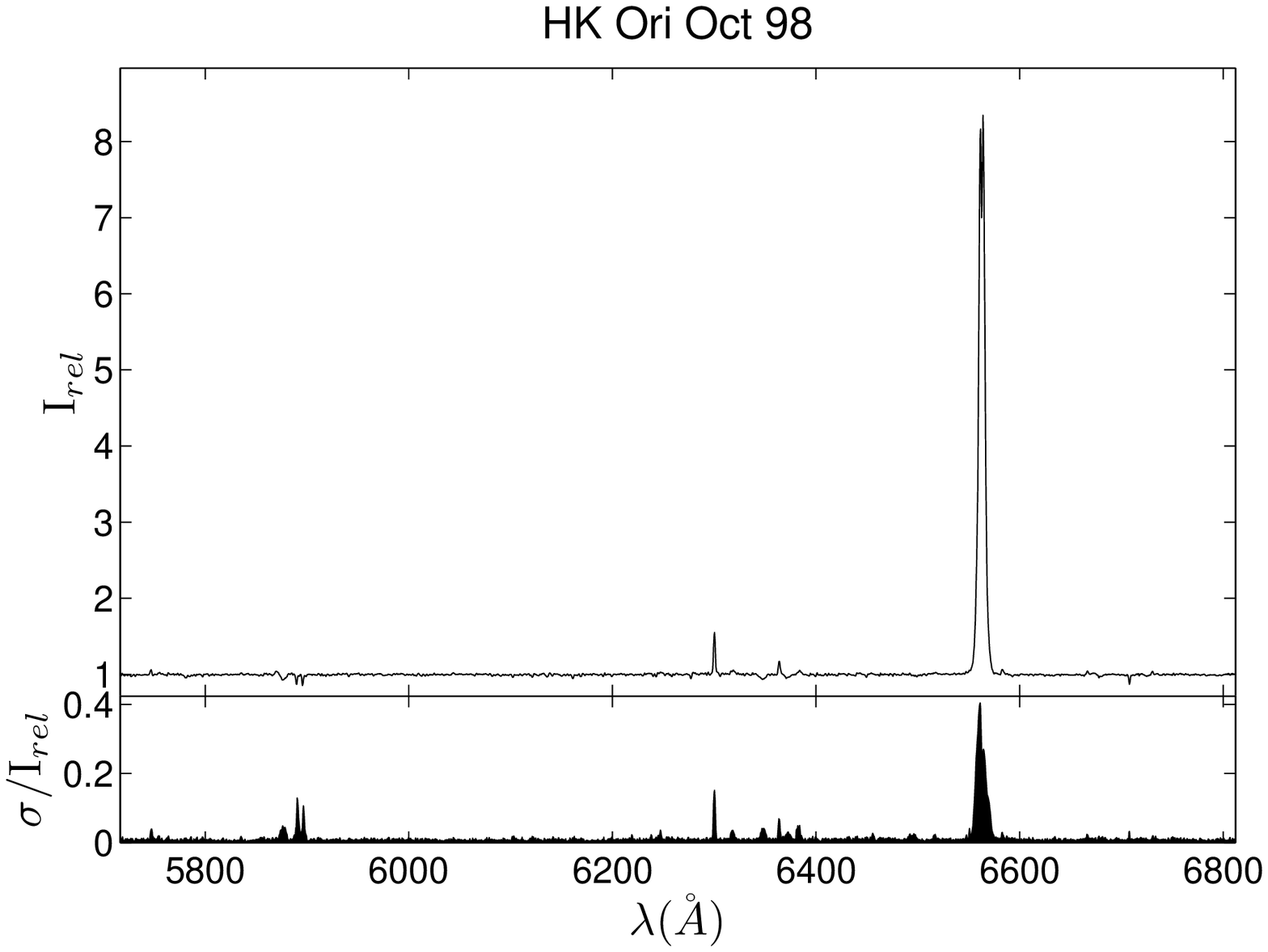}&
\includegraphics[bb=4 77 763 470,height=45mm,clip=true]{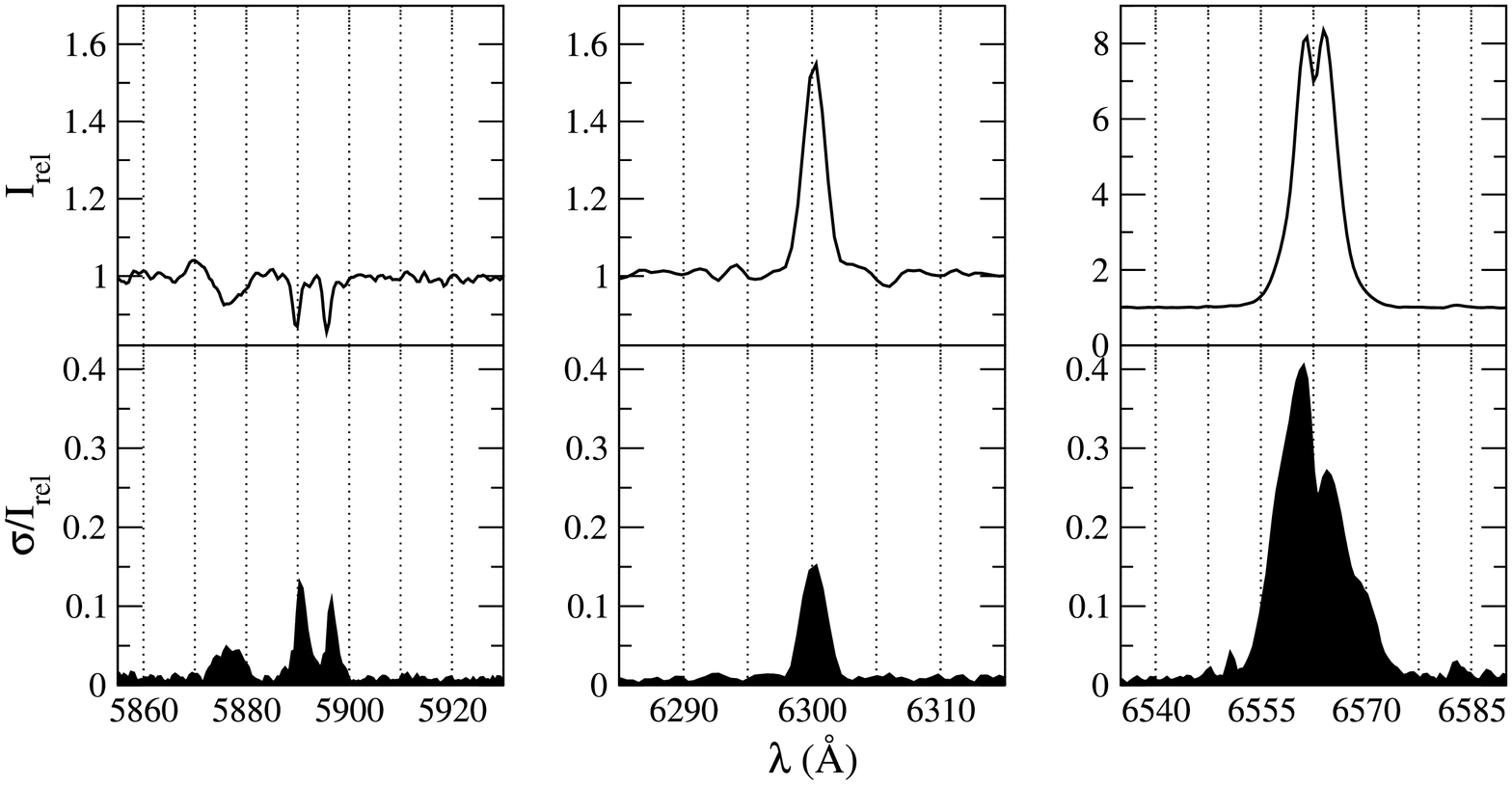} \\ 
\includegraphics[height=47mm,clip=true]{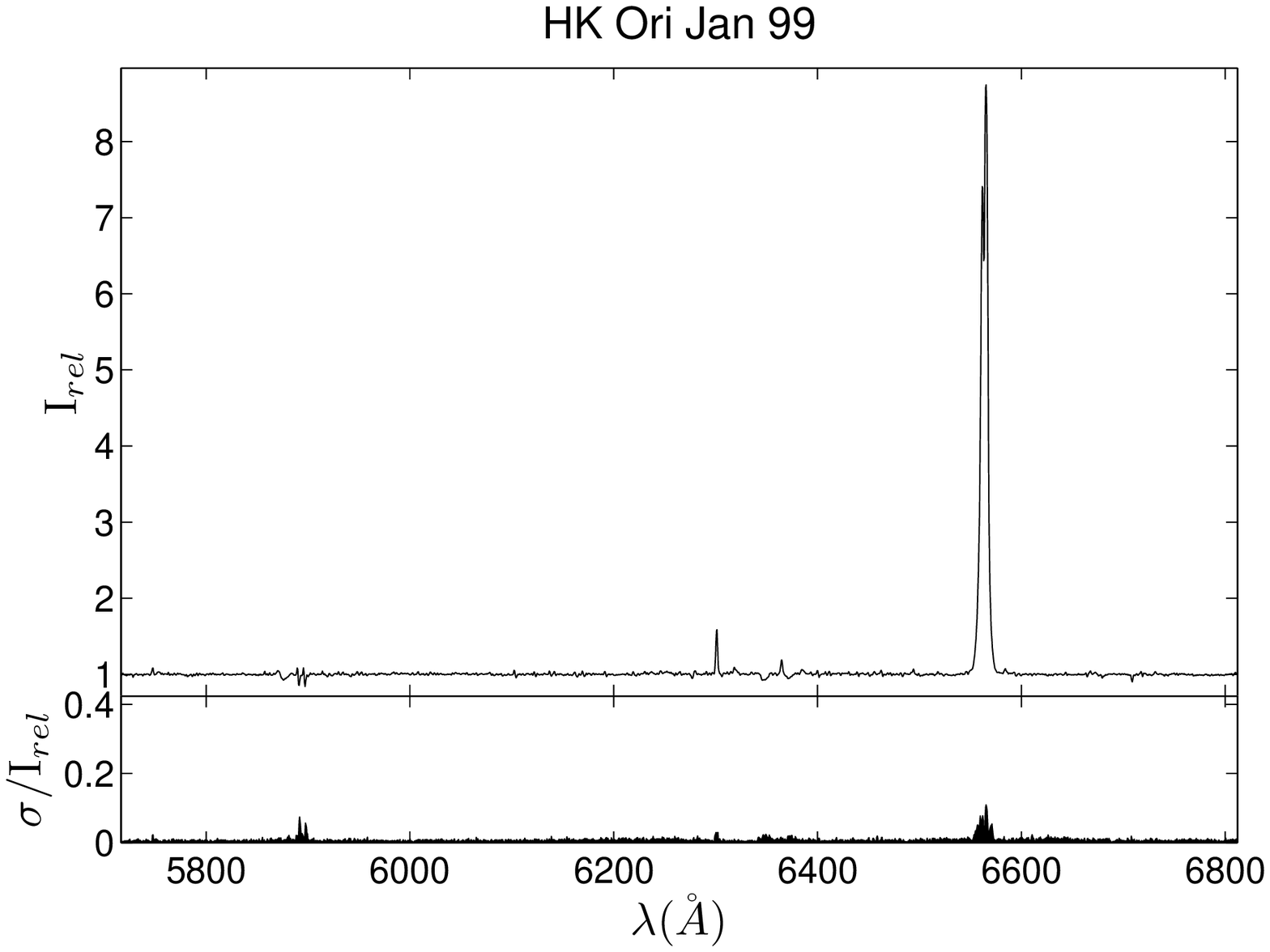}&
\includegraphics[bb=4 77 763 470,height=45mm,clip=true]{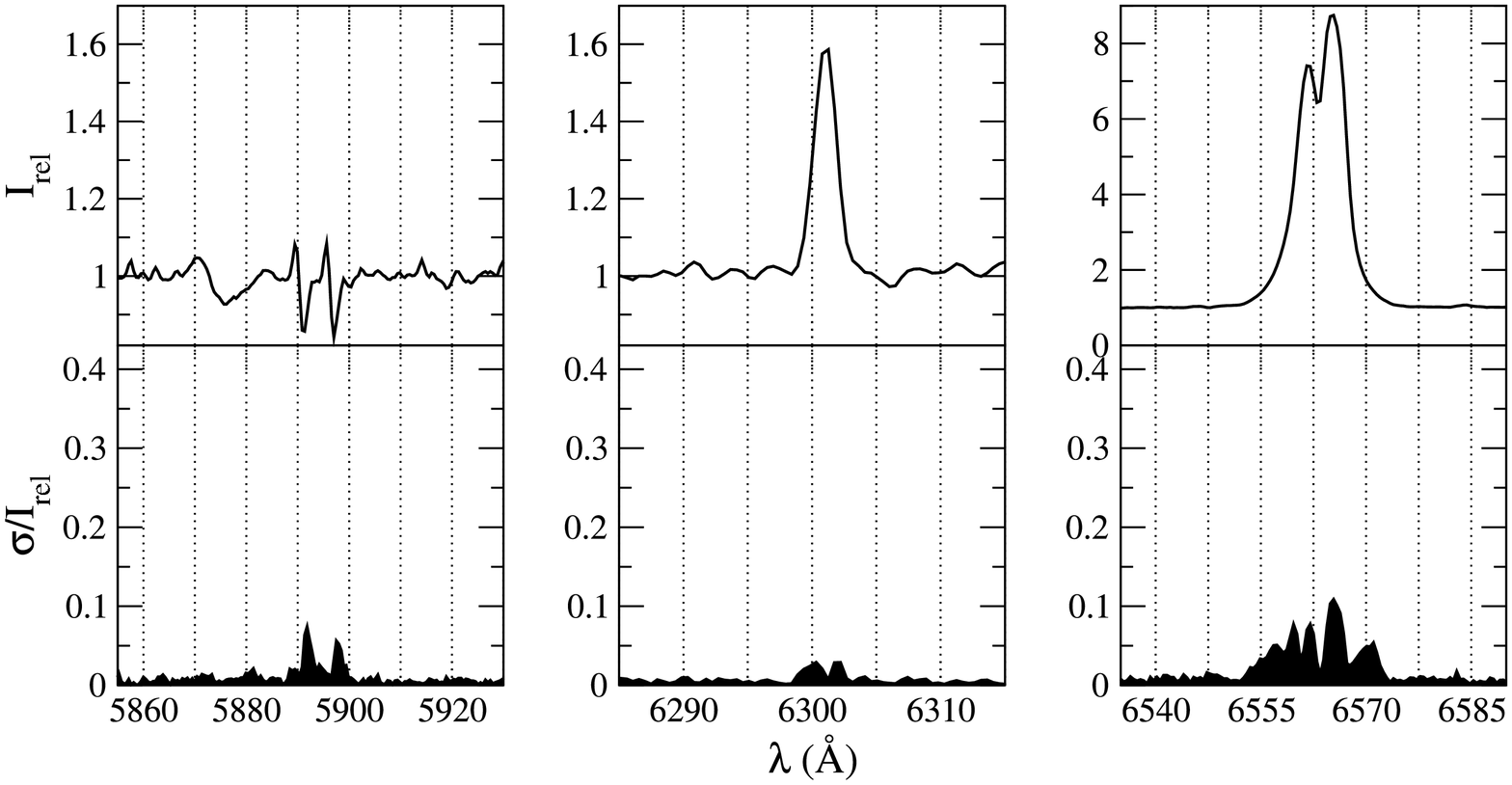} \\ 
\end{tabular}
\end{table}
\clearpage
\begin{table}
\centering
\renewcommand\arraystretch{10}
\begin{tabular}{cc}
\includegraphics[height=47mm,clip=true]{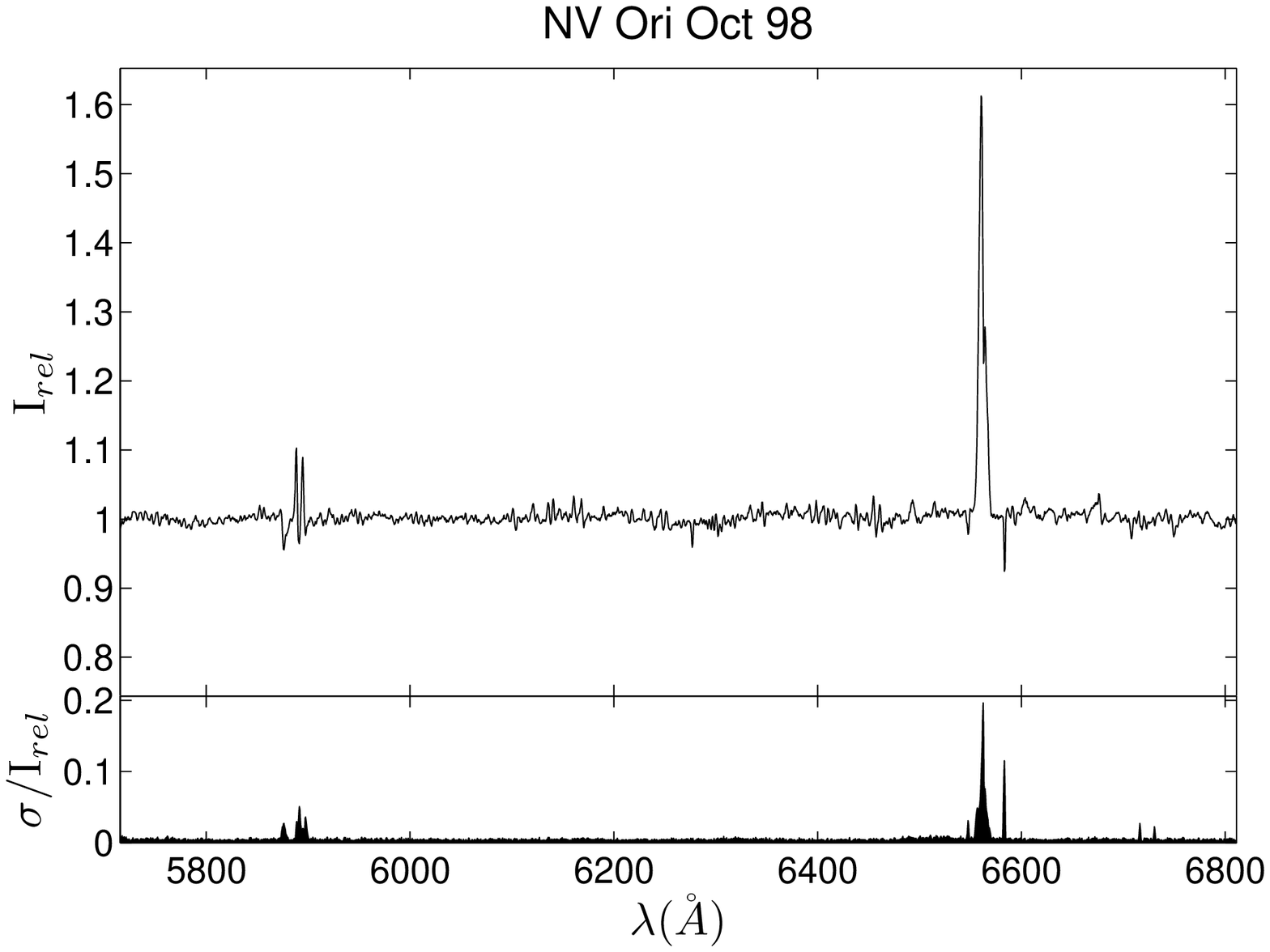}&
\includegraphics[bb=4 77 763 470,height=45mm,clip=true]{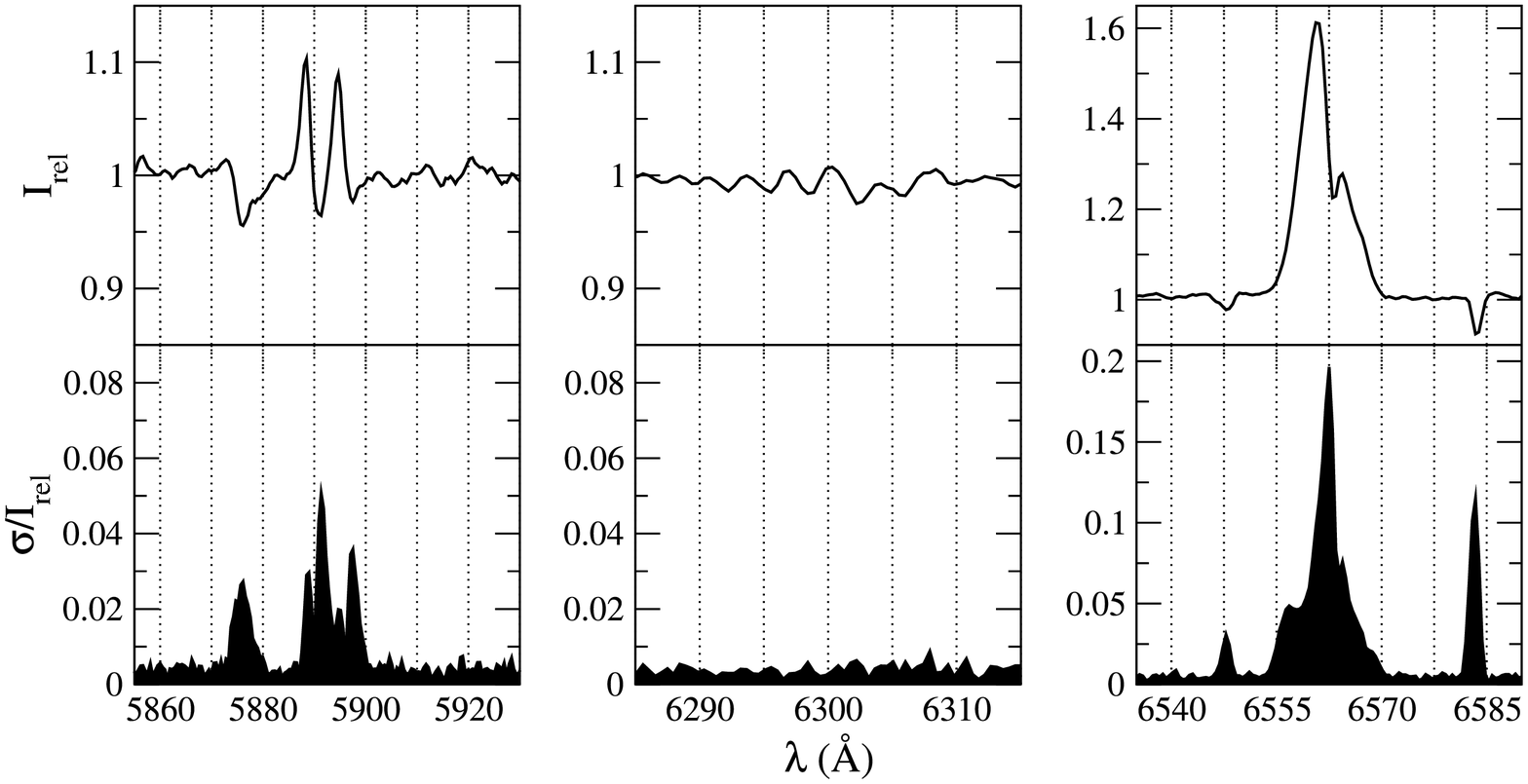} \\ 
\includegraphics[height=47mm,clip=true]{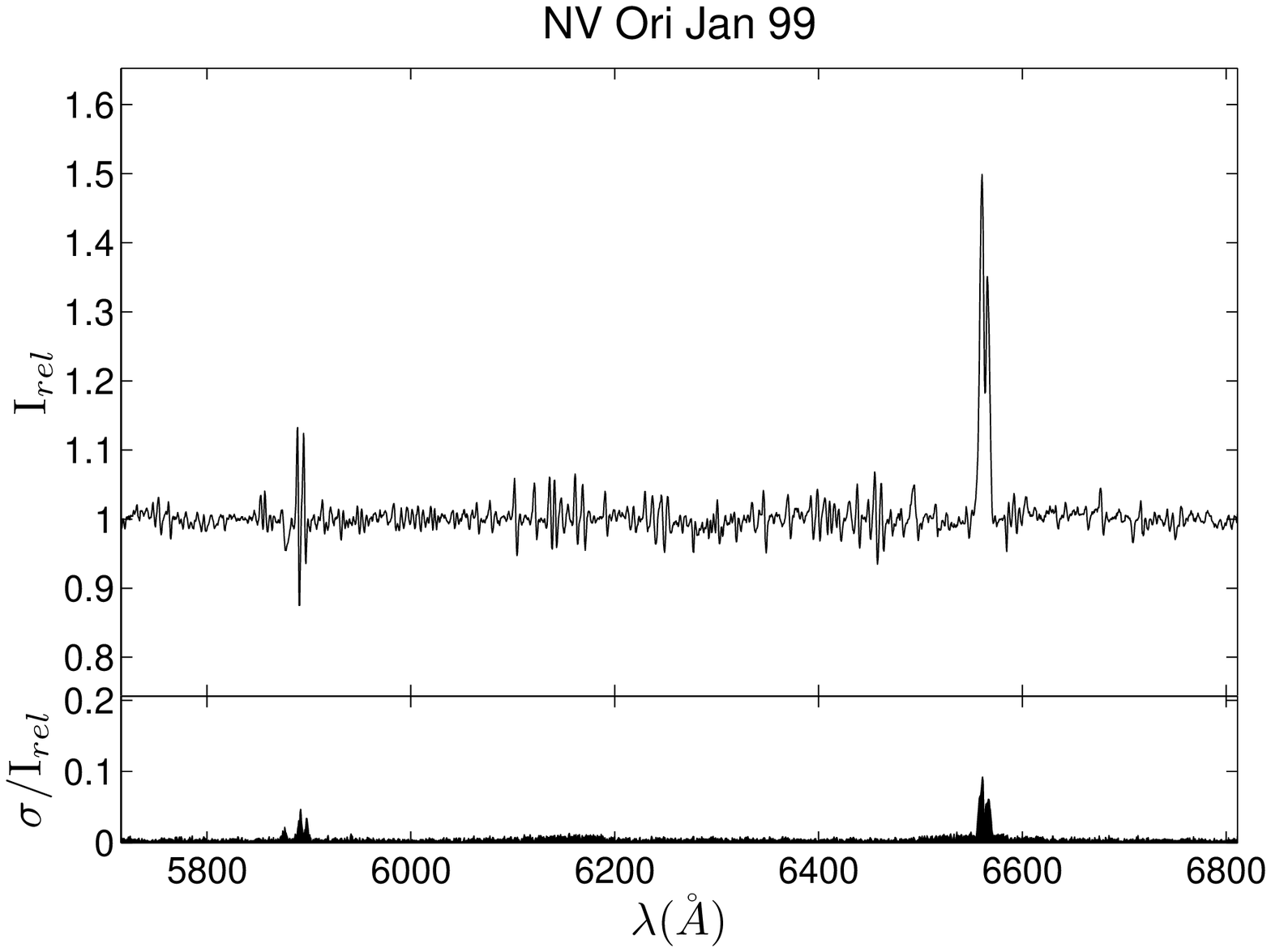}&
\includegraphics[bb=4 77 763 470,height=45mm,clip=true]{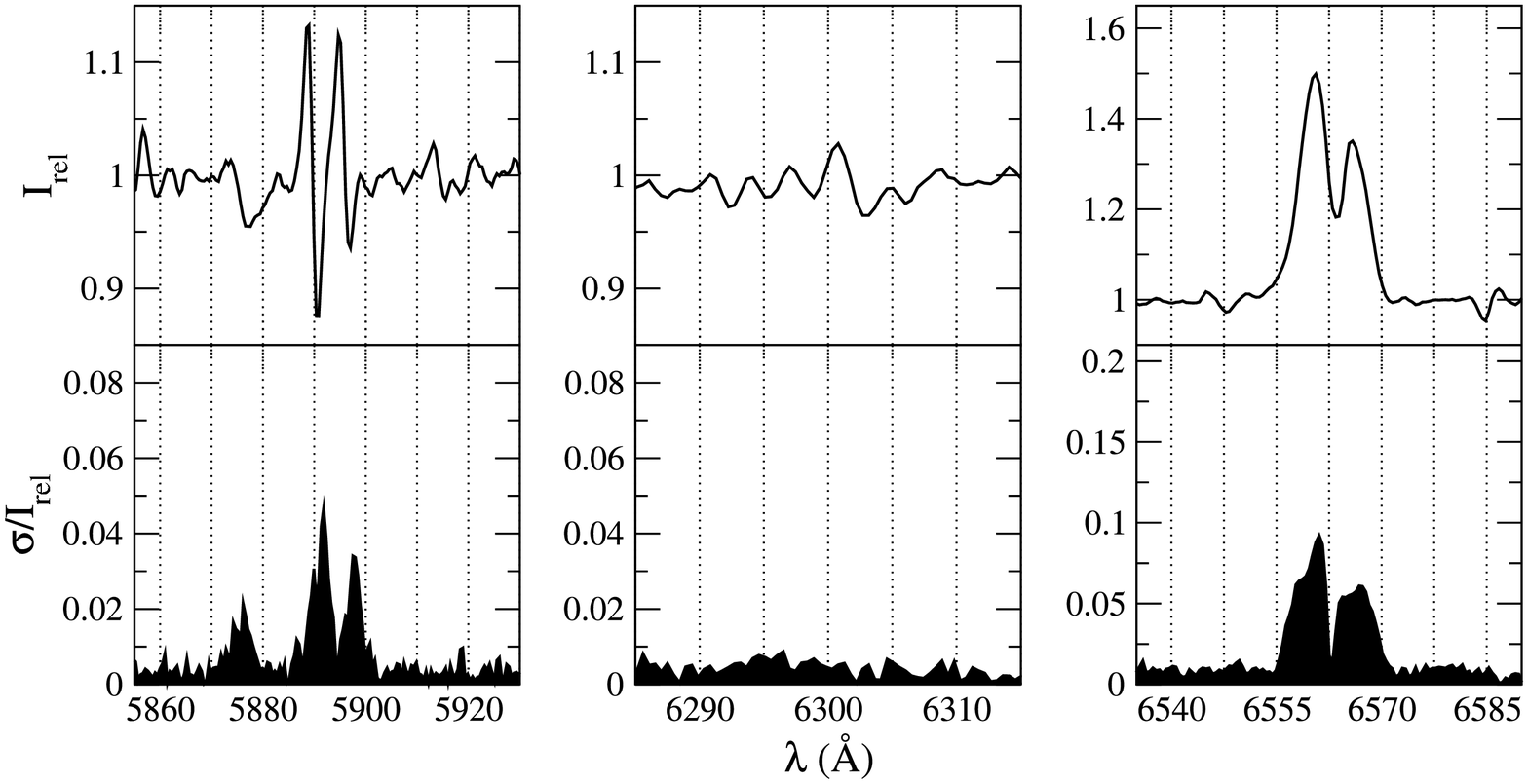} \\ 
\includegraphics[height=47mm,clip=true]{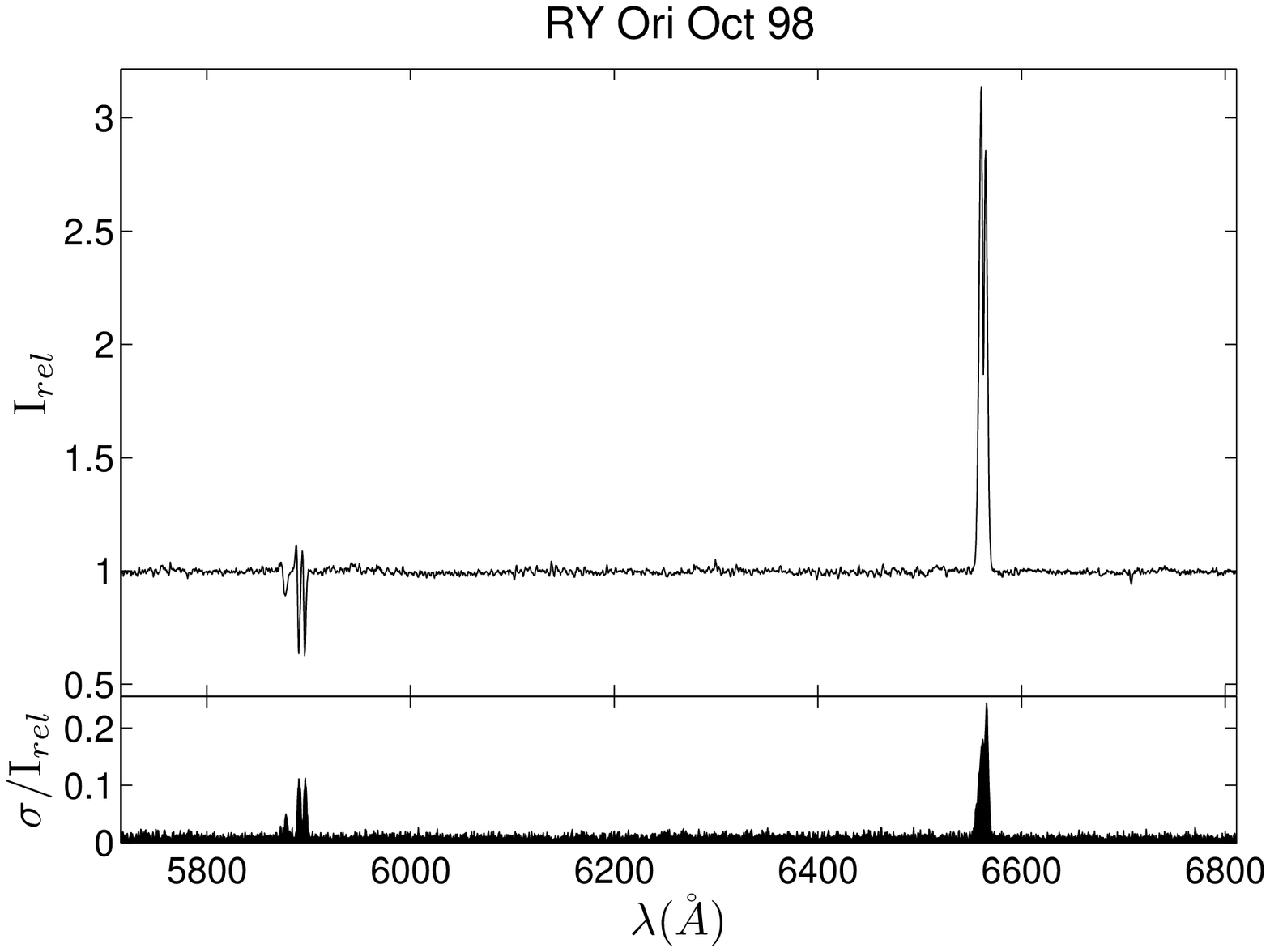}&
\includegraphics[bb=4 77 763 470,height=45mm,clip=true]{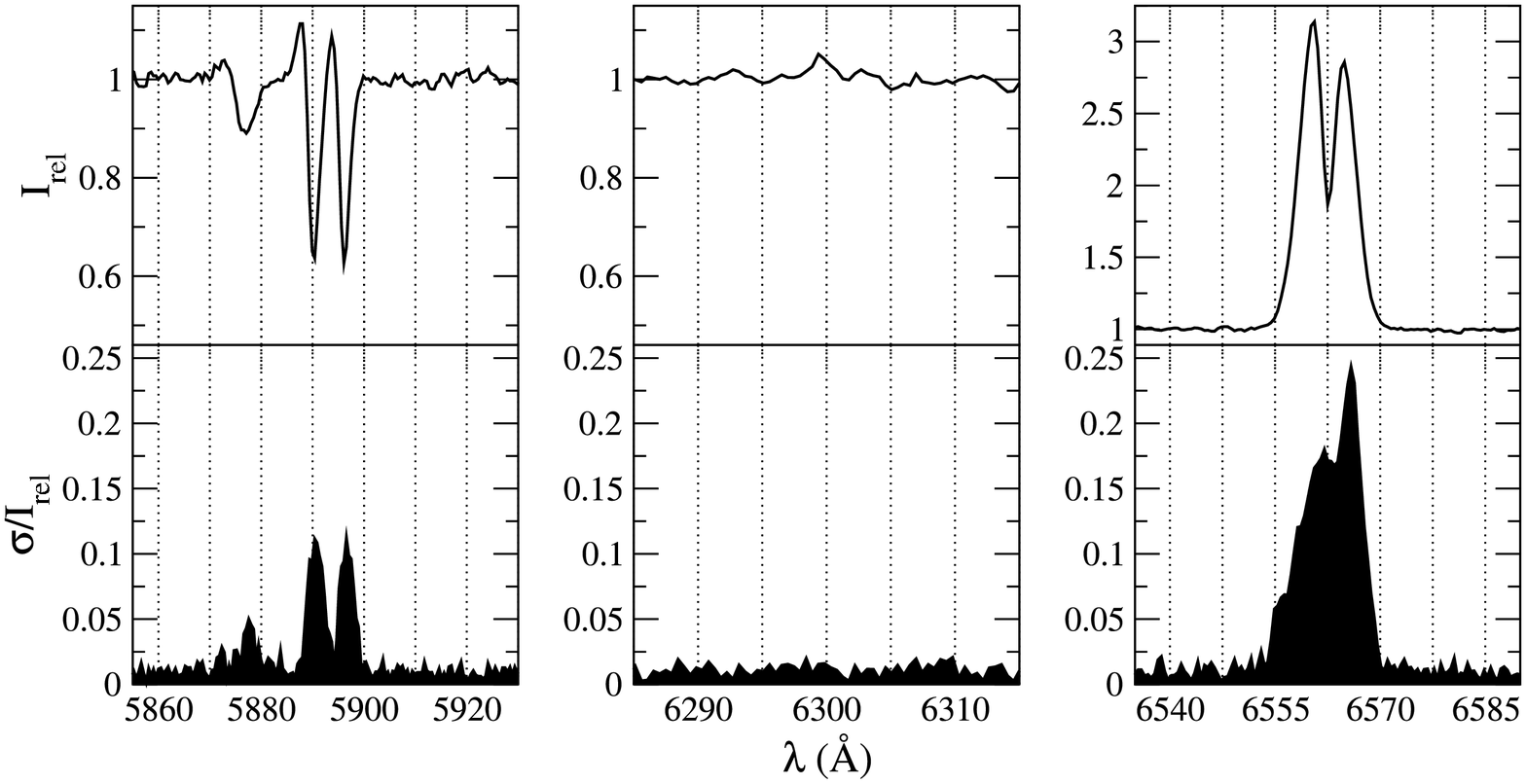} \\ 
\includegraphics[height=47mm,clip=true]{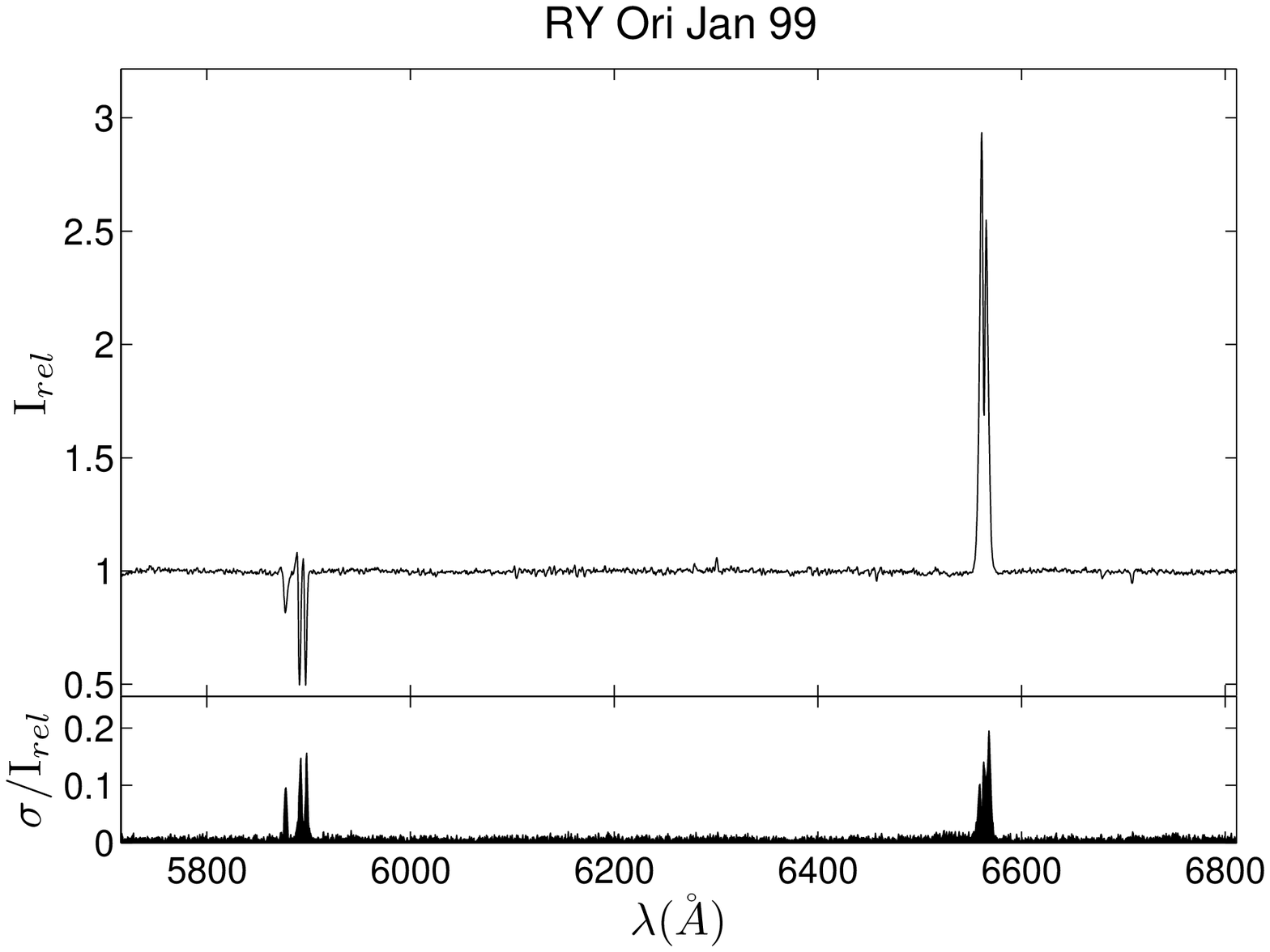}&
\includegraphics[bb=4 77 763 470,height=45mm,clip=true]{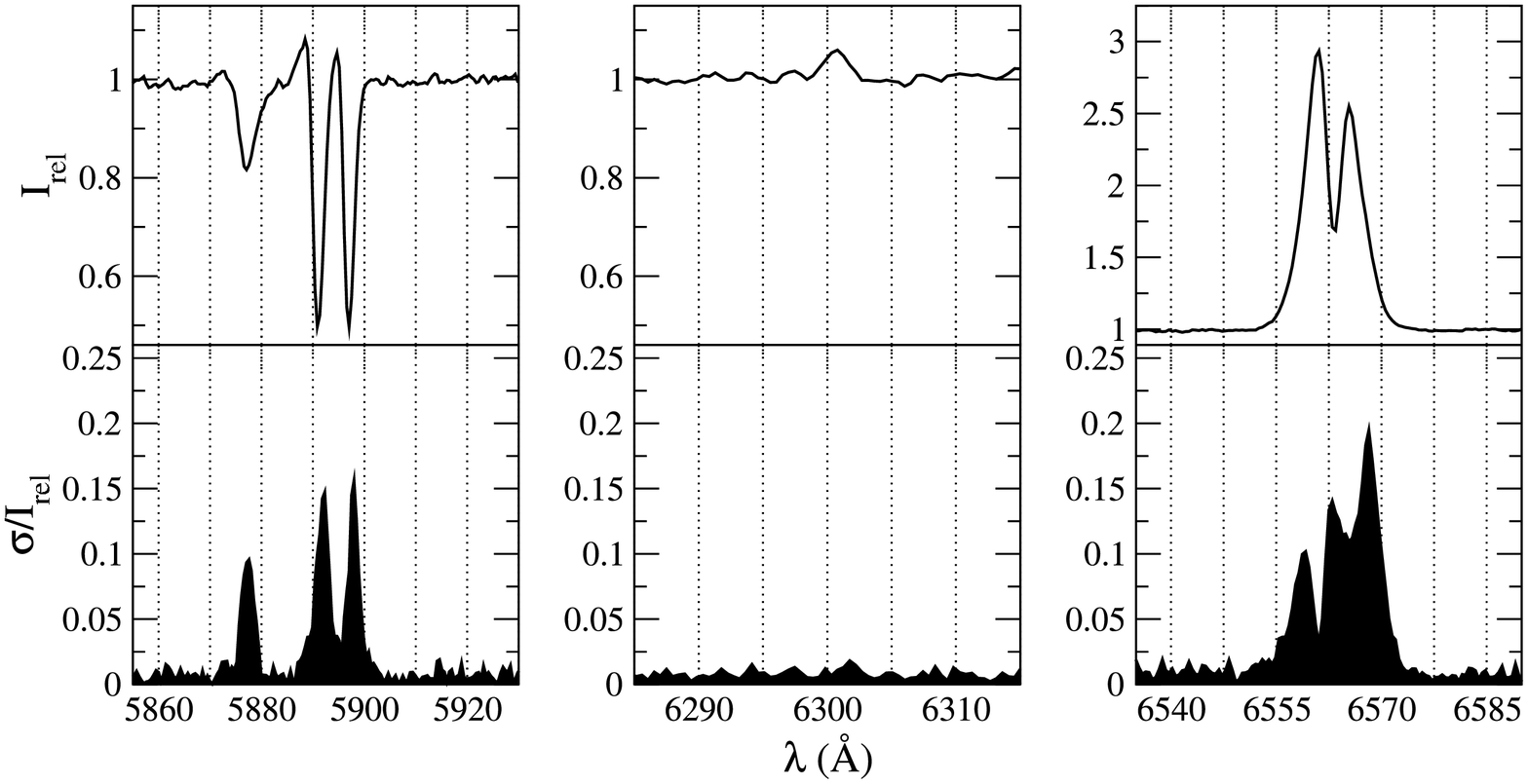} \\ 
\end{tabular}
\end{table}
\clearpage
\begin{table}
\centering
\renewcommand\arraystretch{10}
\begin{tabular}{cc}
\includegraphics[height=47mm,clip=true]{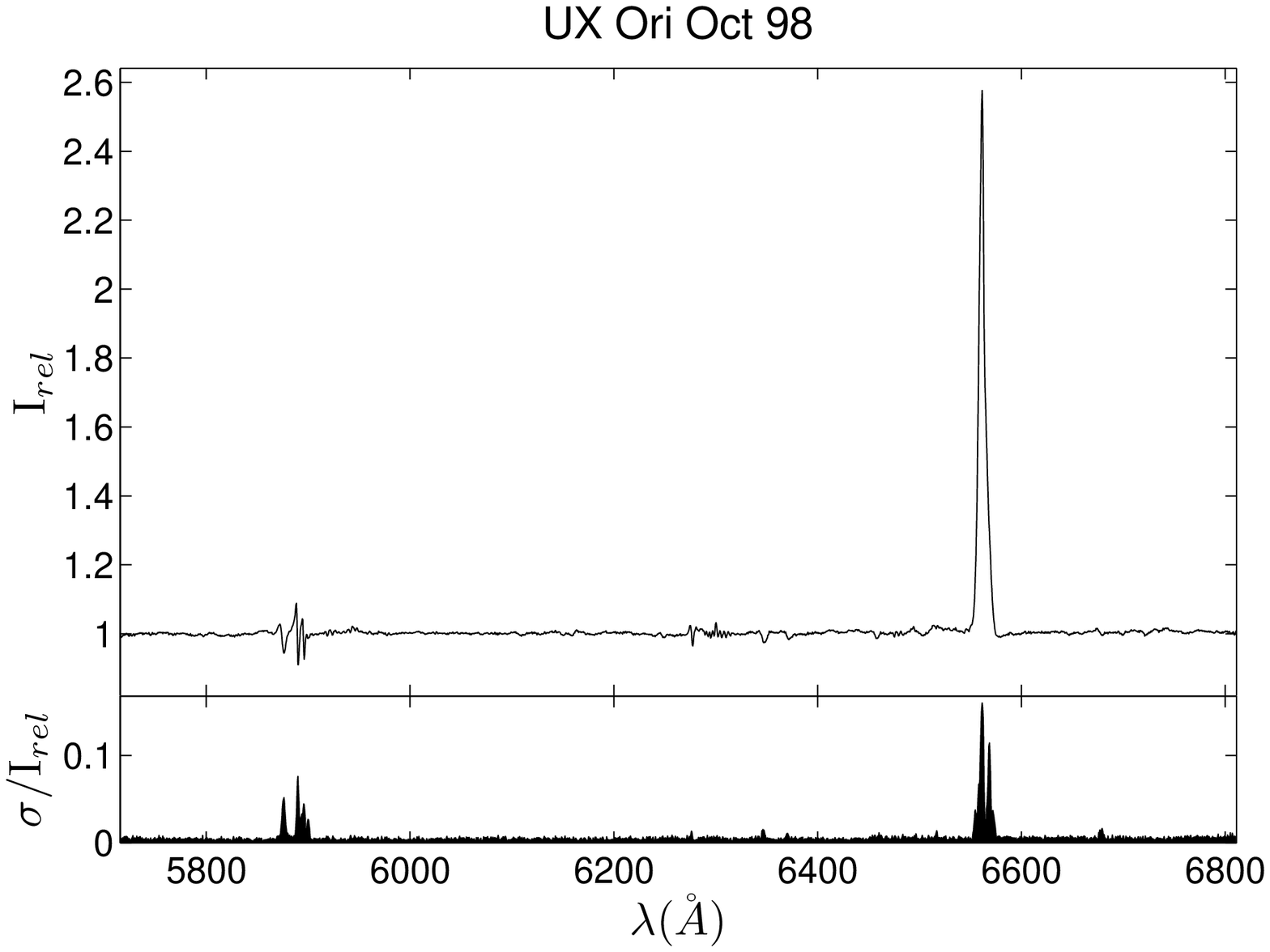}&
\includegraphics[bb=4 77 763 470,height=45mm,clip=true]{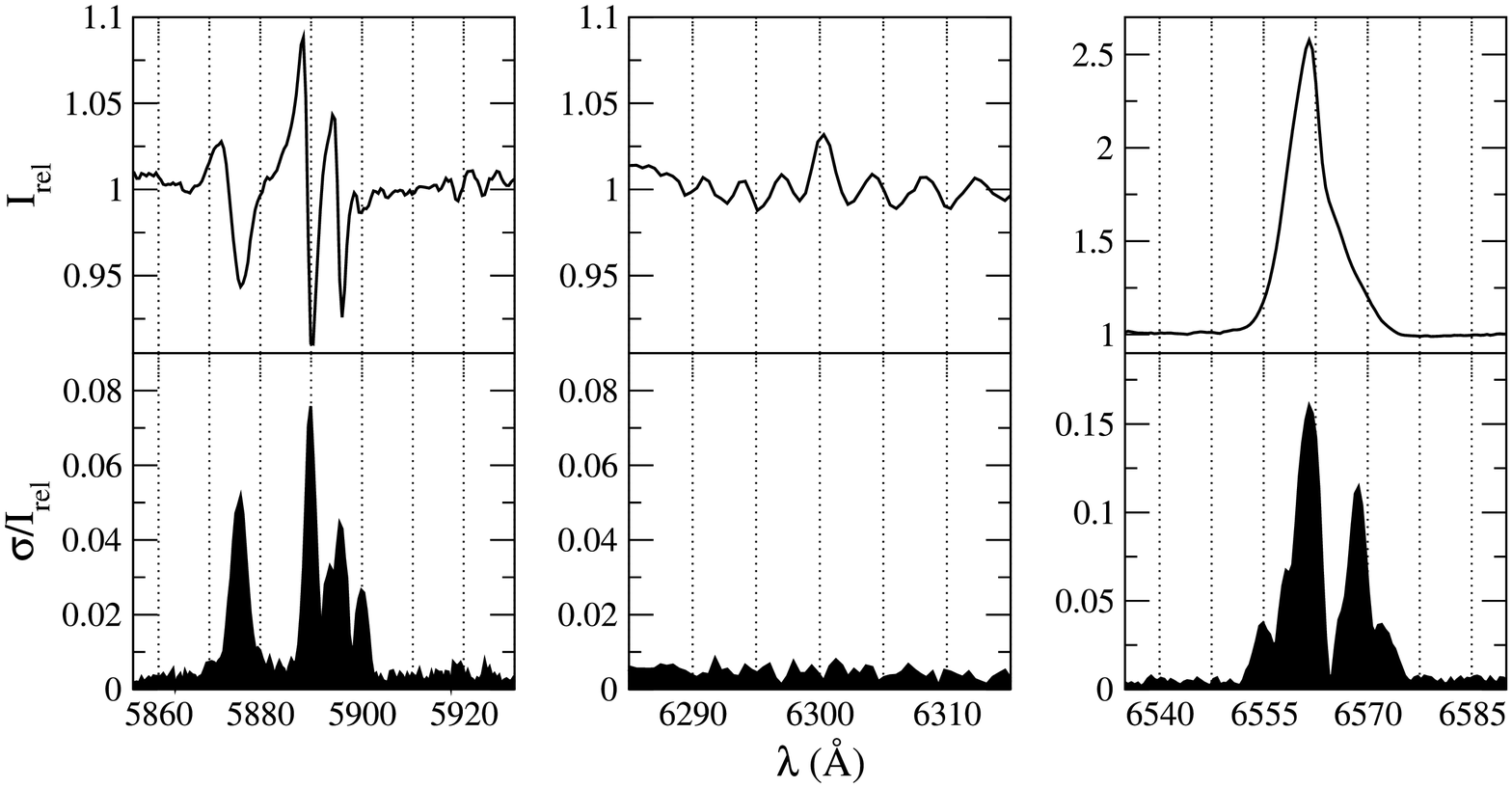} \\ 
\includegraphics[height=47mm,clip=true]{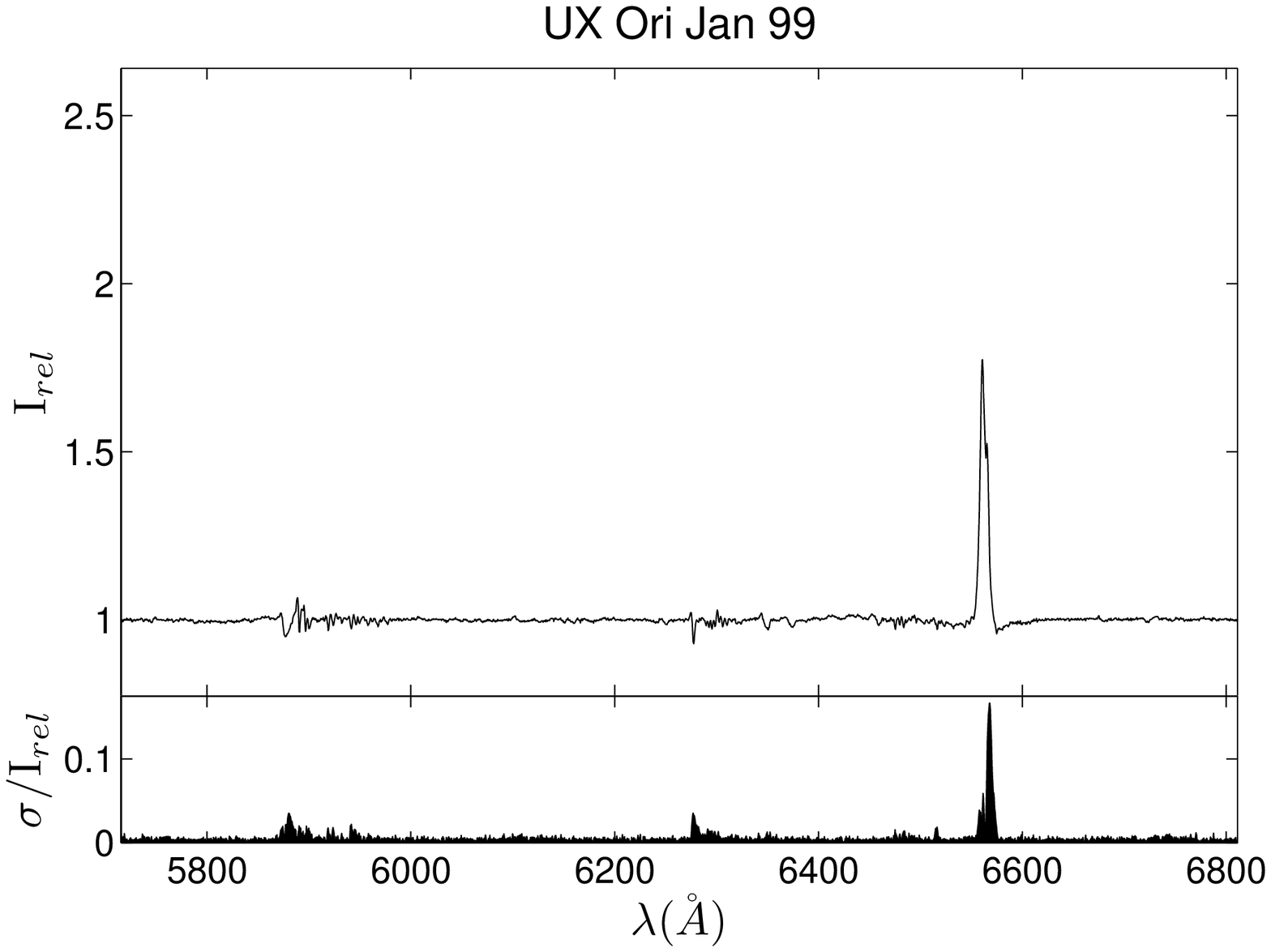}&
\includegraphics[bb=4 77 763 470,height=45mm,clip=true]{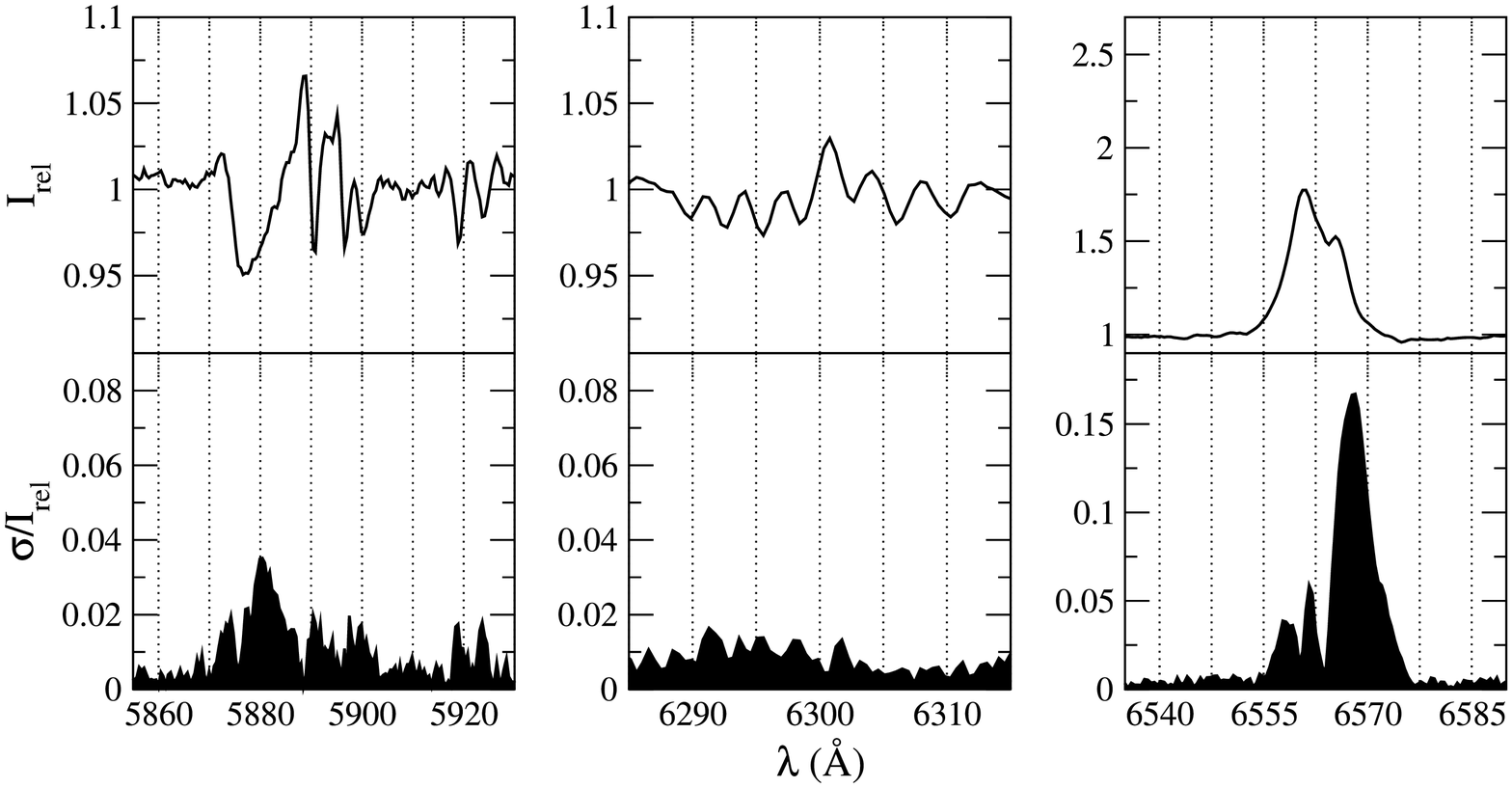} \\ 
\includegraphics[height=47mm,clip=true]{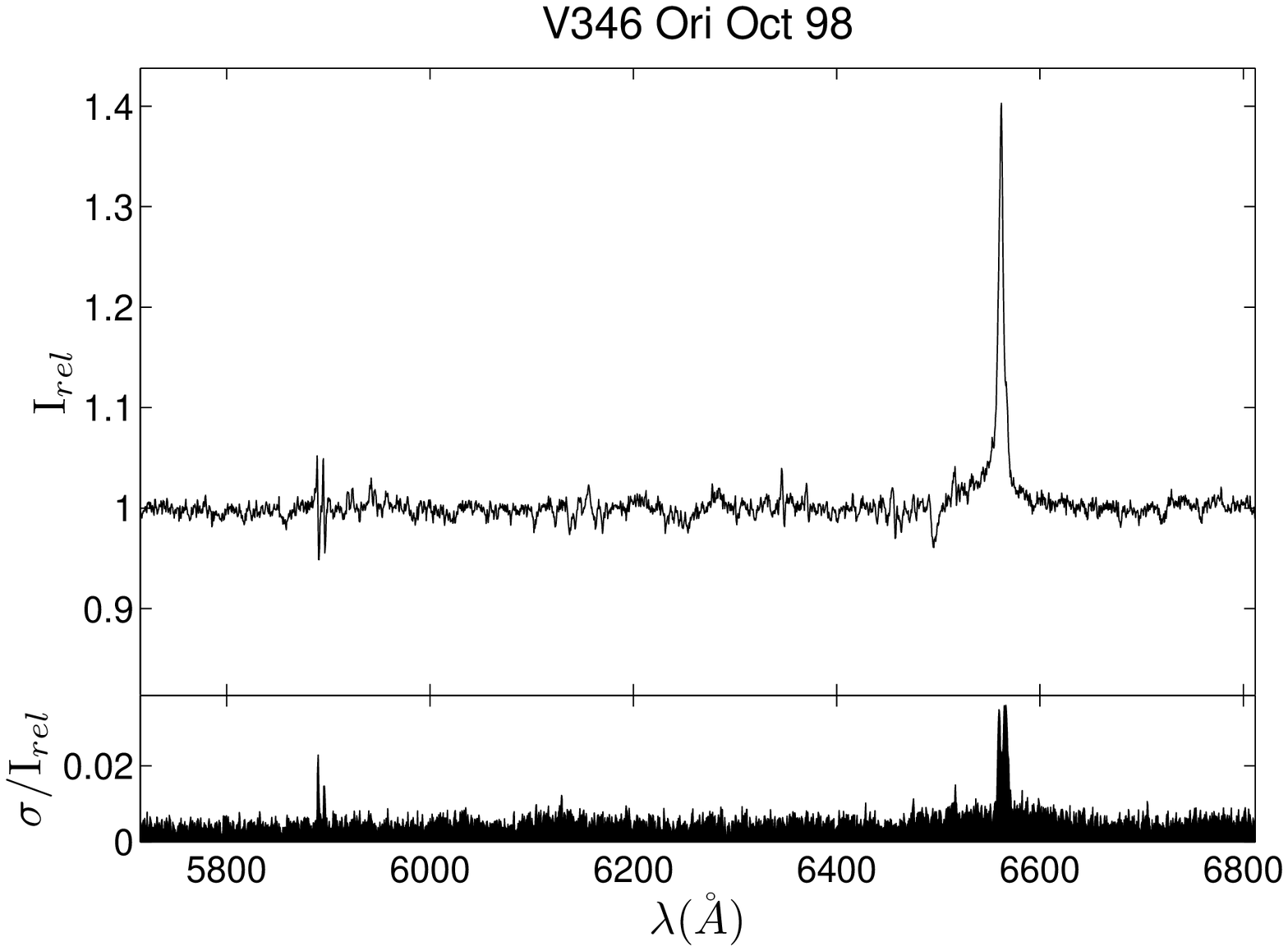}&
\includegraphics[bb=4 77 763 470,height=45mm,clip=true]{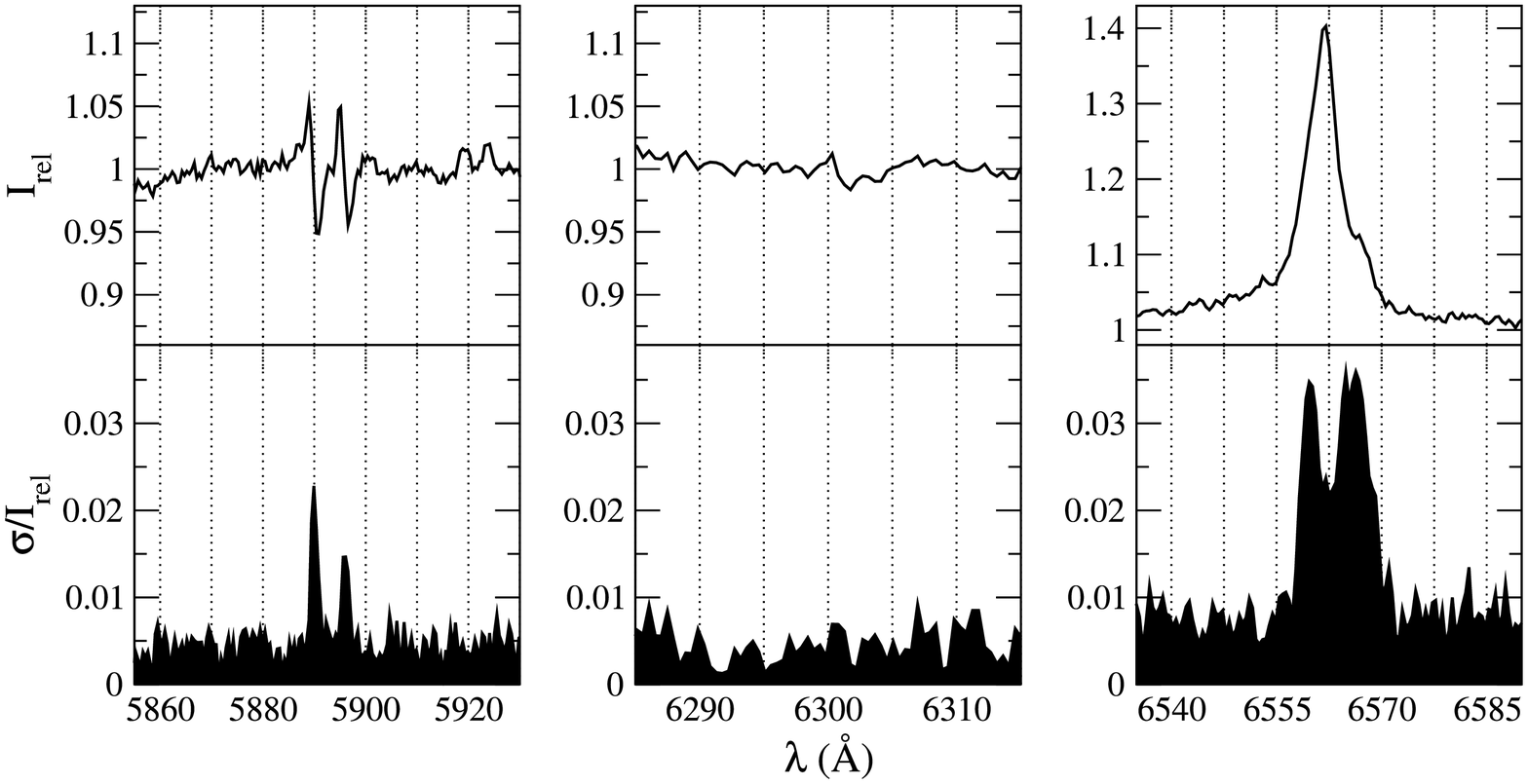} \\ 
\includegraphics[height=47mm,clip=true]{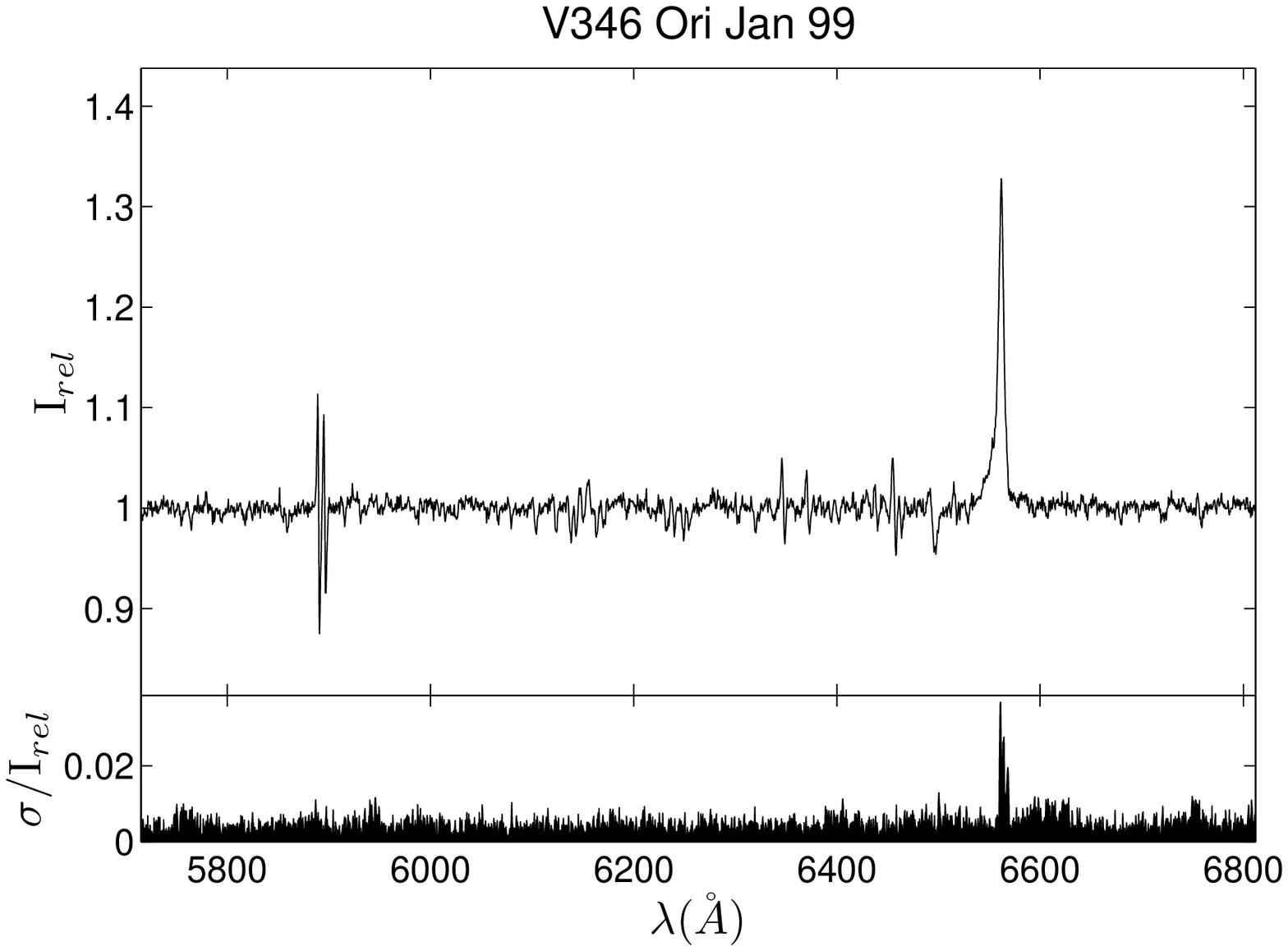}&
\includegraphics[bb=4 77 763 470,height=45mm,clip=true]{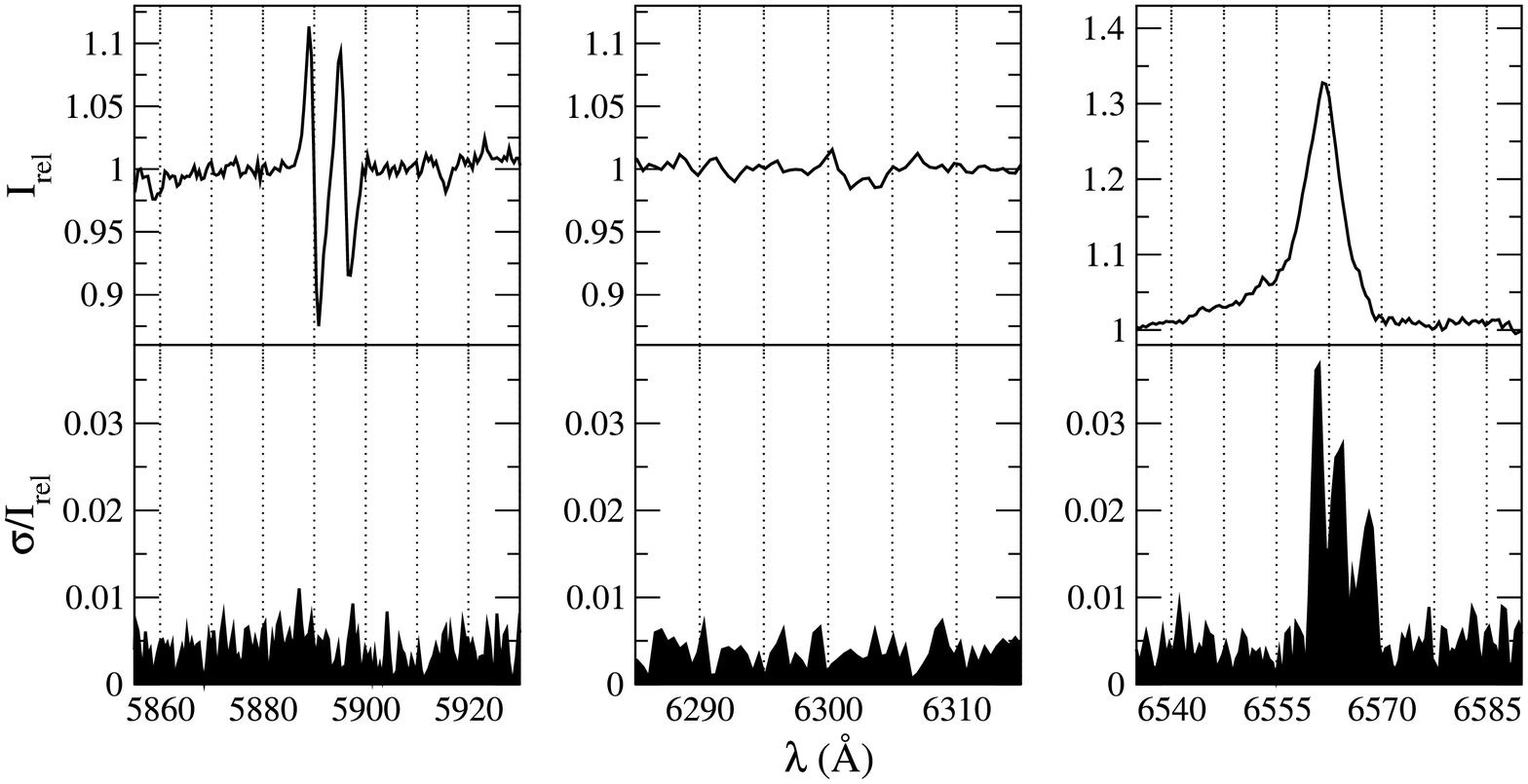} \\ 
\end{tabular}
\end{table}
\clearpage
\begin{table}
\centering
\renewcommand\arraystretch{10}
\begin{tabular}{cc}
\includegraphics[height=47mm,clip=true]{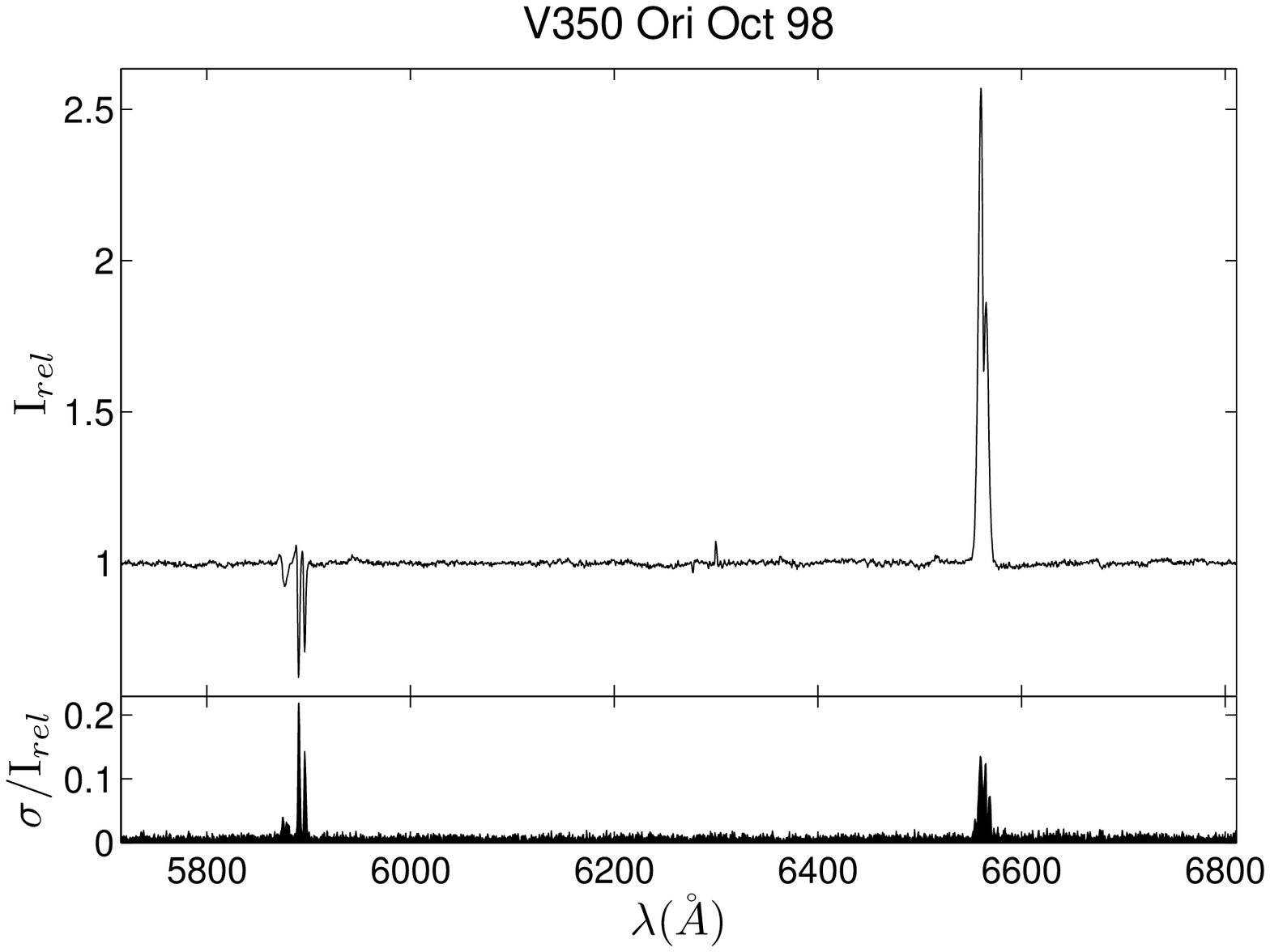}&
\includegraphics[bb=4 77 763 470,height=45mm,clip=true]{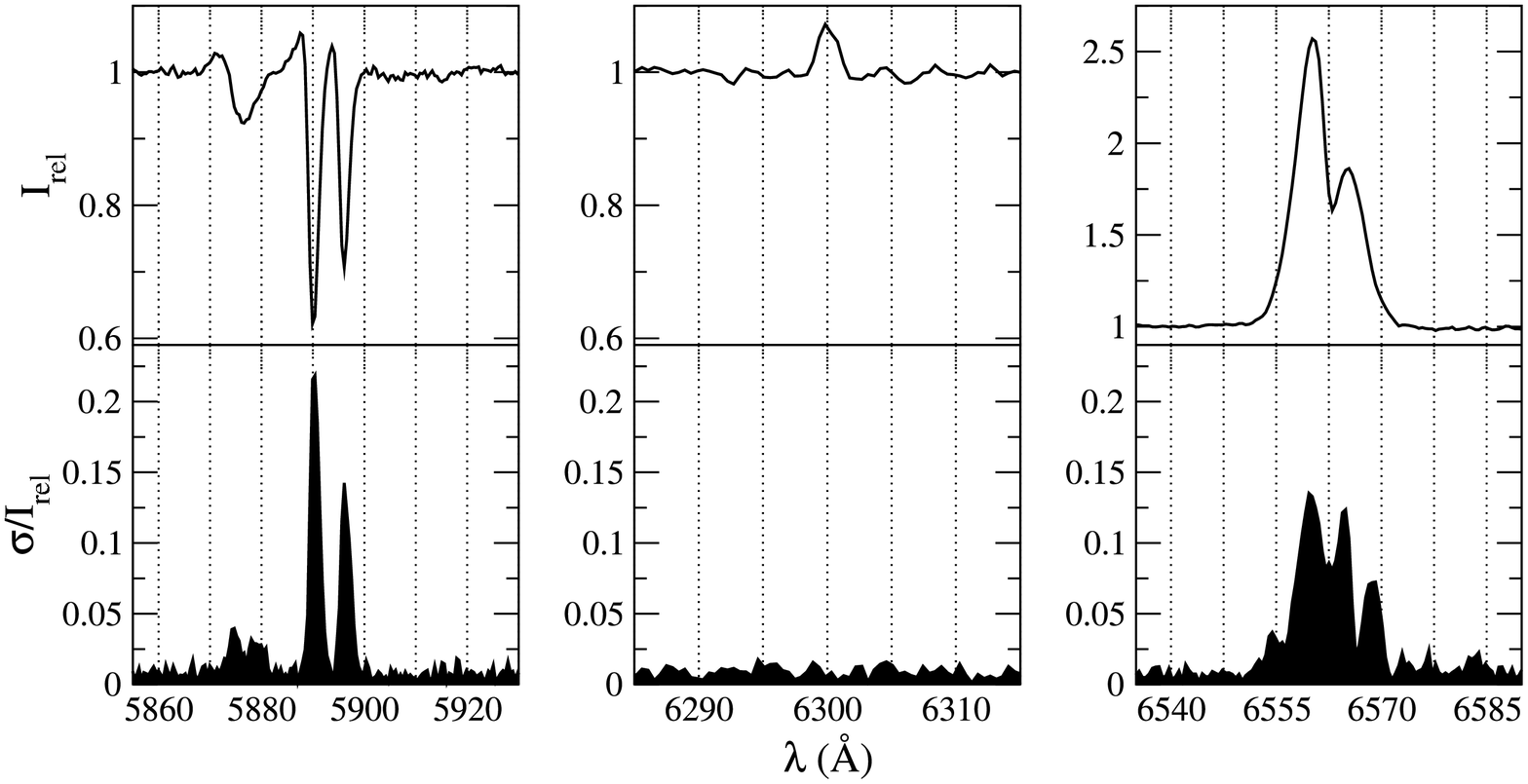} \\ 
\includegraphics[height=47mm,clip=true]{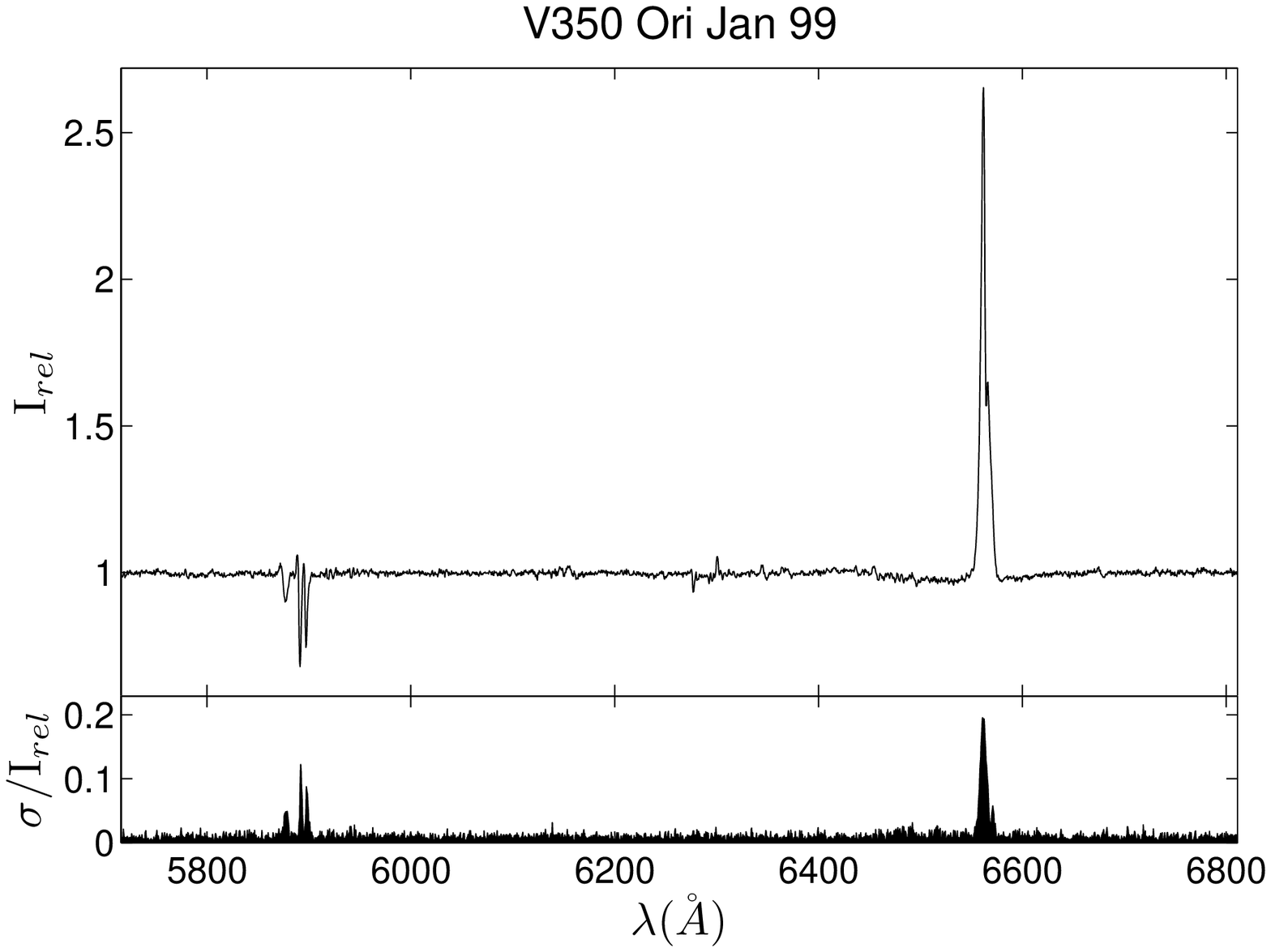}&
\includegraphics[bb=4 77 763 470,height=45mm,clip=true]{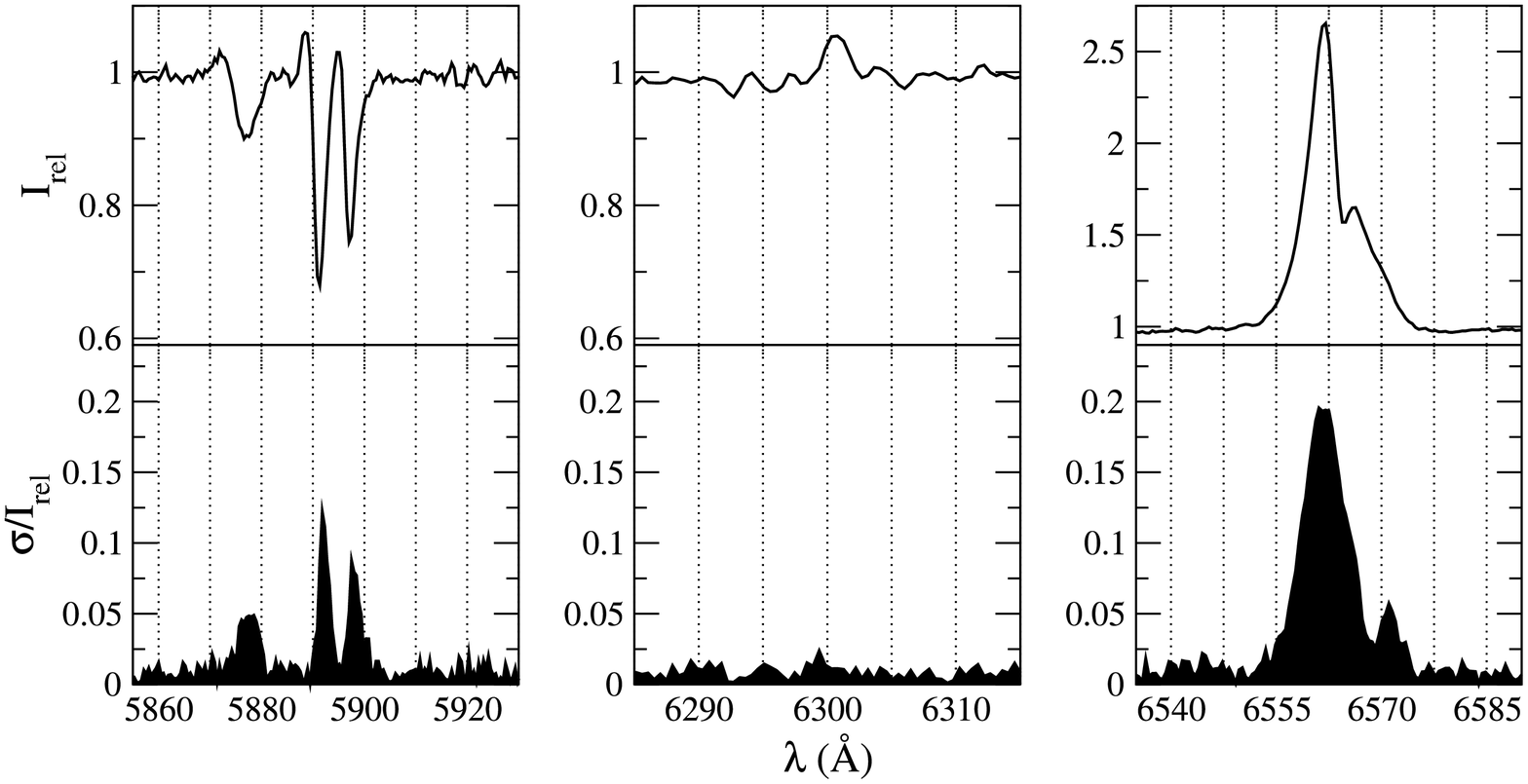} \\
\includegraphics[height=47mm,clip=true]{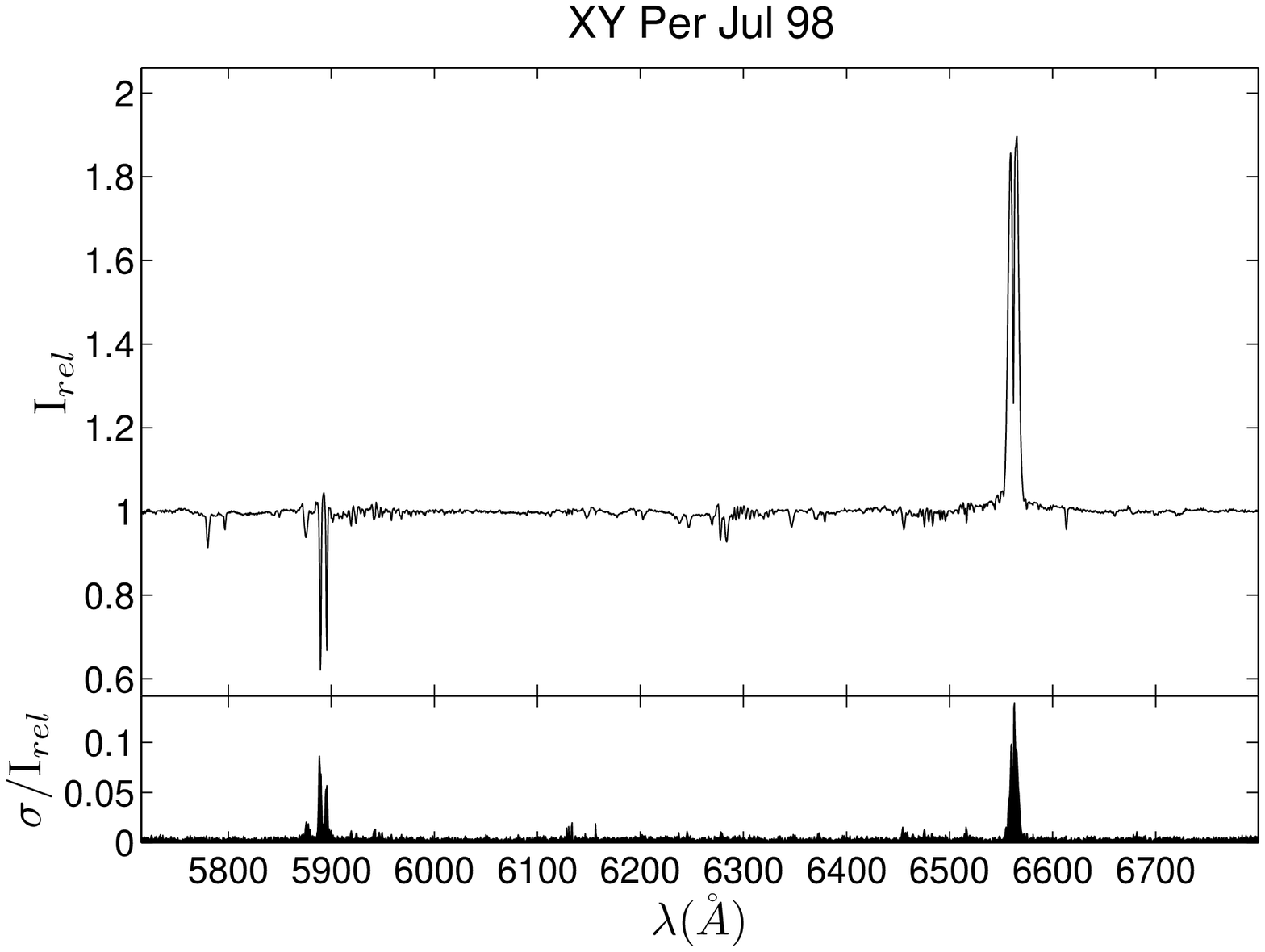}&
\includegraphics[bb=4 77 763 470,height=45mm,clip=true]{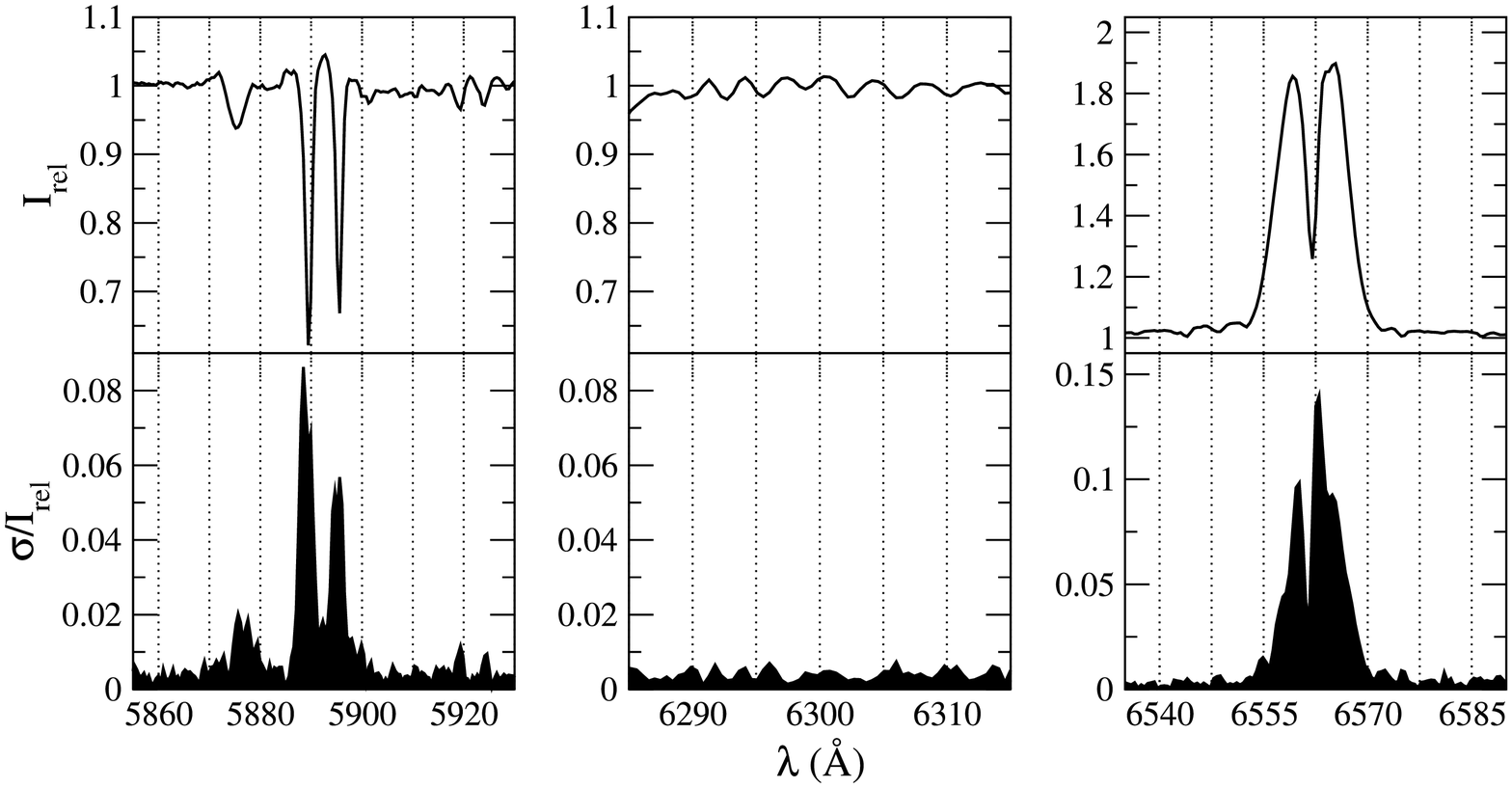} \\
\includegraphics[height=47mm,clip=true]{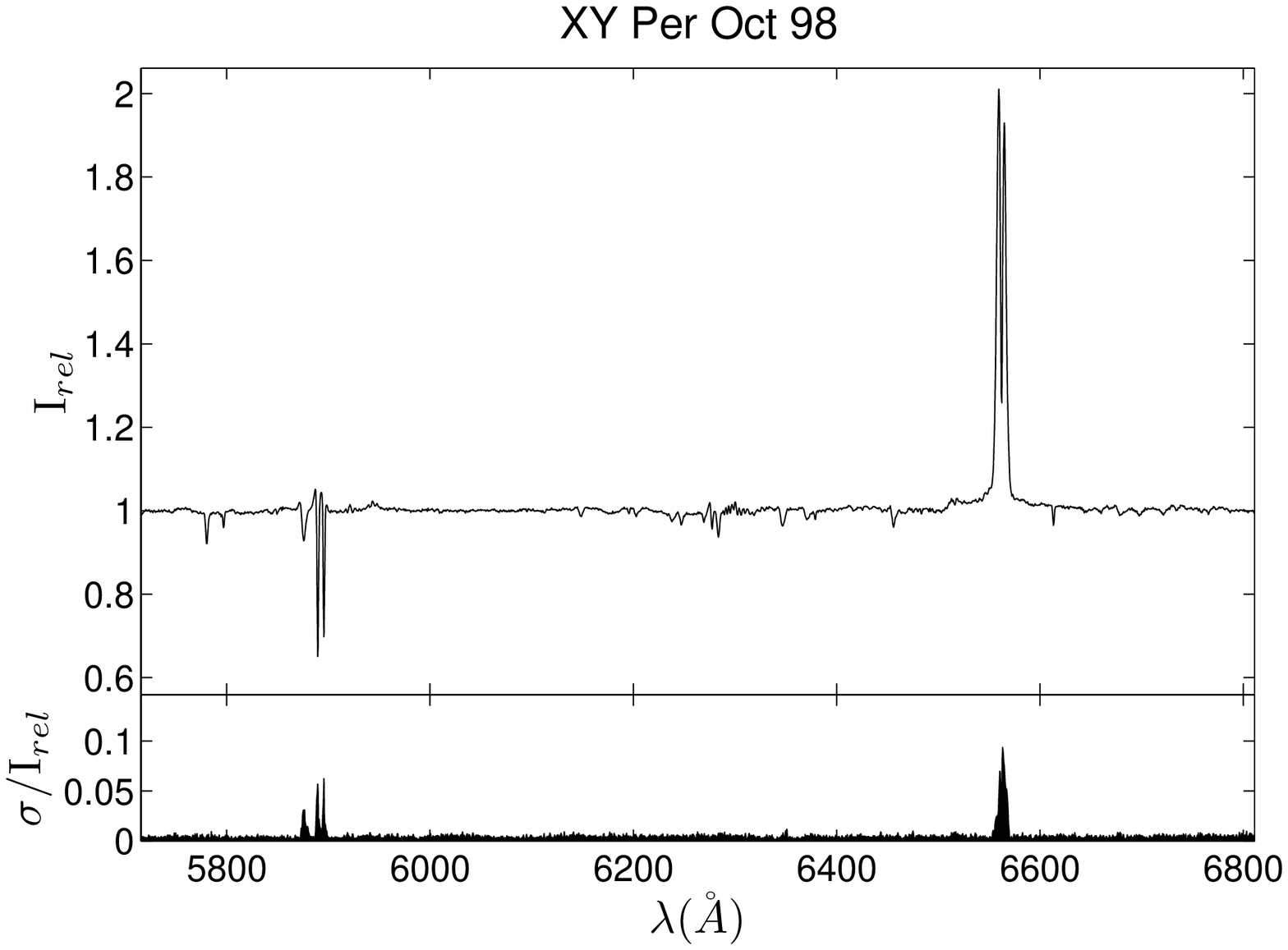}&
\includegraphics[bb=4 77 763 470,height=45mm,clip=true]{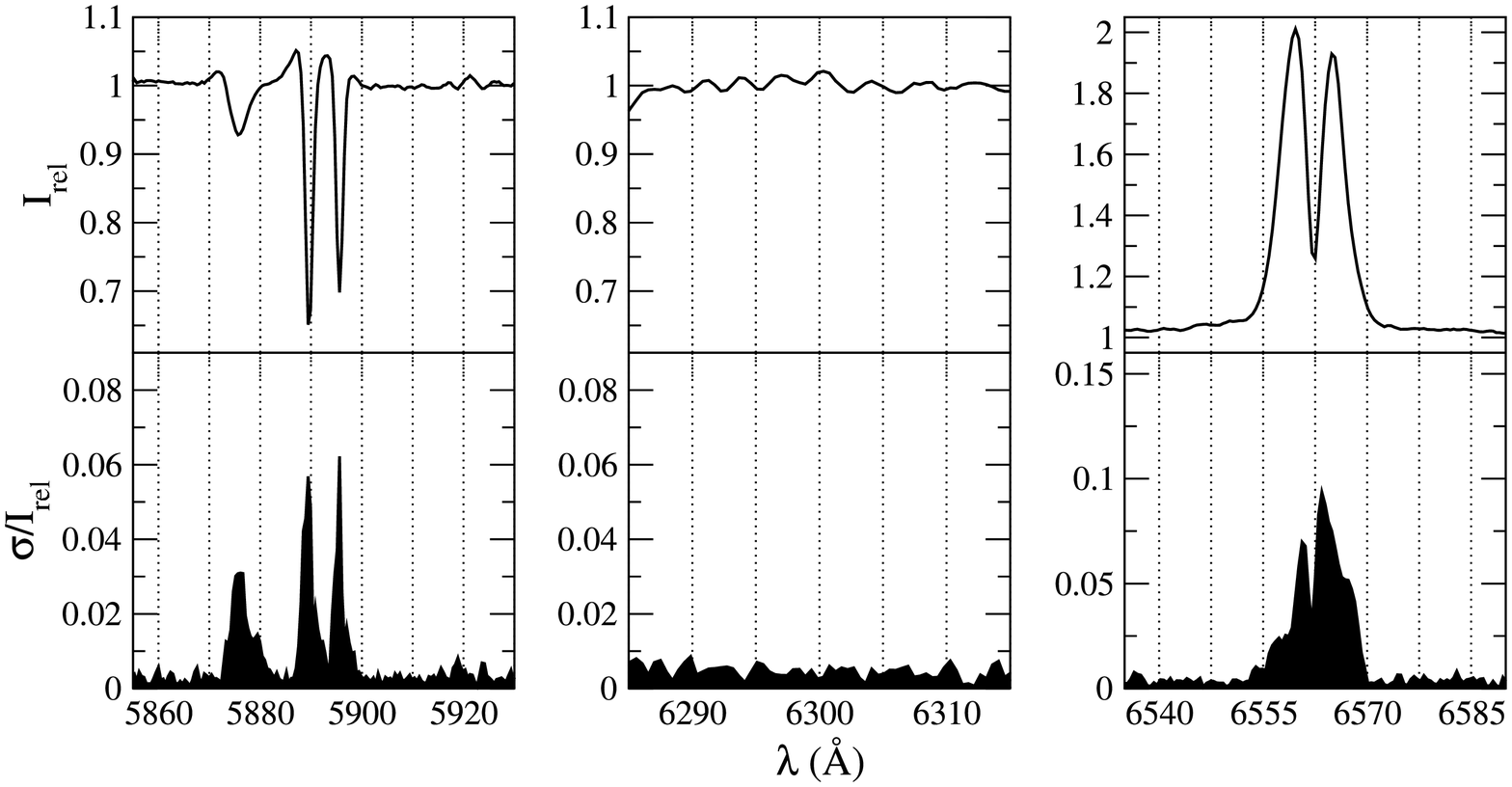} \\
\end{tabular}
\end{table}
\clearpage
\begin{table}
\centering
\renewcommand\arraystretch{10}
\begin{tabular}{cc}
\includegraphics[height=47mm,clip=true]{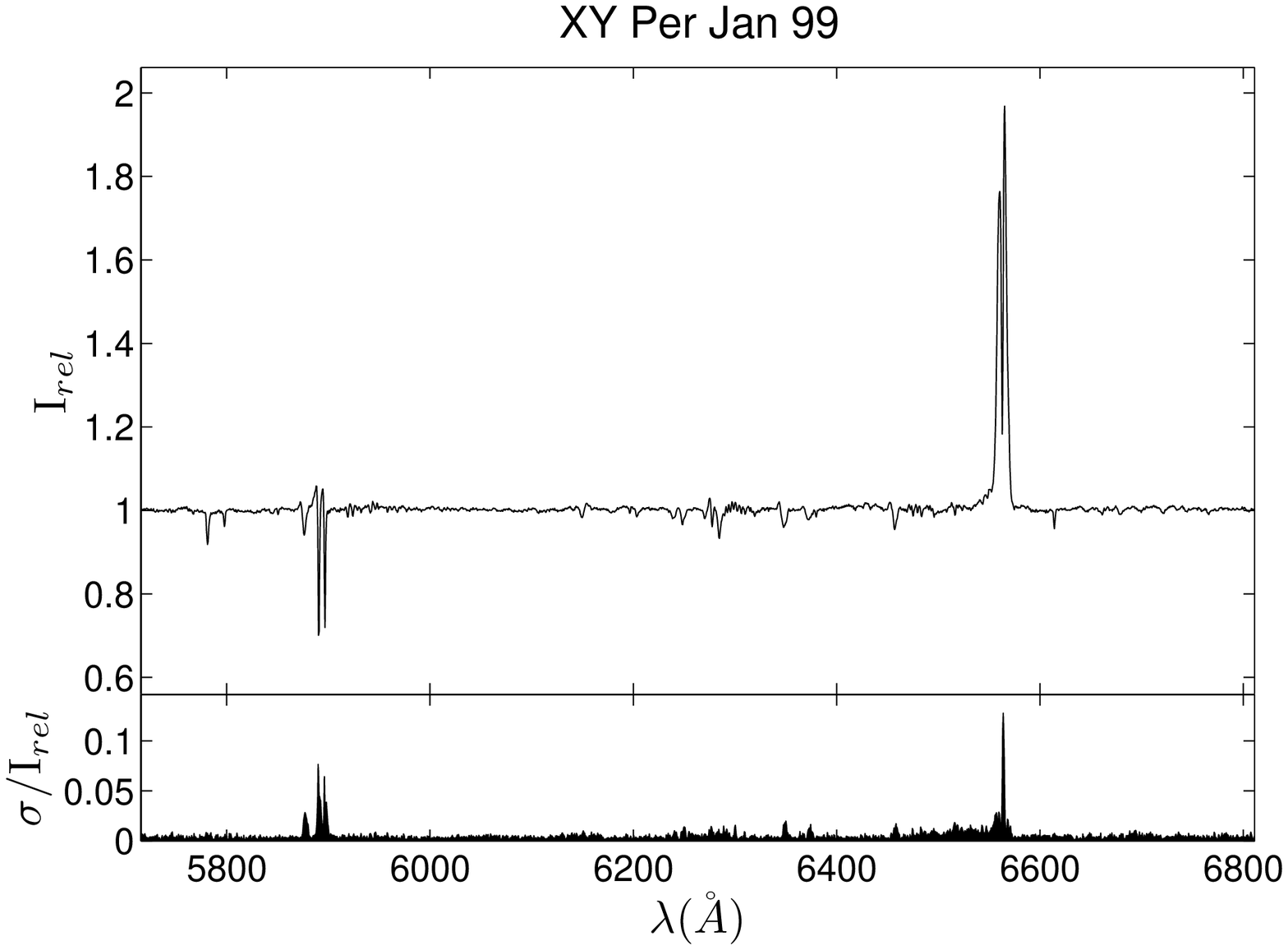}&
\includegraphics[bb=4 77 763 470,height=45mm,clip=true]{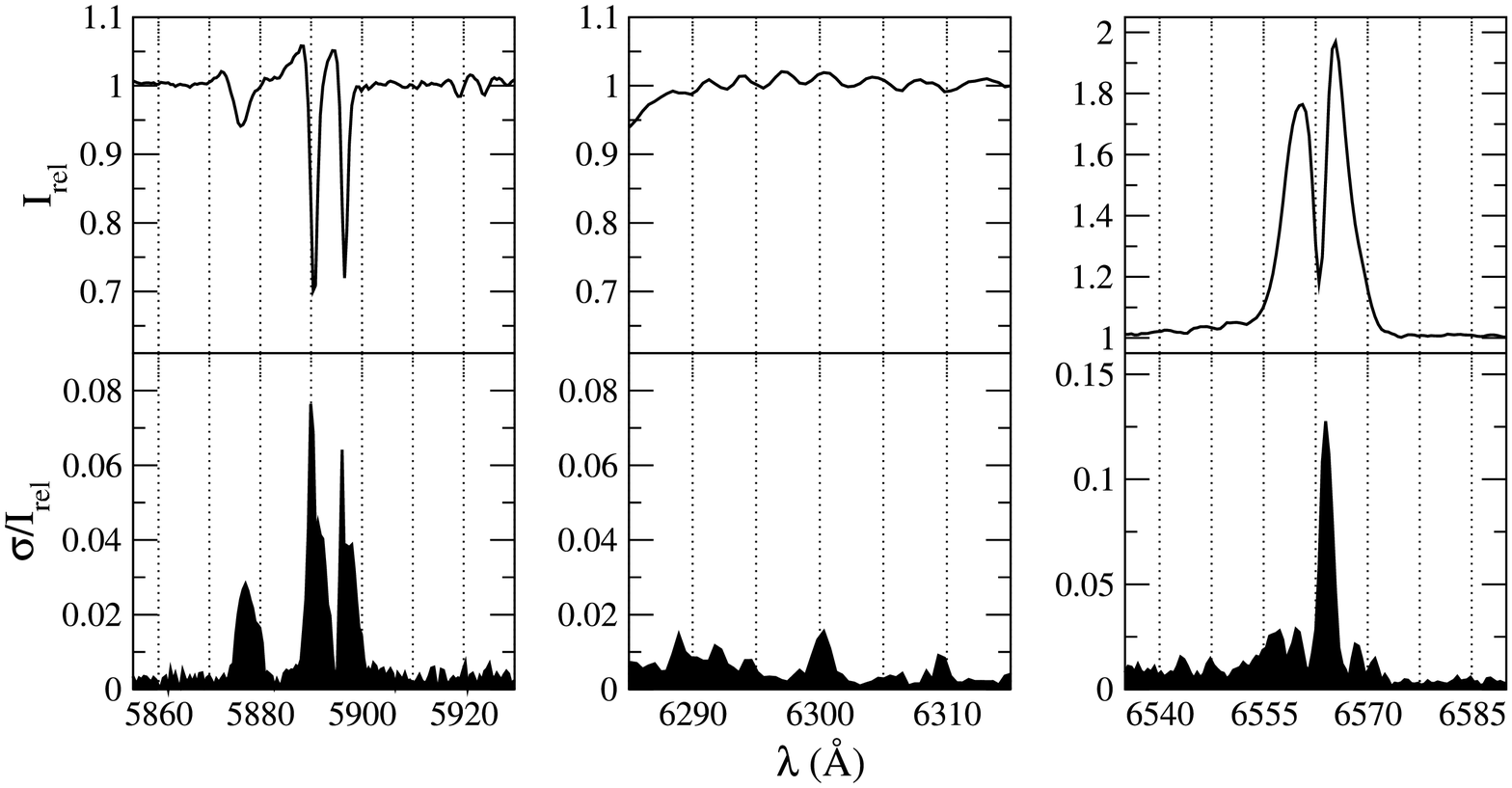} \\
\includegraphics[height=47mm,clip=true]{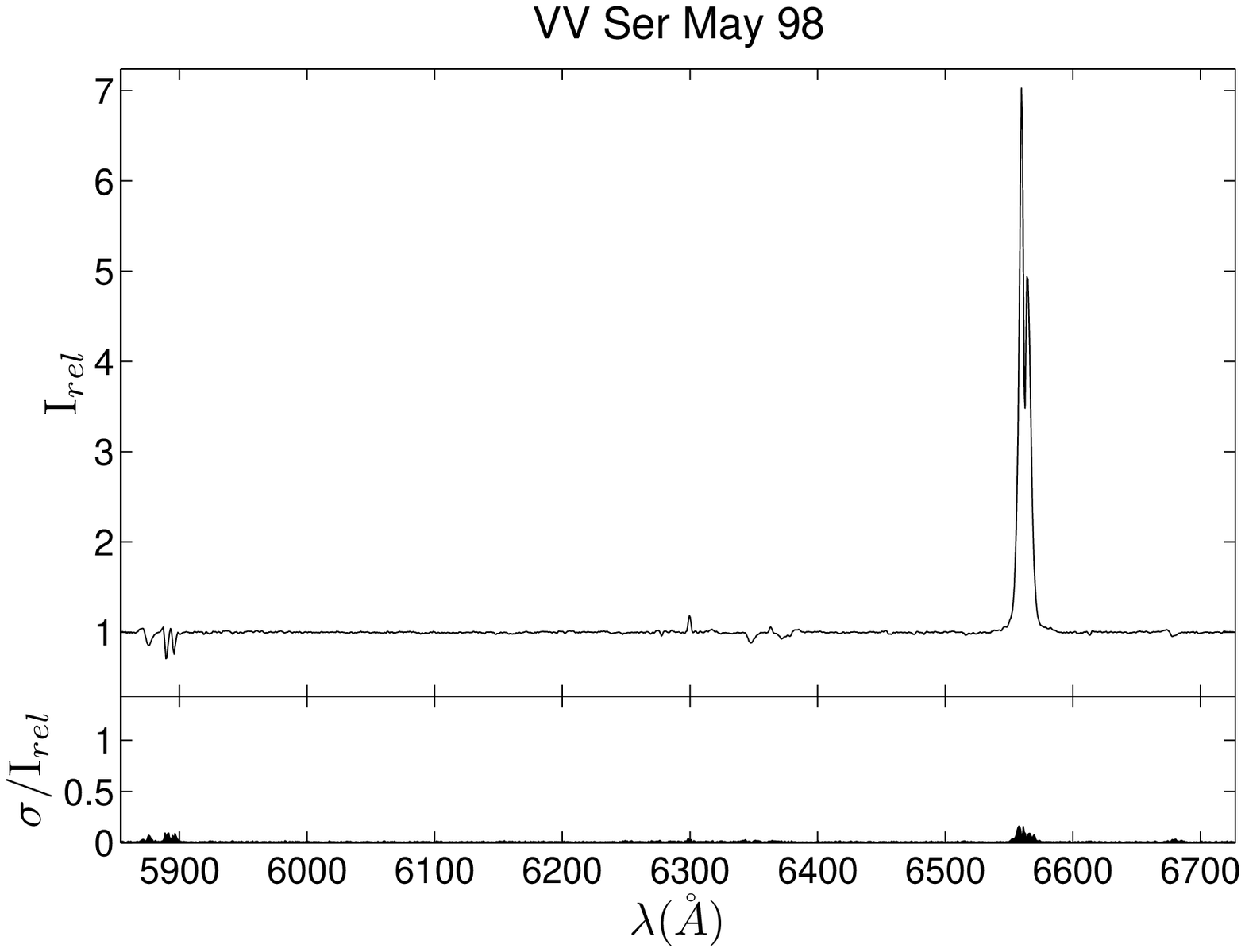}&
\includegraphics[bb=4 77 763 470,height=45mm,clip=true]{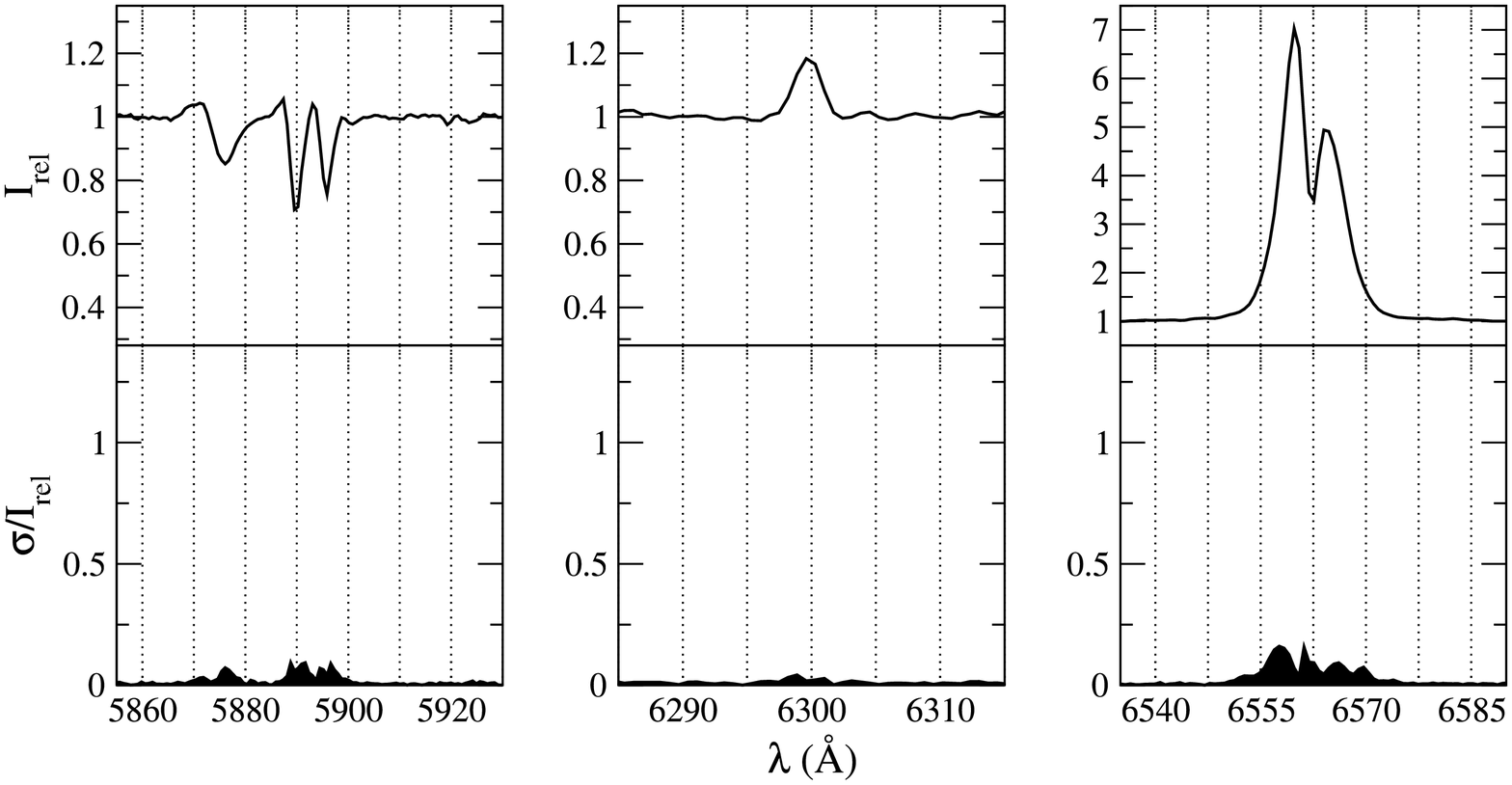} \\
\includegraphics[height=47mm,clip=true]{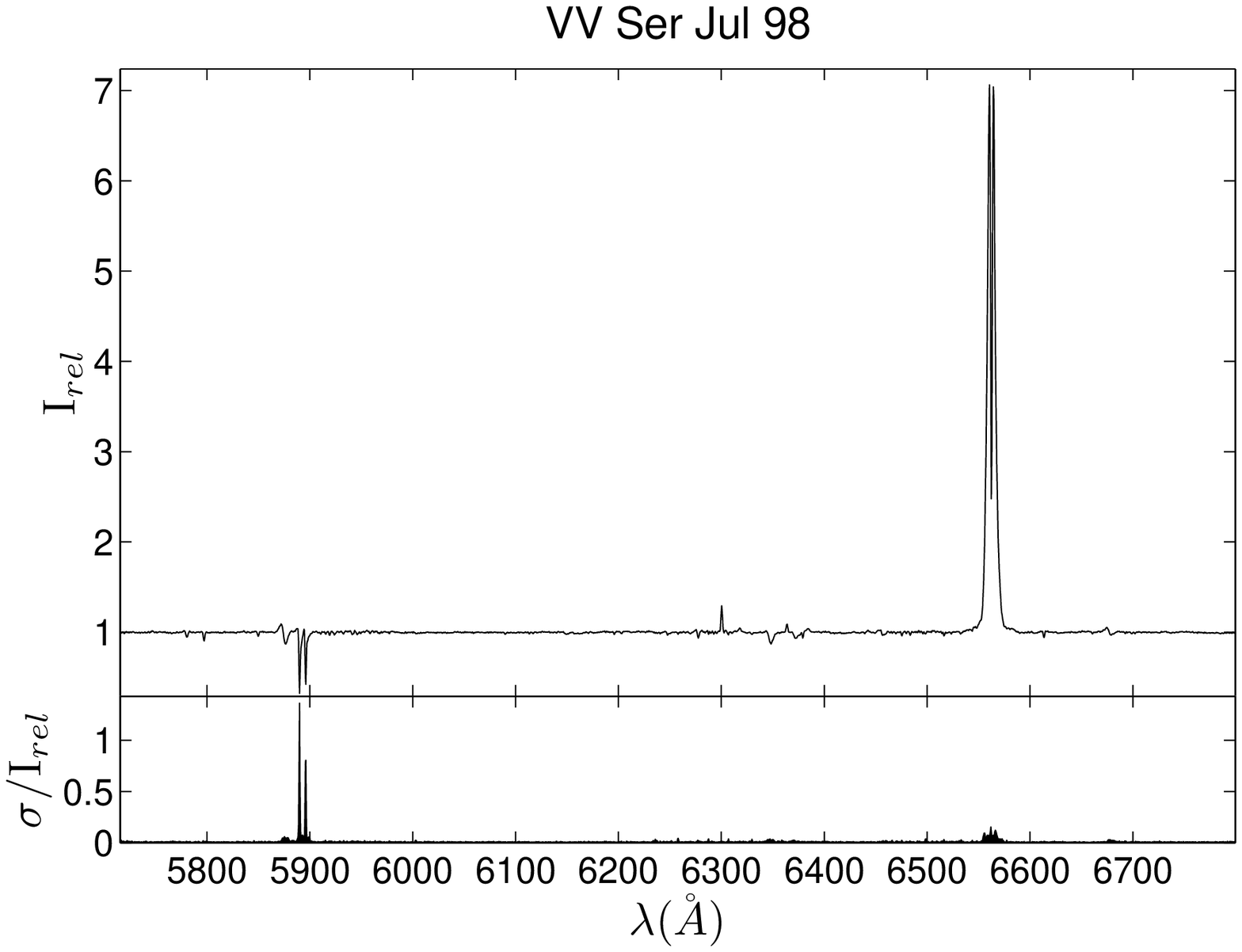}&
\includegraphics[bb=4 77 763 470,height=45mm,clip=true]{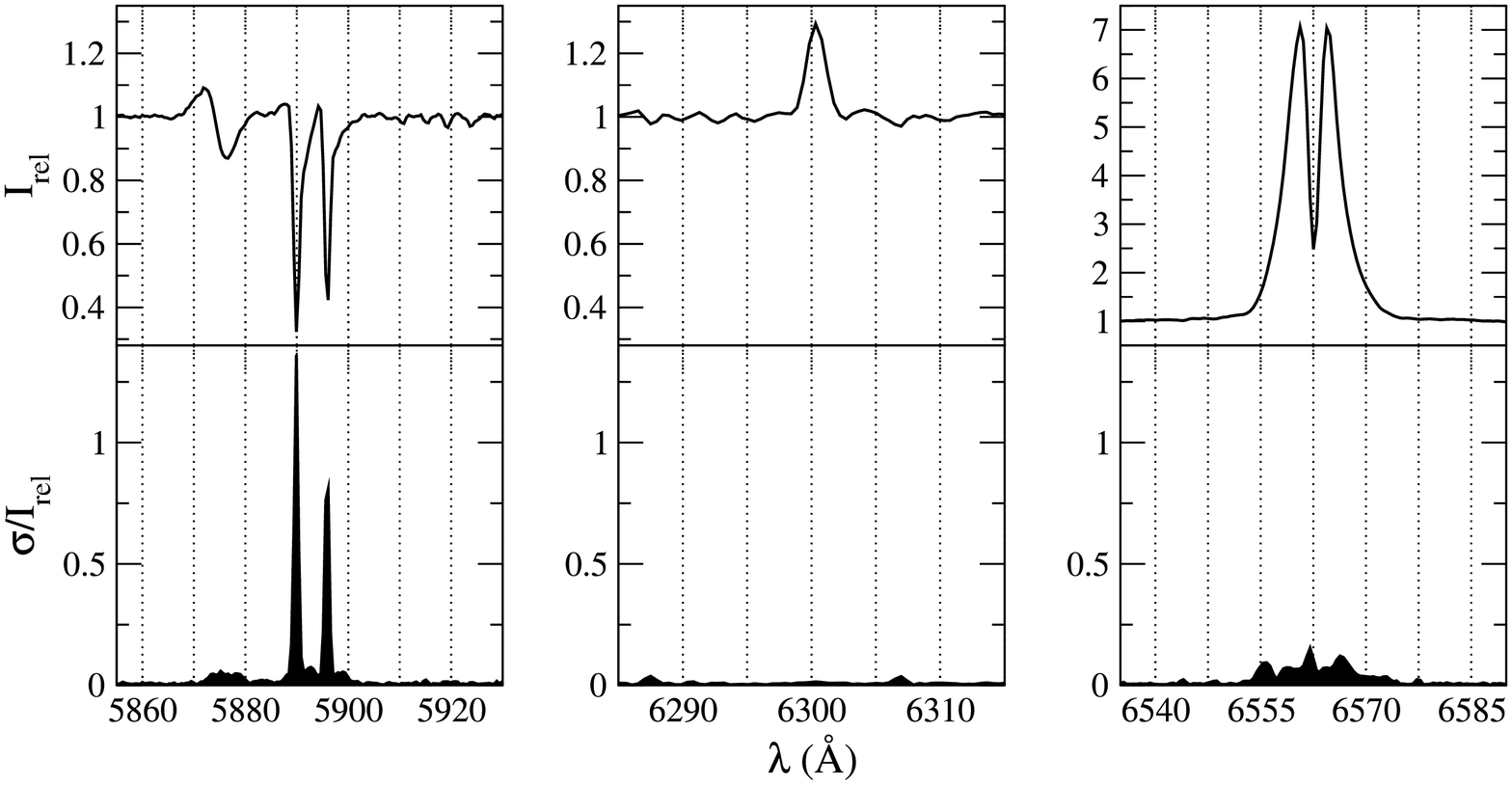} \\
\includegraphics[height=47mm,clip=true]{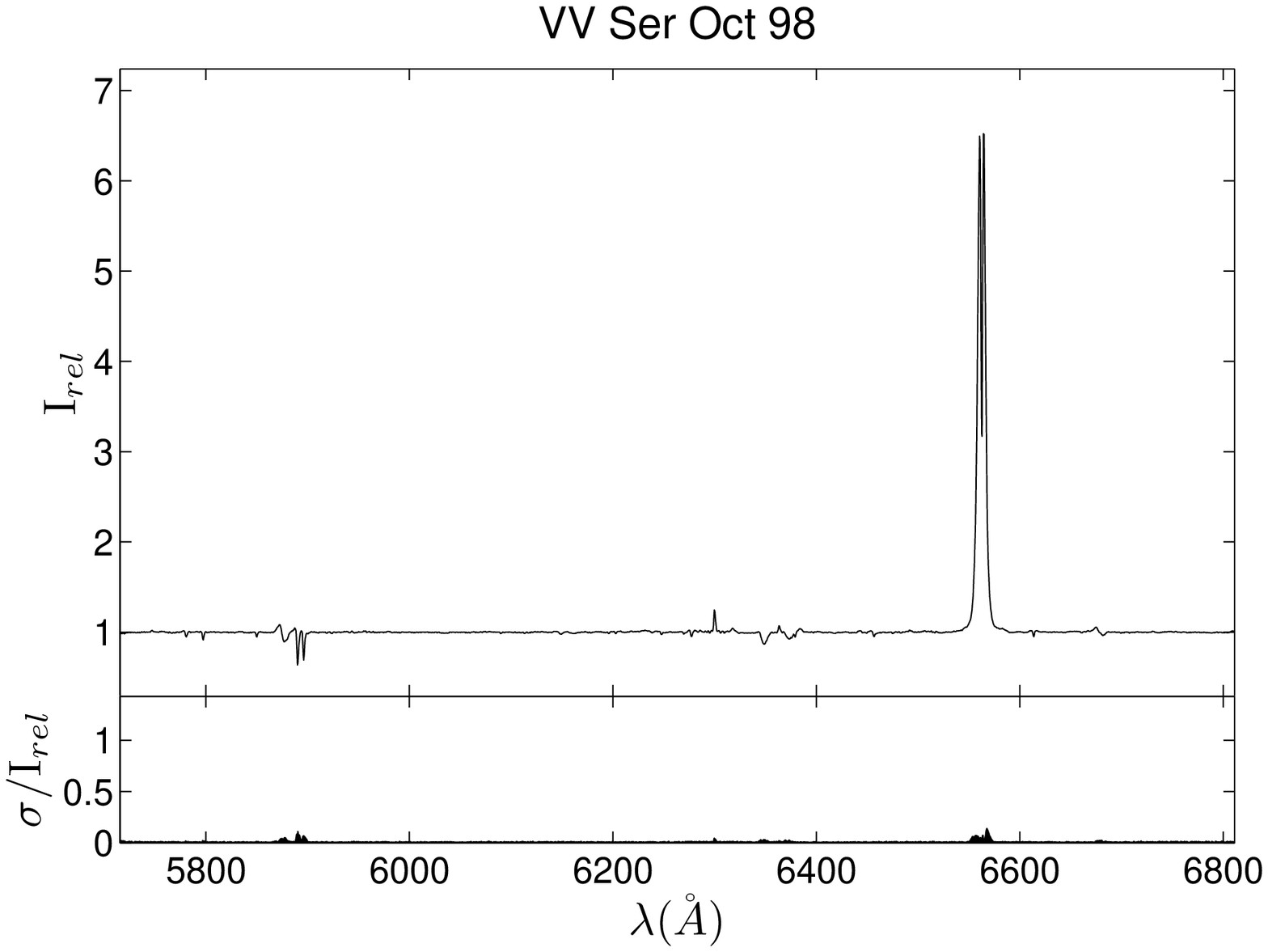}&
\includegraphics[bb=4 77 763 470,height=45mm,clip=true]{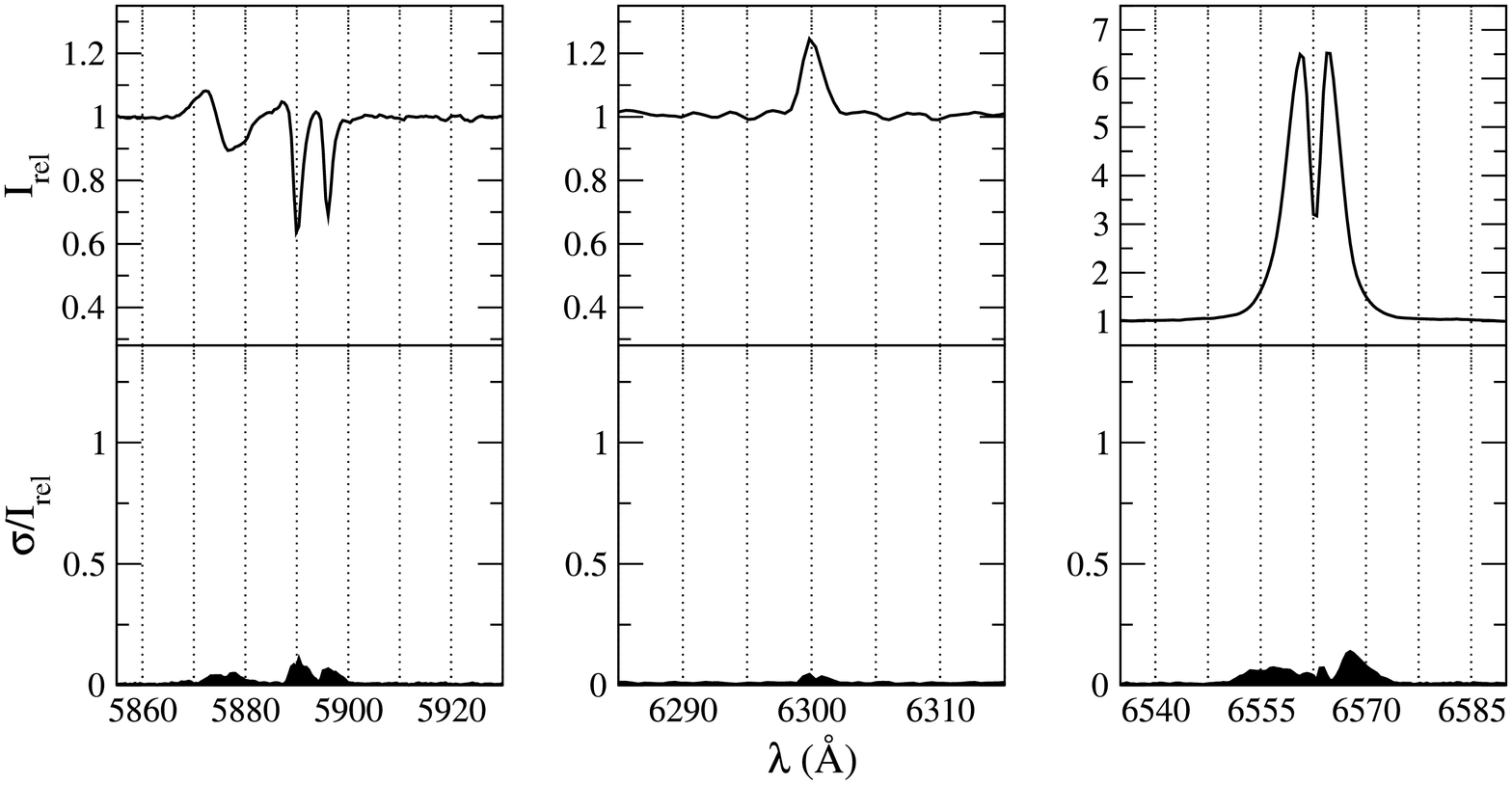} \\
\end{tabular}
\end{table}
\clearpage
\begin{table}
\centering
\renewcommand\arraystretch{10}
\begin{tabular}{cc}
\includegraphics[height=47mm,clip=true]{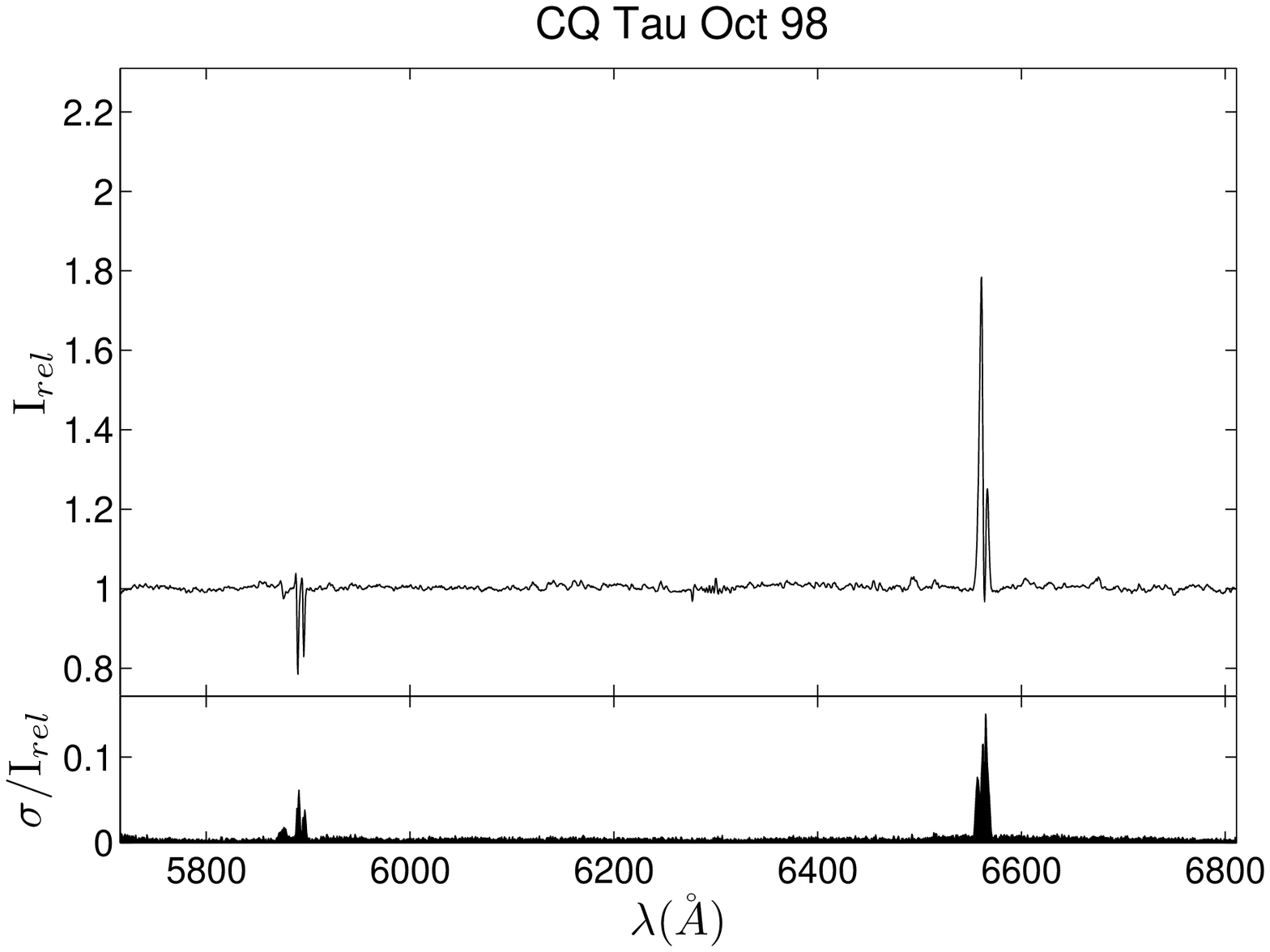}&
\includegraphics[bb=4 77 763 470,height=45mm,clip=true]{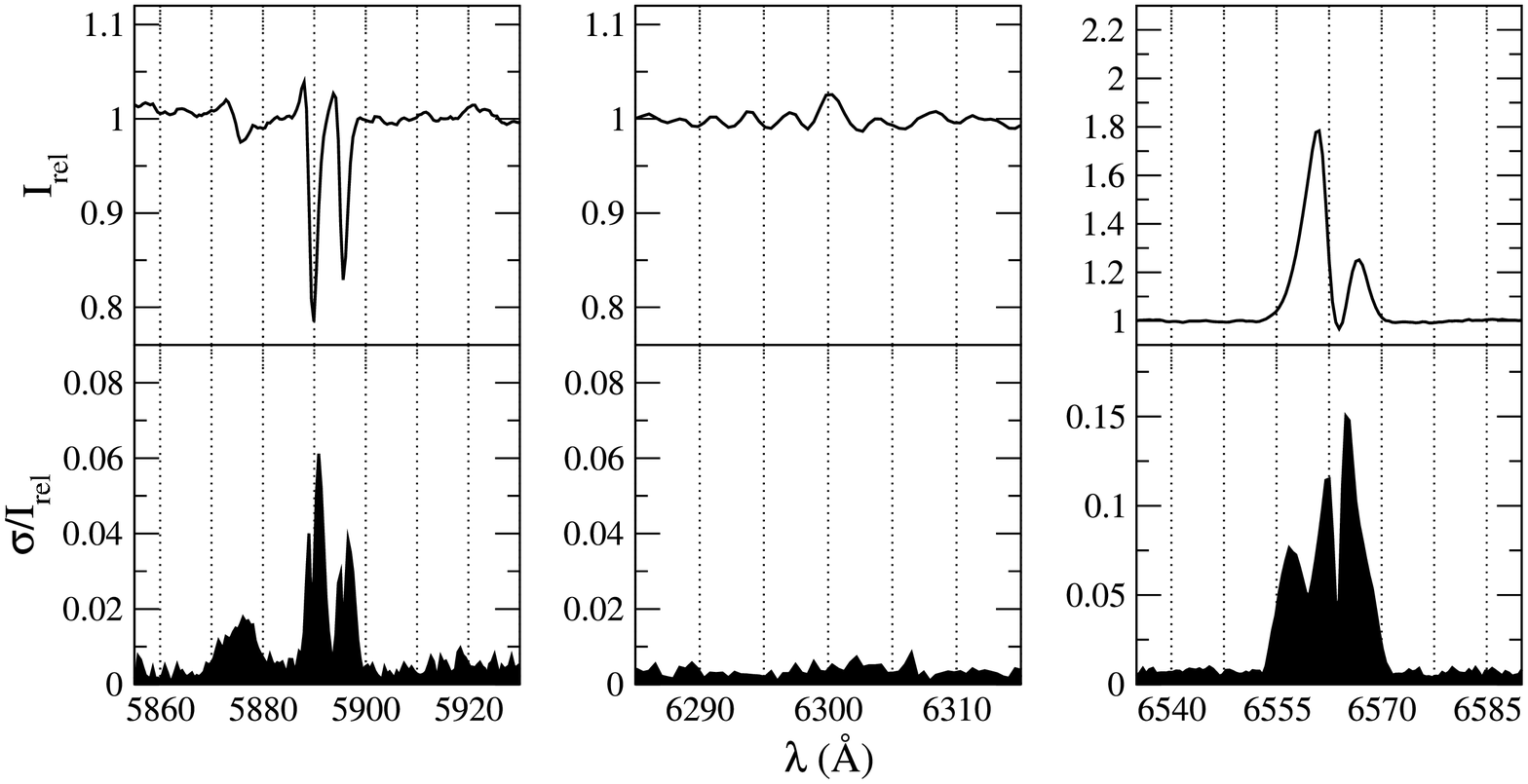} \\ 
\includegraphics[height=47mm,clip=true]{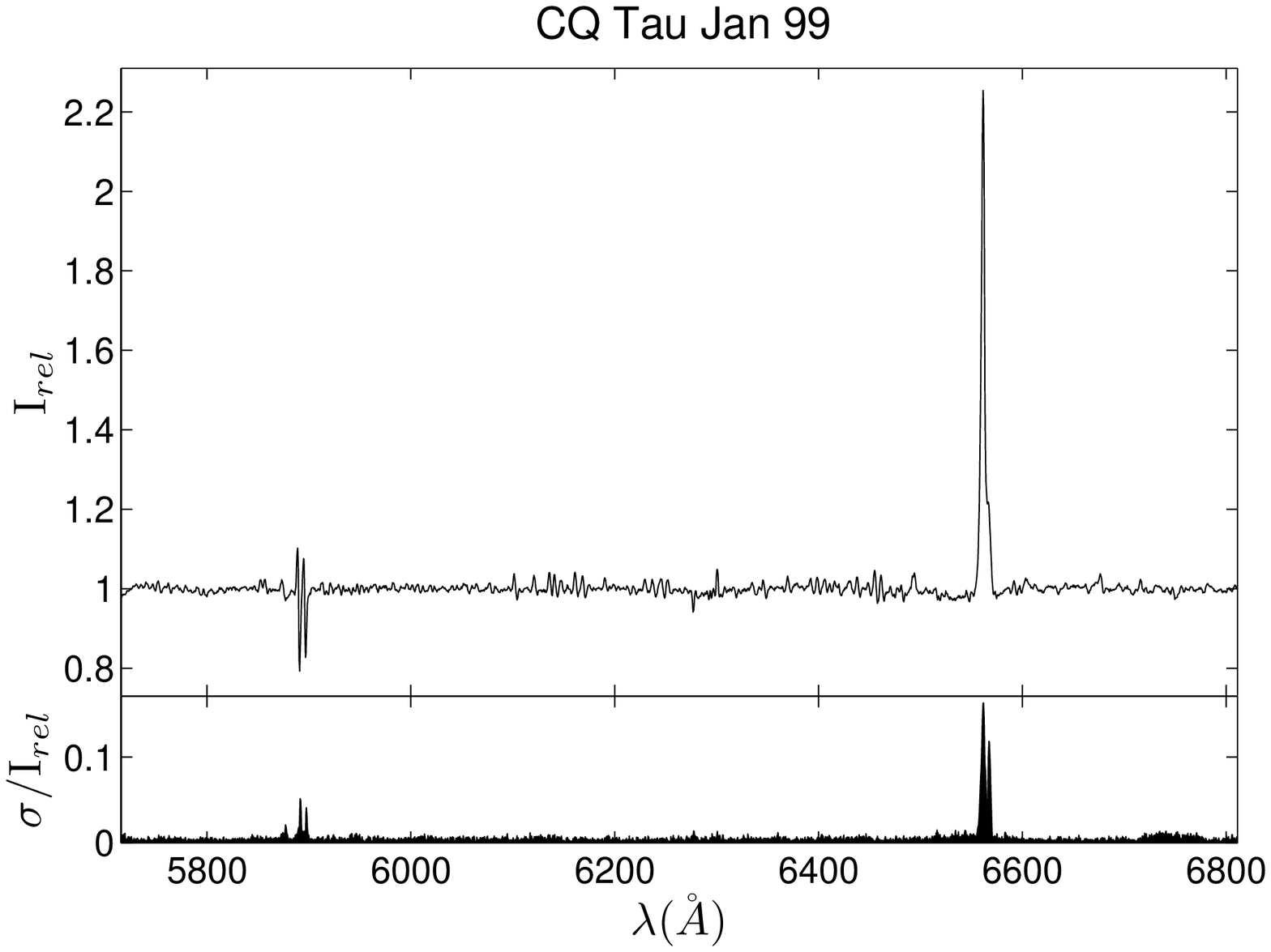}&
\includegraphics[bb=4 77 763 470,height=45mm,clip=true]{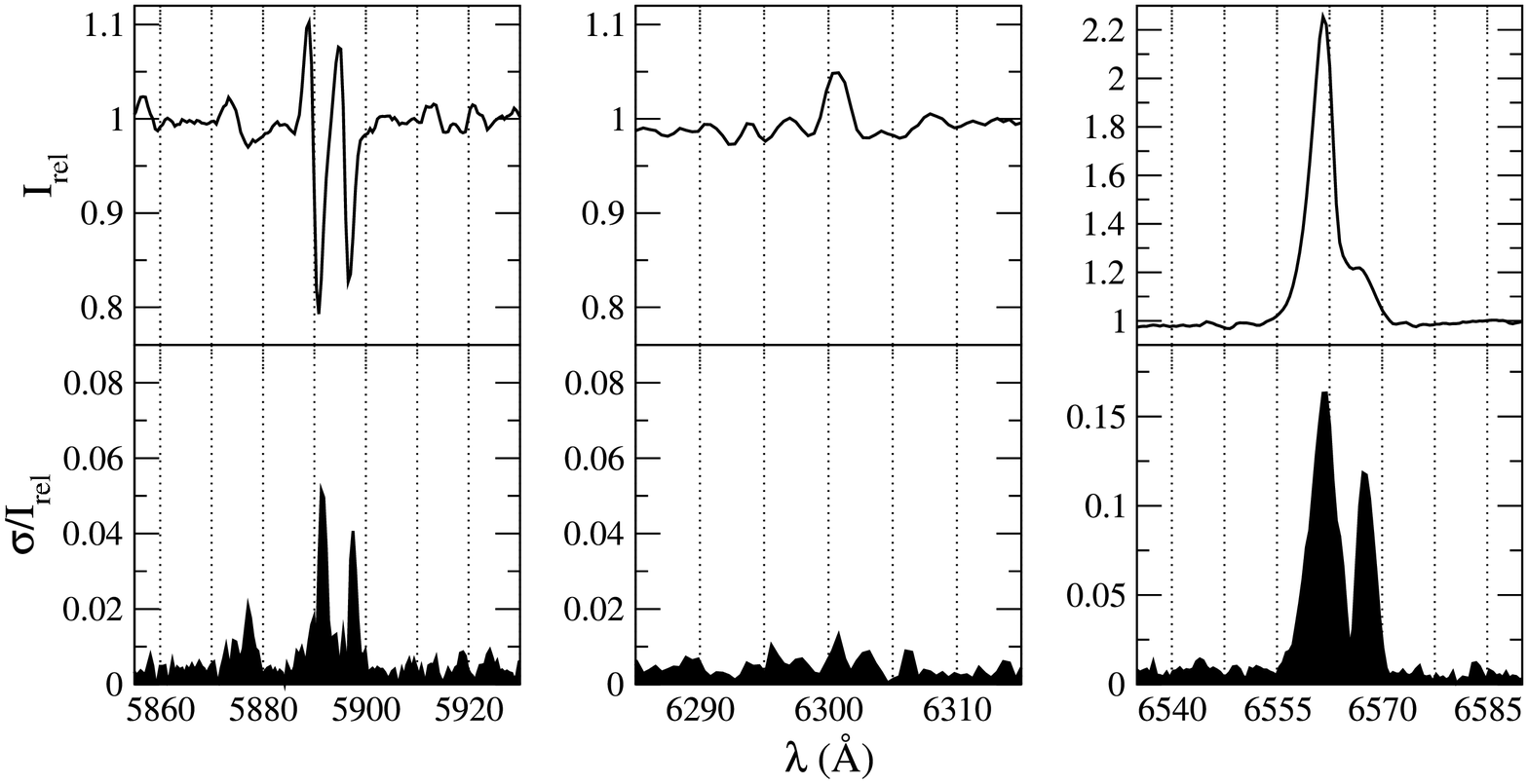} \\ 
\includegraphics[height=47mm,clip=true]{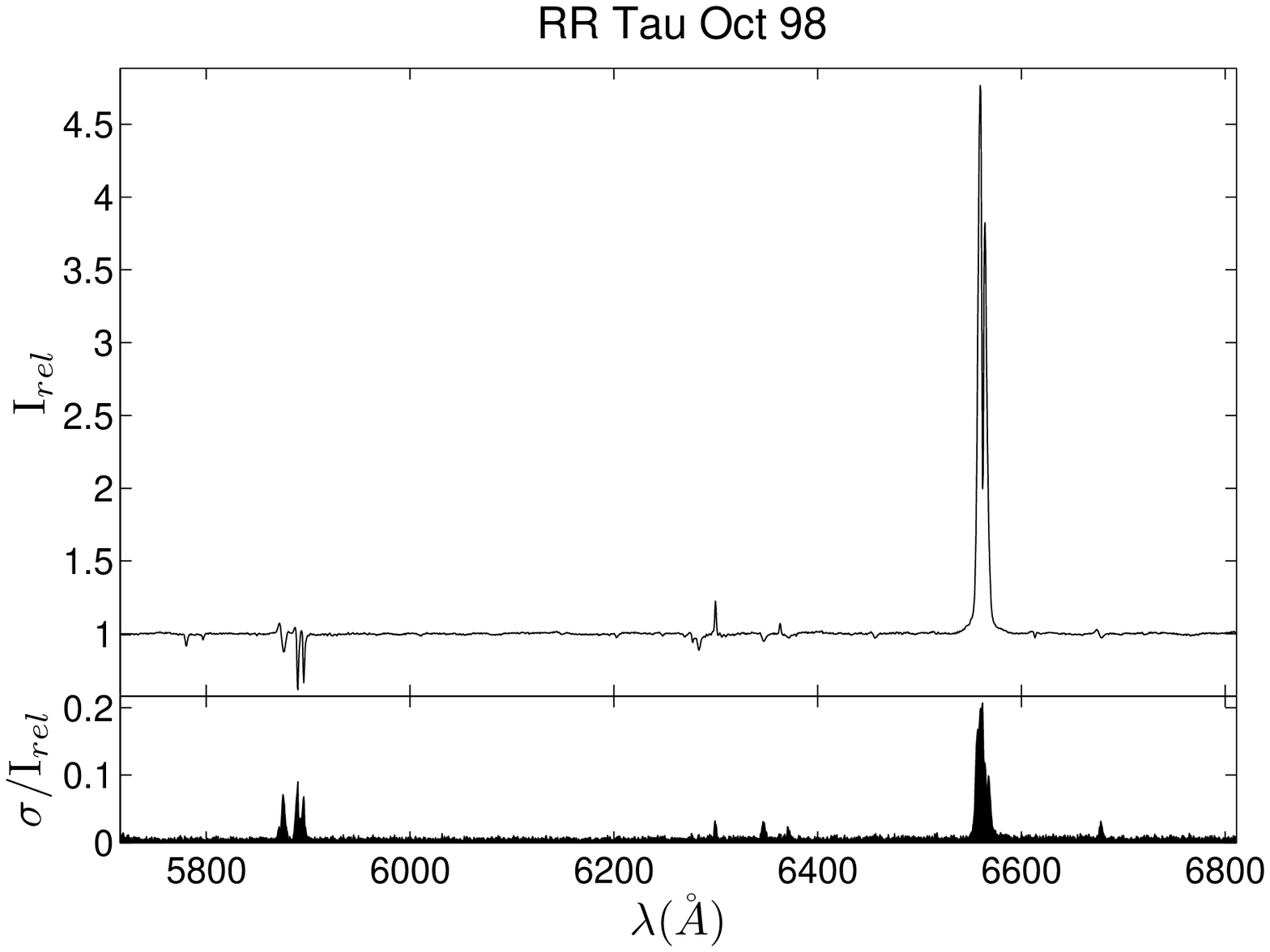}&
\includegraphics[bb=4 77 763 470,height=45mm,clip=true]{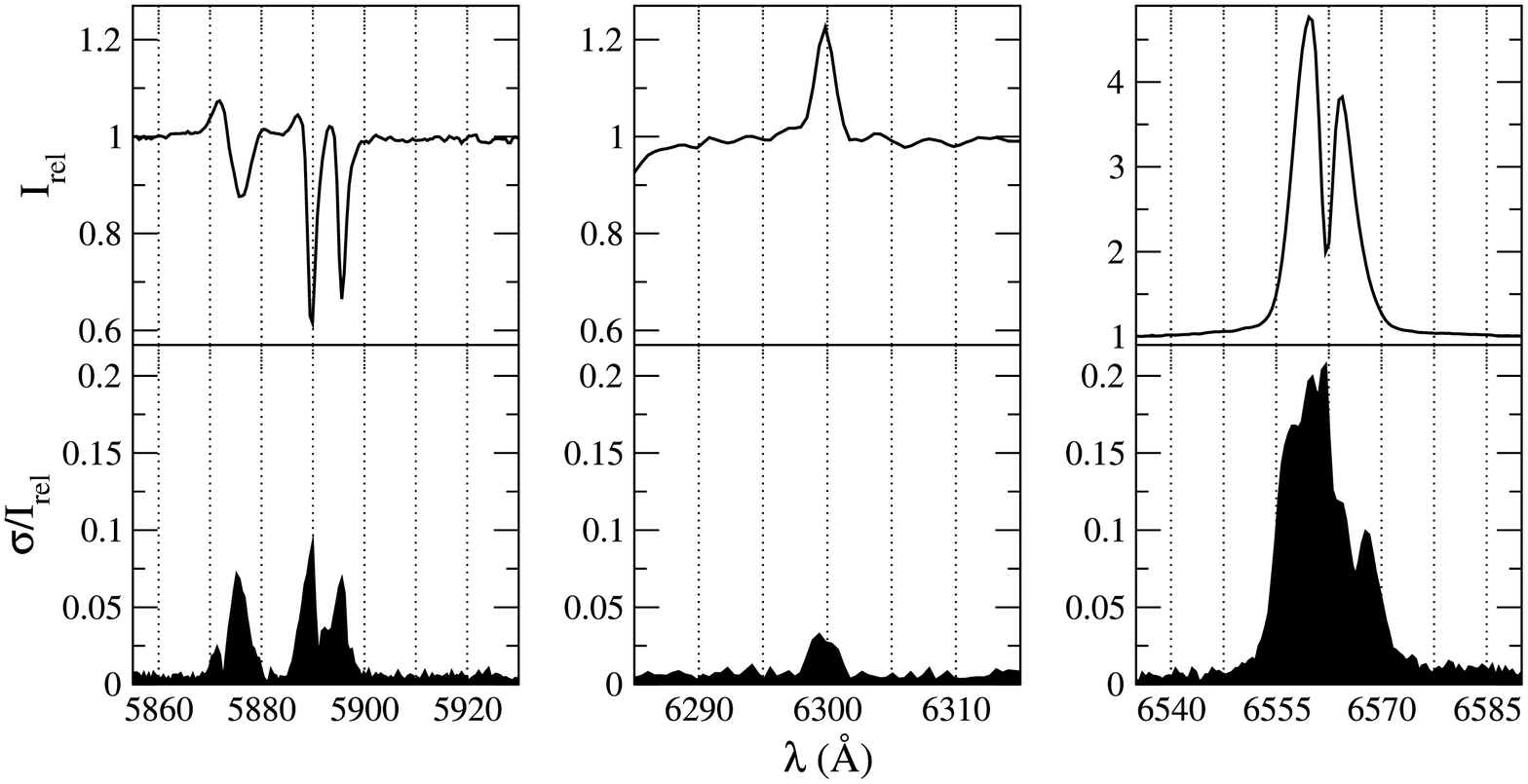} \\ 
\includegraphics[height=47mm,clip=true]{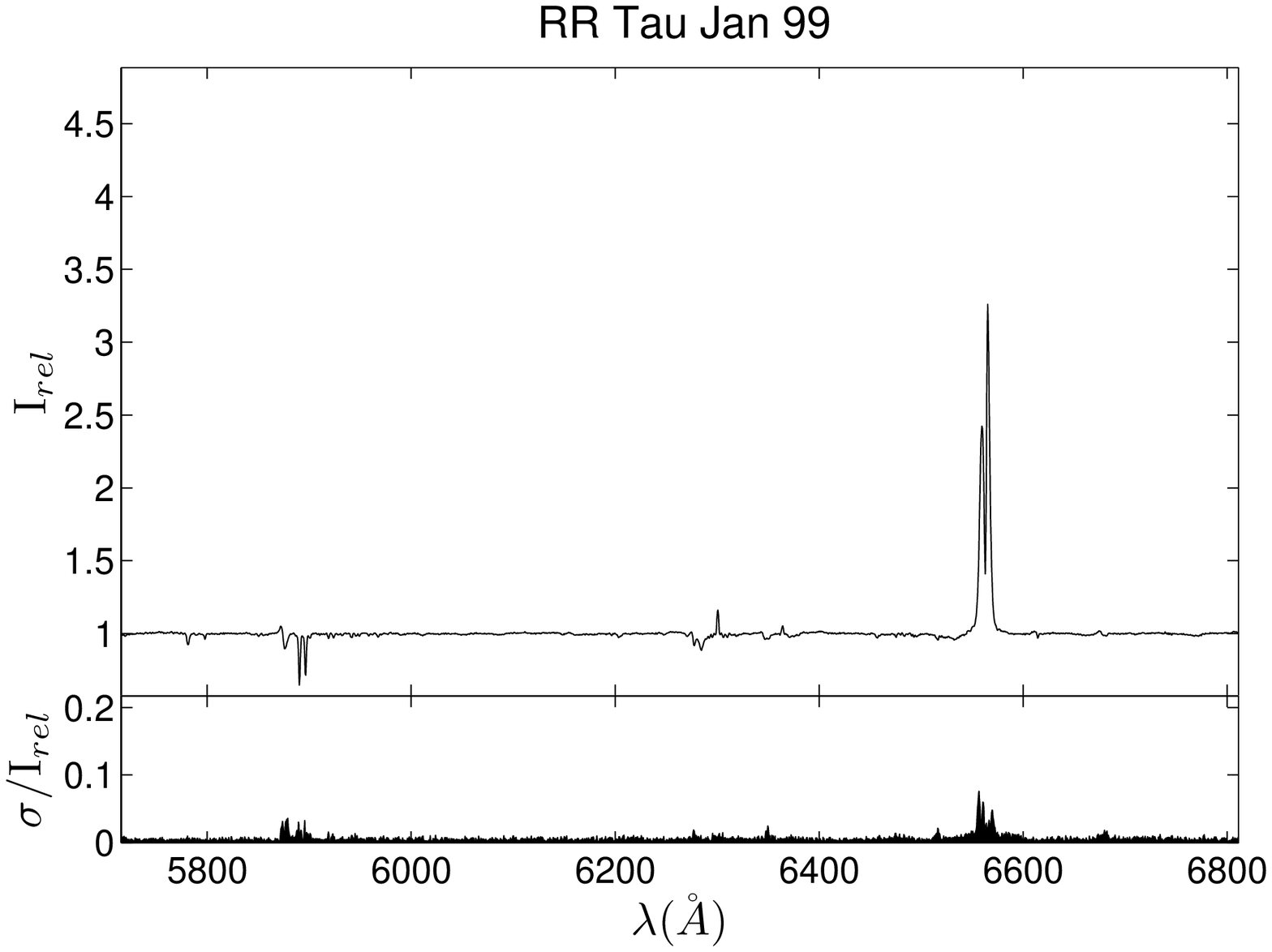}&
\includegraphics[bb=4 77 763 470,height=45mm,clip=true]{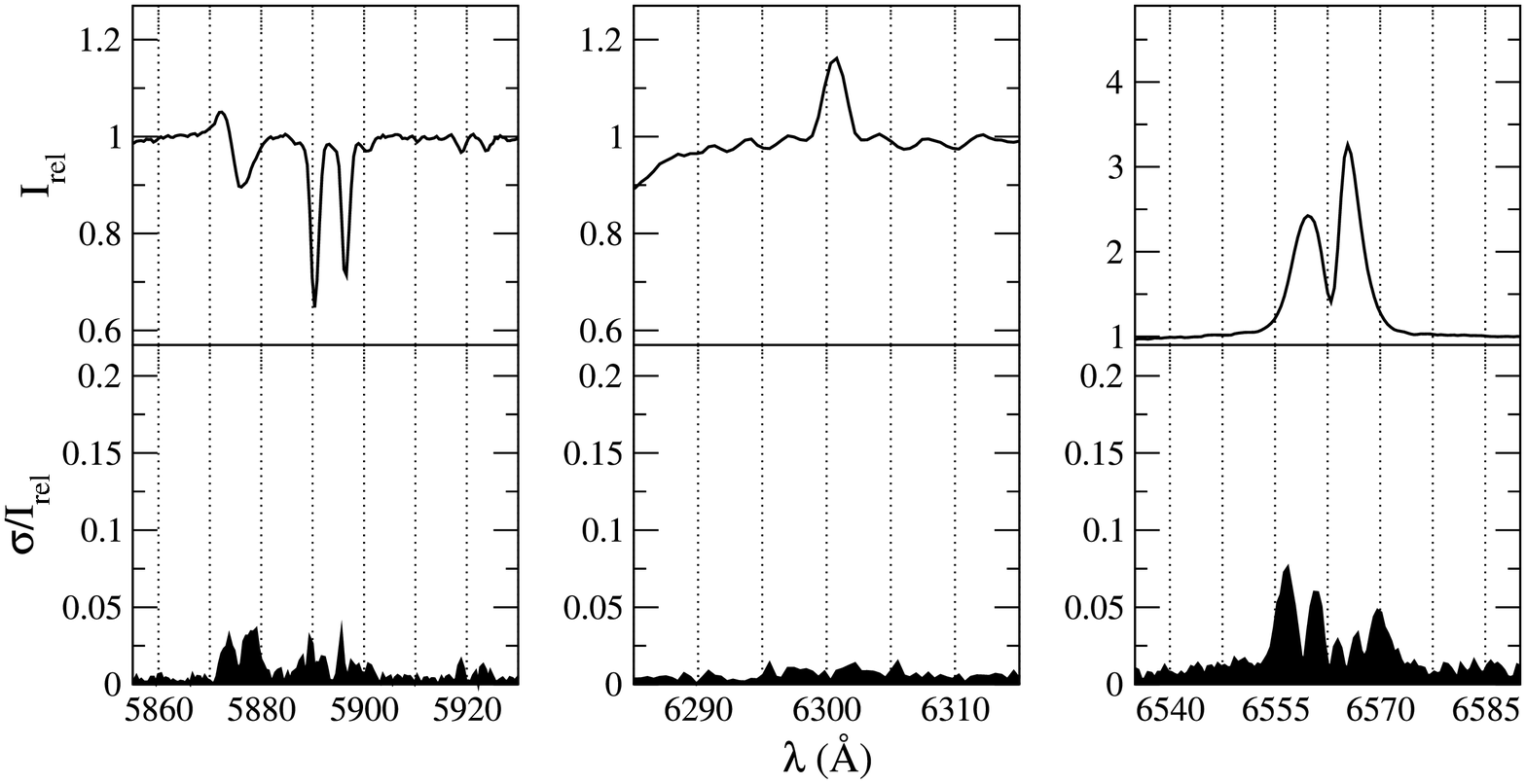} \\ 
\end{tabular}
\end{table}
\clearpage
\begin{table}
\centering
\renewcommand\arraystretch{10}
\begin{tabular}{cc}
\includegraphics[height=47mm,clip=true]{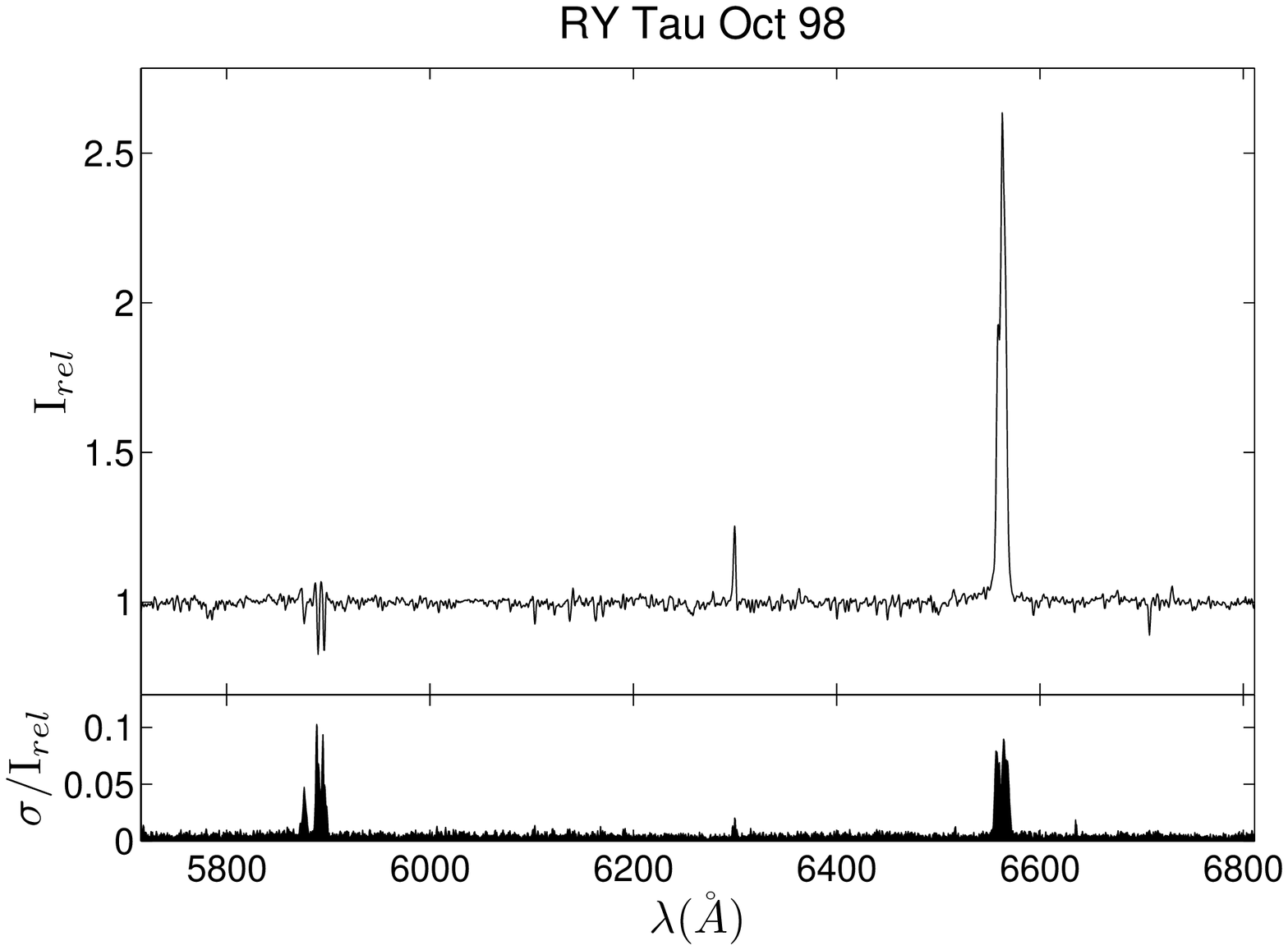}&
\includegraphics[bb=4 77 763 470,height=45mm,clip=true]{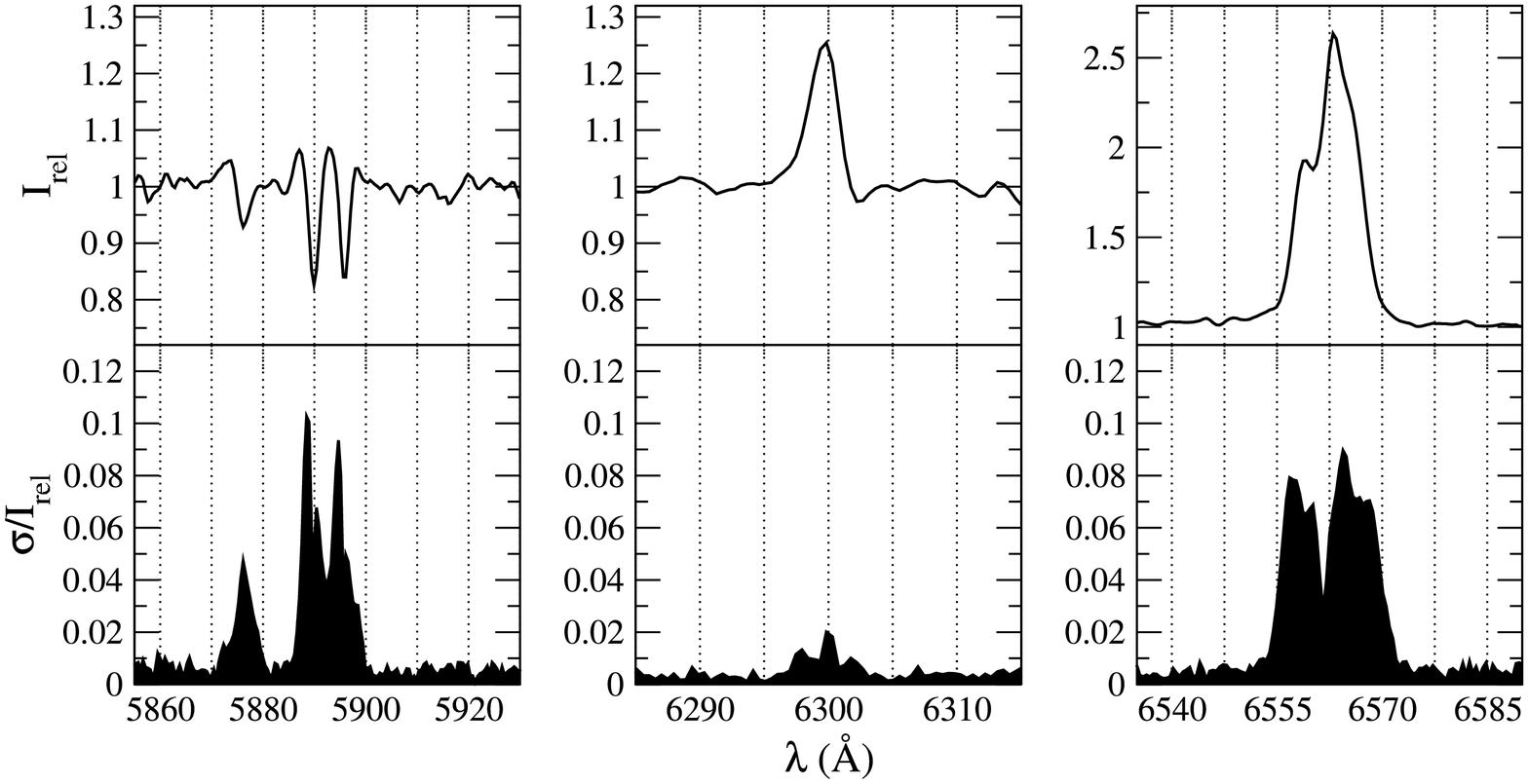} \\ 
\includegraphics[height=47mm,clip=true]{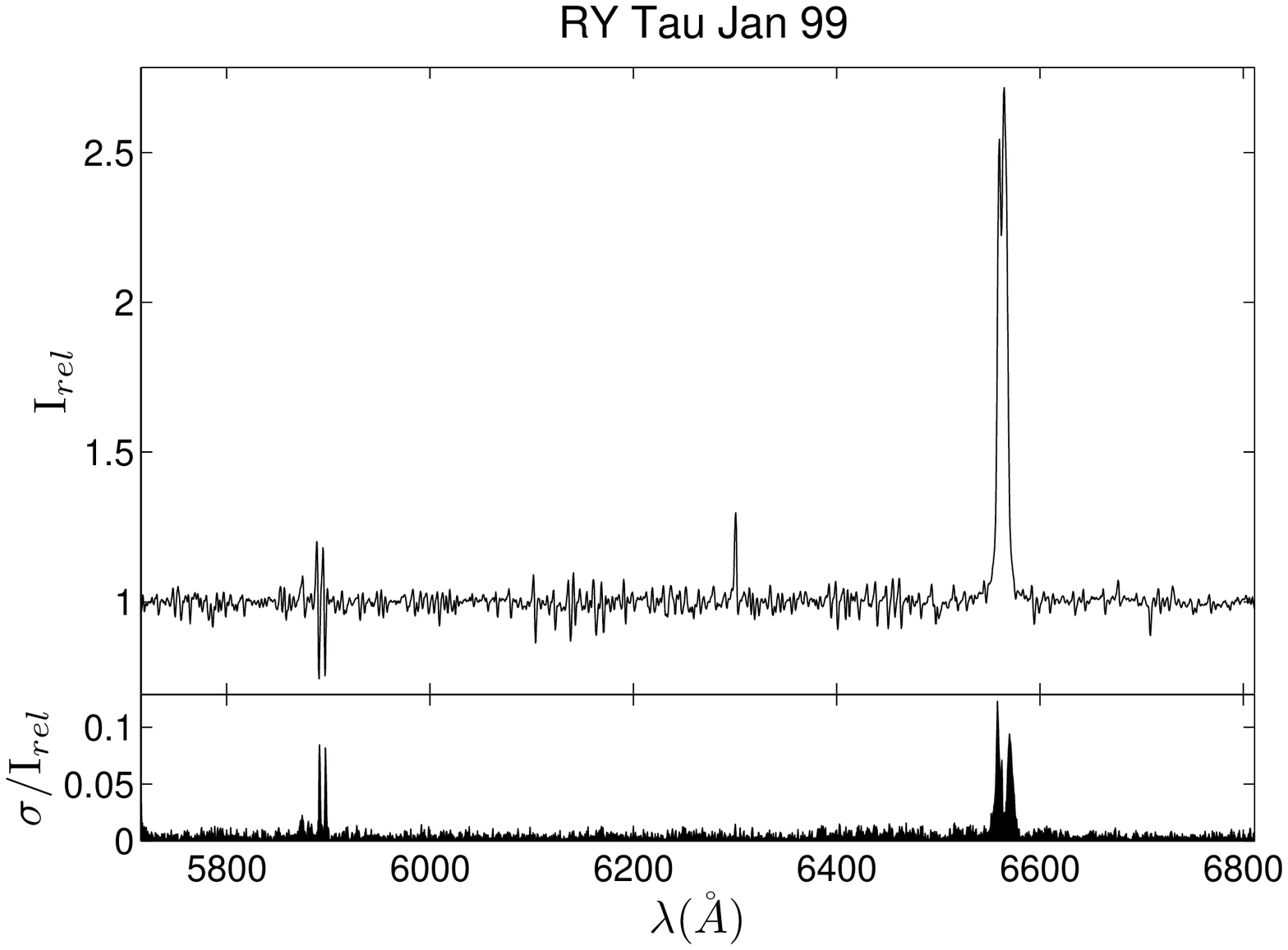}&
\includegraphics[bb=4 77 763 470,height=45mm,clip=true]{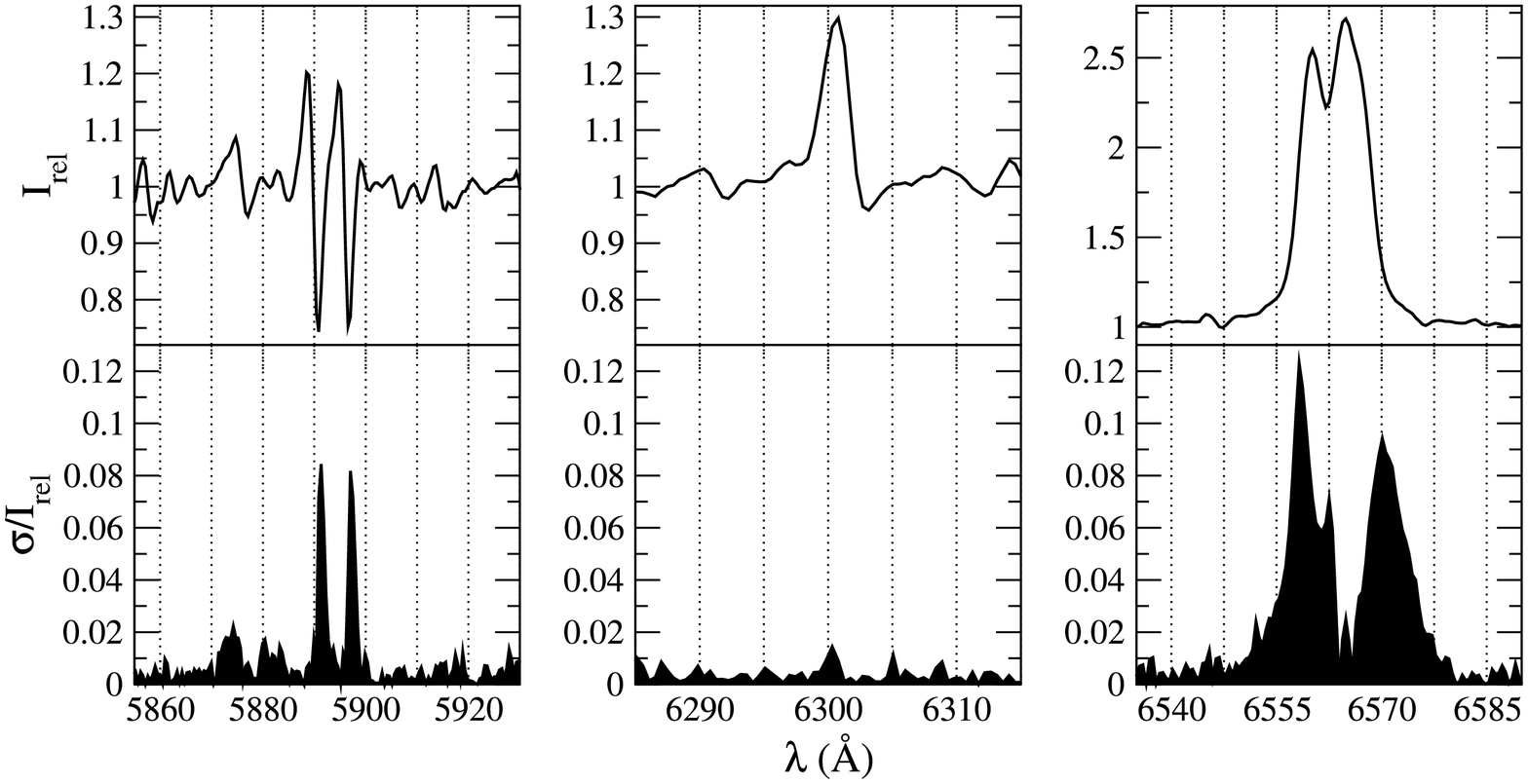} \\ 
\includegraphics[height=47mm,clip=true]{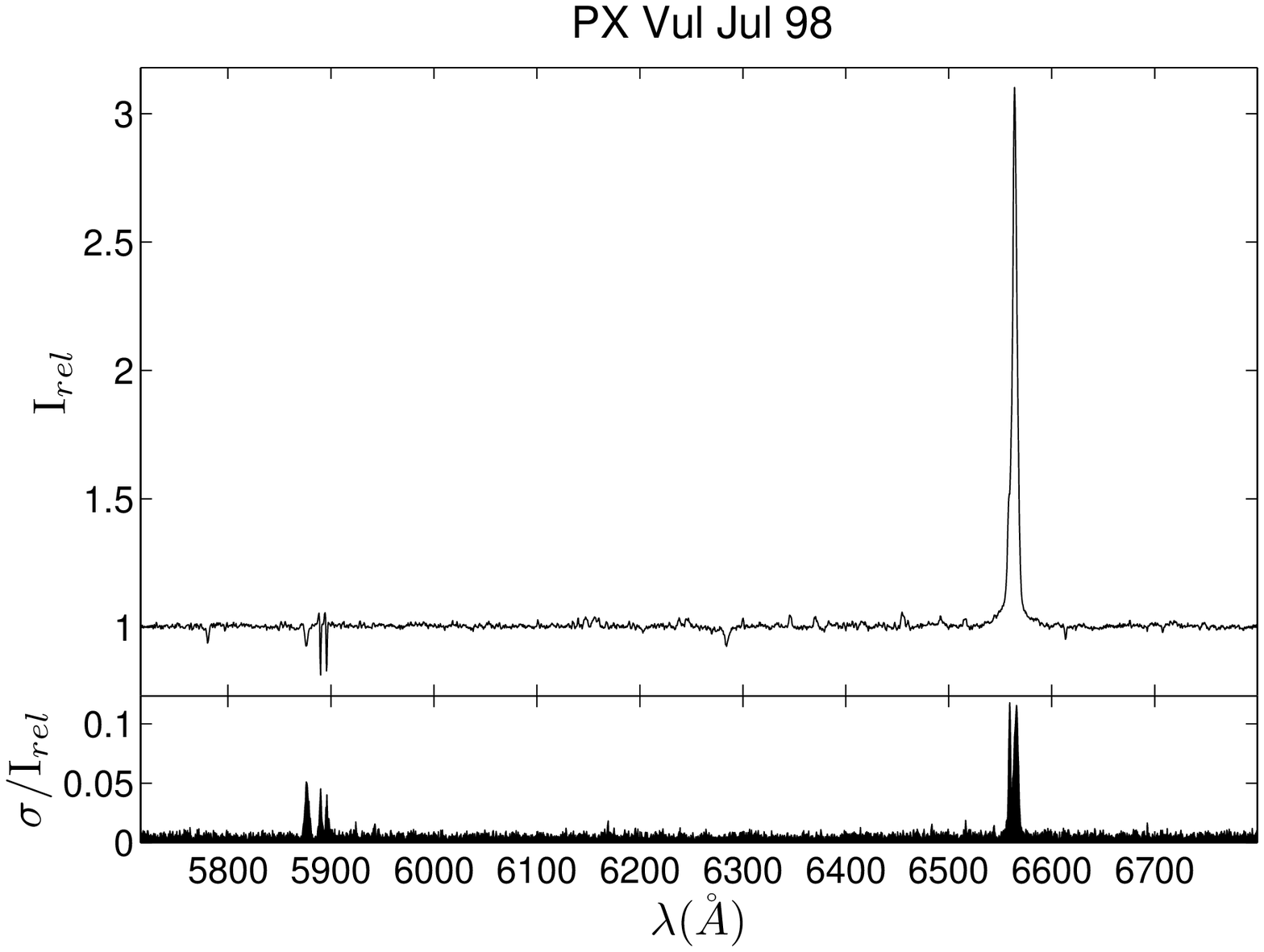}&
\includegraphics[bb=4 77 763 470,height=45mm,clip=true]{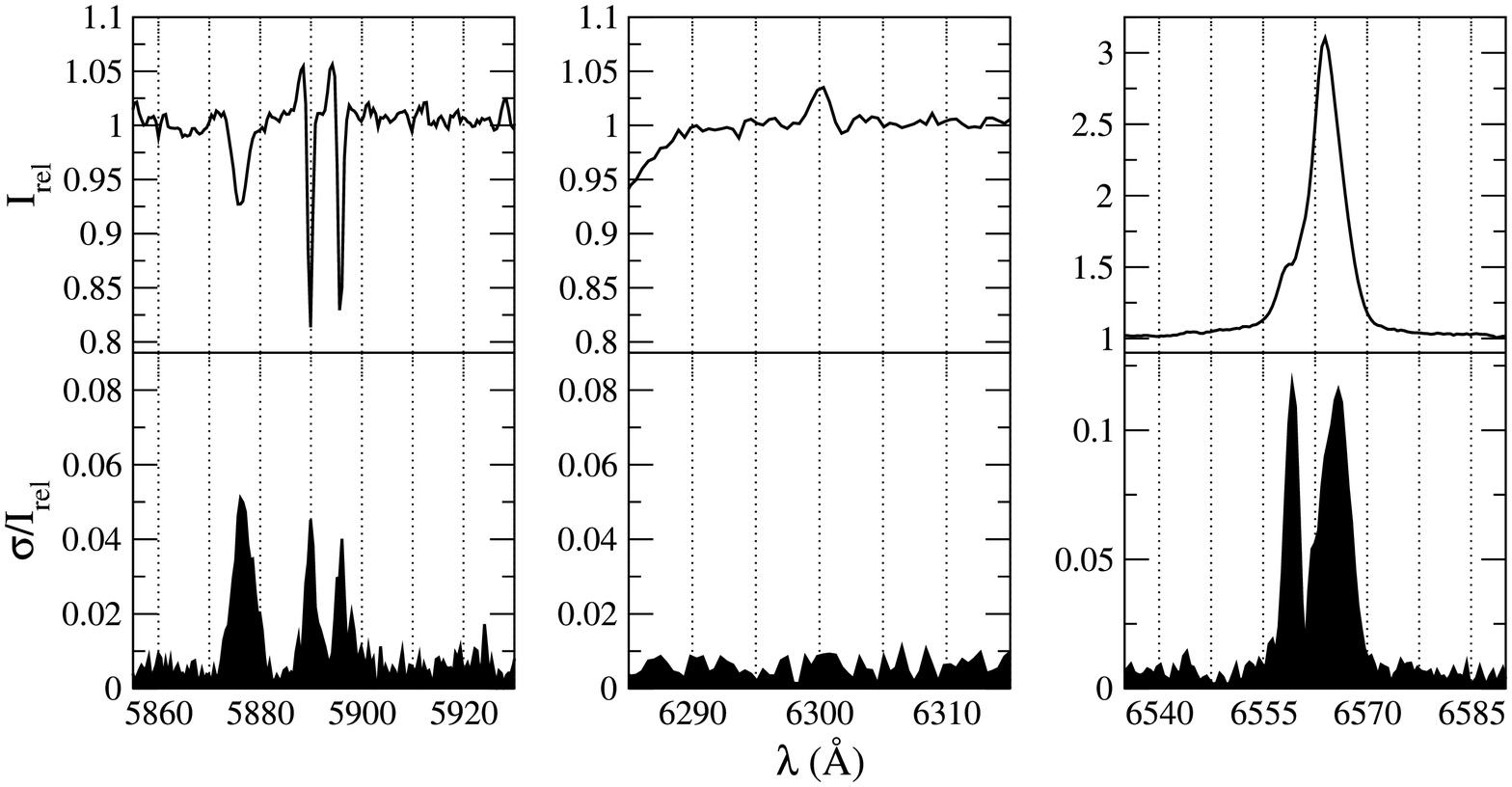} \\ 
\includegraphics[height=47mm,clip=true]{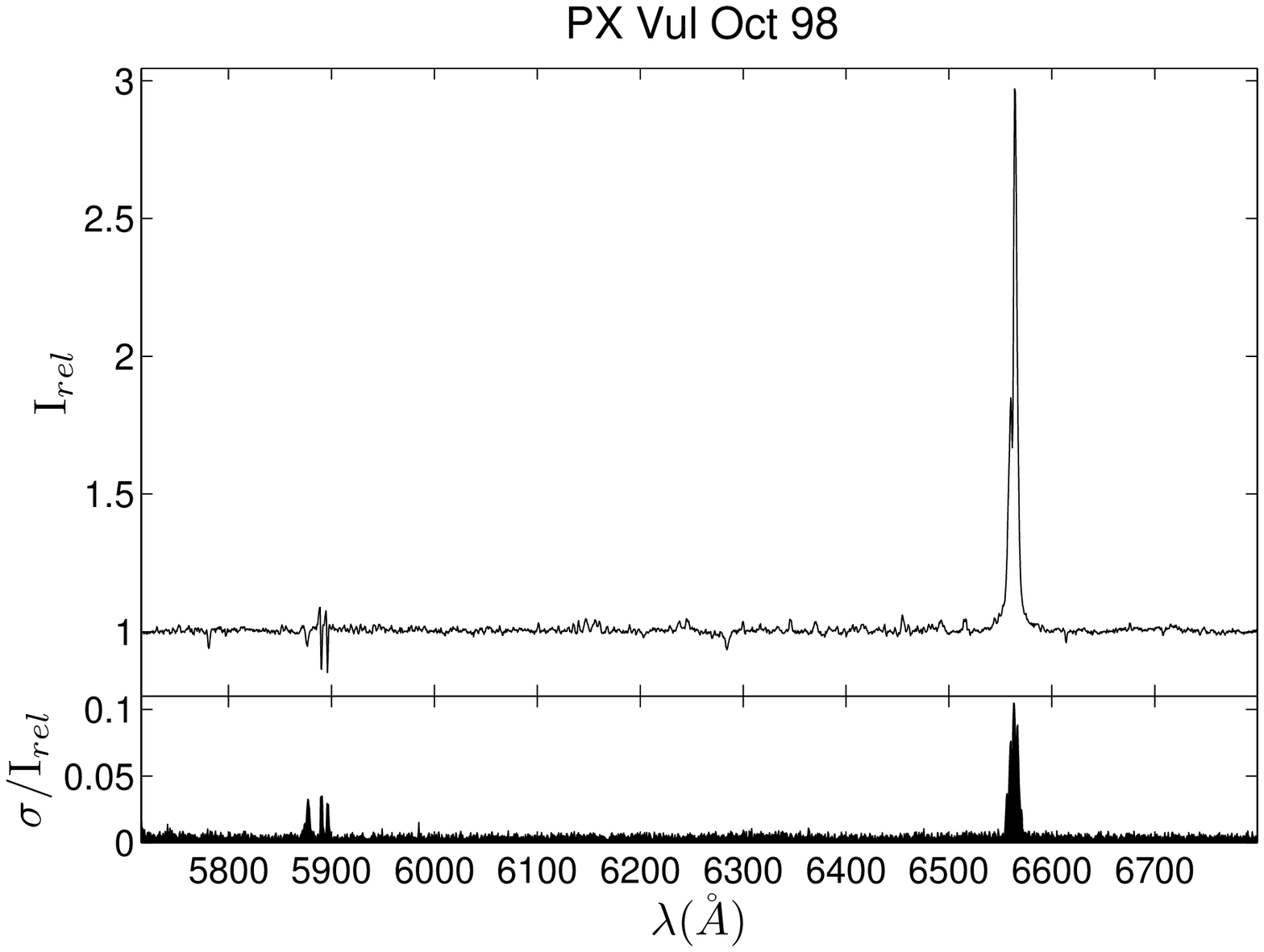}&
\includegraphics[bb=4 77 763 470,height=45mm,clip=true]{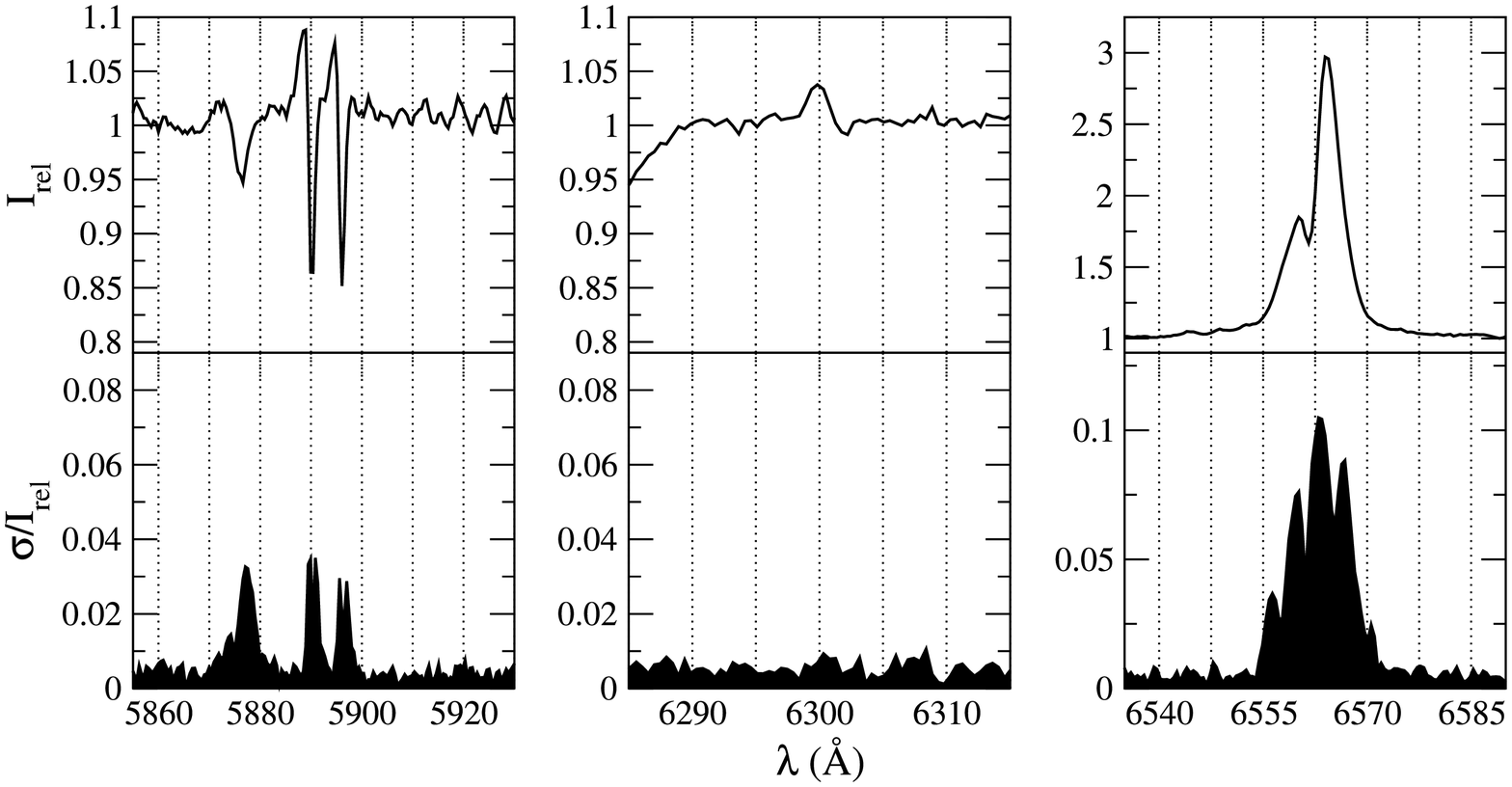} \\ 
\end{tabular}
\end{table}
\clearpage
\begin{table}
\centering
\renewcommand\arraystretch{10}
\begin{tabular}{cc}
\includegraphics[height=47mm,clip=true]{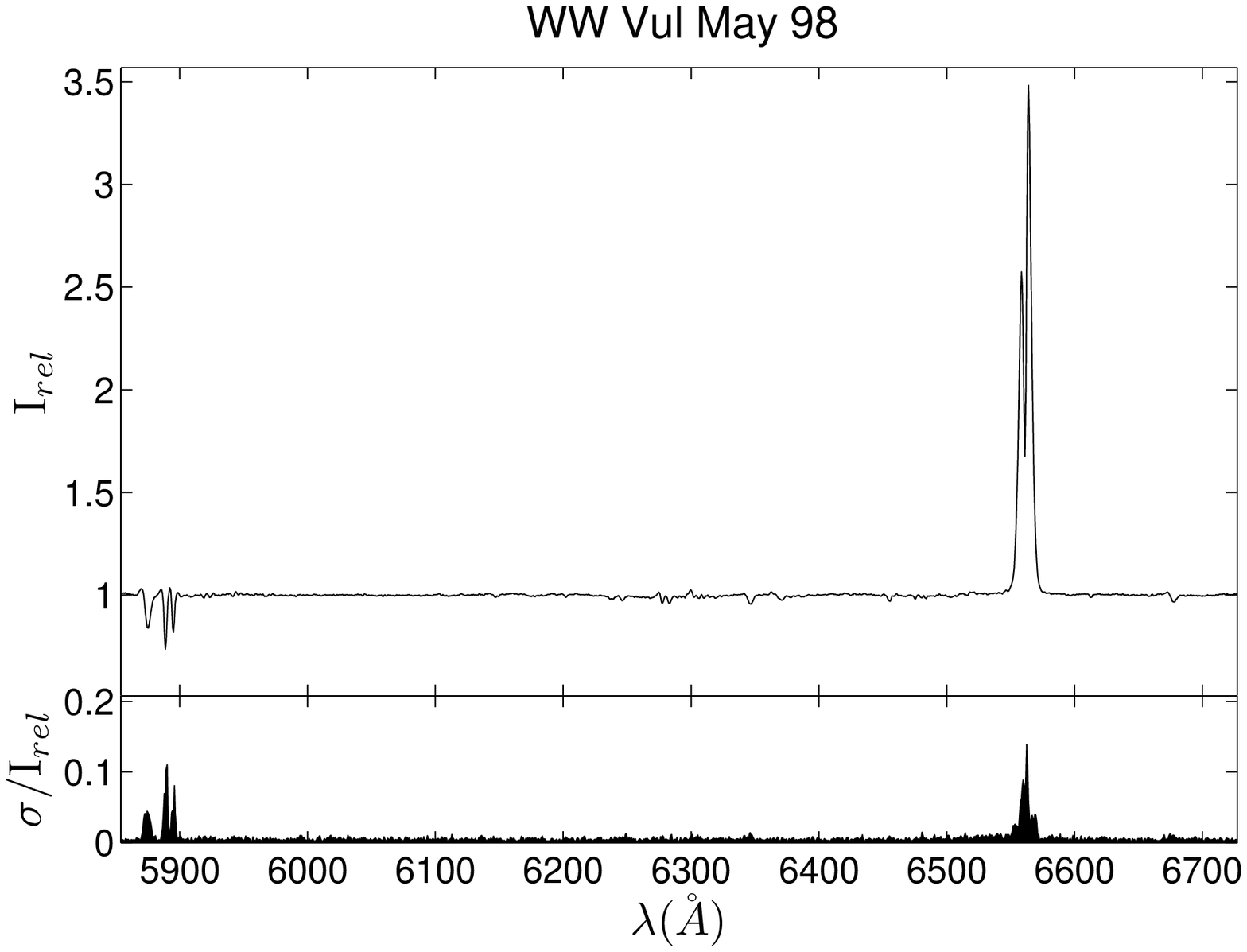}&
\includegraphics[bb=4 77 763 470,height=45mm,clip=true]{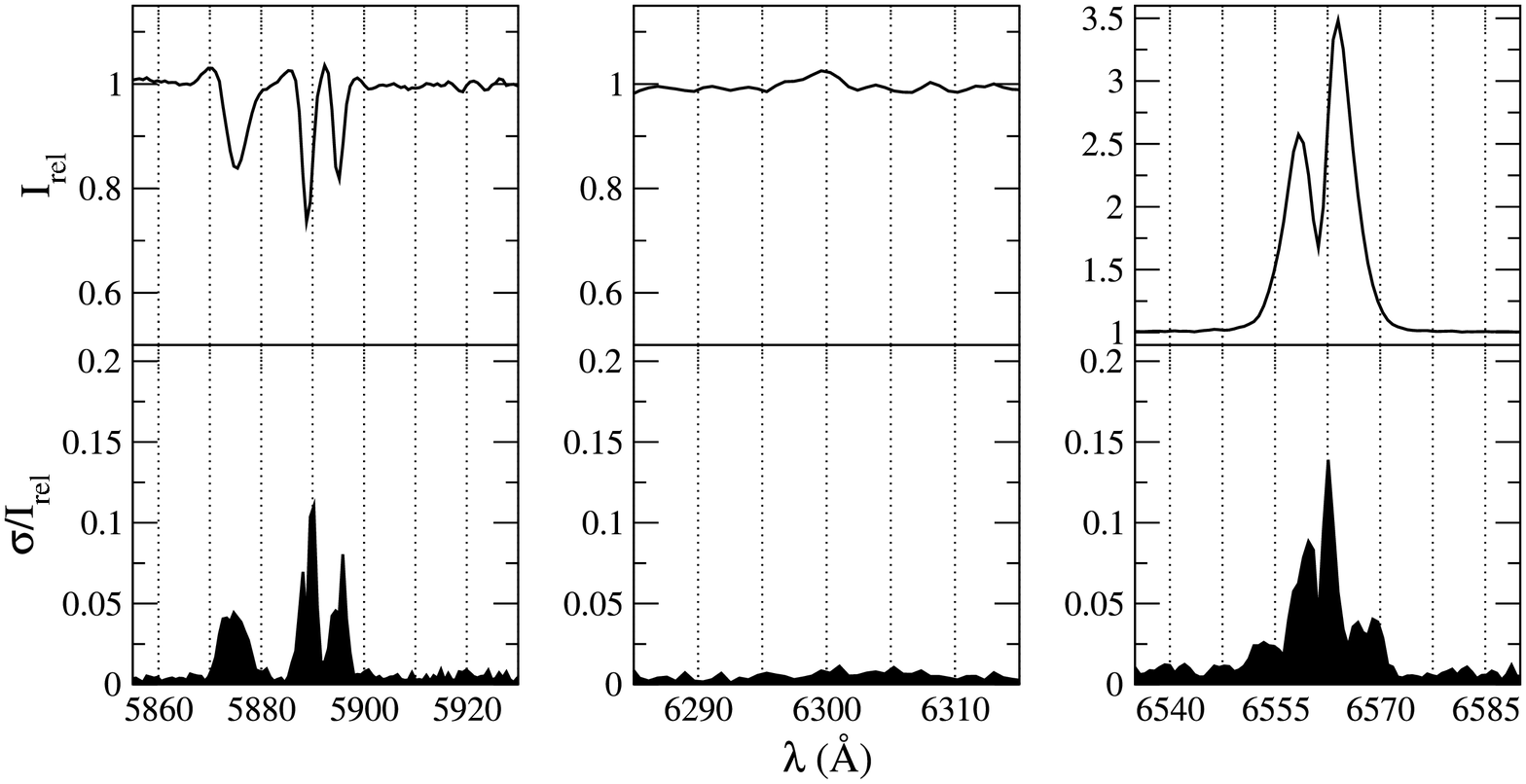} \\ 
\includegraphics[height=47mm,clip=true]{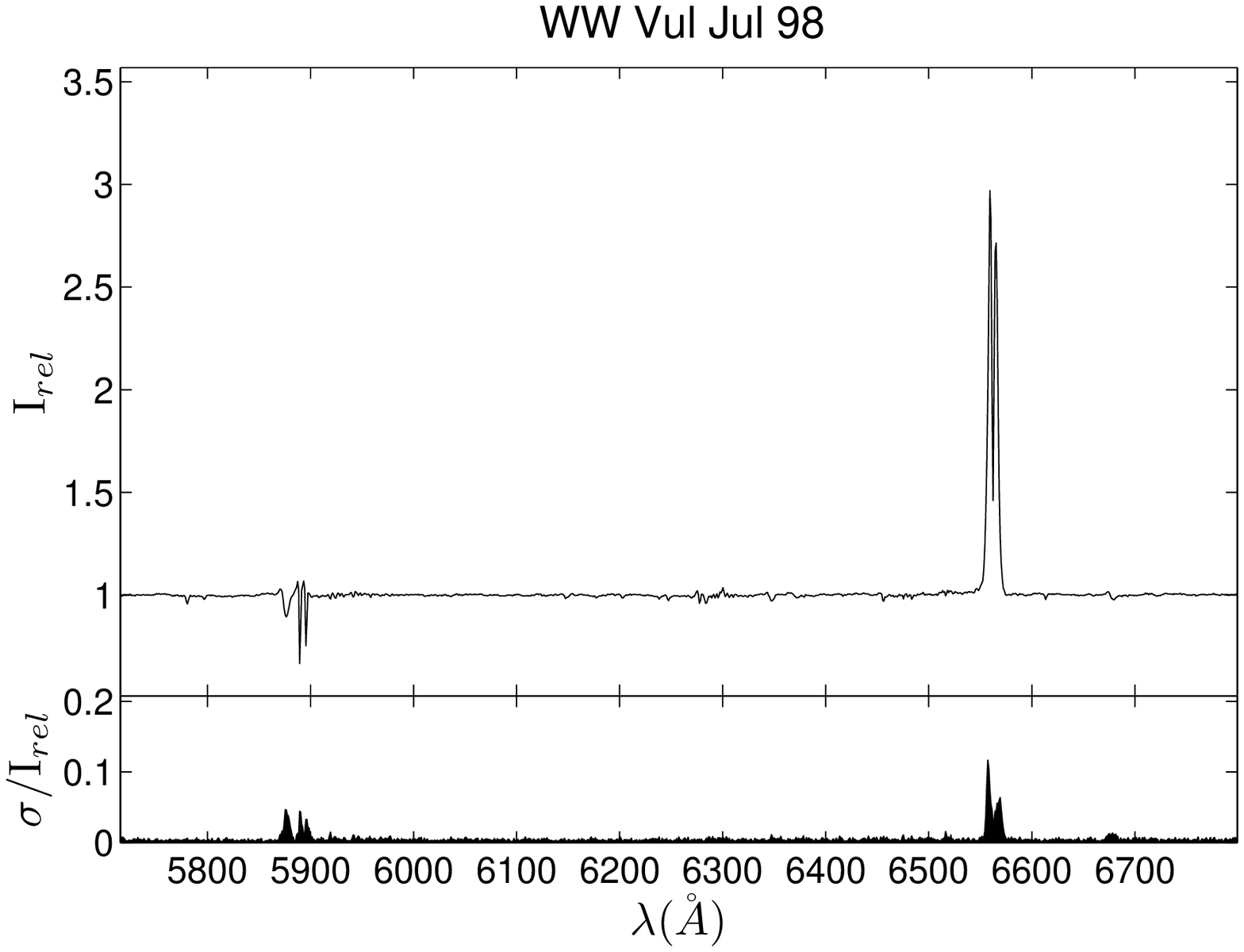}&
\includegraphics[bb=4 77 763 470,height=45mm,clip=true]{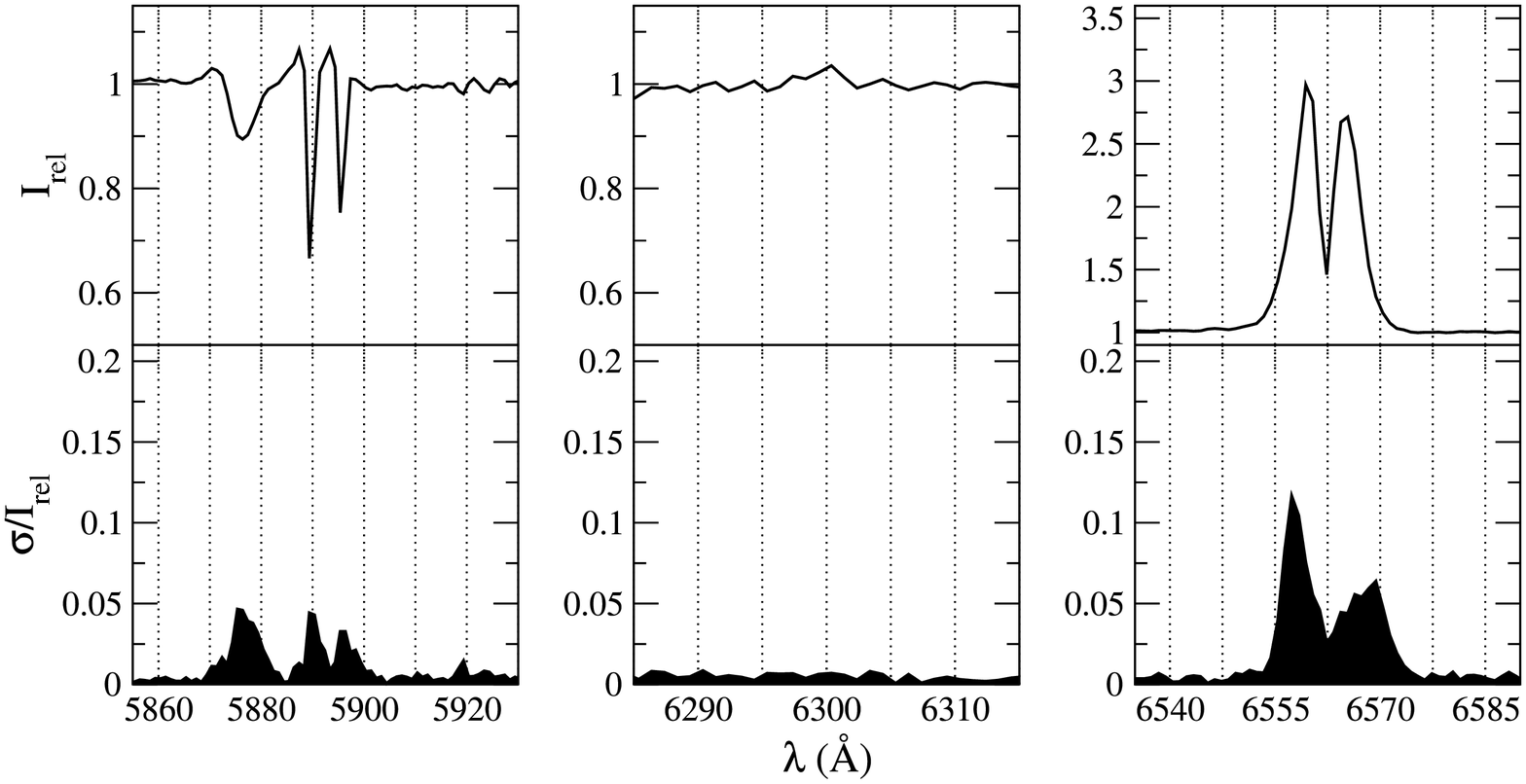} \\ 
\includegraphics[height=47mm,clip=true]{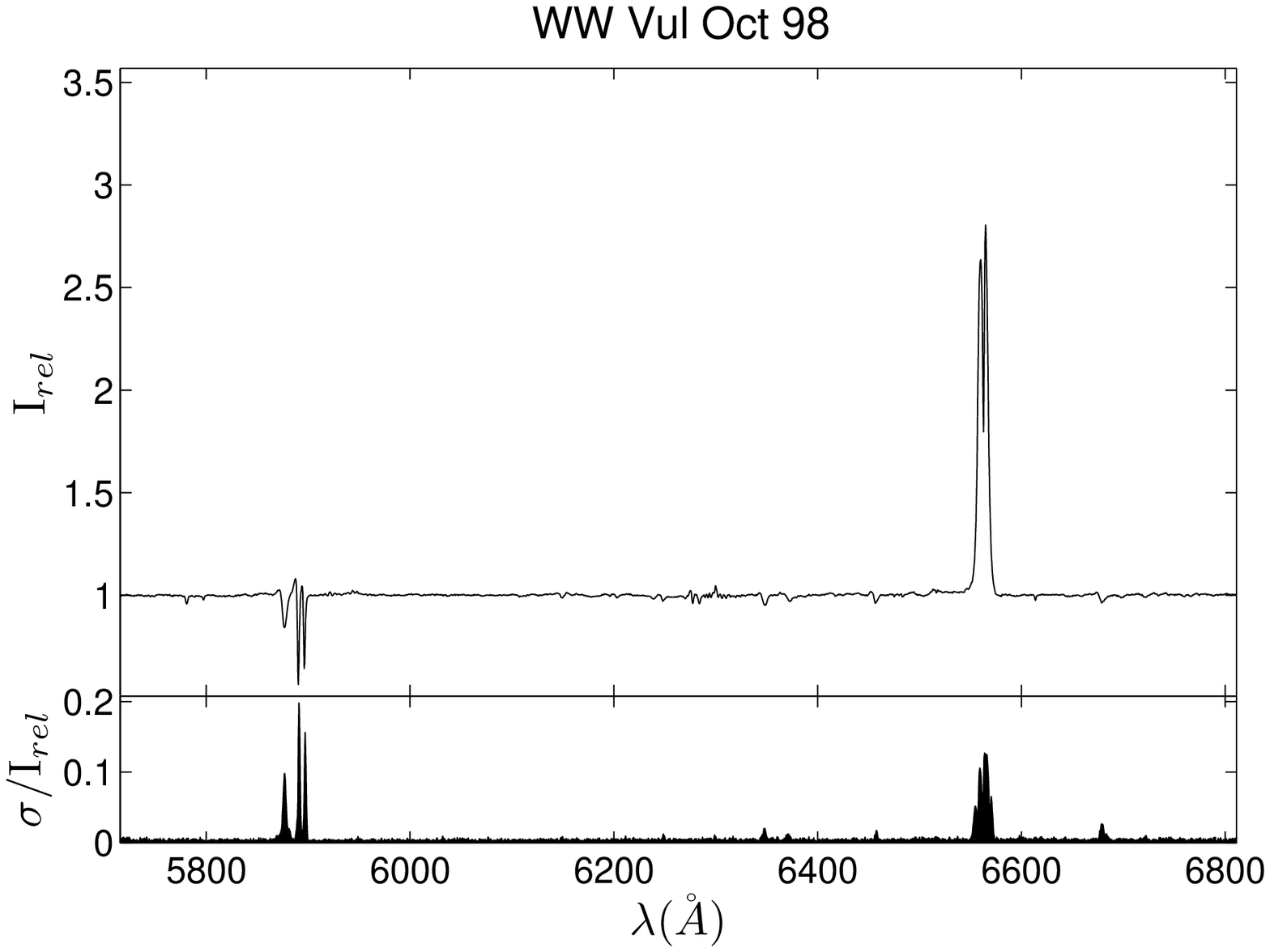}&
\includegraphics[bb=4 77 763 470,height=45mm,clip=true]{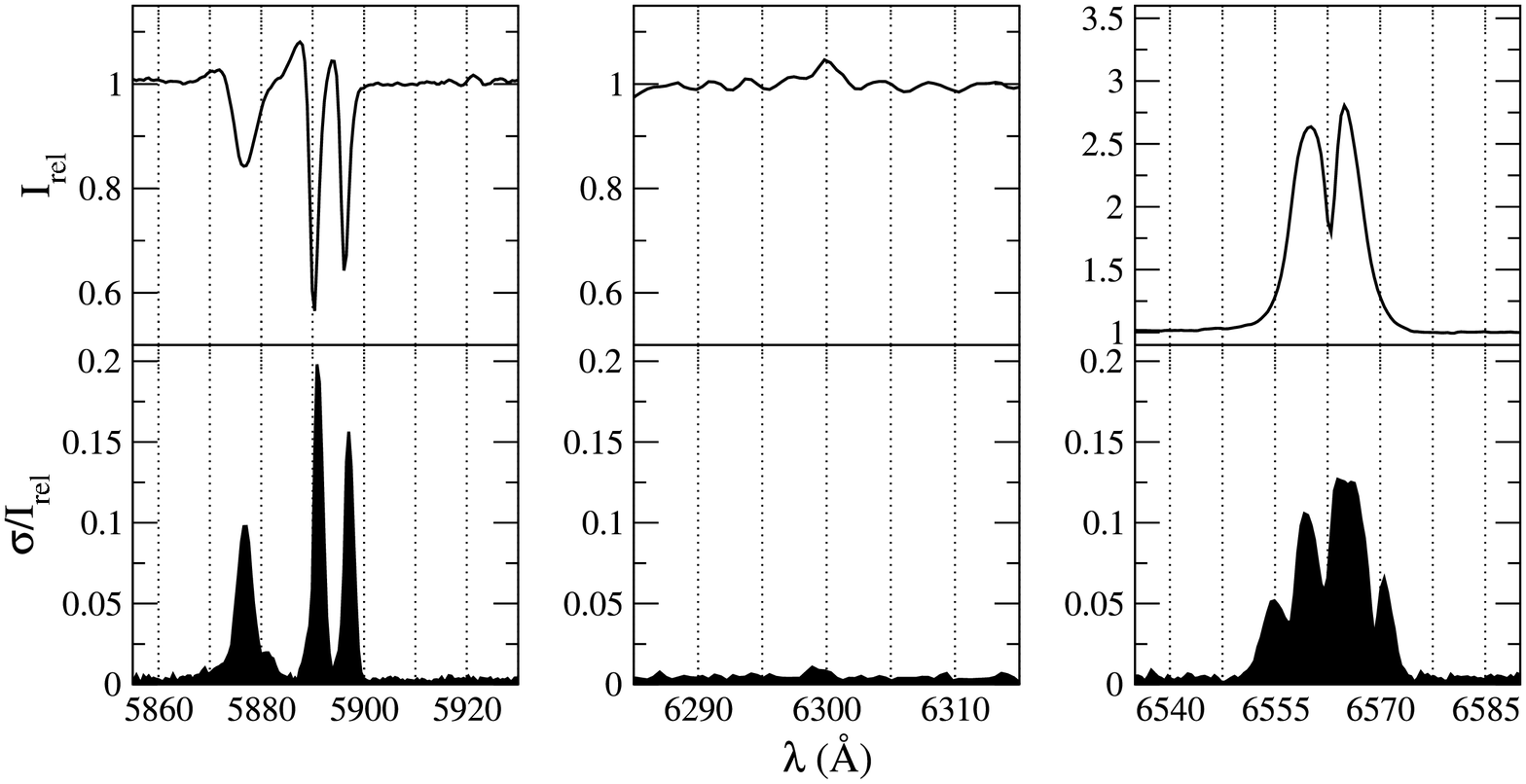} \\ 
\includegraphics[height=47mm,clip=true]{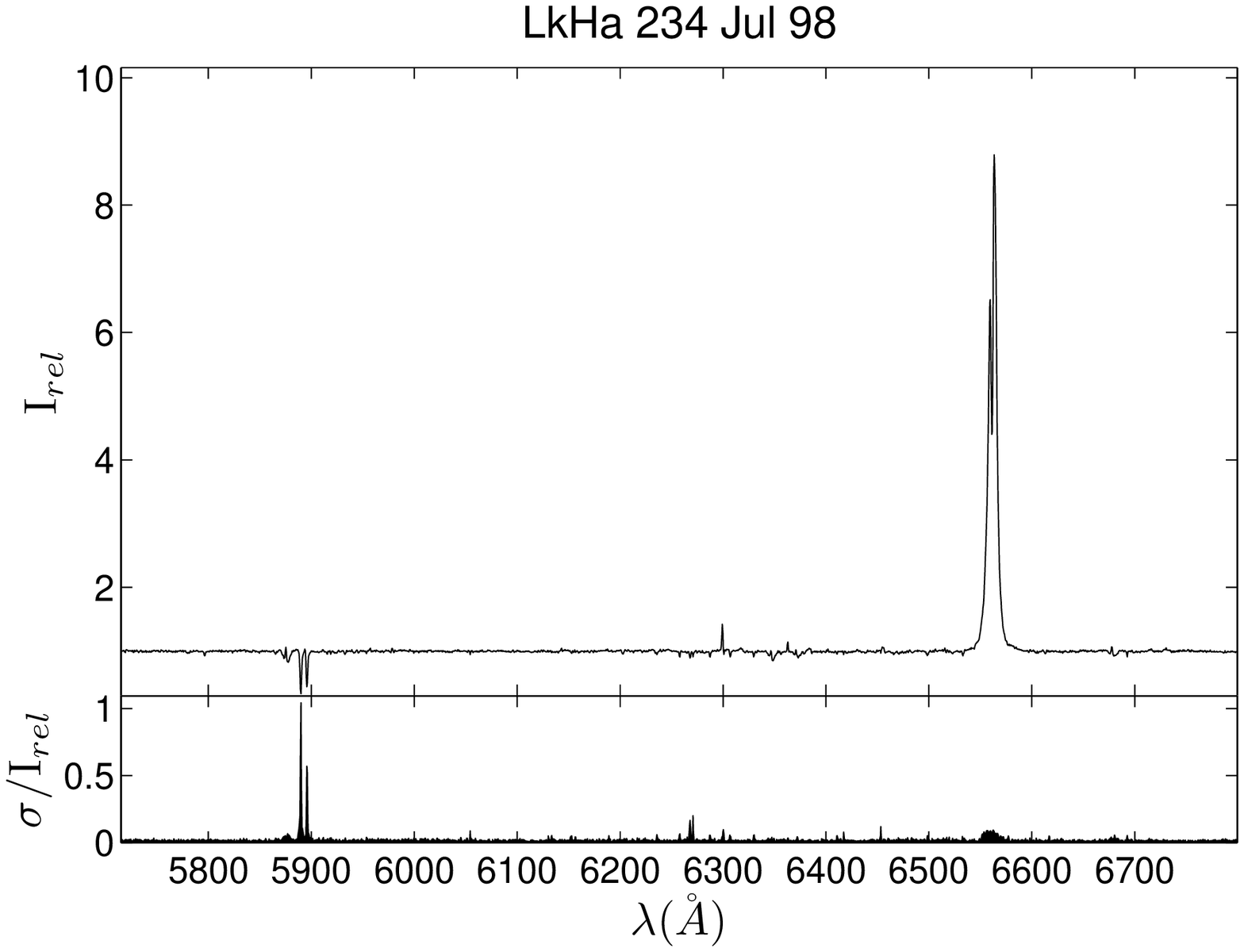}&
\includegraphics[bb=4 77 763 470,height=45mm,clip=true]{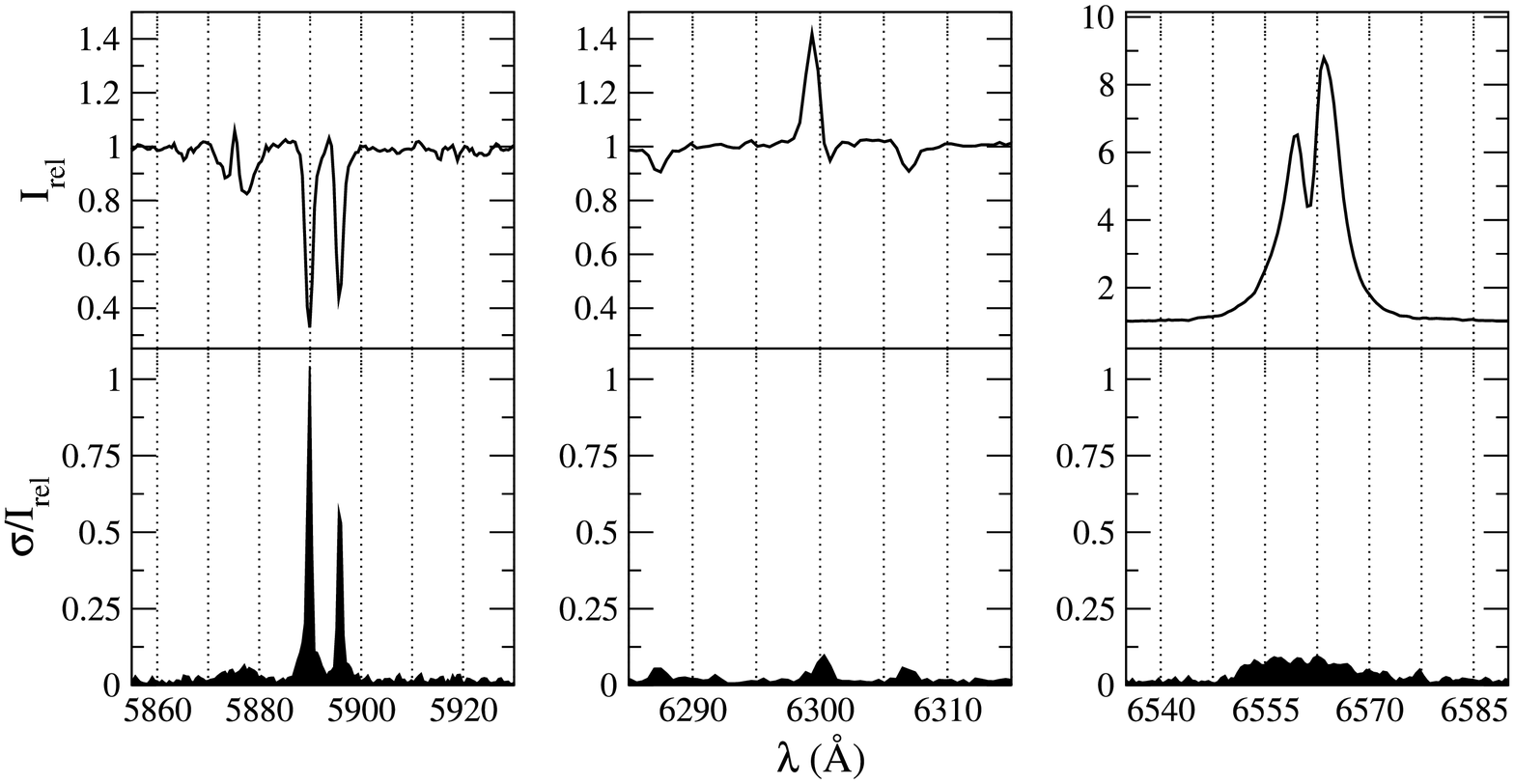} \\ 
\end{tabular}
\end{table}
\clearpage
\begin{table}
\centering
\renewcommand\arraystretch{10}
\begin{tabular}{cc}
\includegraphics[height=47mm,clip=true]{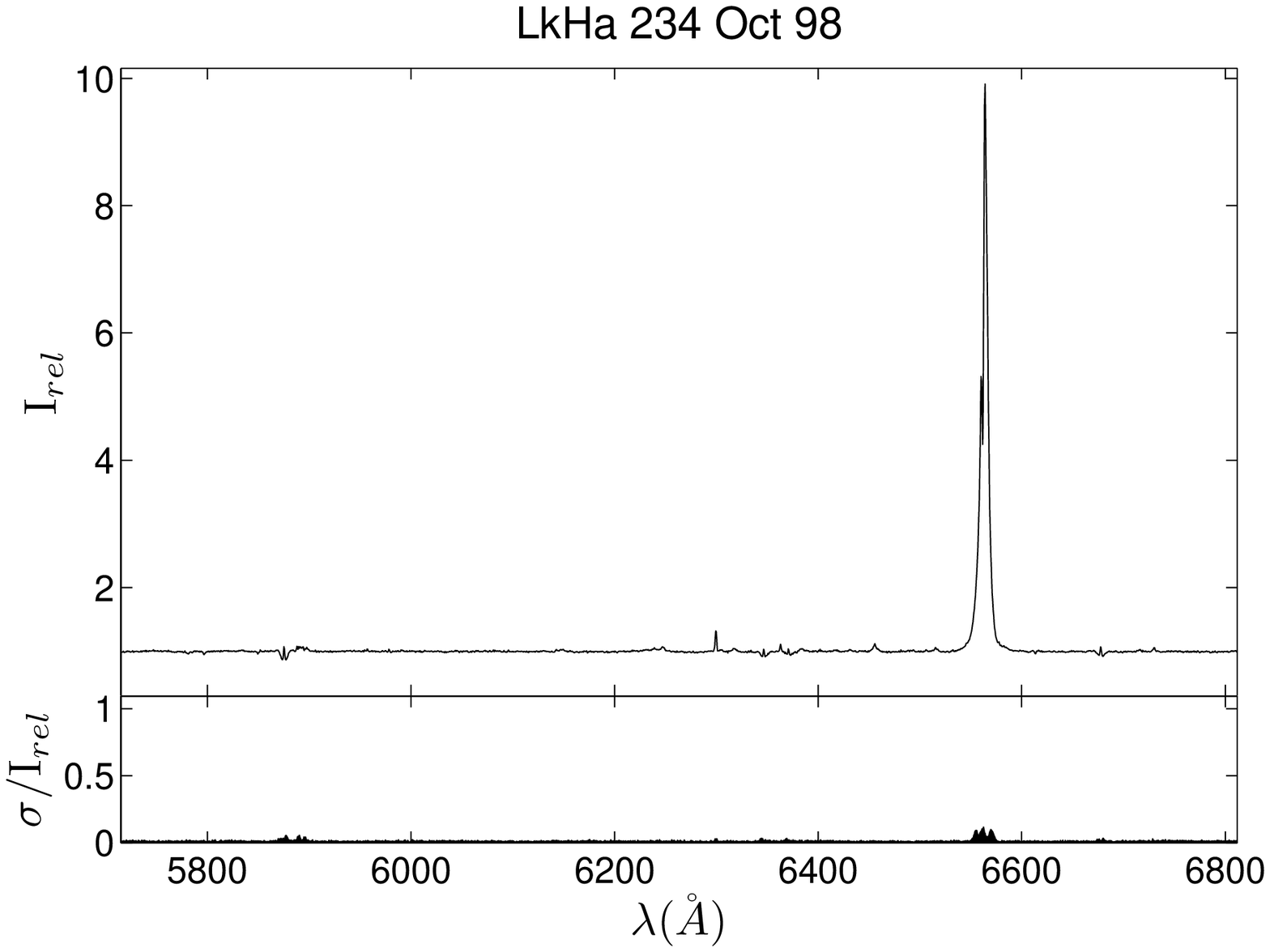}&
\includegraphics[bb=4 77 763 470,height=45mm,clip=true]{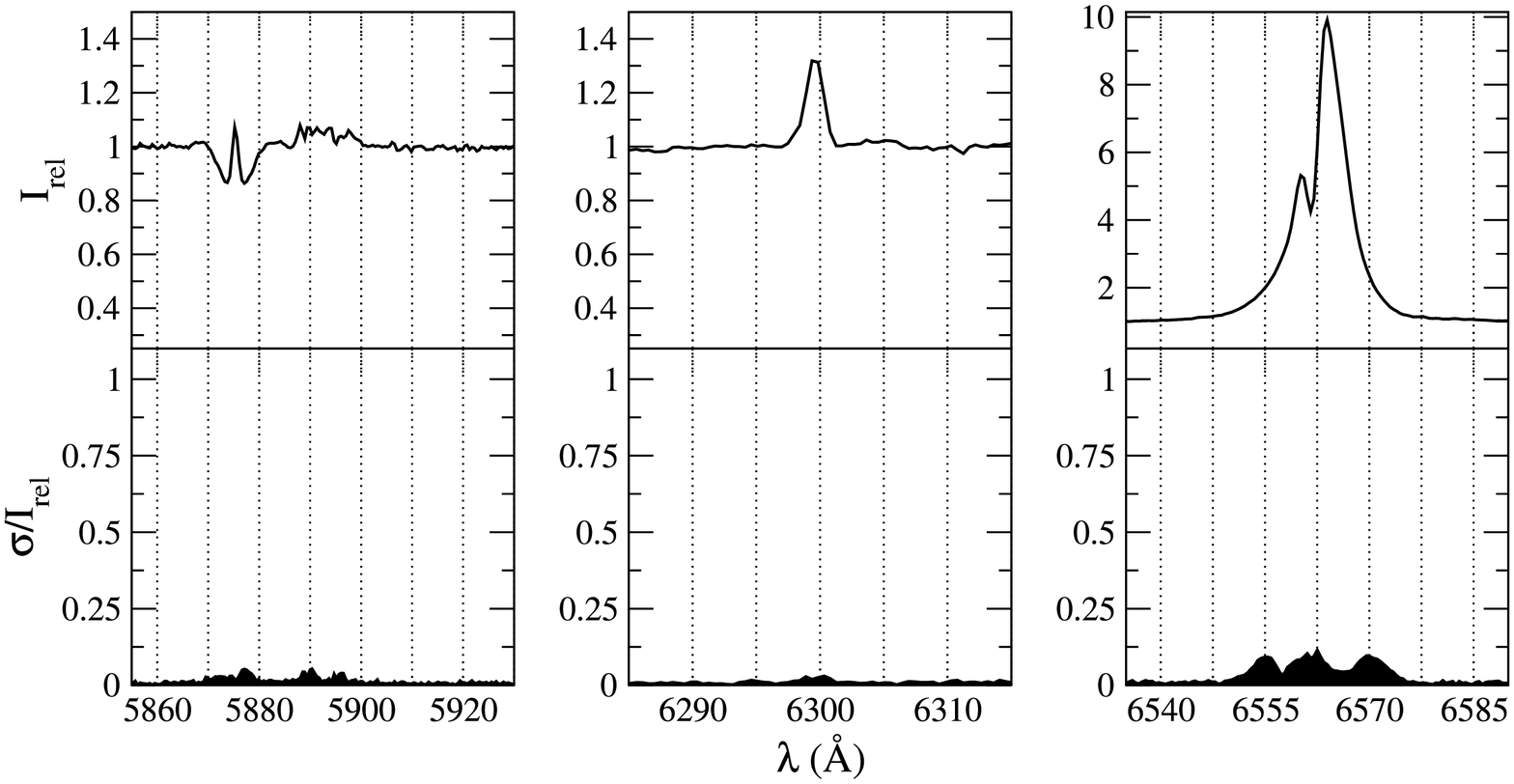} \\
\end{tabular}
\end{table}
\begin{minipage}{18cm}
Fig. B1: For each star and observing campaign, the mean spectra and relative variability distributions are plotted on the left side. The HeI5876, NaID, [OI]6300 and H$\alpha$ regions are enlarged on the right side, for a better visualization. The mean spectrum is given by I$_{rel,k}$ = $\frac{1}{N}$$\cdot$$\sum$$_{i=1}^N${I$_{i,k}$},  and the relative variability by $\sigma$$_{k}$/I$_{rel,k}$, being  $\sigma$$_{k}$ = [$\frac{1}{(N-1)}$$\cdot$$\sum$$_{i=1}^N${(I$_{i,k}$ - I$_{rel,k}$)$^2$}]$^{1/2}$. The subindex $k$ refers to each spectral bin and the subindex $i$ to each one of the $N$ spectra per observing campaign. The relative variability plots provide information on the strength and wavelength position of the changes in the line intensity \citep{JohnsBasri95}. 
\end{minipage}
\clearpage
\begin{table}
\centering
\renewcommand\arraystretch{10}
\begin{tabular}{cc} 
\includegraphics[bb=4 77 763 470,height=45mm,clip=true]{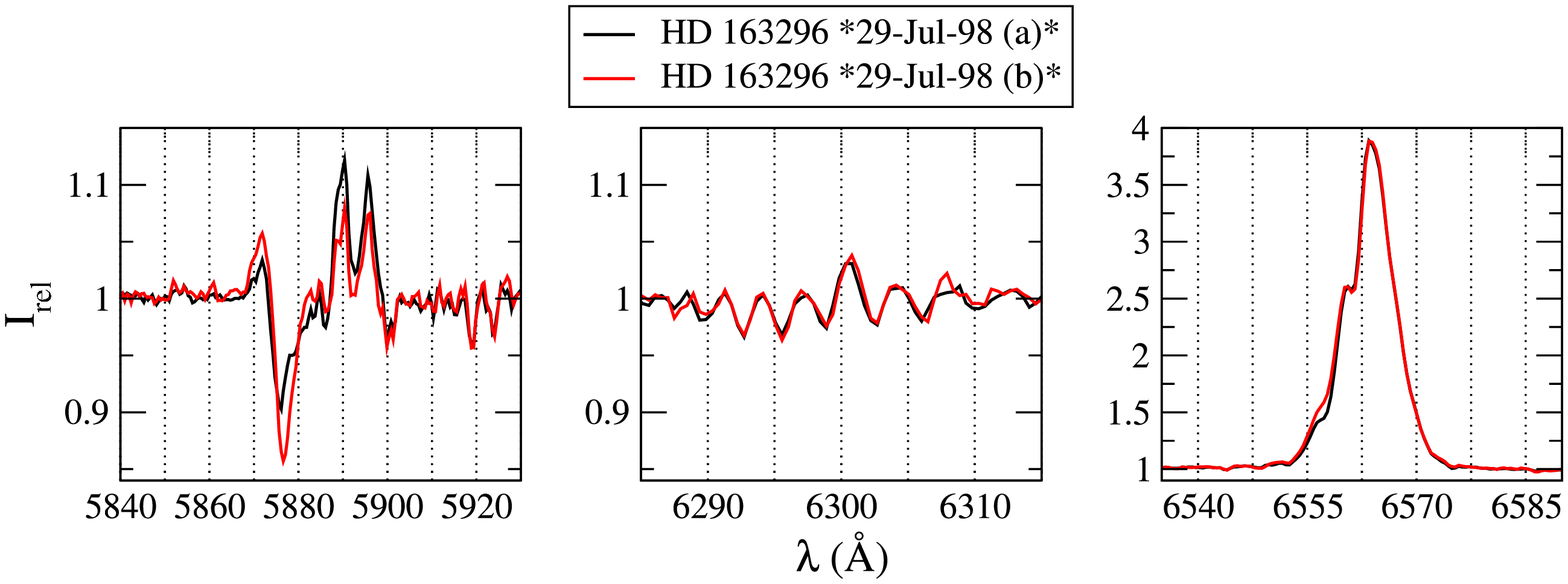}&
\includegraphics[bb=4 77 763 470,height=45mm,clip=true]{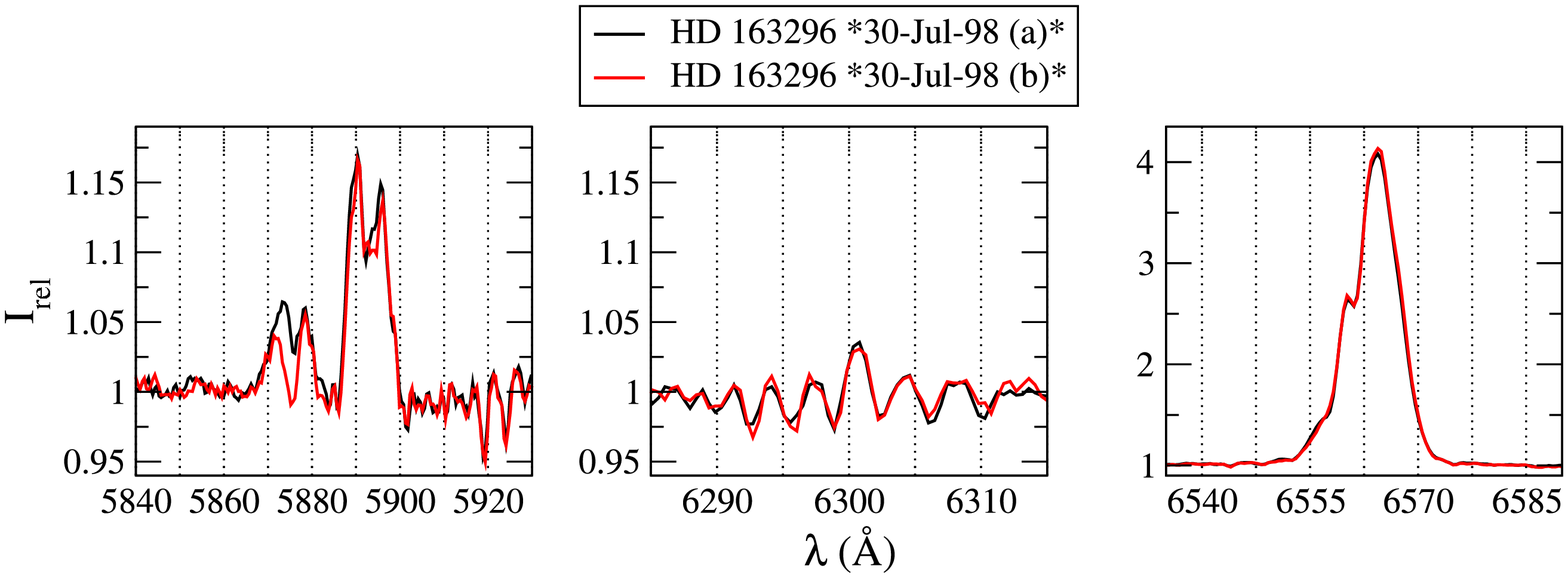}\\
\includegraphics[bb=4 77 763 470,height=45mm,clip=true]{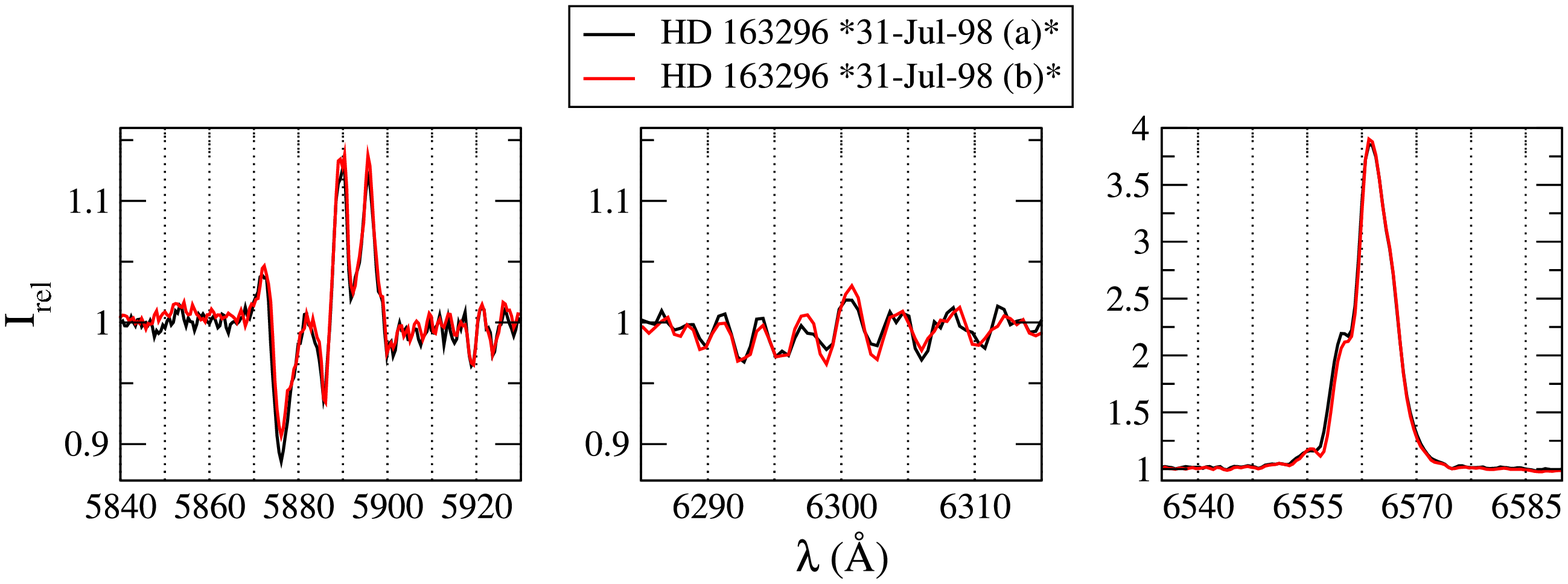}&
\includegraphics[bb=4 77 763 470,height=45mm,clip=true]{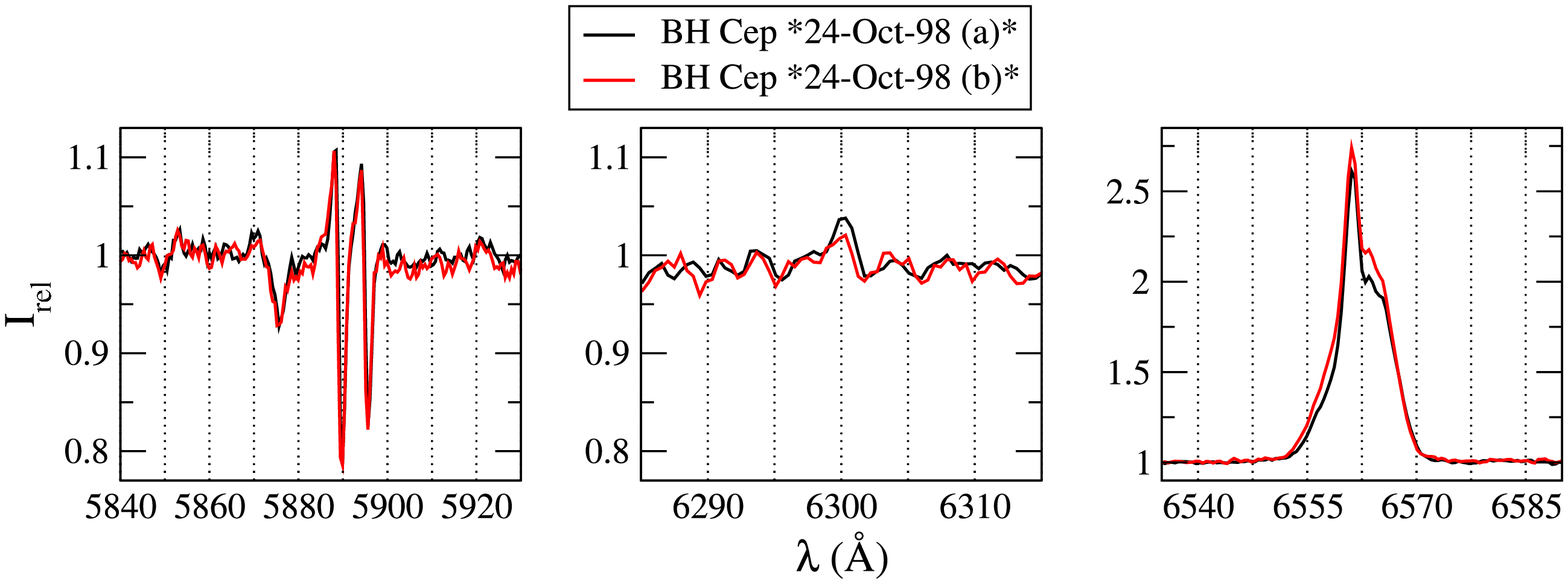}\\
\includegraphics[bb=4 77 763 470,height=45mm,clip=true]{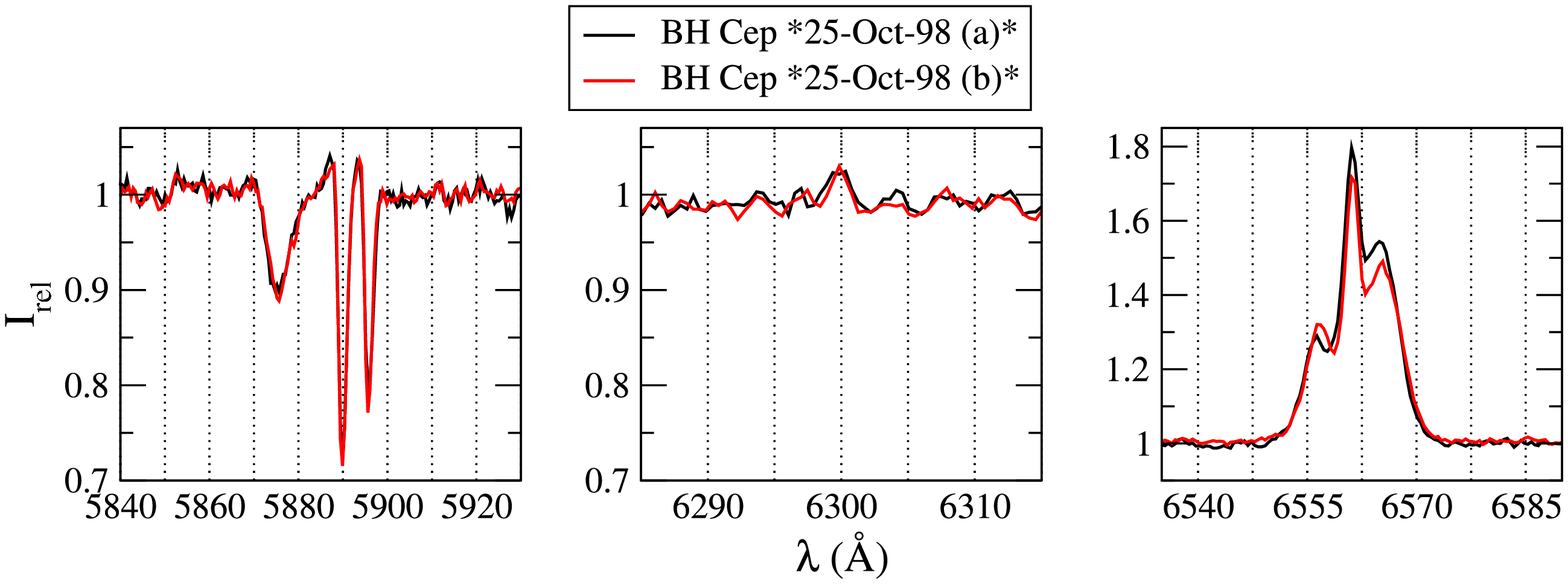}&
\includegraphics[bb=4 77 763 470,height=45mm,clip=true]{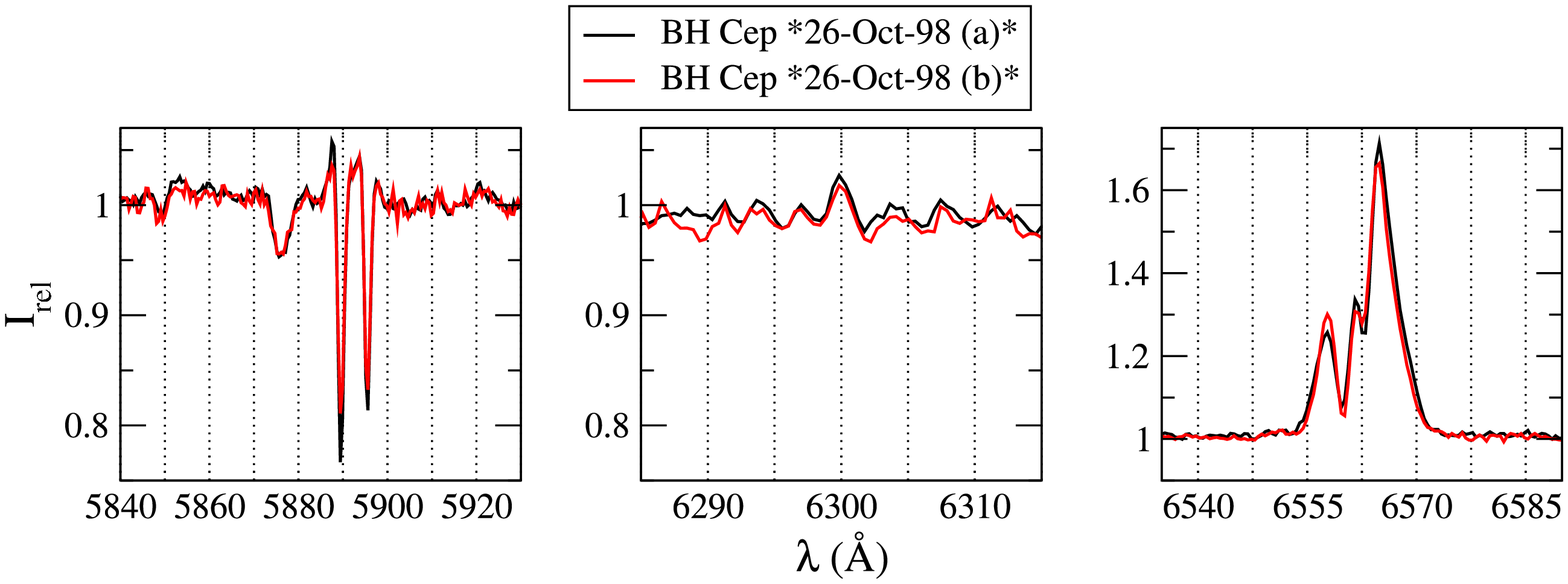}\\
\includegraphics[bb=4 77 763 470,height=45mm,clip=true]{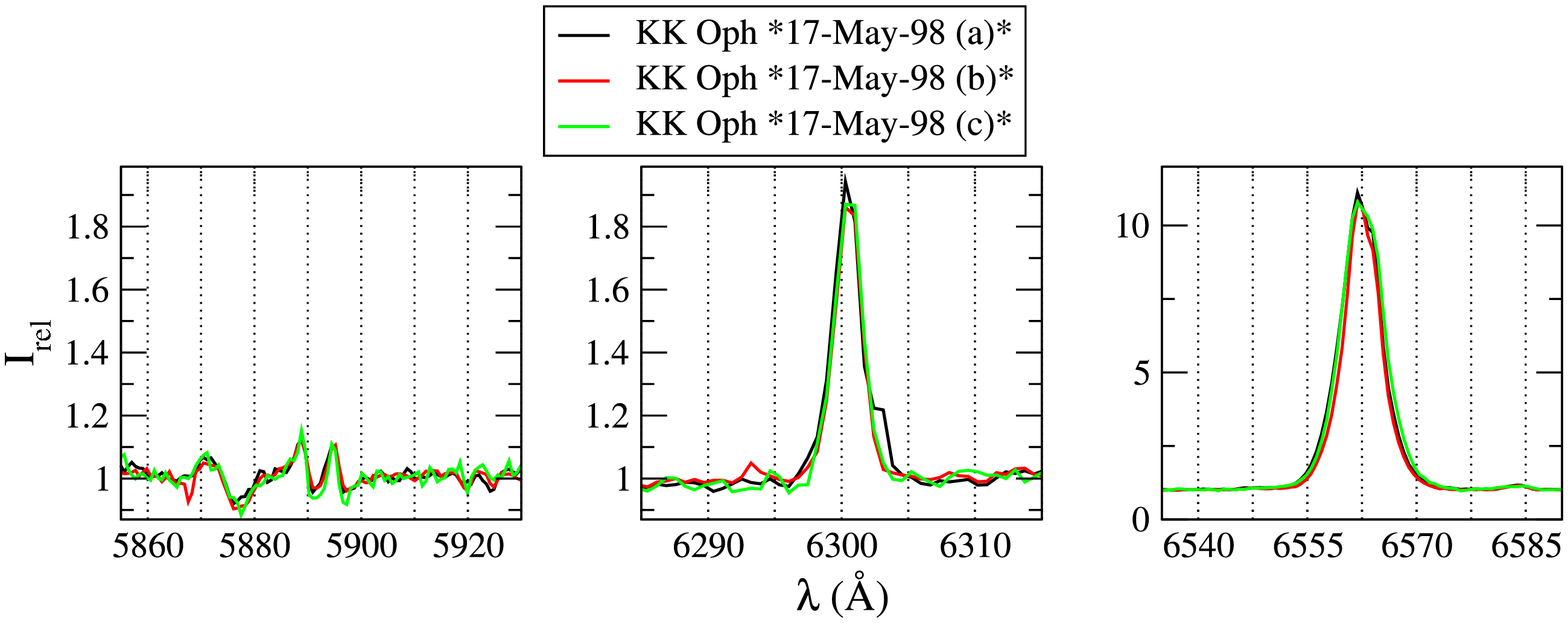}& 
\includegraphics[bb=4 77 763 470,height=45mm,clip=true]{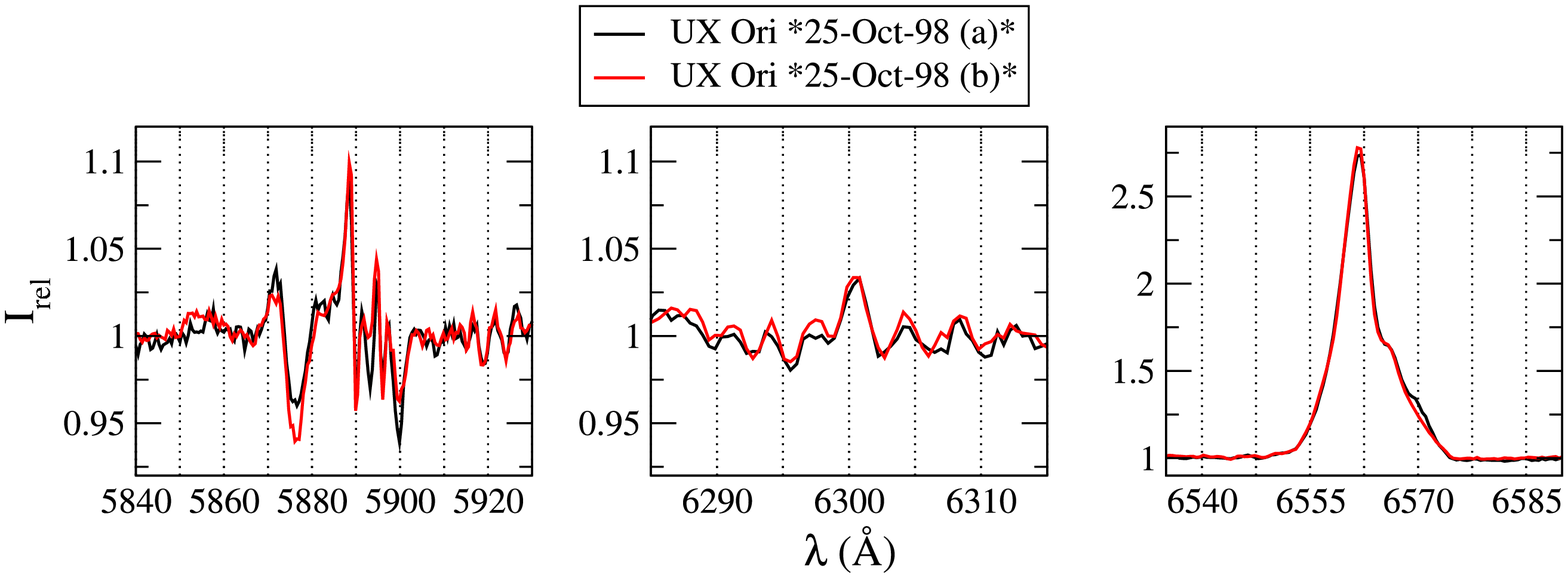}\\
\end{tabular}
\end{table}
\clearpage
\begin{table}
\centering
\renewcommand\arraystretch{10}
\begin{tabular}{cc}
\includegraphics[bb=4 77 763 470,height=45mm,clip=true]{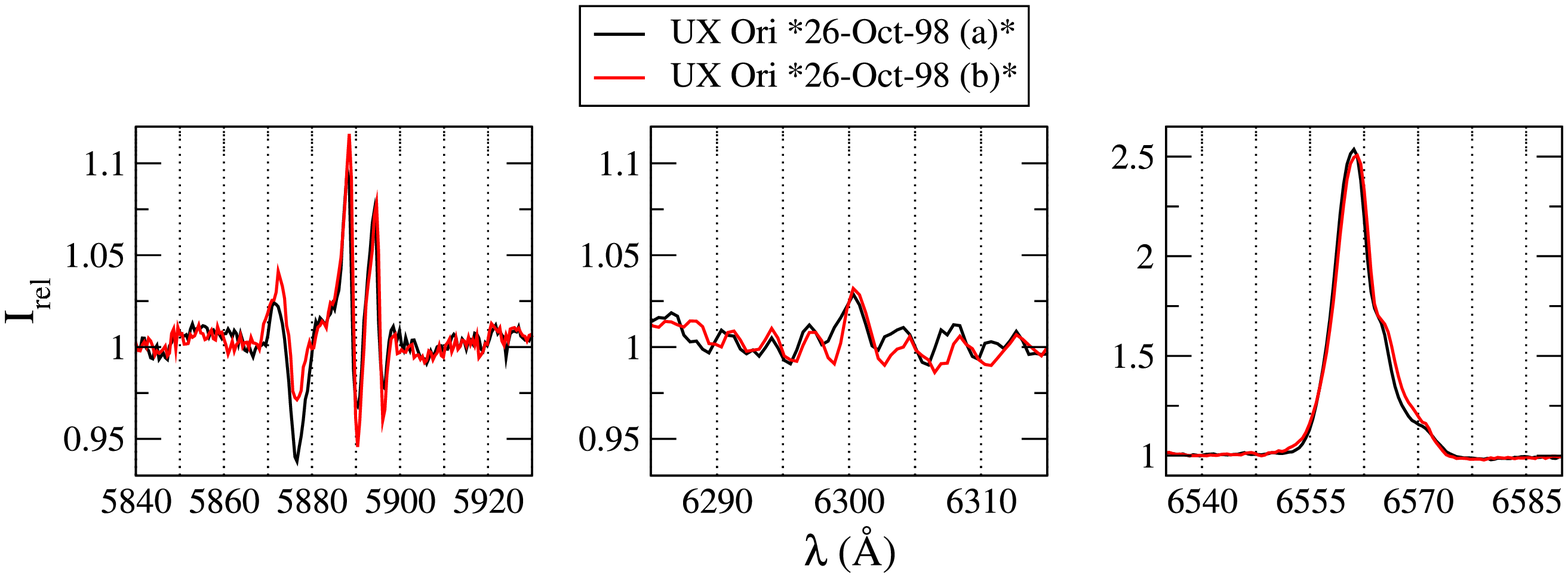}&
\includegraphics[bb=4 77 763 470,height=45mm,clip=true]{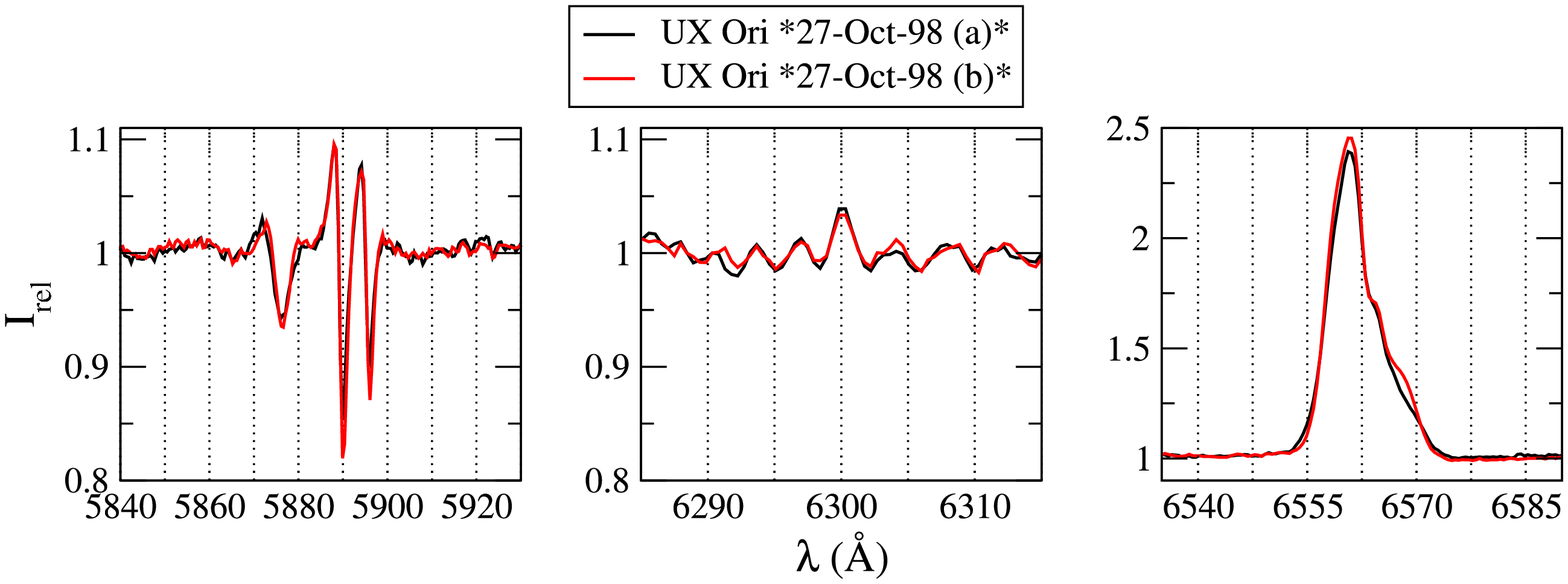}\\
\includegraphics[bb=4 77 763 470,height=45mm,clip=true]{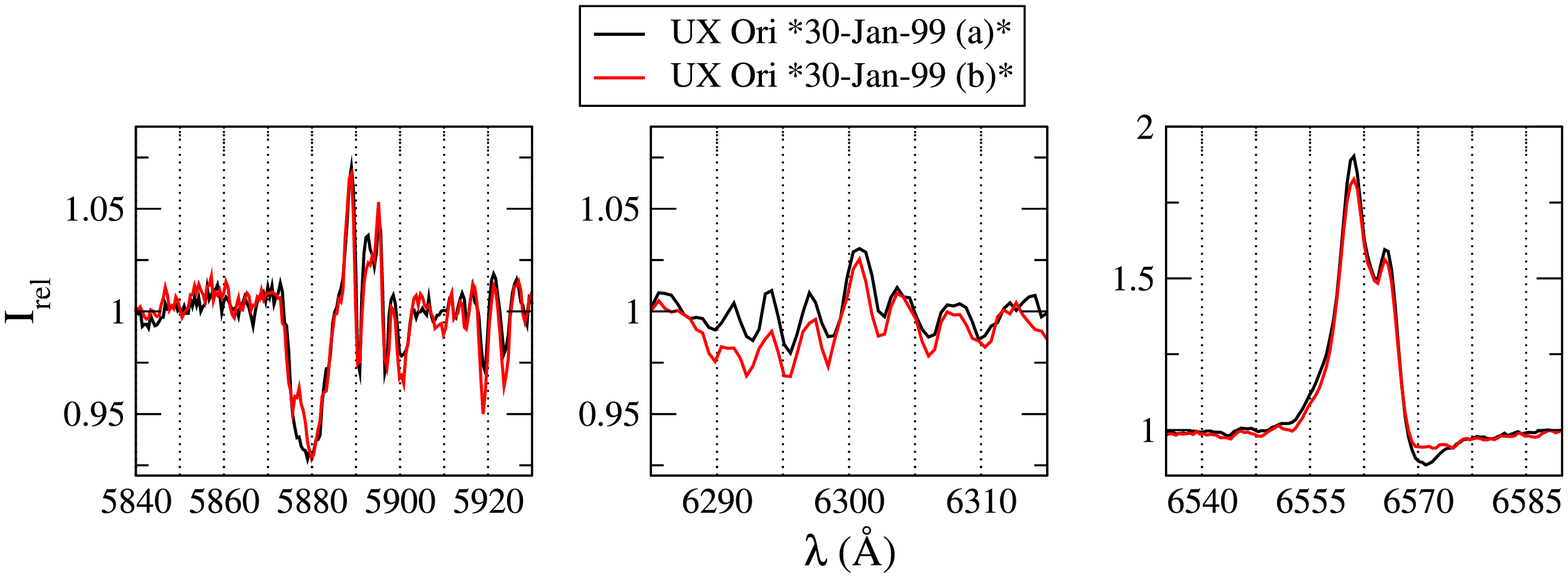}&
\includegraphics[bb=4 77 763 470,height=45mm,clip=true]{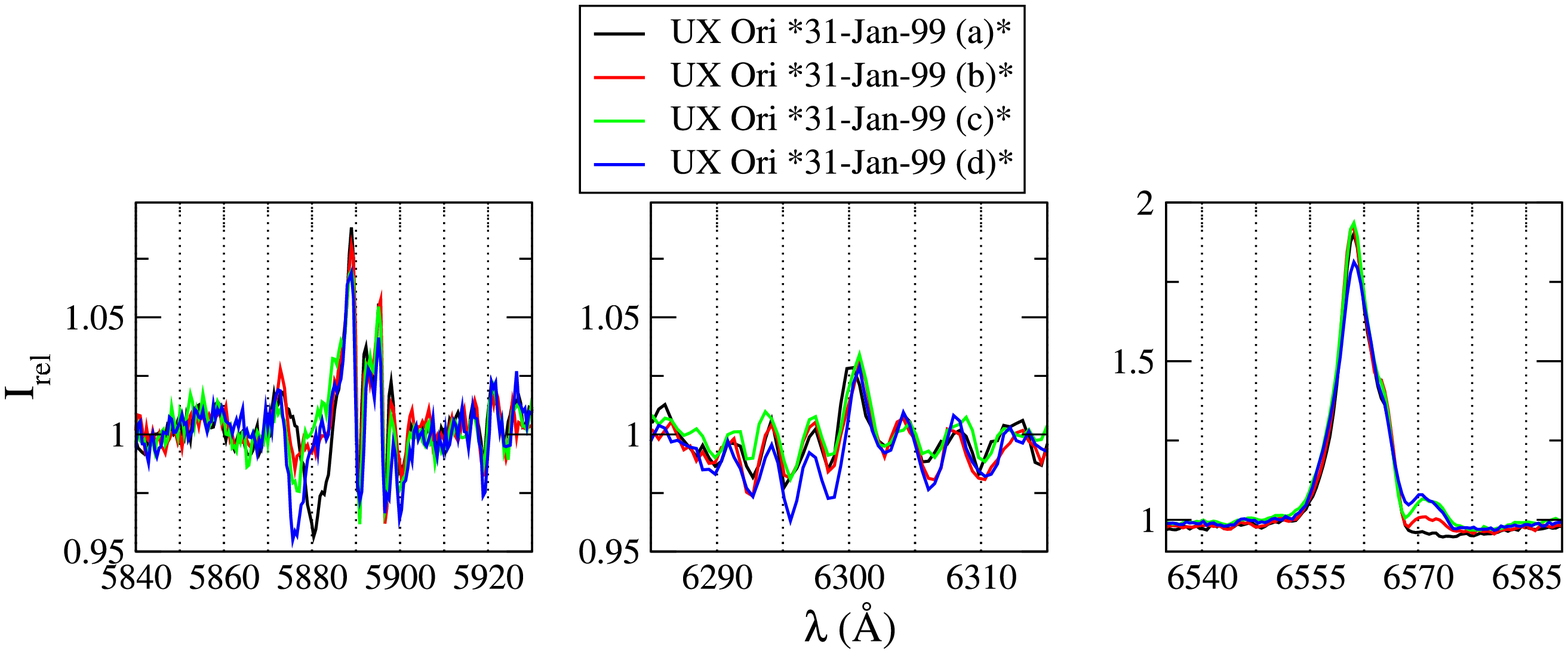}\\
\includegraphics[bb=4 77 763 470,height=45mm,clip=true]{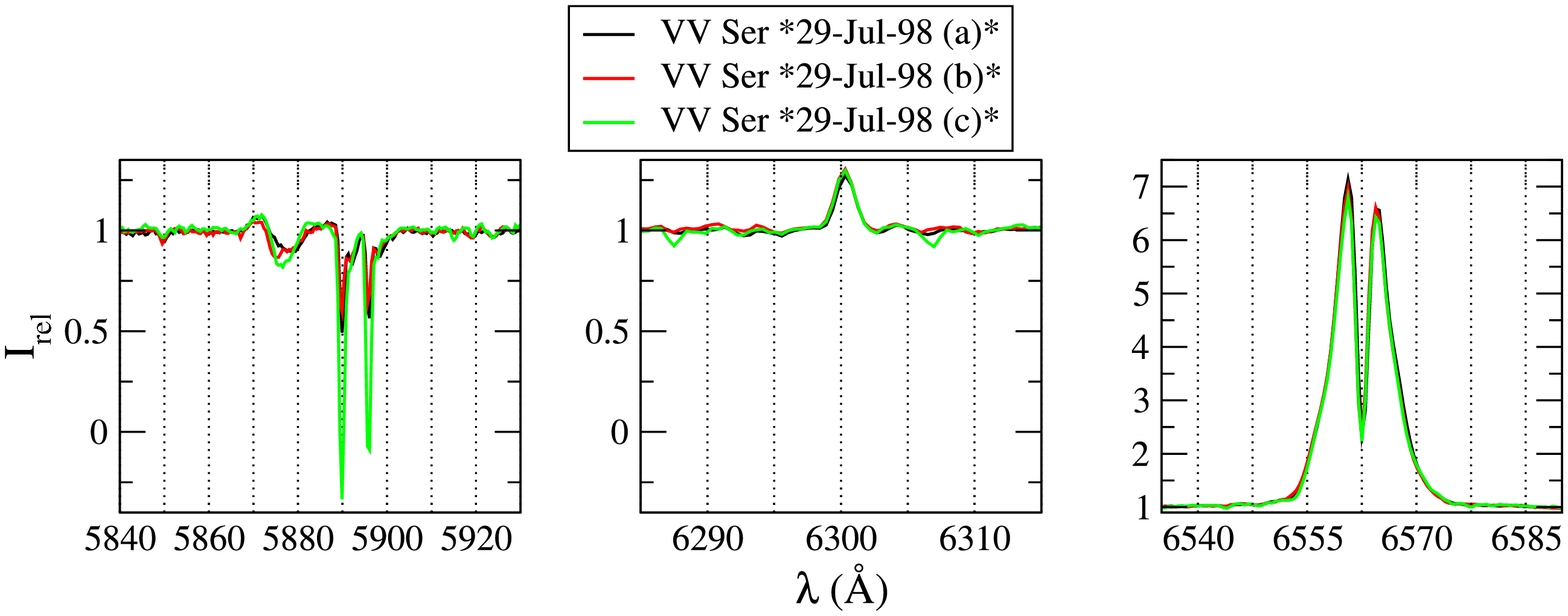}&
\includegraphics[bb=4 77 763 470,height=45mm,clip=true]{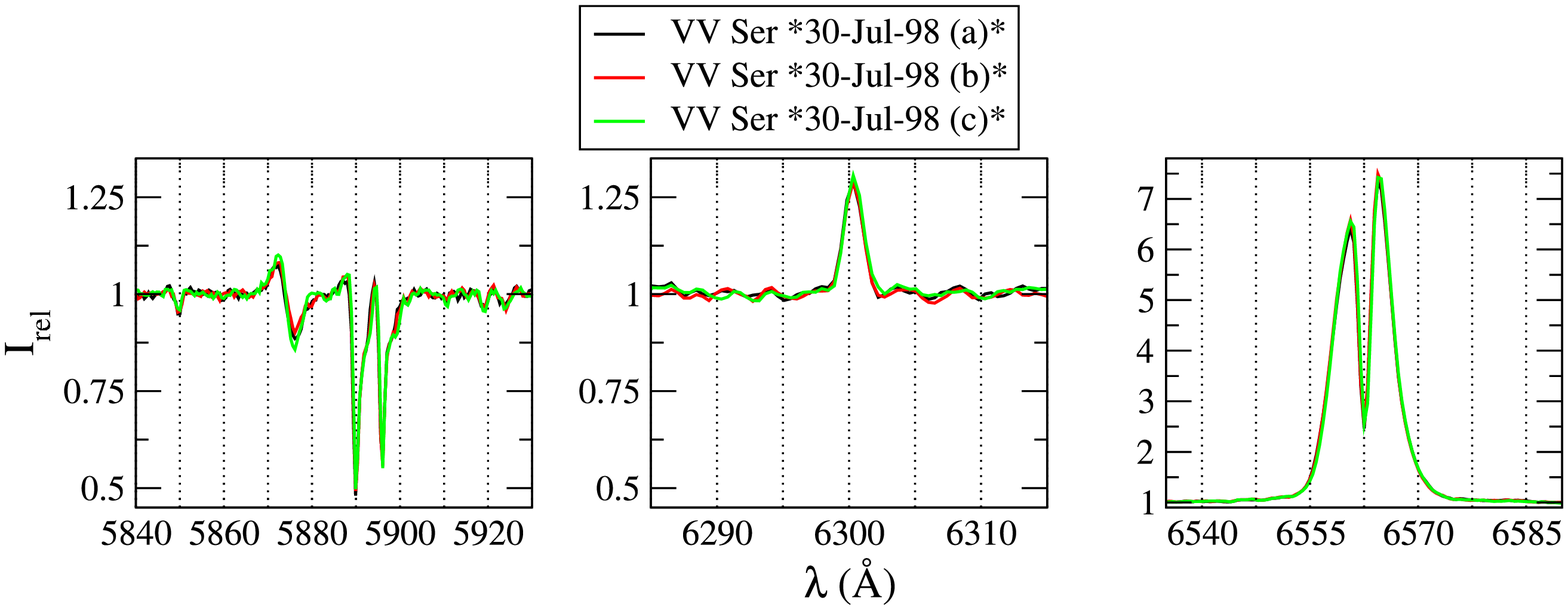}\\
\includegraphics[bb=4 77 763 470,height=45mm,clip=true]{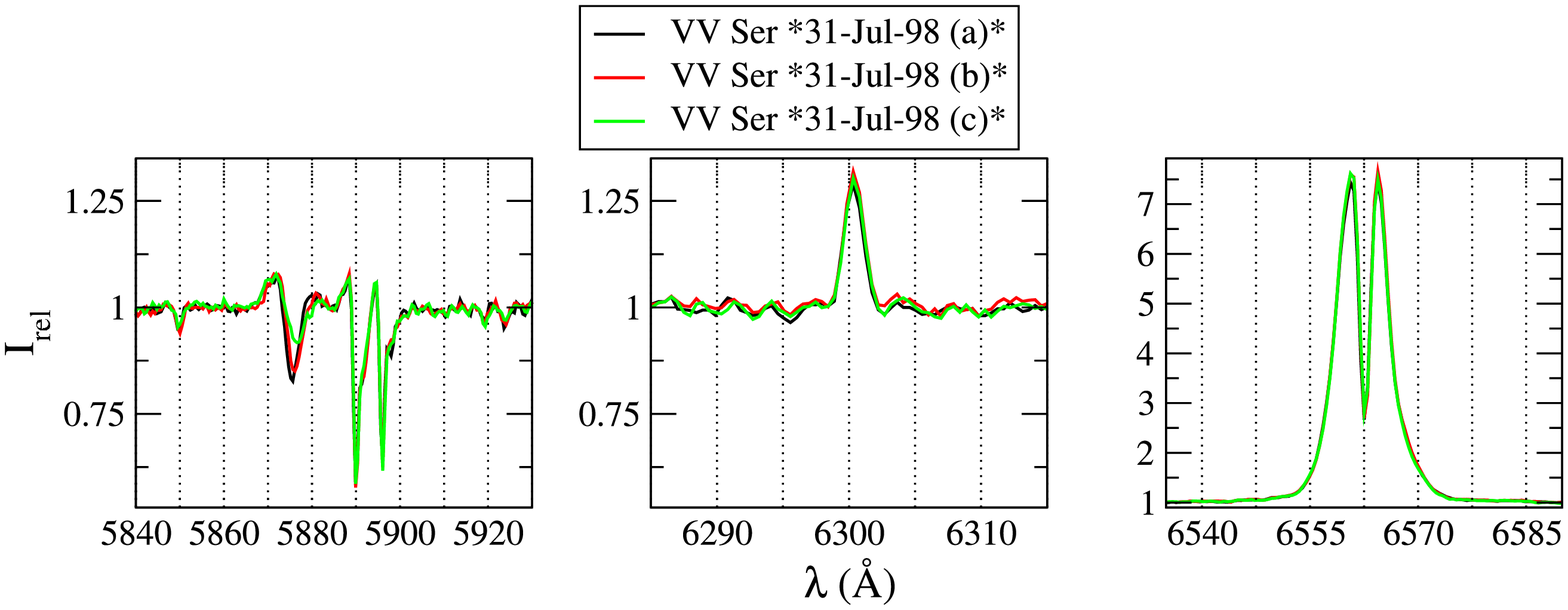}&
\end{tabular}
\end{table}
\begin{minipage}{18cm}
Fig. B2: Individual spectra taken within a time span of hours (i.e. during the same night) for the five objects monitored on this timescale. 
\end{minipage}
\end{appendix}


\end{document}